\documentclass[usegraphicx,useAMS,usenatbib, a4paper]{mn2e}
\usepackage{times}
\usepackage{amssymb}
\usepackage{amsmath}
\usepackage{graphicx}
\usepackage{euscript}
\usepackage{rotating}
\usepackage{epsfig}
\usepackage{url}

\def\ghz{{\rm\thinspace GHz}}

\def\ergps{{\rm\thinspace erg~s^{-1}}}

\def\kpc{{\rm\thinspace kpc}}

\def\kev{{\rm\thinspace keV}}

\def\Ms{{\rm\thinspace Ms}}
\def\pcm2{{\rm\thinspace cm^{-2}}}

\def\gx339{GX~339-4}
\def\pca{{\it PCA}}
\def\hexte{{\it HEXTE}}
\def\rxte{{\it RXTE}}

\newcommand{\eg}{e.g.\thinspace}
\newcommand{\ie}{i.e.\thinspace}

\begin{document}

\title{A Global Spectral Study of Black Hole X-ray Binaries}
\author[Dunn, Fender, K\"ording, Belloni \& Cabanac]
{\parbox[]{6.in} {R.~J.~H.~Dunn$^{1,2}$\thanks{E-mail:
      robert.dunn@ph.tum.de}\thanks{Alexander von Humboldt Fellow}, R.~P.~Fender$^2$,
    E.~G.~K\"ording$^{2,3}$, T.~Belloni$^{4}$ and C.~Cabanac$^2$\\
    \footnotesize
    $^1$Excellence Cluster Universe, Technische Universit\"at M\"unchen, Garching, 85748, Germany\\
    $^2$School of Physics and Astronomy, Southampton, University of Southampton, SO17
    1BJ, UK,\\
    $^3$AIM - Unit\'e Mixte de Recherche CEA - CNRS - Universit\'e
    Paris VII - UMR 7158, CEA-Saclay, Service d'Astrophysique, F-91191
    Gif-sur-Yvette Cedex, France\\
    $^4$INAF-Osservatorio Astronomico di Brera, Via E. Bianchi 46, I-23807 Merate (LC), Italy\\  }}

\maketitle

\begin{abstract}We report on a consistent and comprehensive spectral
  analysis of the X-ray emission of 25 Black Hole X-ray Binaries.  All
  publicly available observations of the black hole binaries in the
  \rxte\ archive were analysed.  Three different types of model were
  fitted to investigate the spectral changes occurring during an
  outburst.  For the population, as well as each binary and each
  outburst from each binary, we construct two diagnostic diagrams.
  The Hardness Intensity/Luminosity Diagram (HID/HLD), the X-ray colour against
  the flux/luminosity of the binary is most useful when studying a
  single binary.  However, to compare between different binary
  systems, the Disc Fraction Luminosity diagram (DFLD) is more useful.
  The DFLD uses the luminosities of the disc and powerlaw components
  to calculate the ratio of the disc luminosity to the total
  luminosity, resulting in a more physical value, which is analogous
  to the X-ray colour calculated for X-ray binaries.  The tracks of
  the outbursts populate the DFLD more evenly that the HLD.  We
  discuss the limitations of both diagnostic 
  diagrams for the study of the X-ray binary outbursts, and we clearly
  illustrate how the two diagrams map onto each other for real
  outburst data.  The similarity of the X-ray colour and Disc Fraction
  behaviour over time during an outburst originally seen in GX~339-4 data is seen in
  other sources' outbursts.  We extract the peak luminosities in a
  single outburst, as well as the luminosities at the transitions away
  from- and returning to the powerlaw dominated state for each
  outburst.  The distribution of the luminosities at the transition
  from the powerlaw to the disc dominated state peaks at around
  $0.3L_{\rm Edd}$, the same as the peak of the distribution of the
  peak luminosities in an outburst.  Using the disc fraction to
  calculate the transition luminosities shows that the distributions
  of the luminosities for the transitions away from- and return to the
  powerlaw dominated state are both broad and appear to overlap.
  Using the change in Disc Fraction to calculate the date a transition
  occurred is not drastically different from the dates obtained from
  changes in the timing behaviour of the X-ray binary.  In addition, we calculate the rate
  of motion of an X-ray binary through the DFLD during an outburst, a
  diagnostic which has the potential to be used as a comparison with
  populations of active galactic nuclei. The fastest rate of motion is
  on the egress and ingress from the powerlaw dominated state.  A
  further region of increased speed through the diagram occurs in the
  disc dominated state on the return to the powerlaw dominated state.
  Finally we compare the measured X-ray luminosities with a small
  number of contemporaneous radio measurements. Overall this is the
  most comprehensive and uniform global study of black hole X-ray
  binaries to date.
\end{abstract}

 \begin{keywords}
accretion, accretion discs - binaries: general - ISM: jets and
outflows - X-rays: binaries
\end{keywords}

\section{Introduction}

Galactic X-ray binaries (XRBs) are the sites of some of the most
energetic and exotic phenomena in the local universe, and may provide
us with insights into the action of supermassive black holes in active
galactic nuclei. Their energetic influence on their surroundings and
the galaxy as a whole are only just becoming clear \citep{Gallo04,
Heinz08}.  

Many black hole X-ray binaries (BHXRBs) spend most of their time in a
quiescent state where both the X-ray (assumed to arise in the
accretion flow) and radio (assumed to arise in a jet-like outflow)
emission are at a very low level.  Most binaries are discovered when
they go through an outburst phase.  As described in detail in
\citet{Fender04} but see also \citet{Nowak95, Done03, Homan05,
  Remillard06, Done07, Belloni09}, the outburst starts in the
``low-hard'' state.  The state is described as ``low-hard'' because
the BHXRB is faint and the X-ray emission is characterised by a hard powerlaw of
$\Gamma\sim 1.5$.  The term is commonly used to indicate the time
during which the X-ray spectrum is hard, regardless of the BHXRBs
brightness. The radio emission in the low-hard state is
characteristic of a steady 
jet emitting synchrotron radiation.  During the early rise of the
outburst, the X-ray and radio luminosities both increase, but the
X-ray colour of the spectrum remains hard \citep{Corbel00,Corbel03}.  As the
outburst progresses the thermal emission from the accretion disc
rapidly becomes more prominent, until it dominates the X-ray emission.
This softening of the X-ray spectrum takes place quickly compared to
the rise up from quiescence.  The state when the disc dominates the
spectrum is called the ``high-soft'' state.
There are two intermediate states between the low-hard and high-soft
states (hard-intermediate and soft-intermediate).  Not all BHXRBs have
been observed to go through both these states during their outbursts.
Some BHXRBs are not observed to do a transition and remain in the
low-hard state until they return to quiescence \citep{Brocksopp04, Capitanio09}.

The short timescale variability characteristics of the X-ray emission
from the BHXRBs also change during the outbursts.  The level of the
root-mean-square (rms) noise and the appearance and frequency of
quasi-periodical oscillations (QPOs) are useful for denoting a more
accurate date of transition between states (see
e.g. \citealp{Homan05,vanderKlis06,Remillard06, Belloni09}).  The
change in spectral information between two neighbouring states can
sometimes be very slight, with no clear distinction between the two.

During the transition to the high-soft state the radio emission has
been observed to flare in some BHXRBs \citep{Fender04, Fender09}.
Even if the flare 
is unobserved (be this the consequence of no suitable radio observations, or
no radio detection during the transition), the radio emission is quenched on the transition to the
high-soft state \citet{Fender99}.  The disc gradually fades in the soft state and the non-thermal
emission recovers.  The source then undergoes the return transition to the
low-hard state at a lower luminosity than the transition from the hard
to the high-soft state.  During this period the radio emission is
observed to recover.  The brightness of the BHXRB continues to
decrease as it returns to quiescence.  See \citet{Fender09} for a
detailed study of the radio emission during BHXRB outbursts and \citet{Corbel00,Corbel03} for the details of the
radio-X-ray correlation in the hard state.

Over the past few years the importance of BHXRBs for the evolution of
the galaxy has become apparent.  The discovery of a jet blown bubble
from Cyg~X-1 \citep{Gallo04} allowed the estimation of the kinetic
energy injection into the inter-stellar medium
\citep{Heinz05,Heinz08}.  Searches have been done for other jet-blown
bubbles, so far with out clear success \citep{Russell06}.  However, it
is reasonable to assume that when exhibiting a steady radio jet, all
BHXRBs inject (mechanical) energy into their surroundings.  This would
have significant effect on the evolution of the galaxy. The radiative
emission from X-ray binaries, of all types, usually dominates the
X-ray emission from non-active galaxies.  The level of this X-ray
binary emission has been used as a proxy for the star formation rate
of the galaxy (e.g. \citealp{Gilfanov03}) as well as the stellar mass
(e.g. \citealp{Gilfanov04}).  The global properties of BHXRBs are also
useful in fully understanding the links between black holes on all
mass scales; in particular studying the coupling between accretion and
feedback (both radiative and kinetic), can help us to understand the
broader picture of how black holes affect their surroundings (see e.g
\citealp{Merloni03,Merloni05,Koerding06,Koerding08,Merloni08}).

For recent reviews on BHXRBs see
\citet{Belloni09,Gilfanov09,Markoff09, Fender09b, Gallo09b}.

Therefore the understanding of the processes which occur during the
life of a BHXRB are important in a wider context.  As the radio power,
and perhaps the energy injection rate, of the jet increases
dramatically during the early stages of an outburst, understanding the
outbursts of the population of BHXRBs is one way of refining our
knowledge of their impact on their surroundings. 

For the study of the X-ray emission from BHXRBs, the \rxte\ satellite
has created an extensive archive of data on a wide variety of these
objects since its launch in December 1995.  This makes a comprehensive
and easily comparable study possible, as the same instruments (though
with changes in calibration over time) have taken all the
observations.  The over 13 year baseline of observations means that a
number of sources have been observed going through outbursts multiple
times.  This allows comparison within as well as between sources.

We present an analysis of a sample of 25 BHXRBs which have been
observed by the \rxte\ satellite.  The sample was not chosen in any
statistical way.  The binaries were selected from well known BHXRBs
(or candidates) as well as those which have been well observed by
\rxte.  We outline the data reduction scheme and model selection in
Sections \ref{sec:data_red} and \ref{sec:model}.  Our initial
comparisons between the binaries using the standard diagnostic
diagrams are presented in Section \ref{sec:pop}.  Sections \ref{sec:Outburst_sim} and
\ref{sec:Trans_L} discuss the outbursts
themselves and the properties of the transitions.  The radio
properties of the binaries during their outbursts are a useful probe
of the jet activity.  The radio observations which are sufficiently
coincident with the X-ray observations are presented in Section \ref{sec:radio}.

\section{Data Selection and Analysis}\label{sec:data_red}

The data reduction scheme is very similar to that used for GX~339-4 in
\citet{Dunn08}.  We briefly recap the method below and highlight any
changes from that scheme.

We used all observations of the binaries we selected which were
publicly available in the \rxte\ archive\footnote{The cut-off date
  used to find public observations was 4 August 2009.}.  This gave
around $12\Ms$ of raw 
\pca\ exposure over an 13 year period.  Not all objects we selected
had similar time coverage, some having large volumes of observations over
all 13 years, whereas others had few.

All the data were reprocessed so that all observations had the same version
of the data reduction scripts applied.  Apart from the \pca\ data, we also analysed the \hexte\
data to constrain the powerlaw slope at high energy which allows for
more detailed fitting at low energies.  We use the data reduction tools from
HEASOFT\footnote{\url{http://heasarc.gsfc.nasa.gov/lheasoft/}} version
6.6.2.  The data reduction and model fitting were automated so that
each observation was treated in exactly the same way.

\subsection{PCA \& HEXTE Data Reduction}\label{sec:data:rxte_red}

The \pca\ and \hexte\ data were reduced according to the procedure in
the \rxte\ 
Cookbook\footnote{\url{http://rxte.gsfc.nasa.gov/docs/xte/recipes/cook_book.html}},
and only a quick summary is given here. 

We used only data from \pca\ PCU-2 as it has always been switched on
throughout the mission, and so can be used
over the entire archive of data.  It is also the best calibrated of the
PCUs on \rxte.  Background spectra were obtained using {\scshape
  pcabackest} from new filter files created using {\scshape xtefilt},
from which updated GTI files were also created.  We use only
  the bright model for the background.  Our aim is to perform a single
  data reduction routine for all sources, and switching between the
  faint and bright models may have introduced jumps into our light
  curves.  We also ignored the {\scshape electron2} selection
  criterion when using {\scshape maketime} to create the GTI files.
  As our analysis concentrates on the outbursts (bright periods) of
  the X-ray binaries, we do not believe that this will bias our
  results.  However, low flux/counts data were therefore treated with
  caution.  The spectra were
extracted and the customary systematic error of 1 per cent was added
to all spectra using {\scshape grppha}.  

In order that the model fitting in {\scshape xspec} was reliable and
relatively quick, we only fitted spectra which had more that 1000
background-subtracted \pca\ counts.  The excluded observations occur
throughout the light curves of the objects, with a concentration in 
the low flux periods.  Our analysis concentrates on the outbursts of
the X-ray binaries rather than the quiescent periods.  Therefore
excluding these low count observations, even if most were taken during
quiescence, is unlikely to bias our conclusions about the outbursts.

Where possible, we
used both \hexte\ Cluster A and Cluster B data.  Background
spectra were obtained using {\scshape hxtback}.  Spectra were extracted
using the routine appropriate to the data-type (Event or
Archive). Dead-time was then calculated using {\scshape hxtdead}.  All
spectra for a given ObsID were then summed using {\scshape sumpha}, and
the appropriate responses and ancillary files were added in as header
key words using {\scshape grppha}.  In order to accurately
determine the slope of the high energy 
power-law we require \hexte\ data to be present when fitting a model.
There have to be at least 2000 background-subtracted counts in one of
the \hexte\ clusters, with the other having a positive number of
counts\footnote{The background subtraction on some observations
  resulted in a negative number of \hexte\ counts in one of the
  clusters.}.  As our analysis concentrates on the outbursts of the
binaries, we also require \pca\ data to be present to fit the soft
energies, where the emission from the accretion disc is found.  We bin
the \hexte\ data up to match the binning found in the Standard-2
\hexte\ data products.

\section{Model Fitting}\label{sec:model}

The spectra were fitted in {\scshape xspec} (v12.5.0an, \citealp{Arnaud96}).  This latest
version of {\scshape xspec} does not allow parameters to be extracted
if there are unconstrained model components present within the fit (an
unnecessary line or disc component for example).  We have used this to
our advantage by stopping fits where a model component is not
necessary (a line component for example) as these would cost time if the code
attempted to extract errors for them, as the parameters in these cases
are usually poorly determined.

Our analysis focuses on the outbursts of the X-ray binaries.  It was
therefore necessary to accurately determine the disc properties during
the outburst, requiring that we
analyse the spectra down to the lowest possible energies.  The
energy boundaries corresponding to channel numbers have changed
during the lifetime of the \rxte\ mission.  The calibration of
channel numbers $\leq6$ is uncertain, and we therefore
ignore all \pca\ channels $\leq6$ (which corresponds to $\sim 3\kev$), which allows a consistent lower
bound to the spectra, extending them to the lowest possible
energies and maintaining calibration.  All PCA data
greater than $25\kev$ were also ignored. The \hexte\ data were fitted between $25$
and $250\kev$.  To investigate the evolution of the thermal and
non-thermal components during an outburst, three types of
models were fitted: Powerlaw ({\scshape power}), Broken Powerlaw
({\scshape bknpower}) and Powerlaw +
Disc ({\scshape power+discbb}).  This allowed us to study the
evolution of the disc and powerlaw over the outburst.  As an iron line
has been detected in a number of X-ray binaries, for each of these
three models, a version
including a Gaussian line fixed  at $6.4\kev$ was also fitted. This gave
a total of six models which were fitted to each spectrum.  

The powerlaws had a soft upper limit of $3\kev$ and a hard upper limit
of $5\kev$.  The break energy was left free throughout the sensitivity
range of \rxte, though this does allow it to mimic a low luminosity
disc (see Section \ref{sec:pop:comp}).  We decided on this approach,
rather than having a higher low-energy bound for the break energy as
this is more flexible in later stages of the analysis.  For the work
presented here we use the method in Section \ref{sec:model:sel}.

Noticeable variations of the hydrogen column density value during
state transitions may be observed in BHBs (see e.g \citealp{Cabanac09,
Oosterbroek96} in GS 2023+338 (V404-Cyg)). However, as RXTE/PCA response
falls under $2\kev$, a systematic study of such variations with this
instrument appears to be difficult. Hence we decided to fix its value
to the commonly accepted value for binaries (see Table \ref{tab:objects}). Galactic
absorption was modelled using the {\scshape wabs} photoelectric
absorption code, with values fixed to the accepted values for the
binaries (see Table \ref{tab:objects}).  To
obtain fluxes outside of the \rxte\ observing band, for the disc for example, dummy responses
were created within {\scshape xspec}.  

Although GX~339-4 had a correction applied for the galactic ridge
emission (GRE) in \citet{Dunn08}, we do not apply any such correction to any
of the sources analysed here.  We did investigate whether it was
possible to determine the level of any such correction from the
\rxte\ data itself.  In some cases (4U~1543-47, XTE~J1550-564, XTE~J1650-500, GX339-4 and
H~1743-322) this was indeed possible - in some cases even being able to
constrain the line parameters.  However in all the other sources,
there did not appear to be any \rxte\ observations in periods of
quiescence where the source was not detectable over the background (in
terms of background subtracted counts per second per PCU).  Rather
than applying the GRE correction to only those sources where it was
possible to determine the level of correction from the \rxte\ data, or
even those where measures using other observatories have been
published, we decided to not apply any correction at all.  The level
of the GRE during an outburst of an BHXRB is very low.  Comparing our
results from GX~339-4 in this analysis to that presented in
\citet{Dunn08} the average flux difference is 2.27 per cent, with 95 per
cent of the observations having a flux difference of less than 5 per cent.  Therefore our
decision to standardise the spectral 
fitting across all objects by not applying a GRE correction should not
affect any conclusions drawn about the outbursts of the BHXRBs.  Any future
work on the decays of the outbursts will need to take the GRE into account, and as such
any correction will be applied then.  

We encountered some difficulty in fitting discs reliably, especially
when the sources are in the intermediate states.  The
response of \rxte\ is only reliable down to $\sim 3\kev$, whereas the
discs we are trying to measure have temperatures of around $1\kev$ or less.  We
set the minimum
disc temperature to $k_{\rm B}T=0.1\kev$ in {\scshape xspec} to
prevent discs from being fit at very low temperatures.  It is likely
that in cases where this occurred, the disc was being fit to take into
account of any curvature in the powerlaw slope rather than to a true
disc component.  Before selecting the best fitting
model we penalised the $\chi^2$ of disc models which
had $k_{\rm B}T<0.4\kev$ as in early results from the fitting
procedure indicated that few discs are detected in \rxte\ data with
temperatures lower than $k_{\rm B}T<0.4\kev$.  In doing this we note
that we are likely to have excluded a few disc fits that were reliable.



Although a more complicated model, {\scshape comptt} for example, would
give more information on the state of the accretion disc and its
surroundings there are difficulties with using these models.  The
models used in this analysis are simple, and so can be fit to
observations with low numbers of counts.  More complicated models
require more counts to enable a complete fit to be made.  In our
analysis of GX~339-4 we find that although {\scshape comptt} does fit
well to the high luminosity hard-state observations with the highest
number of counts \citep{Dunn08}.  But in the soft state, even the observations with
the highest number of counts are unable to constrain all the
parameters.  To use more complicated models
restrict the numbers of observations we could use in our analysis, and
so the picture arising from this work would be less complete.  

The {\scshape diskbb} model is a comparatively simplistic model and
does not include effects on the disc spectrum from the gravitational
potential or from the disc atmosphere.  In the \pca\ pass-band, when
the discs are at low temperatures, this is unlikely to be of concern.
However for higher temperature and brighter, more dominant discs, the
resulting distortion of the disc emission may not be well fit by the
{\scshape diskbb} model.  But, for the reasons noted above, we still
use the {\scshape diskbb} model rather than adding in further model parameters.

\subsection{Selecting the best fitting model}\label{sec:model:sel}

\begin{figure}
\centering
\includegraphics[width=1.0\columnwidth]{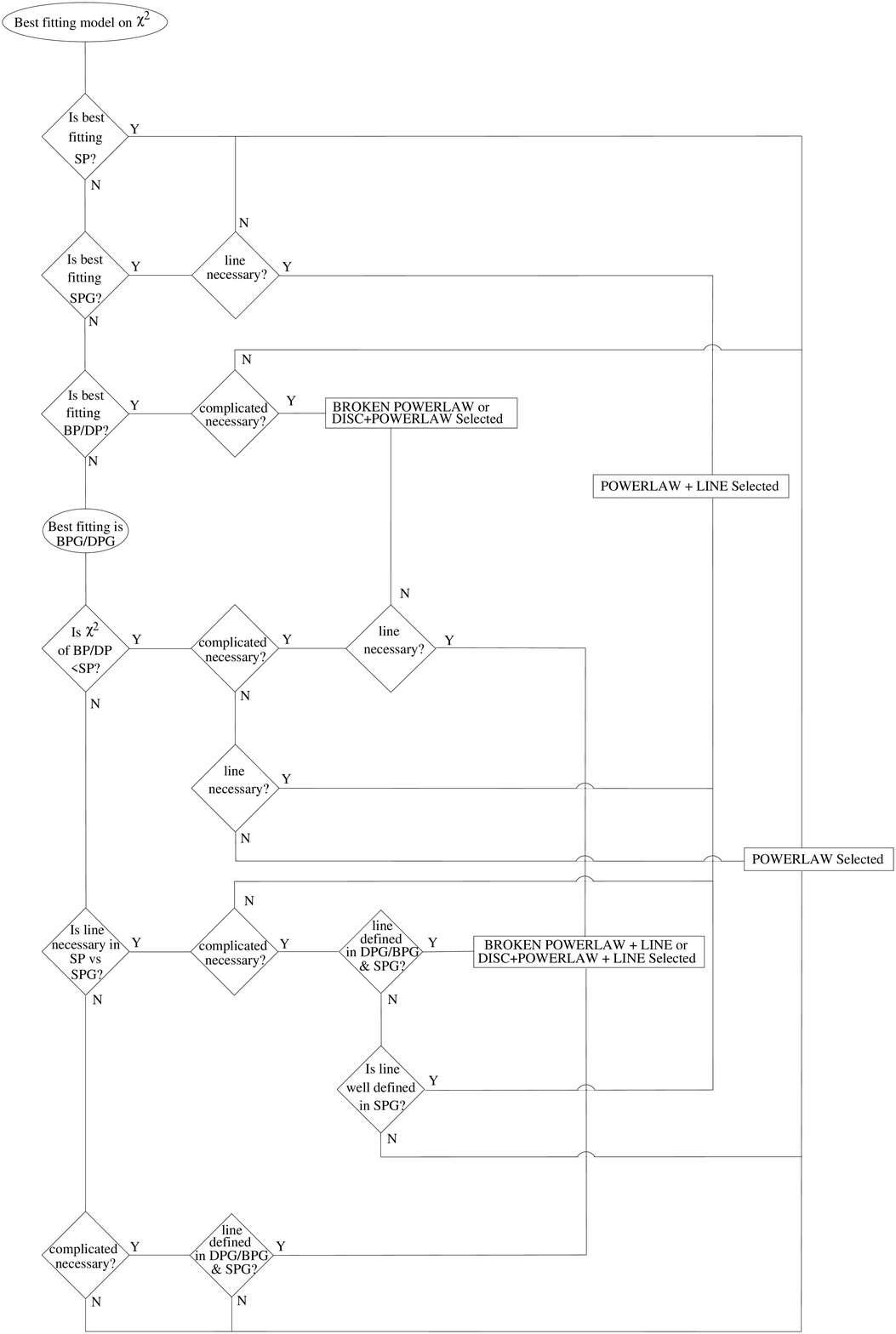}
\caption{\label{fig:flowchart} Decision Tree for selecting the best
  model. }
\end{figure}

Out best-fitting model-selection routine has changed subtly from that
outlined in \citet{Dunn08}.  We show the flowchart used in
Fig. \ref{fig:flowchart} and outline the procedure here. We still initially select the model with
the lowest reduced $\chi^2$.  If this is the simple powerlaw (SP), then
that is the best fitting model, as it is also the simplest.  If it is the simple powerlaw +
gaussian line (SPG) then we test whether the gaussian component is an
accurate description of a true iron line in the data.  The scheme for
this ``line test'' is identical to that described in
\citet{Dunn08}\footnote{An $F$-test using 
  $\mathcal{P}<0.001$ as the significance level combined with the
  normalisation of the line and its uncertainty ($\mathcal{N}_{\rm
    line}>3\sigma_{N_{\rm line}}$).  Both criteria have to be satisfied
  for the line to be taken as real.  Also, if the line width,
  $\sigma>2\kev$ then this was taken to mean that the line was not
  well constrained, and the line deemed to be not real.}.

If this best fitting model is a broken powerlaw (BP) or a
disc+powerlaw model (DP), we
test whether using this more complex continuum model is a significant
enough an improvement over a simple unbroken powerlaw, by performing an $F$-test.
If the $F$-statistic probability $\mathcal{P}<0.001$ then we select the more
complex continuum model.  

However, if the best fitting model
is BP (or DP) and contains a gaussian (BPG or DPG respectively), then
the following is done. 

If the $\chi^2$ of the BP (or DP) without the line is less than that of the SP, then
these continuum models are tested against each other to see if the more complicated
model is necessary.  Then subsequently the addition of the line
component is tested on the result of the model testing.

If, however,  the $\chi^2$ of BP (or DP) is greater than that of the SP,
then firstly the existence of the line is tested between SP and SPG.
Then if the line is necessary in SPG, the BPG (or DPG) are tested
against SPG. If SPG is the best 
fitting model, then we use the SPG model fit as the best fitting.  If the best fitting
model is BPG (or DPG), then if the line is well defined
($\sigma<2\kev$) in both the complicated (BPG or DPG) and SPG models,
then the BPG (or DPG) are chosen.  If the line is only well defined in
the SPG model, then this is chosen, else the SP model is taken as the
best fit.

However, if the line is not necessary in the SP models, then although it is not a fair
test, then BPG (or DPG) are tested against SP.  If SP is the best
fitting model, then this is a reliable result.  However, if BPG (or DPG)
are the best fitting model, then only if the line is well defined
($\sigma<2\kev$) in {\it both} the BPG (or DPG) and SPG models,
then the BPG (or DPG) are chosen.  Else the best fitting model is SP.
All observations with this final set are noted so they can be
investigated at a later stage if necessary.

The new feature of {\scshape xspec} where fits are terminated if there
are unnecessary model components present, allows us to stream-line the
model selection procedure.  These model ``fits'' are automatically
removed from those available to be selected as they are set to have a
very high chi-squared value and so are not selected by the routine as
a best fit.

We note that we may be missing non-dominant disc components in,
for example, the hard state.  Even though we fit all observations with
a powerlaw + disc model, the lack of sensitivity of the
\rxte\ \pca\ below $\sim 3\kev$ makes detecting non-dominant discs
difficult.  Therefore even if no significant disc is detected in our
analysis, there may be discs present at a very low level in the hard
and intermediate states -- absence of evidence does not imply evidence
of absence.  

We also note that we are attempting to fit a line component with a
gaussian at a fixed energy, whereas deeper spectra with higher
resolution (from \eg {\it Chandra} or {\it XMM-Newton}) are better fit with
relativistically blurred line whose peak energy varies (\eg
\citealp{Miller04}).  The results from the line parameters will appear in
a forthcoming publication.   However, we do not select models where
the line parameters (usually the line width) are not well
constrained\footnote{Those observations where the line width is
  unconstrained or if the value of the line width is
greater than $2\kev$.}, even if these are the best fitting.  

Any observation
with a $3-10 \kev$ flux from the best fitting model of less than
$1\times 10^{-11} \ergps$ was discarded from further analysis, as were
ones where the flux was 
not well determined (the error on the $3-10 \kev$ flux was larger than
the flux itself).  It was noted that in some cases, although the flux
was high enough and well determined, some of the model parameters were
not, especially in the soft state, were the powerlaw parameters may
be difficult to determine if the disc is very strong.  We therefore
excluded these observations from the analysis presented here, however
as the disc was fitted successfully, they may be included in a future
paper on the disc properties.

Even after this level of selection some spectra still were not well
fit by any of the models.  To remove the parameters obtained from
these poor fits we cut at $\chi^2<5.0$.  

This selection procedure resulted in a final list
of 3919 observations, corresponding to $\sim 10 \Ms$, with well fitted
spectra and high enough fluxes and counts.
We extract a range of parameters and fluxes for different energy bands and
model components which are presented and analysed further the
following sections.  

\begin{figure}
\centering
\includegraphics[width=1.0\columnwidth]{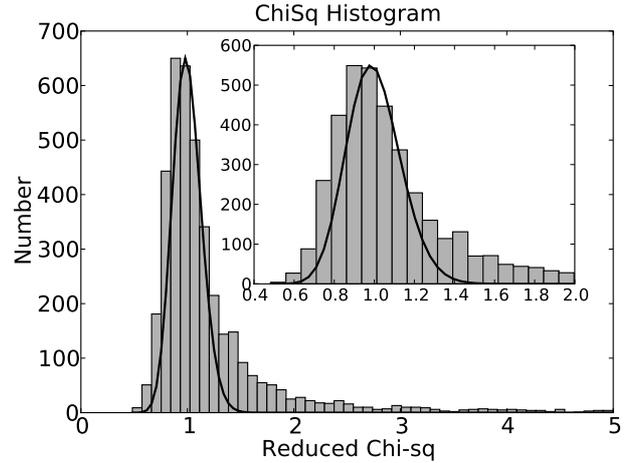}
\caption{\label{fig:chi_dist} The distribution of reduced $\chi^2$
  from all the observations.  The inset shows an enlargement around
  the peak of the distribution.  The selection of $\chi^2<5.0$ has
  been applied when creating this plot.  We show on top a the
  $\chi^2$ distribution using 111 degrees of freedom.}
\end{figure}

The distribution of the reduced $\chi^2$ from all the observations is
shown in Fig. \ref{fig:chi_dist}.  There is a clear peak at $\chi^2=1$
with a tail extending to higher values of $\chi^2$, 60.1 per cent of the
model fits have $0.8\leq \chi^2 \leq 1.2$ and 90.2 per cent between $0.7\leq \chi^2 \leq 2.0$.  Our average
number of degrees of freedom per spectrum is 111.4.  We show on Fig. \ref{fig:chi_dist} the expected
distribution for 111 degrees of freedom.  Given the automatic nature
of the data reduction procedure used in this analysis, this tail to
higher values of $\chi^2$ is expected.  A manual investigation into some of
the models with a high $\chi^2$ showed that these observations are a
mix of low counts data and occasions where the automated fitting
routine failed to find the best fitting model, leaving a poorly-fit
simple powerlaw (+gaussian) as the only model with fitted parameters.  However, the fraction of high
$\chi^2$ values is reasonably small.

\section{Selected Objects}\label{sec:obj}

In our analysis we primarily  aimed to select those objects which
appear to have been observed regularly with \rxte.  However, some of
the objects only have a few observations taken with \rxte.
In some cases, although there are many observations, only few pass
the counts, flux and model fitting criteria to be analysed further.
This sample of X-ray binaries was not intended to
be complete in any way, however, we have attempted to include all BHXRBs
with well studied outbursts.  

\begin{table*}
\centering
\caption{\label{tab:objects} {\sc X-ray Binary Parameters}}
\begin{tabular}{lrlrlll}
\hline
\hline
Object & $M_{\rm BH}$ & $D$ & $N_{\rm H}$ & $P_{\rm orb}$& $M_*$\\
&$M_{\odot}$&$\kpc$&$\times 10^{22}\pcm2$ & $h$ & $M_{\odot}$\\
\hline
4U 1543-47    	&$9.4\pm2.0~(1,2)	$&$7.5\pm0.5~(3,4)	$&$0.43~(2,4)	$&$26.8~(4)	$&$2.45~(1)	$\\
4U 1630-47	&$[10]	 		$&$10.0\pm5.0~(5)	$&$>6~(6)	$&$	-	$&$-		$\\
4U 1957+115	&$[10]	 	 	$&$[5]			$&$0.15~(7)	$&$9.3~(8)	$&$1.0~(9)	$\\
GRO J1655-40	&$7.0\pm0.2~(10,11)	$&$3.2\pm0.2~(4,12)	$&$0.8~(13)    	$&$62.9~(4)	$&$2.35~(10)	$\\
GRS 1737-31	&$[10]	 	 	$&$[5]			$&$6.0~(14)     $&$	-	$&$-		$\\
GRS 1739-278	&$[10]	 	 	$&$8.5\pm2.5~(15) 	$&$2~(15)	$&$	-	$&$-		$\\
GRS 1758-258	&$[10]	 		$&$[5]			$&$1.50~(16)	$&$18.5 ~(17)	$&$-		$\\
GS 1354-644	&$>7.8=10.0\pm2.0~(1)	$&$>27=33\pm6~(18)	$&$3.72~(18,19)	$&$61.1~(18)	$&$1.02~(1)	$\\
GS 2023+338	&$10\pm2~(1)	 	$&$4.0\pm2.0~(4) 	$&$0.7~(4)     	$&$155.3~(4)	$&$0.65~(1)	$\\
GX 339-4	&$5.8\pm0.5~(20)	$&$8.0\pm4.0~(21)	$&$0.4~(22)     $&$42.1~(4)	$&$0.52~(20)	$\\
H 1743-322	&$[10]		 	$&$[5]		 	$&$2.4~(23)     $&$	-	$&$-		$\\
XTE J1118+480	&$6.8\pm0.4~(1,24)	$&$1.7\pm0.05~(25,26)	$&$0.01~(25)    $&$4.08~(4)	$&$0.28~(1)	$\\
XTE J1550-564	&$10.6\pm1.0~(3)	$&$5.3\pm2.3~(4)  	$&$0.65~(27)    $&$37.0~(4)	$&$1.30~(3)	$\\
XTE J1650-500	&$<7.3=6\pm3~(28)	$&$2.6\pm0.7~(29)	$&$0.7~(30)     $&$7.7~(28)	$&$-		$\\
XTE J1720-318	&$[10]~(31)	 	$&$>8=8\pm6~(31)	$&$1.24~(31)    $&$	-	$&$-		$\\
XTE J1748-288	&$[10]		 	$&$>8=10\pm2~(32)	$&$7.5~(33)     $&$	-	$&$-		$\\
XTE J1755-324	&$[10]		 	$&$[5]		 	$&$0.37~(34)    $&$	-	$&$-		$\\
XTE J1817-330	&$<6=4\pm2~(35)	 	$&$>1=[10]~(35)	 	$&$0.15~(35)   	$&$	-	$&$-		$\\
XTE J1859+226	&$10\pm5~(36)	 	$&$6.3\pm1.7~(4)  	$&$0.34~(36)    $&$9.17~(4)	$&$0.9~(36)	$\\
XTE J2012+381	&$[10]		 	$&$[5]  	 	$&$1.3~( 37)    $&$	-	$&$-		$\\
LMC X1		&$10\pm5~(38)	 	$&$52\pm1.0~(39)  	$&$0.5~(13)	$&$93.8~(40)	$&$-		$\\
LMC X3  	&$10\pm2~(41)	 	$&$52\pm1.0~(39)  	$&$0.06~(42)    $&$40.8~(43)	$&$6~(41)	$\\
SAX 1711.6-3808	&$[10]	 	 	$&$[5]	   	 	$&$2.8~(44)	$&$	-	$&$-		$\\
SAX 1819.3-2525	&$10\pm2~(46)	 	$&$10\pm3~(46) 	 	$&$0.1~(47)     $&$67.6~(46)	$&$-		$\\
SLX 1746-331	&$[10]		 	$&$[5]   	 	$&$0.4~(45)     $&$	-	$&$-		$\\
\hline
\hline
\end{tabular}
\begin{quote}
Many of the objects do not have well determined distances or masses.
In this case we have taken the distances to be $5\kpc$ and the masses
$10~M_{\odot}$, shown in square brackets.  A recent critical look at the distance estimates for
GRO~J1655-40 by \citet{Foellmi08} indicates a revised estimate of the
distance of $<2.0\kpc$.  

(1) \citet{Ritter03}
, (2) \citet{Park04}
, (3) \citet{Orosz02}
, (4) \citet{Jonker04}
, (5) \citet{Augusteijn01}
, (6) \citet{Tomsick05}
, (7) \citet{Nowak08}
, (8) \citet{Thorstensen87}
, (9) \citet{Shahbaz96}
, (10) \citet{Hynes98}
, (11) \citet{Shahbaz99}
, (12) \citet{Hjellming95}
, (13) \citet{Gierlinski01}
, (14) \citet{Cui97}
, (15) \citet{Greiner96}
, (16) \citet{Pottschmidt06}
, (17) \citet{Smith02}
, (18) \citet{Casares04}
, (19) \citet{Kitamoto90}
, (20) \citet{Hynes03}
, (21) \citet{Zdziarski04}
, (22) \citet{Miller04}
, (23) \citet{Capitanio05}
, (24) \citet{Wager01}
, (25) \citet{Chaty03}
, (26) \citet{Gelino06}
, (27) \citet{Gierlinski03}
, (28) \citet{Orosz04}
, (29) \citet{Homan06}
, (30) \citet{Miniutti04}
, (31) \citet{CadolleBel04}
, (32) \citet{Hjellming98}
, (33) \citet{Kotani00}
, (34) \citet{Revnivtsev98}
, (35) \citet{Sala07}
, (36) \citet{Hynes02}
, (37) \citet{Campana02}
, (38) \citet{Hutchings87}
, (39) \citet{DiBenedetto97}
, (40) \citet{Orosz08}
, (41) \citet{Cowley83}
, (42) \citet{Haardt01}
, (43) \citet{Hutchings03}
, (44) \citet{IntZand02}
, (45) \citet{Wilson03}
, (46) \citet{Orosz01}
, (47) \citet{IntZand00}.
\end{quote}
\end{table*}

In Table \ref{tab:objects} we list the masses, distances, $N_{\rm H}$
values of the binaries studied in this work.  We also summarise the
number and total exposure of the X-ray observations used (Table \ref{tab:obstimes}).  Many of the
masses and distances are not well determined, and in these cases we
have assumed the distances to be $5\kpc$ and the masses
$10M_{\odot}$.

The uncertainties in distances and masses affect the calculation of
the Eddington Luminosity for these sources.  As the majority of
relations and diagrams in this work use $L_{\rm Edd}$ to scale the
observables, allowing the different binaries to be compared with one
another.  Without well determined distances and masses then any
relation obtained for the ensemble of binaries will be uncertain to
some degree.

We note that there is a large range in the number of observations per
binary (from almost 700 to less than 10).  Therefore in parts of this
analysis where the population as a whole is studied, the results will
be strongly influenced by those BHXRBs with many observations.
However, those BHXRBs with the most observations also tend to be the
ones which go through clear outbursts.  We also look at the
lightcurves to ``pre-identify'' outbursts, which can then be compared,
either to other outbursts from the same source, or to ones from
different sources.

\begin{table}
\centering
\caption{\label{tab:obstimes} {\sc Observation Numbers and Times}}
\begin{tabular}{lllll}
\hline
\hline

Object &Total Obs& Exposure & Selected Obs$^*$ & Exposure\\
&&$\Ms$&&$\Ms$\\
\hline
4U 1543-47    	&$101	$&$0.210	$&$61   $&$0.147$\\
4U 1630-47	&$868	$&$1.629	$&$704  $&$1.371$\\
4U 1957+115	&$98	$&$0.458	$&$59   $&$0.260$\\
GRO J1655-40	&$572	$&$2.254	$&$484  $&$1.829$\\
GRS 1737-31	&$5	$&$0.045	$&$5    $&$0.045$\\
GRS 1739-278	&$10	$&$0.022	$&$6    $&$0.017$\\
GRS 1758-258	&$10	$&$0.007	$&$9    $&$0.007$\\
GS 1354-644	&$9	$&$0.049	$&$8    $&$0.049$\\
GS 2023+338	&$4	$&$0.005	$&$0    $&$0.000$\\
GX 339-4	&$971	$&$1.954	$&$709  $&$1.682$\\
H 1743-322	&$364	$&$1.030	$&$346  $&$0.998$\\
XTE J1118+480	&$97	$&$0.186	$&$81   $&$0.170$\\
XTE J1550-564	&$397	$&$0.868	$&$365  $&$0.833$\\
XTE J1650-500	&$175	$&$0.271	$&$108  $&$0.191$\\
XTE J1720-318	&$97	$&$0.247	$&$63   $&$0.125$\\
XTE J1748-288	&$23	$&$0.092	$&$21   $&$0.074$\\
XTE J1755-324	&$2	$&$0.006	$&$2    $&$0.006$\\
XTE J1817-330	&$154	$&$0.383	$&$123  $&$0.329$\\
XTE J1859+226	&$127	$&$0.300	$&$121  $&$0.292$\\
XTE J2012+381	&$25	$&$0.045	$&$15   $&$0.036$\\
LMC X1		&$72	$&$0.351	$&$69   $&$0.349$\\
LMC X3  	&$684	$&$1.607	$&$471  $&$1.048$\\
SAX 1711.6-3808	&$17	$&$0.040	$&$13   $&$0.029$\\
SAX 1819.3-2525	&$58	$&$0.132 	$&$48   $&$0.114$\\
SLX 1746-331	&$58	$&$0.153	$&$28   $&$0.091$\\
\hline
Totals		&4998	&12.34	&3919	&10.09\\
\hline
\hline
\end{tabular}
\begin{quote}
Differences between the exposure stated here for GX~339-4 and in
\citet{Dunn08} arise from the use of the true exposure rather than the
difference in start and stop times of the observation. $^*$ After
removing high $\chi^2$ observations, low flux observations et cetera.
\end{quote}
\end{table}

\begin{figure*}
\centering
\includegraphics[width=1.0\textwidth]{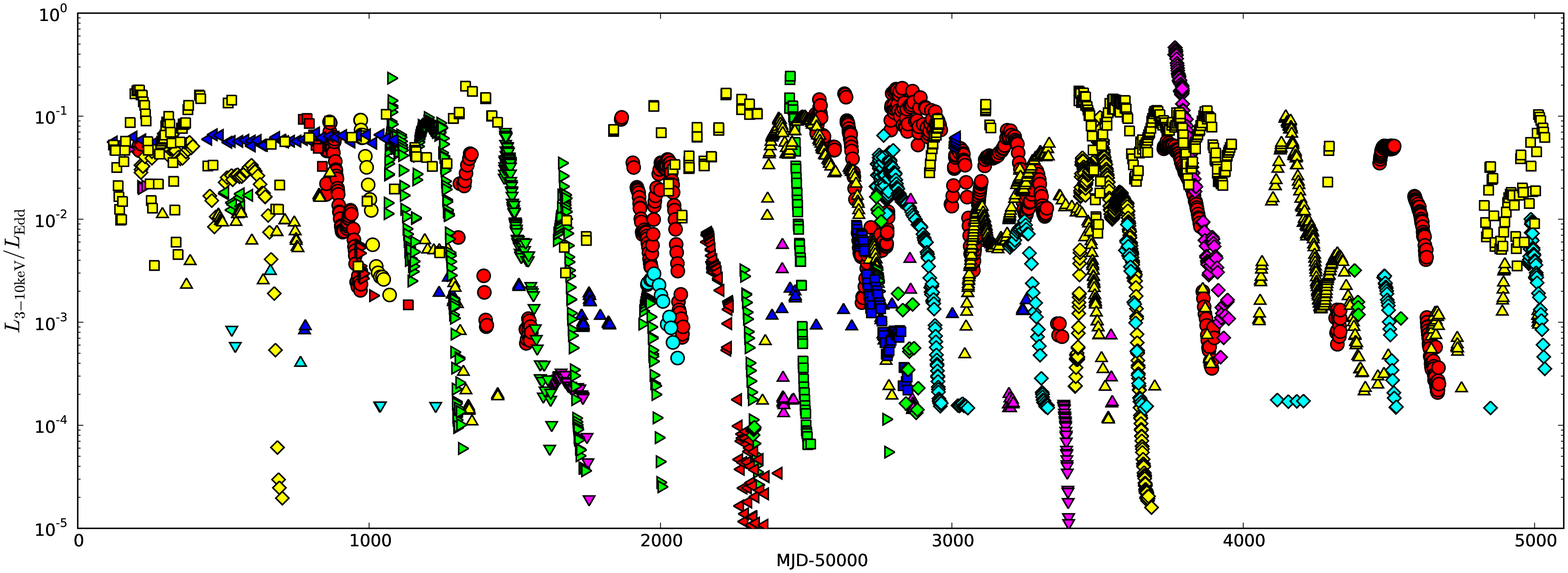}
\includegraphics[width=0.98\textwidth]{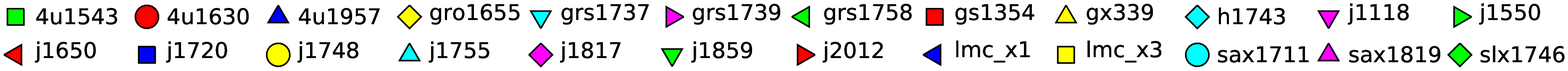}
\caption{\label{fig:lc} {\scshape top} The light curve for all the objects in the
  sample. Each object has a different symbol.  {\scshape bottom }
  Legend for the different symbols used for the plots showing different objects.}
\end{figure*}

The full \pca\ lightcurve of from all the objects over the full 13
years of the \rxte\ mission is shown in
Fig. \ref{fig:lc}.  The level of activity from X-ray binaries can
easily be seen.  Although this plot is in terms of the Eddington
ratio, most sources are between $5$ and $10\kpc$ away, so the overall
variation in X-ray flux is a factor of four greater.  Even with
this incomplete sample and the restrictions on which observations are
analysed in this work, there are few periods when no BHXRB is
radiating at a significant fraction of $L_{\rm Edd}$.

\section{Population Study: Global properties of BHXRBs}\label{sec:pop}

Having analysed this large sample of X-ray binaries we now look at
their global properties using well characterised methods.

\subsection{Hardness Luminosity Diagram}\label{sec:pop:hid}

\begin{figure}
\centering
\includegraphics[width=1.0\columnwidth]{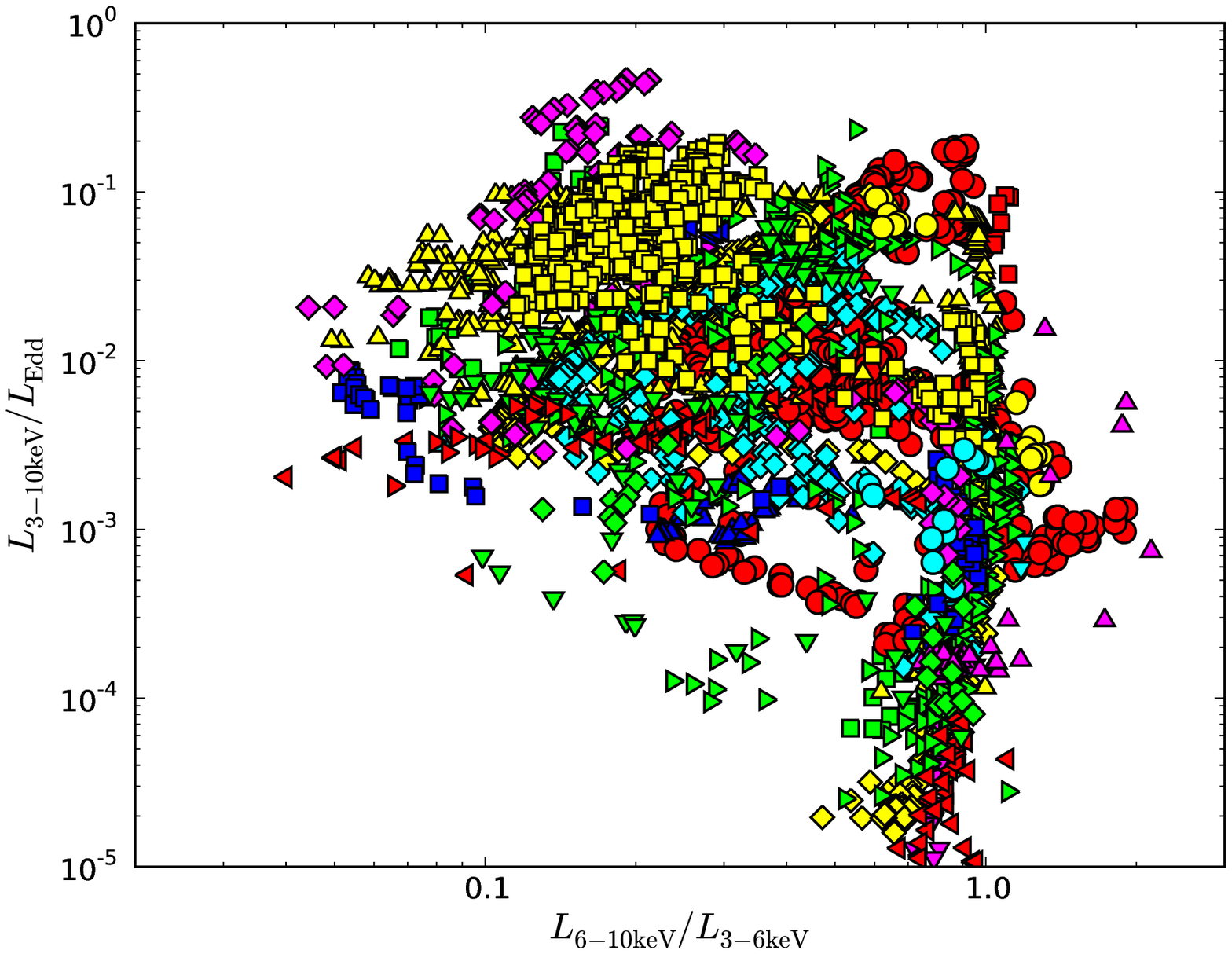}
\includegraphics[width=1.0\columnwidth]{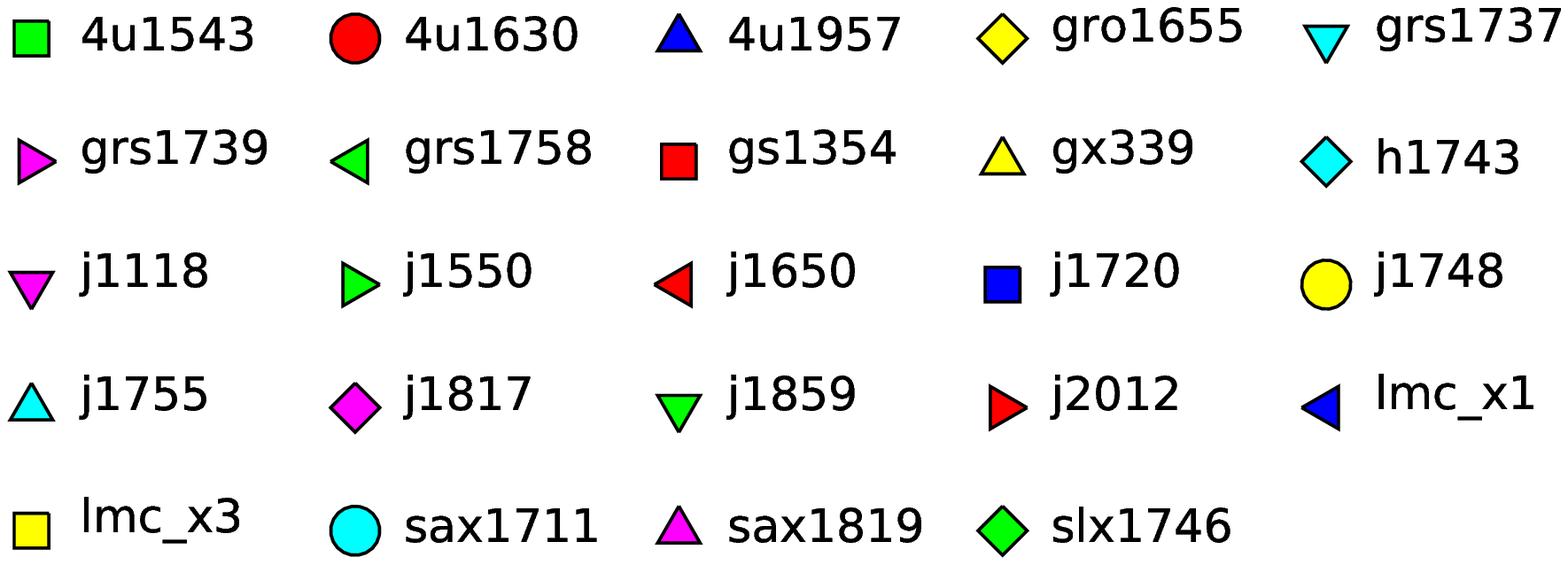}
\caption{\label{fig:HID} {\scshape top} The HLD for all observations of all the
  objects.  Each object has a different symbol.  {\scshape bottom }
  Legend for the different symbols used for 
  the plots showing different objects.}
\end{figure}

The natural first step is to follow \eg \citet{Homan01, Belloni04,
  Fender04}, and plot the X-ray observations 
in a Hardness-Intensity Diagram (HID) - in this case a Hardness-Luminosity
Diagram (HLD) to show the state changes of
the BHXRBs during their outbursts.  We extract the $3-6$, $6-10$ and
$3-10\kev$ luminosities from the spectra.
The X-ray colour was calculated from $L_{6-10\kev}/L_{3-6\kev}$ and plotted
against $L_{3-10\kev}$ in Fig. \ref{fig:HID}.

If all the X-ray binaries behave in a similar way during their
outbursts, they would be expected to overlap in a well defined region
of the diagram.  However the overlap is not expected to be perfect
(see later in this section).  For the Hardness-Intensity diagrams for
the individual outbursts from individual sources see Appendix
Fig. \ref{fig:curves} and Fig. \ref{fig:remainder}.

Most of the objects do overlap in the hard-state ``stalk'' and in the
soft-state region.  However there are some outliers which fall below
the majority of the soft state, and some which appear harder than most
of the other hard state observations.  The two objects which have soft
X-ray colours but are at lower 
luminosities than most (XTE~J1550-564, XTE~J1859+226), have two separate
explanations.  In XTE~J1550-564 the X-ray colour softens towards the end
of the outburst in 1998 and so the track
reappears below the majority of the soft state points.  Fainter
observations which were cut from our analysis show the XTE~J1550
returns to the hard state shortly after our points cease \citep{Homan01}.  For XTE~J1859
the peak of the outburst is not far below the rest on
Fig. \ref{fig:HID}, so in this case, the decay in luminosity while in
the soft state is just more than in the other binaries.

The two objects which have observations at higher X-ray colours
($\gtrsim 1$) than
most others in the sample are 4U~1630-47 and SAX~J1819.3-2525.  SAX~J1819.3-2525 is known to be
a peculiar source with strong and rapid variability with a possible
shrouding of the black hole \citep{Maitra06}.  4U~1630-47 has a high $N_{\rm
  H}$, which is a possible cause for its ``shift'' of the canonical
track through the HLD to harder values.  In the Appendix, the
individual HLDs and curves show that almost all of the individual
outbursts are shifted towards harder X-ray colours.  However, both
GRS~J1737-31 and XTE~J1748-288 also have high $N_{\rm H}$ values but neither
has the equivalent extreme shift in X-ray colours.  XTE~J1748-288 shows a
slight hardening compared to the rest of the binaries on its return to
the hard state.  There are only a few observations of GRS~J1737-37 but
these are also fractionally harder than the norm.  This effect of the
values of $N_{\rm H}$ is discussed below. 

There are two main problems with the HLD as it has been used in
Fig. \ref{fig:HID}.  The main problem is that as the distances of
the BHXRBs uncertain to probably a
factor of two, the relative luminosities are not that well constrained.  Secondly,
as we are using the $3-10\kev$ luminosities in the construction of the
HLD, the effects of galactic absorption are different for each object
and affect the values of the measured luminosities.  

The HLD does well when studying outbursts of individual objects, especially when in
combination with X-ray timing and variability information, to determine the state of the
XRB.  However, as the X-ray colour is used, then with the
inability of being able to accurately determine both the disc emission
and the effect of any absorption from the spectrum alone limits the usability of the HLD to
compare quickly and easily between sources.  We therefore turn to
another diagnostic diagram for the study of the evolution of the outbursts of
XRBs.

We do note, however, that when using a wide band luminosity
(e.g. $0.1-100\kev$) rather
than just $3-10\kev$ there is better agreement between the soft states
of the binaries as this a more accurate measure of the total
luminosity of the source.  However, this cannot account for the effects of
absorption on the X-ray colour.

\subsection{Disc Fraction Luminosity Diagram}\label{sec:pop:dfld}

The HLD compares the soft X-ray luminosity to the hard X-ray
luminosity to give a rough characteristic of the spectral state.  The
soft X-ray band is dominated by the disc when in the soft state, and
the hard X-ray band comes mainly from the non-thermal X-ray emission
which has been suggested comes from a corona.  The X-ray colour is
therefore a proxy for the extent to which the thermal component is
dominating the X-ray emission from the binary.

We therefore construct a diagram to compare the relative strengths of
the disc and the powerlaw components.  Previous studies which have
used the relative strengths of the disc and powerlaw components to
study the outburst properties include
(\citealp{Kalemci02,Kalemci04,Tomsick05,Kalemci06,Koerding06,Dunn08}).
\citet{Koerding06} formulated the ``Disc 
Fraction Luminosity Diagram'' (DFLD) for AGN, as an HLD from the X-ray
spectrum alone does not work as
the disc emission peaks in the UV.  The soft X-rays give information
on warm absorbers rather than the accretion disc  They therefore calculated the disc
and powerlaw luminosities, and combined them in a way to emulate the HLD:
\begin{eqnarray}
{\rm Powerlaw\, Fraction}={\rm PLF}=\frac{L_{1-100\kev,\ {\rm
      PL}}}{L_{0.001-100\kev,\ {\rm Disc}}+L_{1-100\kev,\ {\rm PL}}}
\nonumber \\
{\rm Disc\, Fraction}={\rm DF}=\frac{L_{0.001-100\kev,\ {\rm Disc}}}{L_{0.001-100\kev,\ {\rm Disc}}+L_{1-100\kev,\ {\rm PL}}}.\nonumber
\end{eqnarray}

\noindent We use the unabsorbed powerlaw luminosity, but only
determine the flux down to $1\kev$ to prevent the low energy end from
being overly dominant.  We are unable to determine both the
absorption and the low-energy cut-off of the powerlaw from the
\rxte\ data, and so use a fixed value of $N_{\rm H}$ for the fitting,
and a standard low energy bound to the powerlaw flux for all objects.
Although this may introduce excess powerlaw luminosity into our
calculations, it is standard across all objects, as the true low
energy behaviour of the powerlaw cannot be determined from the
\rxte\ spectra.    The disc luminosity is also the
unabsorbed luminosity to full capture the radiative output of the
disc.  The energy range used is purposefully large to ensure we cover
as much of the disc emission as possible.  This diagnostic diagram is therefore less sensitive to the effects of
$N_{\rm H}$ (the powerlaw luminosity still includes the effects of
$N_{\rm H}$ instead of a low energy cut off), as in the outbursts, the
low energy emission is dominated by the unabsorbed disc emission.

\begin{figure}
\centering
\includegraphics[width=1.0\columnwidth]{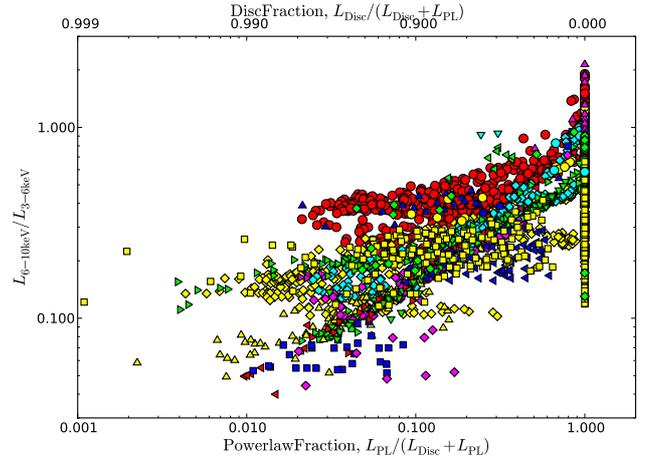}
\caption{\label{fig:DFXrCl} The correlation between the X-ray colour
  and the disc fraction.   See
  Fig. \ref{fig:HID} for the legend to the symbols used.}
\end{figure}

We show in Fig. \ref{fig:DFXrCl} the relation between the disc
fraction and the X-ray colour of the observation.  The arrangement of
the points shows that the calculated disc fractions are well
determined.  In fact, for many objects
there is a constant relation between the two diagnostic values when
the disc fraction is not zero.  The approximate relation is quadratic,
where the powerlaw fraction is proportional to the square of the
hardness ratio.  There
are some which do not appear to lie on the relation.  The clearest are
4U~1630-47 and LMC-X3.  4U~1630-47 has a large value for $N_{\rm H}$, which
hardens the spectrum, so moving the soft state to harder X-ray
colours.  The same can be said, to varying degrees, of H~1743-322 and
XTE~J1748-288.  In LMC-X3, there is very little variation in the X-ray
colour for a large range in disc fraction.  In this source, the
powerlaw is not always well determined underneath a very dominant
disc, likely to result from the distance of the source.

GRO~J1655-40 and XTE~J1550-564 also appear at harder
X-ray colours than the majority of the other sources at the same disc
fractions.  The $N_{\rm H}$s for these two sources are not
particularly high, so this is unlikely to be the reason for their
offset.  They do both have atypical outbursts, with XTE~J1550-564
performing a complete loop in the HLD, and GRO~J1655-40 having a very
strong very high state.  Whether these features also manifest
themselves in the relative strengths of the powerlaw and disc
components is not clear.

\begin{figure}
\centering
\includegraphics[width=1.0\columnwidth]{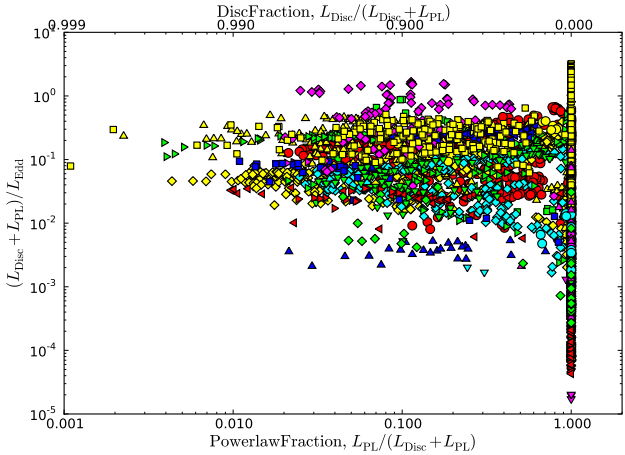}
\caption{\label{fig:DFLD} The DFLD for all observations of all the
  objects.  Each object has a different symbol.   See
  Fig. \ref{fig:HID} for the legend to the symbols used.}
\end{figure}

On our Disc Fraction Luminosity
diagrams (DFLDs), we plot both the
Powerlaw and the Disc Fractions as axes.  We retain the powerlaw
fraction for the main $x$-axis for similarity with the HLDs, and use the disc fraction as the
secondary (upper) $x$-axis.  We hope that this will minimise confusion
when talking about the disc fraction.  We note that uncertainties
in the luminosity of the observation arising from uncertainties in the
masses and distances of the object have not been accounted for in Fig. \ref{fig:DFLD}.

Rather than using the canonical state descriptions (hard, soft,
intermediate etc.) when discussing the
DFLD, we will use disc- and powerlaw-dominated states.  The reason for
this is that the intermediate and
hard states are compressed together.  This makes distinguishing the
standard states as determined from the X-ray colours difficult on the DFLD.  Therefore, we use the two-state
description as mentioned above.  We define the two states as follows,
a powerlaw-dominated state is one where the powerlaw fraction is
greater than 0.8 (disc fraction less than 0.2).  The disc dominated
state is therefore the converse.  For the moment we do not consider
the observations which are intermediate between the disc dominated and
powerlaw dominated states.

Some of the objects (XTE~J1817-330 and LMC~X-3) appear to have
luminosities above the Eddington luminosity.  In the case of XTE~J1817-330
the distance to the source is not well known and there are large
uncertainties on the mass of the compact object.  In the assumption
that XTE~J1817-330 is not a super-Eddington source, we can use the Eddington
luminosity to place limits on the distance of the source for different
values of the mass of the compact object.  We have currently used
a value of $10\kpc$, which results in a maximum Luminosity of
$1.7L_{\rm Edd}$.  Therefore, assuming that the mass of the central
object is $\sim4M_{\odot}$, the upper limit on the distance is $\sim8\kpc$.
For the upper mass limit of $\sim6M_{\odot}$, the upper limit on the
distance is $\sim10\kpc$ and for the lower mass limit of
$\sim2M_{\odot}$ $\sim5.6\kpc$.  

For LMC~X-3 the situation is less clear.  The distance to the source
is well known and the mass is also comparatively well determined.  The
DFLD of LMC~X-3 alone is strange.  The ``stalk'' extends well beyond
the most luminous disc dominated state.  Investigating the individual
fits more closely, we found that not all of the best fit results were
well fit.  Although there are sufficient counts in the \hexte\ band to
pass our selection criteria, they are not sufficient to constrain the
powerlaw at high energies.  At low energies, in the \pca\ band, the
data are dominated by the disc, preventing a powerlaw component to be
fitted to the data.  As the powerlaw parameters cannot be accurately
determined, {\scshape xspec} may not extract values for this (disc +
powerlaw) model.  This model fit is then penalised in the model
selection routine, and results in a best fitting model with only a
powerlaw component.  In some cases, the powerlaw is well-fitted, but
has a very steep slope.  The steep slope at the low energies, where the
powerlaw fits the rise of the disc component results in a high
powerlaw normalisation, from which the powerlaw luminosity, and hence
total luminosity, is calculated.  Fitting only a disc model, although
it would allow the study of the disc parameters (and may be done for a
future publication) would not allow the calculation of an accurate
disc fraction.

The DFLD constructed from the SDSS\footnote{Sloan Digital Sky Survey,
  \citet{Adelman-McCarthy08}} by \citet{Koerding06} is shown in their
Fig. 10.  The DFLD presented in Fig. \ref{fig:DFLD} appears similar in
appearance to the expected DFLD simulated for a sample of BHXRBs
undergoing outbursts by \citet{Koerding06}.  There are some outlying
points, but most of the observations lie in a swathe between $1$
and $50$  per cent of $L_{\rm Edd}$.  The outlying points may arise
from the uncertainty in the values of the distances and masses of the
XRBs.  The radio properties across the DFLD are described in Section
\ref{sec:radio}.  We now discuss the properties of the DFLD in more detail.

\subsection{Cross mapping of HID and DFLD}\label{sec:pop:comp}

\begin{figure*}
\centering
\includegraphics[width=0.49\textwidth]{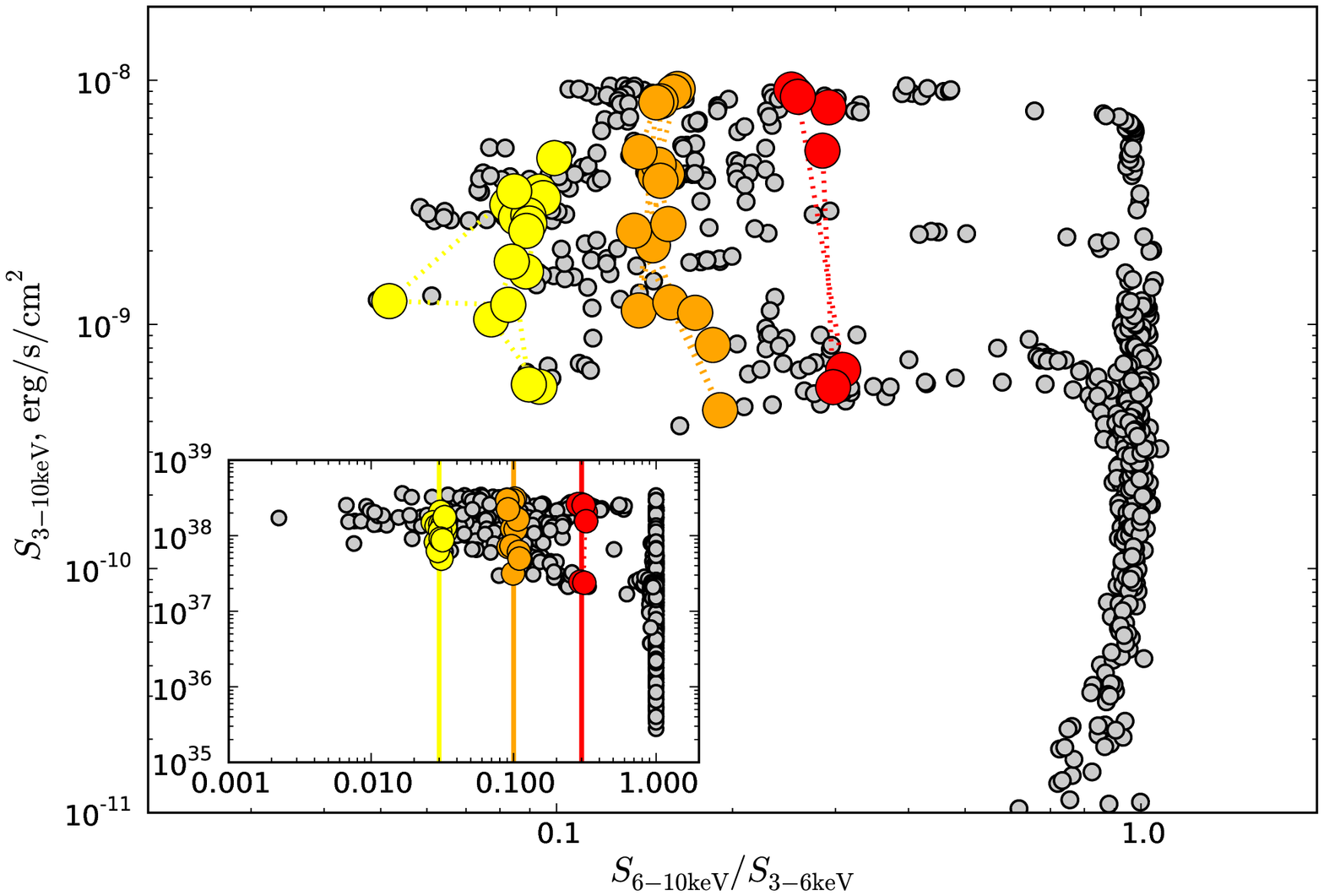}
\includegraphics[width=0.49\textwidth]{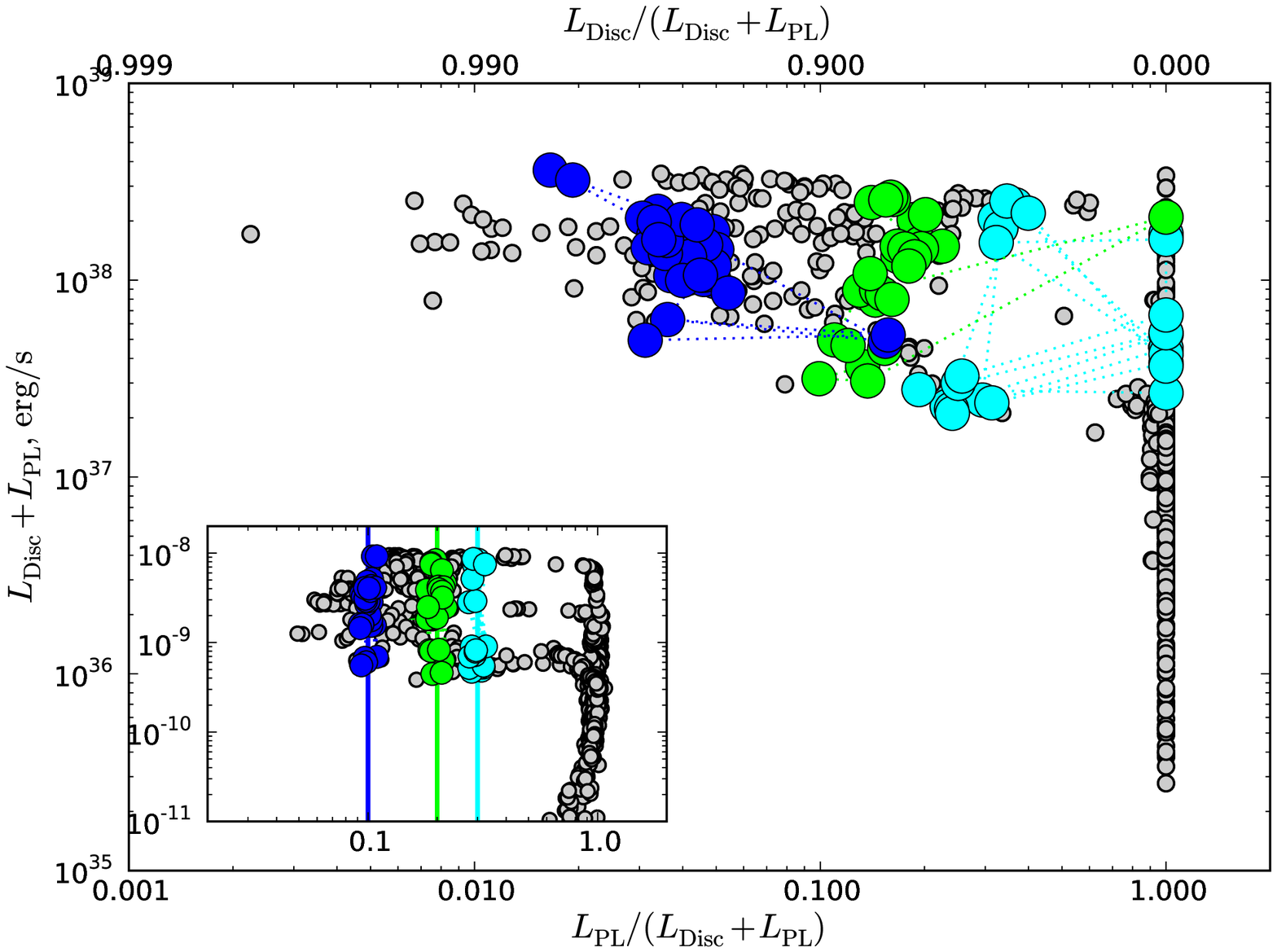}
\includegraphics[width=0.49\textwidth]{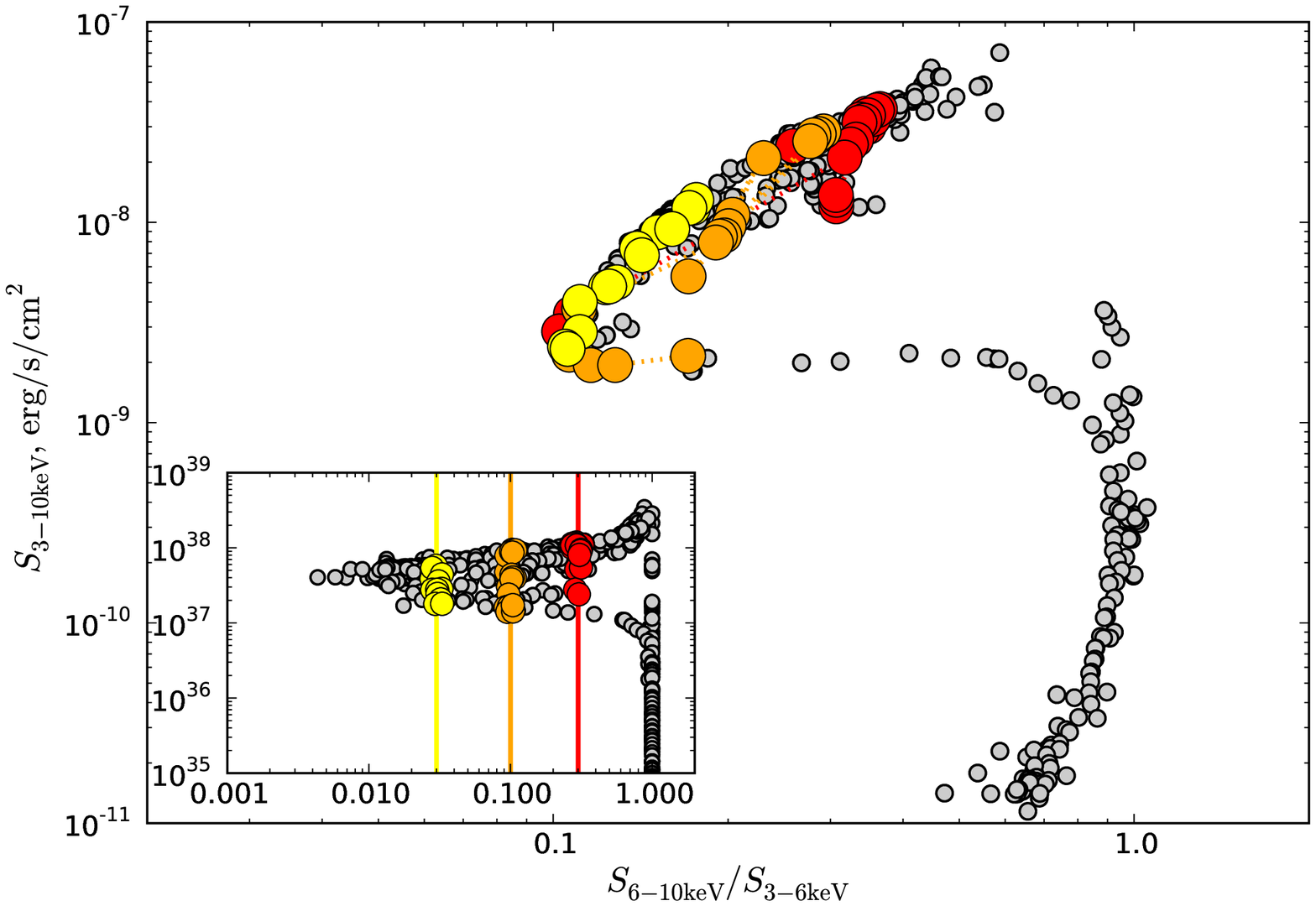}
\includegraphics[width=0.49\textwidth]{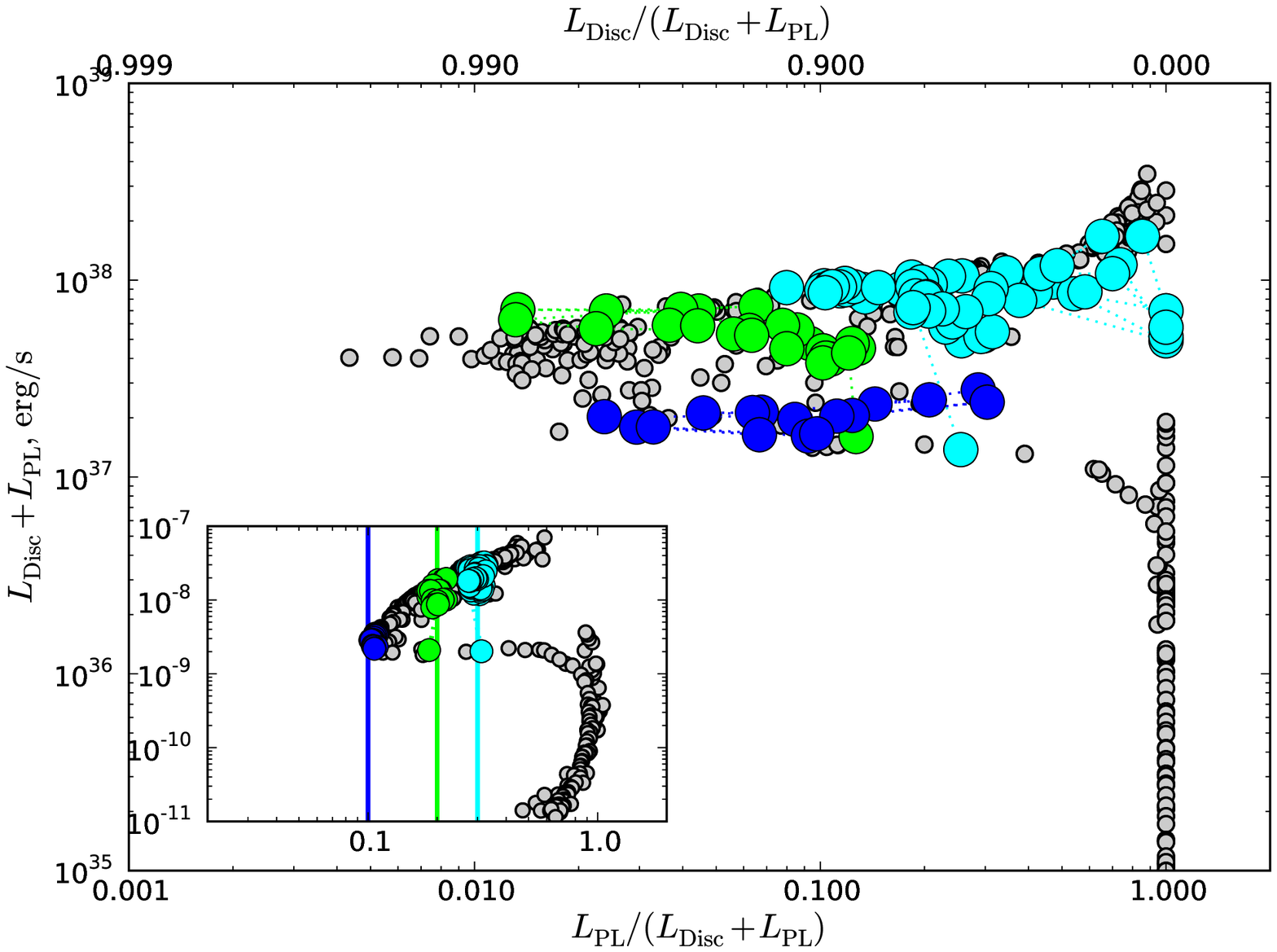}
\includegraphics[width=0.49\textwidth]{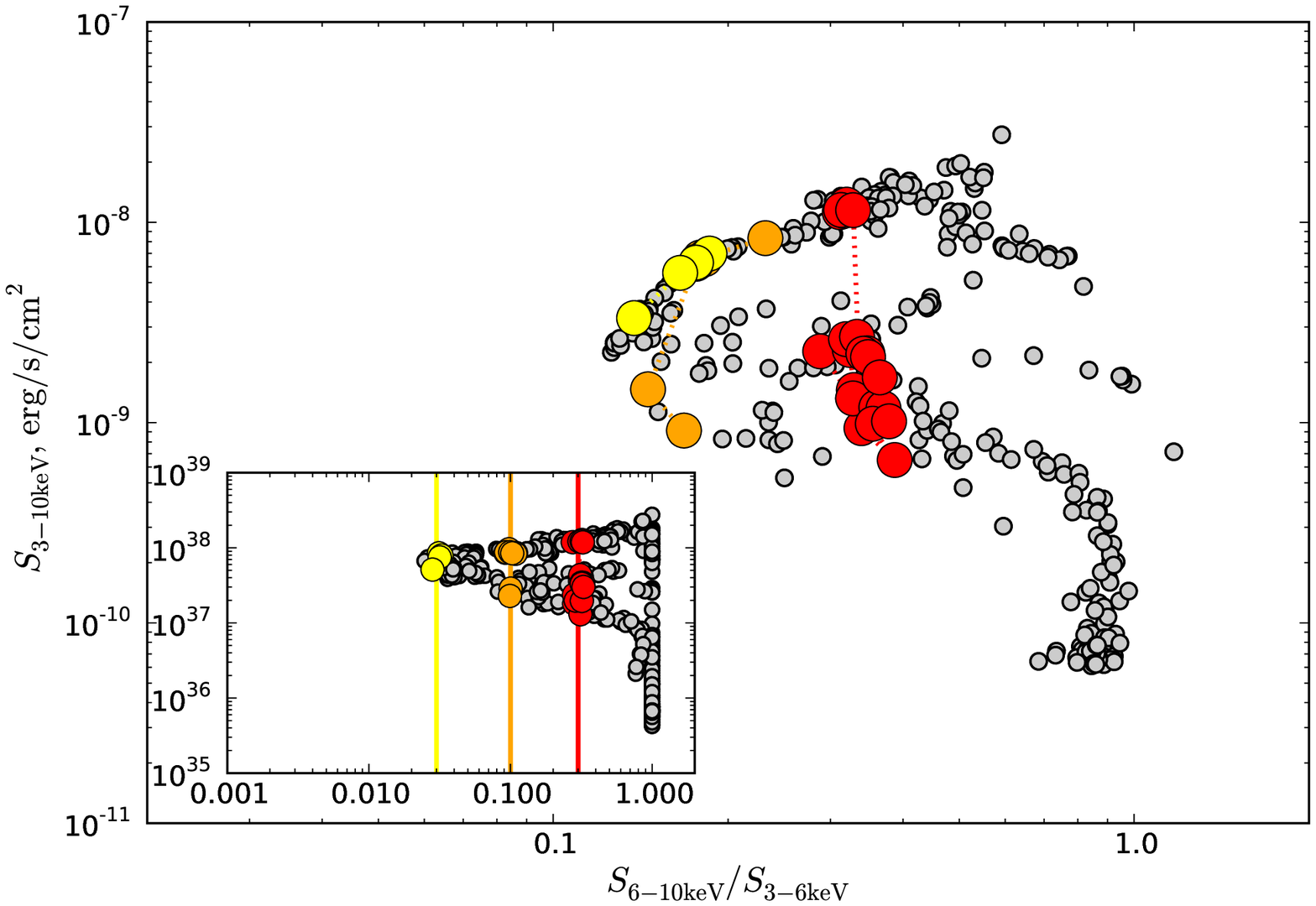}
\includegraphics[width=0.49\textwidth]{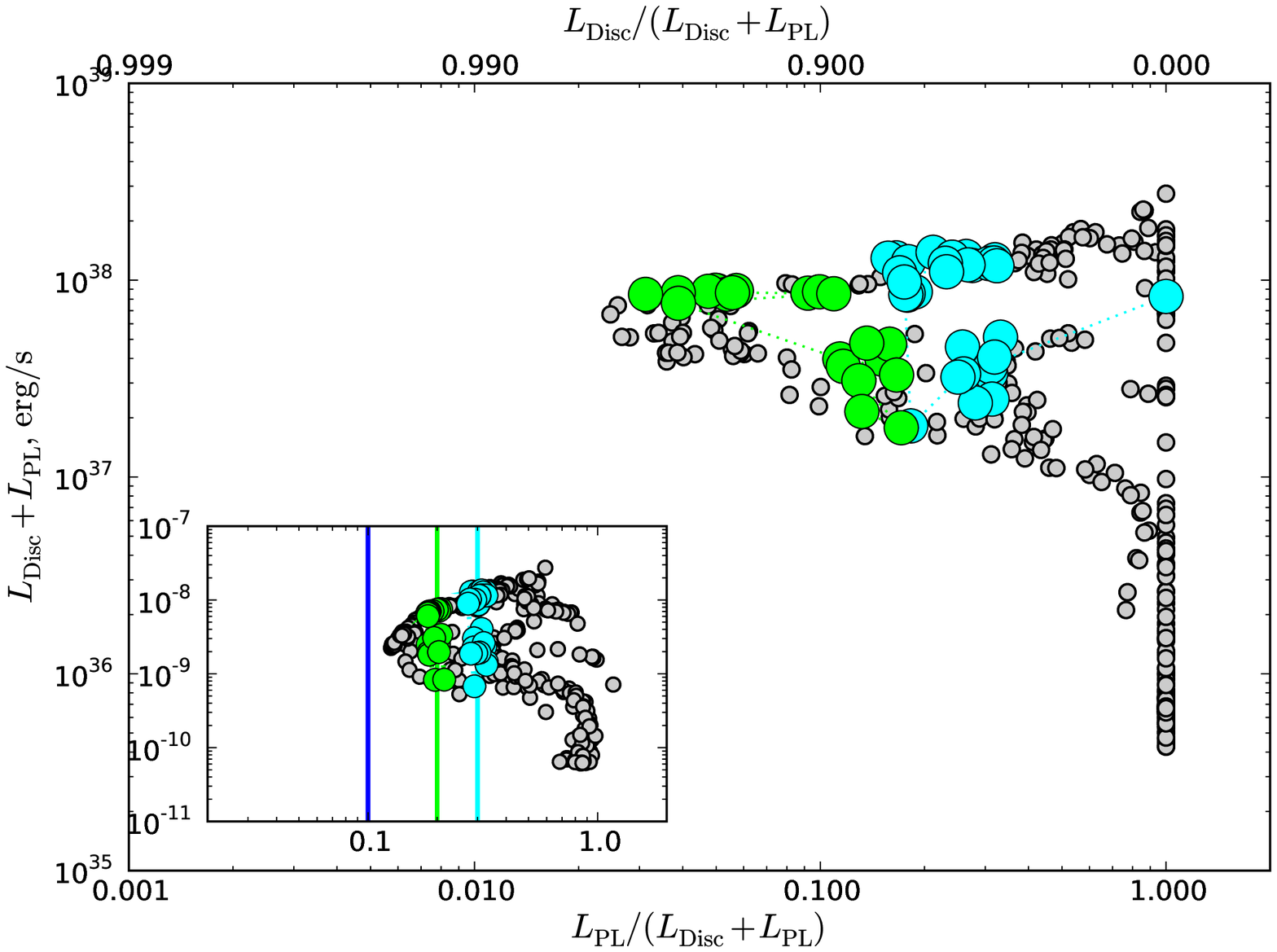}
\caption{\label{fig:DFLD_on_HID} The mapping of one diagram onto the
  other.  {\scshape left}: HID with specific disc fractions highlighted, and
  {\scshape right}: DFLD
  with specific X-ray colours highlighted.  The highlighted disc fractions
  are red 0.3, orange 0.1, yellow 0.03; and the highlighted
  X-ray colours are 
  cyan 0.3, green 0.2, blue 0.1.  {\scshape top}: GX~339-4, {\scshape
  middle}: GRO~1655-40 and {\scshape bottom}: H~1743-322}
\end{figure*}

The mapping of the HID to the DFLD and vice-versa was not investigated
clearly in \citet{Dunn08}.  Although we now have more populated
diagram, the mapping from HID to DFLD and vice versa is not easily
determined for the complete population.  This arises because of the problems
with the HID outlined in Section \ref{sec:pop:hid}.  The effects of
absorption on the soft end of the spectrum re-appear when converting
Disc Fractions for all the sources back onto the HID.  It is therefore
easiest to see the mapping for individual objects.  As every object,
and even every outburst (see Section \ref{sec:Outburst_sim}) is different we show the
mapping for three representative outbursts in
Fig. \ref{fig:DFLD_on_HID}.   Correcting for the hardness when used in
the HID would more easily allow the comparison between sources
\citep{Done03}.  

In Fig. \ref{fig:DFLD_on_HID} we show the HID and DFLD for GX~339-4, GRO
J1655-40 and H~1743-322.  To highlight
the variation across the diagram, we select a number observations with
specific X-ray colours or disc fractions.  These have been given a different colour to
show the pattern of variation between the two diagnostic measures.

Roughly, the
most disc dominated states are the soft states, and
the powerlaw dominated states are the hard and intermediate states.
However the exact mapping, especially the HID intermediate states,
appears to vary from source to source.  The state transitions will be easier to
map when they can be clearly identified from changes in the timing
properties of the sources, rather than from X-ray colours at present.
For further discussions on the transitions between states see Section \ref{sec:Trans_L}. 

The apparent lack of points between powerlaw fractions of $1$ and
$\sim 0.3$ result from the low energy limitations of
the \rxte\ \pca.  The discs we are attempting to analyse are
$\lesssim1\kev$ whereas the lowest calibrated spectral energy bin is
around $3\kev$.  Therefore as the disc emission strengthens as the
source moves into the soft state, the X-ray colour softens.  However,
our fitting algorithm may take a broken powerlaw over a disc and
powerlaw fit for these intermediate states.  Only when the curvature of
the multicolour disc is clearly evident will a disc and powerlaw be a
better fit, by which time the source is almost fully into the soft
state.  There is very little difference in the $\chi^2$ values of the
fits, but as both models have the same number of degrees of freedom,
finding out which is the appropriate one to choose is not
straightforward.  Rather than choosing disc models where they may not
have been fitted well, or may not be appropriate, we currently
continue to use the model selected on $\chi^2$ terms.  This has on a
handful of occasions resulted in observations where the best fit is a
broken powerlaw, but all the neighbouring observations are disc model
fits (see Appendix Fig. \ref{fig:curves} - 4U~1630-47 Outburst 7 \& 8).  Rather than adjusting these model fits by hand, and therefore
possibly biasing our results, we leave these observations as they are,
especially as they stand out clearly.


Therefore in some cases the DFLD as determined from
\rxte\ compresses some of the intermediate states
into the powerlaw dominated state.  The extent to which this occurs
varies on a source-by-source basis.  As can be seen in
Fig. \ref{fig:DFLD} the area between disc fractions of $0.3$ to $0.9$ is
fairly well populated whereas in Fig. \ref{fig:DFLD_on_HID} for GX~339-4,
there is a gap, but in GRO~J1655-40 there is not.  However above a certain
X-ray hardness, all the points are compressed on to a single line.
The DFLD is therefore not ideally suited to the study of the hard and
intermediate states.

The fan-like mapping of lines of constant X-ray colour onto the DFLD,
indicates that while the source remains in the soft state, the disc
begins to cool and decay.  Although this can be seen as the total luminosity in
the soft state decreases, the level to which this is the disc or the
powerlaw is less clear from using the X-ray colour and a $3-10\kev$
luminosity.  In the DFLD the level to which the disc dominates the
X-ray emission from the binary is clearly seen to decrease before the
source re-enters to the hard state.  

This mapping of the X-ray colour onto the Disc Fraction was
expected from the simulation of an BHXRB DFLD by \citet{Koerding06} as
shown in their Fig. 11.  Although also intended to show the positions
of AGN classes on the DFLD, the split into only hard, hard
intermediate and soft states is similar to the distribution of the
lines of constant X-ray colour (from which the states can usually be
determined). 

The transition across from the powerlaw to the disc dominated state
in the DFLD can be at a single or decreasing luminosity (horizontal or sloping
line in Fig. \ref{fig:DFLD_on_HID}), mimicking that what happens in the
HID.  However, on the return, the HID transition appears to horizontal
and rapid, whereas the decline to the powerlaw dominated state in the
DFLD shows the disc dominance gradually declining as the luminosity
decreases.  There are more observations where the return transition
has been monitored, only one of which is close to horizontal
(XTE~J1650-500, see Appendix Fig. \ref{fig:curves}).  

This comparison between the HID and DFLD shows that for individual
outbursts or objects the HID is most suited as the different states
are not compressed together.  The intermediate, hard and soft states
are easily discernable.  When using the DFLD, the intermediate states
are compressed with the hard states.  However, for the reasons
outlined in Section \ref{sec:pop:hid}, it is not best suited for comparing
outbursts from different objects.

\subsection{Speed of motion}\label{sec:pop:speed}

We investigated the speed of motion of an BHXRB through the HID and
DFLD.  Under the assumption that AGN are just scaled-up versions of
BHXRBs, then as they are going to evolve much more slowly than BHXRBs,
a population study will definitely be required to investigate them.
Under the assumption that the temporal evolution of AGN are the same
as for BHXRBs, the speed of motion of the BHXRBs through the DFLD
shows where AGN are most likely to be found.  When the BHXRBs move
rapidly through the DFLD, they do not spend long in that area of the
diagram, and so are not observed often in that area of the diagram.
Hence for a population of AGN, they will be preferentially be observed
in the areas of the DFLD where the BHXRBs move more slowly.

To allow the easy comparison of the speed at different points
within these diagrams, the relative changes were calculated:
\begin{equation}
{\rm
  Rate}=\left(\frac{x_i-x_{i-1}}{\frac{(x_i+x_{i-1})}{2}}\right)/(t_i-t_{i-1})~
{\rm day ^{-1}},
\end{equation}
where $x_i$ is the quantity whose rate of change we calculate at the
observations of interest, $x_{i-1}$ is the value at the previous
observation, and $t$ are the dates of the observations.  We do not
take values for the rates of motion when the two observations are
separated by more than five days.  This is to remove large values when
an outburst is badly sampled on the rise or decline.  This process
removed around 200 observations from these plots.  

\begin{figure}
\centering
\includegraphics[width=0.49\textwidth]{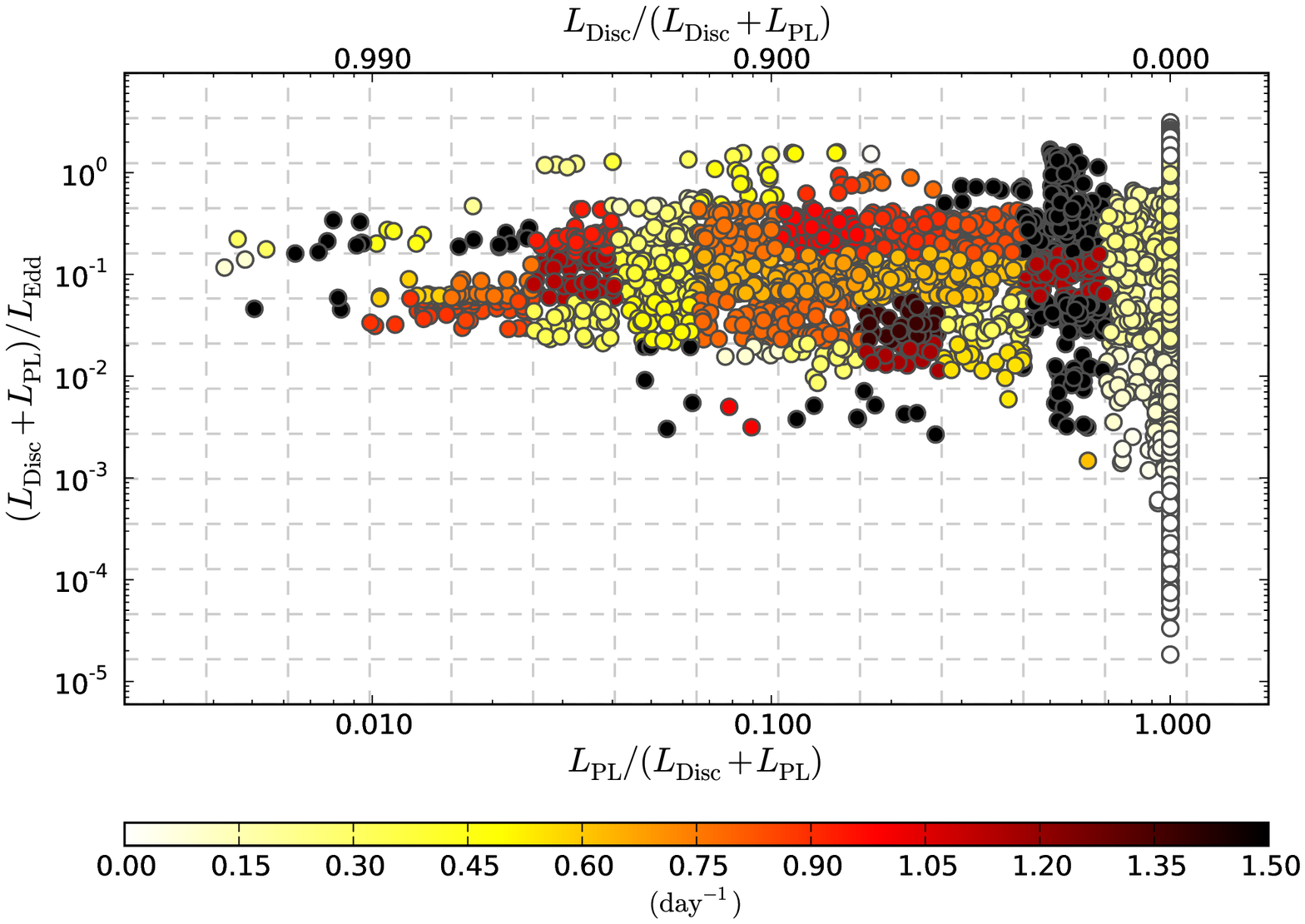}
\includegraphics[width=0.49\textwidth]{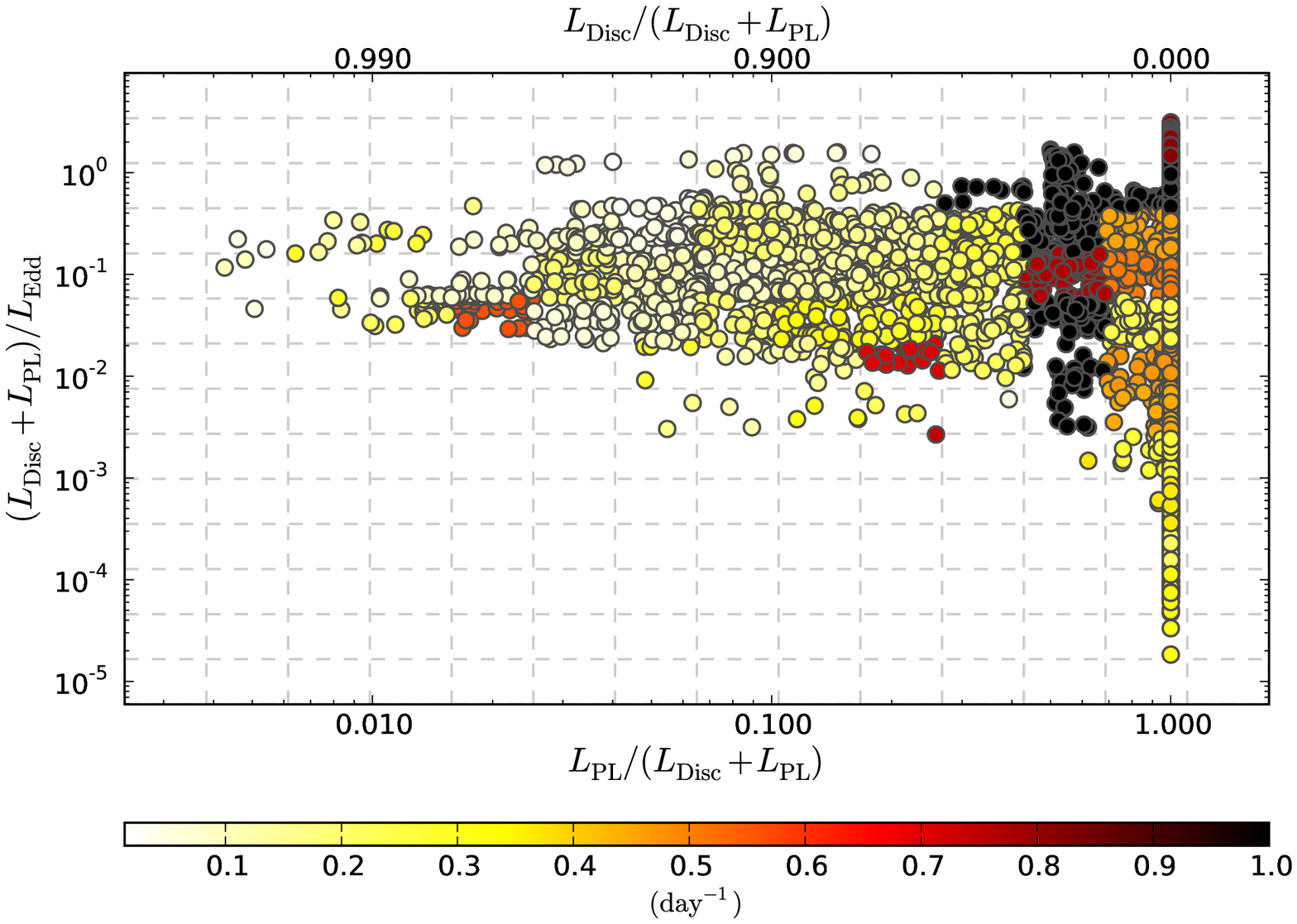}
\includegraphics[width=0.49\textwidth]{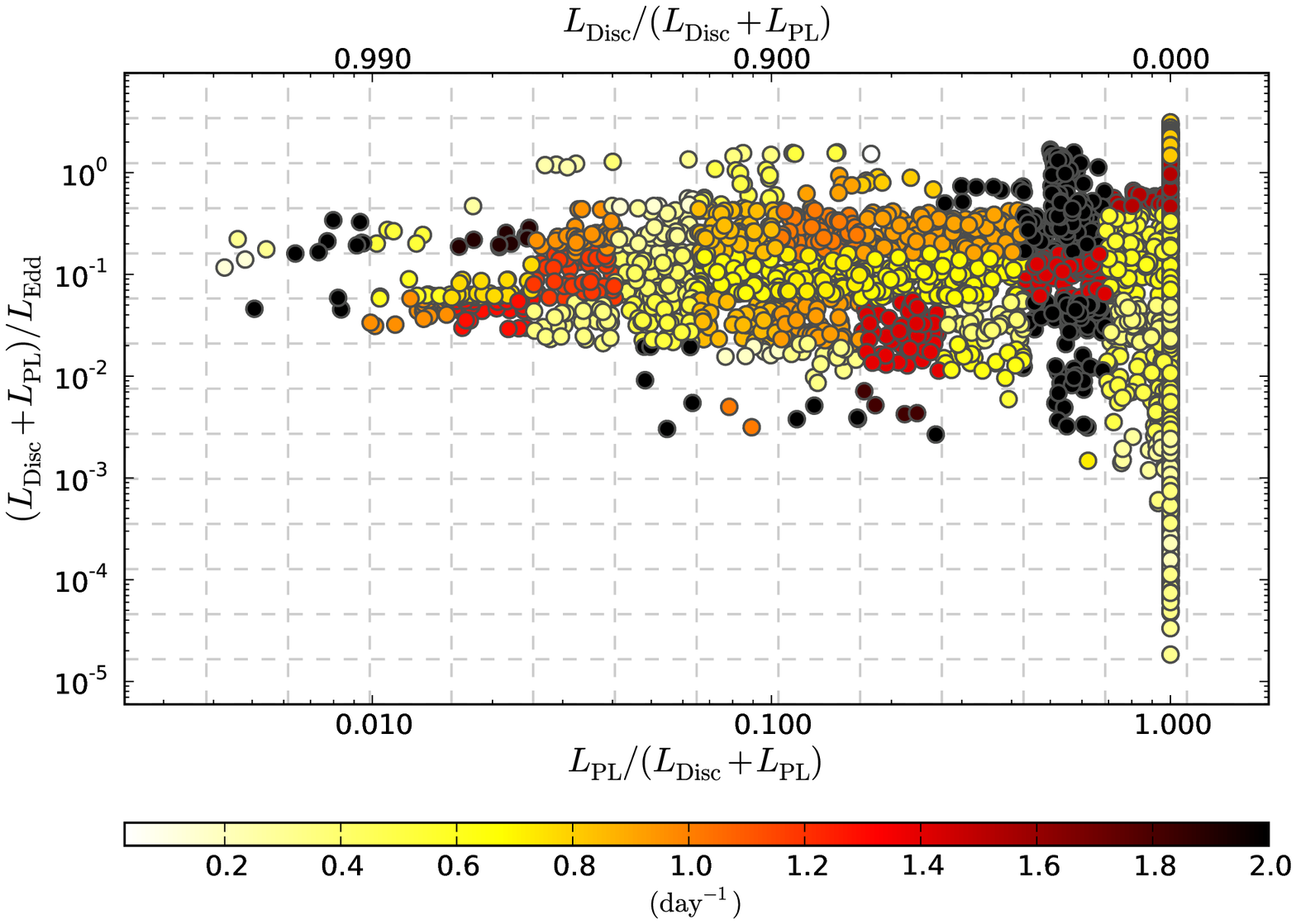}
\caption{\label{fig:rate} Rate of motion through the DFLD
  in Disc Fraction ({\scshape top}), Luminosity ({\scshape middle})
  and both combined ({\scshape bottom}).  The colour scales have been
  truncated for clarity.}
\end{figure}

We calculate the rates in Disc Fraction and Luminosity separately.
From these two rate calculations we calculate the overall rate of
motion through the plane using Pythagoras' theorem.  However we do not
currently calculate the direction of the overall motion.  To show the
variation of the rate of motion through the plane, we average within the
grid squares shown by the dotted lines in Fig. \ref{fig:rate}.  To
further clarify the diagrams, we also truncated the colour scales. 

It is obvious that the slowest change in disc fraction would be in
the powerlaw dominated state (Disc Fraction $<0.2$).  This slow rate
of change of disc fraction can be seen at the base of the ``stalk'' in the low
luminosity part of the diagram.  The fastest rate of change of disc
fraction might be expected on the transition across to the disc dominated
state.  This is seen on the transition to- and the transition from- the
disc dominated state at all luminosities as the dark band around a disc fraction
of $\sim 0.4$.  There is another minor rise in rate mid-way through the
disc dominated state (disc fractions $\sim 0.8$) at the lower end of
the luminosity range at these disc fractions.    This may be an
indication of an increase in the rate of motion back towards the
powerlaw dominated states.  This feature may become clearer once the
distances and masses of the BHXRBs are more certain, and the levels
of their transitions, and tracks in the DLFD are more precisely positioned.

The variation in change of rate of the luminosity is much clearer.
There is a small peak at the top of the powerlaw dominated state, but
the greatest rate of change is found in the same region as for the
rate of change of disc fraction - on the departure from and
entry to the powerlaw dominate state (disc fractions
of $\sim 0.4$).  In the most disc dominated
state there is little variation in the luminosity - and the rates of
change are uniform across the region.  There appears to be a slight
increase in the 
rate during the decline, as the source approaches the powerlaw
dominated state.

When combining the rates, the effect of both the disc
fraction and the luminosity rate distributions are seen.  The disc
fraction rates 
dominate the diagram being much stronger.  What is clear, however, is
that the rate of motion through the softest, most disc dominated
states is comparatively slow.  The final brightening of the hard
state along with the beginnings of the transition across towards the
disc dominated state, as well as the final parts of the decline back to the hard
state are the regions where the BHXRBs change their characteristics
the fastest.

The region marking the transition from the powerlaw dominated to the
disc dominated states with an increased rate of motion may result from
our model selection procedure.  The differences in $\chi^2$ between the
broken powerlaw (DF $=0$) and disc+powerlaw (DF $\ne0$) are small, but
the differences in disc fraction are large.  As the BHXRB
rejoins the hard state there is some fluctuation as to which model is
chosen.  This effect occurs on both transitions,
and at a period where a high duty cycle of observations is likely.
Therefore it is possible that the sudden change from pure powerlaw to disc +
powerlaw, possibly with some oscillation, contributes to this region
of increased rate of motion.  

Some of the grid squares include very few points.  The small numbers of
points in these squares some cases leads to extreme rates, usually
high.  We ignore 
these grid squares in our interpretation above as the few numbers of
points do not reflect the global properties.

\section{Similarity of Outbursts}\label{sec:Outburst_sim}

As noted in \citet{Dunn08} two outbursts of GX~339-4 were remarkably
similar, even though they had different tracks in the HID and were
separated by two years.  This had been pointed out for the bright part
of the outbursts by \citet{Belloni06}.  Of the binaries in this sample there are many
which do not show much of an outburst.  This can result from the fact
that no strong outburst has been observed by \rxte.  However in a
number of cases, only the tail end of the outburst has been seen, \ie
after the source has reached the soft state.  The brightening of the
binary in the hard state was missed.  The rise from the soft state is
likely to have been monitored by the {\it ASM} on \rxte.  However
pointed \pca\ observations will only have started after an interval,
by which time these sources have already reached the soft state.  This
allows us to conclude that the rises and transitions of these sources
was extremely rapid, occurring over a couple of days or so.  Even with
this missing information we are still 
able to extract parameters from the lightcurves and their shape
to be compared to those outbursts which have been fully sampled.

In the Appendix in Fig. \ref{fig:curves} we show the
outbursts disc fraction curves, as well as the X-ray colour curves, with their luminosity as colour
scale.  The basic shape is the same as in the two outbursts from GX~339-4 \citep{Dunn08}.  There is a
``hardening'' (disc becomes less dominant) in the middle of the outbursts
which is associated with an increase in flux.

In some cases the ``hardening'' occurs without an increase in flux,
and in others vice versa.    In some cases we do not observe the
beginning of the outburst, but still can see some of the spectral and
flux variation within the outburst - 4U~1543-47, 4U~1630-47, XTE~J1817-330 and
XTE~J1859+226 all clearly show some of these features.  For comparison
we also show the same 
curves using the X-ray colour in the Appendix, Fig.
\ref{fig:curves}.

Using the All Sky Monitor (ASM) on \rxte\ \citet{Gierlinski06}
investigated the behaviour of the transition from the hard to the soft
state.  They found two categories for the transitions - bright/slow
and dark/fast, with separations in the transition luminosities ($\sim
0.06$ vs $<0.01~L_{\rm Edd}$ respectively), the duration of the
transition and the shape of the transition on the HID.  In the BHXRBs'
spectral evolution presented here, as we are using the \pca, in many
cases the beginning of the outburst has been missed, and we are
therefore unable to accurately comment on the type of outburst
(e.g. 4U~1543-47).  Unsurprisingly we do agree on the classifications
from \citet{Gierlinski06}.  Only in 4U~1630-47 and GRO~J1655-40 can we
investigate additional transitions.  The outburst sequence in
4U~1630-47 is very complex, and therefore very difficult to classify,
and we do not attempt to.  GRO~J1655-40 is likely to be a bright-slow
transition as it takes 100 days to reach its most disc
dominated/softest state, makes the transition at around $\sim
0.06L_{\rm Edd}$ and has a strong very high state.

\subsection{Peak Luminosities}

On the curves in Fig \ref{fig:curves} we show the lines used to calculate the transitions
between the different states.  The vertical lines show the dates where the
transition between disc and powerlaw dominated states took place as
calculated by the changes in disc fraction.  We also show as the two
horizontal lines the powerlaw fractions used to determine the date of
the transition.  The
date given is when the source crossed the powerlaw fraction $=0.8$ line if
this was able to be interpolated reliably.  We use a powerlaw fraction
$=0.8$ when calculating the transitions to make sure that we clearly
identify the tracks away from- and return to the powerlaw dominated/hard state.  If not, then the first/last
disc dominated state is used to estimate the luminosity on the entry
to and exit from the disc dominated state (see Section
\ref{sec:Trans_L} for more complete details of the procedure).

The difficulty of using \rxte\ data in this analysis is apparent in
some of the Disc Fraction curves (see Appendix Fig. \ref{fig:curves}).
There are some points which have a 
powerlaw fraction of 1.0 even though most of those
which surround them are in the disc dominated part of the diagram
(e.g. GX~339-4, Outburst 4, and 4U~1630-47).  In these cases the disc
is not strong 
enough to be the best fit model given the statistical cuts we have
made.  These observations are easily identified in the Disc Fraction curves
in the Appendix.  Using an X-ray observatory which has a better low energy
response than \rxte\ would allow the more gradual change from pure
powerlaw to powerlaw + disc to be observed in all objects.  In some
objects this discrete nature is not observed, presumably because
either the disc or powerlaw parameters are such that the disc is
easily identified at low disc luminosities.

\begin{figure}
\centering
\includegraphics[width=1.0\columnwidth]{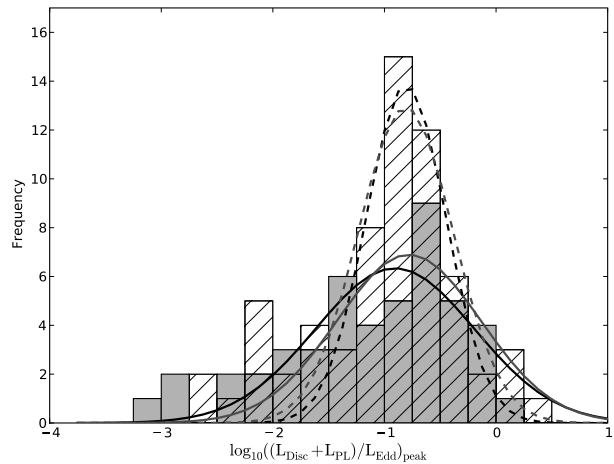}
\caption{\label{fig:PeakL} The distribution of the peak luminosities
  within each outburst for the objects in our study (solid grey) and
  that from \citet{Chen97} (``/'' hatched).  Least squares fitting in
  grey, the downhill simplex
fitting in black lines.  The solid curves are the fits to our
distribution.  We fit the data from \citet{Chen97} using both the
fitting methods, and show the results with the dashed lines.}
\end{figure}

We compare the distribution of peak Luminosities achieved in the
outburst with those analysed in \citet{Chen97}.  Their best fit to the
distribution is a gaussian with a mean of $-0.7$ and a
Full Width Half Maximum (FWHM) of $0.82$
in logarithmic space\footnote{We fit gaussian profiles to the data
  from \citet{Chen97} in
  order to compare the two distributions statistically.  Our best fit
  values are $\mu\sim-0.81$ and $\sigma\sim 0.36$ in logarithmic space.}.  Our distribution, see Fig. \ref{fig:PeakL}, shows a
very similar distribution ($\mu\sim-0.91$, $\sigma\sim 0.71$, the
average of the least squares and downhill simplex algorithms).  Our
distribution has a very similar mean, but is broader.  We perform a
Kolmogorov-Smirnov (K-S) and a $t$-test on the two samples, and it is unlikely that they are drawn
from different populations and unlikely that they have different means ($P_{\rm
  K-S}=0.38$ and $P_t=0.40$).

The excess at
low Eddington fractions is likely to arise from identified
outbursts\footnote{We pre-identify times when the source was bright to
make the automatic transition detection easier.}  -
periods of emission in the \rxte\ lightcurve - where the source did not
reach a true peak in emission.  It was either monitored by chance in
quiescence for a length of time, the full outburst was not observed
by \rxte\ (as in XTE~J1118+480) or the outburst was a hard state only
outburst.  We therefore ignore this tail.  This excess was also seen 
in \citet{Chen97}, who put it down to 
arising from uncertainties in the masses, distances and spectral
shapes of their sources.

As was already noted in \citet{Chen97} amongst others, the peak of the peak luminosity
distribution does not occur at 100 per cent of the Eddington ratio,
but at around $0.2L_{\rm Edd}$.  This will be explored further with
the study of the transition luminosities in Section \ref{sec:Trans_L}.

From disc instability models, the peak outburst luminosity is expected
to correlate with the orbital size of the binary \citep{King98,
  Shahbaz98}.  As shown in Table \ref{tab:objects}, we have the orbital
parameters for about half of our sample of BHXRBs.  We extracted the
peak luminosities for outbursts which showed a transition (either away
from or return to the powerlaw dominated/hard state, see Section
\ref{sec:Trans_L}) and plotted these against the orbital period.
There was no clear correlation, neither when using the X-ray colour,
nor when using the Disc Fraction when determining the transitions.
However, there are significant limitations to our calculations of the
transitions (see Section \ref{sec:Trans_L}) and unmeasured orbital
periods, and so unfortunately we cannot conclude anything from the 
apparent lack of a correlation.

\section{Transition Luminosities}\label{sec:Trans_L}

Having produced a large number of HLDs and DFLDs we can extract the
luminosity of the BHXRB during the transition between different
states.  If the BHXRBs behave in a very similar way, then these should
occur at very similar fractions of $L_{\rm Edd}$.  Although we
calculate the transitions for each outburst where possible, to study
their distribution, we only take those where both transitions have
been observed.

In those outbursts when only the return to
the canonical hard state is observed we do not attempt to interpolate a
beginning to the outburst.  If the first observation of an outburst is sufficiently
soft (X-ray colour or Powerlaw fraction $<0.3$) then no attempt is made to
find the transition away from the powerlaw dominated state.  If the first observation is in the
powerlaw dominated state (X-ray colour or Powerlaw fraction $>0.8$) then
to calculate a transition each observation is stepped through to find
the first which has an X-ray colour/Powerlaw Fraction $<0.3$.  Then the
observations are stepped through in reverse to find the last
observation with X-ray colour/Powerlaw Fraction $>0.8$.  The observations
which bracket an X-ray colour/Powerlaw Fraction $>0.8$ are used to
interpolate when the transition occurred and what luminosity the
source was at.  If the bracketing observations are more than 10 days
apart, then the softer/more disc dominated observations' date and luminosity are used.
This process is reversed to find the return transition.  These dates
are shown on the curves in Appendix Fig. \ref{fig:curves}.

We chose the values of $0.3$ and $0.8$ to ensure we have a complete
transition from the hard to the soft state.  In \citet{Dunn08} we used
values of the X-ray colour of $0.22$ and $0.87$ derived from the
timing properties in \citet{Belloni06} to select purely soft- or
hard-state observations.  When looking at the HIDs from the BHXRBs in
this work, using these
values would have been too restrictive, not detecting transitions in
sources where the HID looks similar to the canonical HID, but is for
some reason truncated or offset (e.g. 4U~1630-47).  We therefore chose
a more ``inclusive'' set of values to determine transitions.  

The exact value of the
Disc/Powerlaw Fraction at the point of transition may vary from source
to source.  For a few of the sources, we investigate the results of
the previous studies on 
the timing properties of the sources as they make the transition to
obtain the true dates.  We show in Table \ref{tab:transitions} a
selection of the transition dates for the source which have had them
determined from timing properties.

We do not look for transitions including the intermediate states, as
these are difficult to determine from spectral information alone.  We
also excluded those sources where the data although being very sparse,
would give a transition date and flux, and those where
no clear outbursts are apparent in the light curves, but the spectral
variations would have resulted in transitions being detected.  This
was accomplished using the pre-identification of outburst periods in
the lightcurves.  

\begin{table*}
\centering
\caption{\label{tab:transitions} {\sc Transitions from Timing Analyses}}
\begin{tabular}{llll@{$\to$}rlll}
\hline
\hline
Object &\multicolumn{2}{c}{Timing Date Range} &
\multicolumn{2}{c}{State Transition}& X-ray Colour Date & Disc Fraction Date &Ref.\\
&MJD&MJD&\multicolumn{2}{c}{}&MJD&MJD\\
\hline
4U 1543-47    	& 52473.25& 52474.17&HSS &IMS &&&(1)\\
		& 52479.74& 52480.66&IMS &LHS &52483 & 52478&(1) \\
4U 1630-47	& 50855& 50856& LHS& VHS&&&(2)\\
		& 50874& 50884& VHS& HSS&&&(2)\\
		& 50888& 50890& HSS& LHS&&&(2)\\
		& 50951& & HSS& LHS&&&(3)\\
		& 51395& & HSS& LHS&51351&&(4)\\
		& 52057.96& 52059.38& HSS& LHS&& 52049&(4)\\
GRO J1655-40	& 50663.7& 50674.4& HSS& LHS& 50676 & 50672 & (4)\\
		& 53440& & LHS& HSS&53440&53440&(5)\\
		& 53442& & LHS& HSS&&&(6)\\
		& 53501& & HSS& VHS&&&(5)\\
		& 53506& & HSS& IMS&&&(6)\\
		& 53520& & VHS& HSS&&&(5)\\
		& 53625& & IMS& LHS & 53633 & 53628 & (6)\\
GX 339-4	& 51200.9& 51220.8& HSS& LHS& 51201& 51201&(4,7)\\
		& 52394.4& 52398.7& LHS& HIMS& 52401& 52411&(8)\\
		& 52410.5& 52411.6& HIMS& SIMS&&&(8)\\
		& 52558.6& 52560.4& SIMS& HSS&&&(8)\\
		& 52693.7& 52694.9& HSS& HIMS& 53734 & 53719&(8)\\
		& 52740.0& 52741.7& HIMS& LHS&&&(8)\\
		& 53233& & HIMS& SIMS& 53228 & 53234& (9)\\
H 1743-322	& 52929.98& 52930.90& HSS& IMS&&& (10)\\
		& 52937.51& 52938.00& IMS& LHS&52941 & 52938& (10)\\
XTE J1550-564	& 51305.12& 51308.32& HSS& LHS&& 51290&(4,11)\\
		& 51661& & LHS& IMS& 51659&51673&(12)\\
		& 51672.4& 51673.0& IMS& HSS&&&(4)$^a$\\
		& 51673.4& 51674.7& HSS& IMS&&&(4)\\
		& 51675.5& 51676.4& IMS& LHS&&&(4)\\
		& 51677& & IMS& LHS& 51683&51664&(12)$^a$\\
XTE J1650-500	& 52231& 52232& HSS& LHS& 52238& 52234& (4)\\
XTE J1720-318	& 52652& & LHS& HSS&&&(13)\\
		& 52715.47& 52728.58& HSS& LHS& 53735& 52735& (14)\\
XTE J1748-288	& 50975& 50977& VHS& HSS&&50975&(15,16)\\
		& 51007& 51012& HSS& LHS&&51006&(15,16)\\
XTE J1755-324	& 50661& 50758& HSS& LHS&&&(4)\\
XTE J1817-330	& 53885& & HSS& IMS& 53906& 53886& (17)\\
		& 53889& & IMS& LHS&&&(17)\\
XTE J1859+226	& 51524& & HSS& IMS& 51613& 51597& (4)\\
\hline
\hline
\end{tabular}
\begin{quote}
LHS - Low Hard State, HSS - High Soft State, SIMS - Soft Intermediate
state, HIMS - Hard Intermediate Sate, IMS - Intermediate State. Where
possible we give the two dates which bracket the change in state 
as determined in the other studies indicated.  Where only a single
date was given or able to be determined, we give that date.  The
nomenclature for the states is that of the study in which transition
dates are presented, and may not reflect the current view.  $^a$~We note that the
two different transition determinations for XTE~J1550-564 do not fully
coincide.  We also show the transition dates obtained in this work, as
determined from the X-ray colour and disc fraction changes.  Those
transitions where no date was obtained from this work are still shown
for comparision to Appendix Fig. \ref{fig:curves}. \\
(1) \citet{Kalemci05}
, (2) \citet{Trudolyubov01}
, (3) \citet{Tomsick00}
, (4) \citet{Kalemci02}
, (5) \citet{Brocksopp06}
, (6) \citet{Debnath09}
, (7) \citet{Nowak02}
, (8) \citet{Belloni05}
, (9) \citet{Belloni06}
, (10) \citet{Kalemci06}
, (11) \citet{Homan01}
, (12) \citet{Miller01}
, (13) \citet{Remillard03}
, (14) \citet{Brocksopp05}
, (15) \citet{Revnivtsev00a}
, (16) \citet{Revnivtsev00b}
, (17) \citet{Gierlinski08}.
\end{quote}
\end{table*}

The locations of the dates where transitions occurred are show in the
figures in the Appendix (Fig. \ref{fig:curves}).  The cross-matching
between the \rxte\ observations shown here, and
those determined from timing information was done to a tolerance of
$\pm1$ day.  In most cases the timing-determined transitions and those
from X-ray colour or Disc Fraction values are not wildly
dissimilar (see Table \ref{tab:transitions}).  Although it would be
better to use the transition values 
as determined from timing information, this is not possible for all
sources and all outbursts.  We therefore use the method outlined above.

In some sources the accurate determination of the transition date from
the observations was not possible from gaps or noise in the
light, X-ray colour or Disc Fraction curves (see e.g. 4U~1630-47
Outburst 3 and GX~339-4 Outburst 1).  This leads to
uncertainties when calculating the 
distribution of the transition luminosities.

\begin{figure}
\centering
\includegraphics[width=1.0\columnwidth]{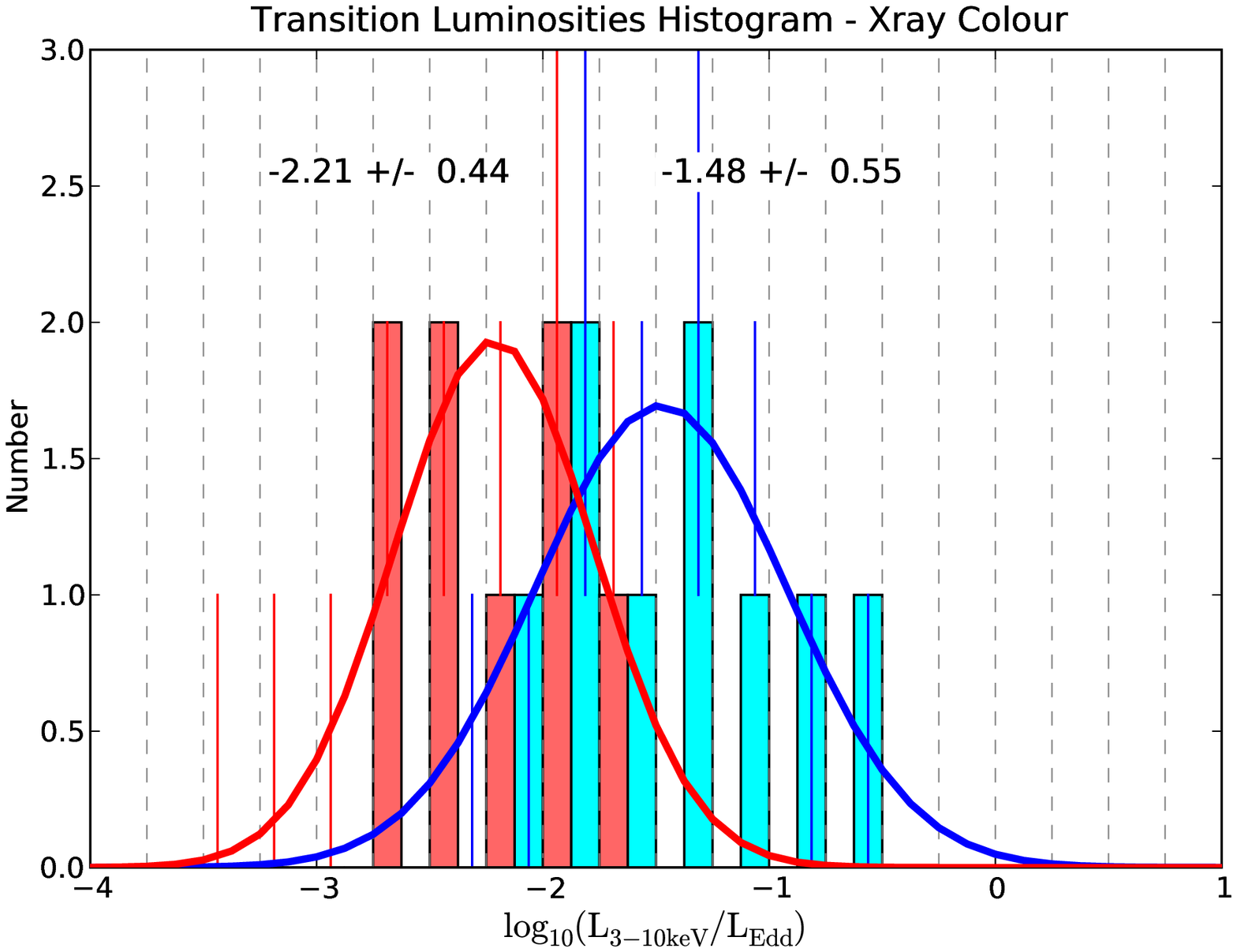}
\includegraphics[width=1.0\columnwidth]{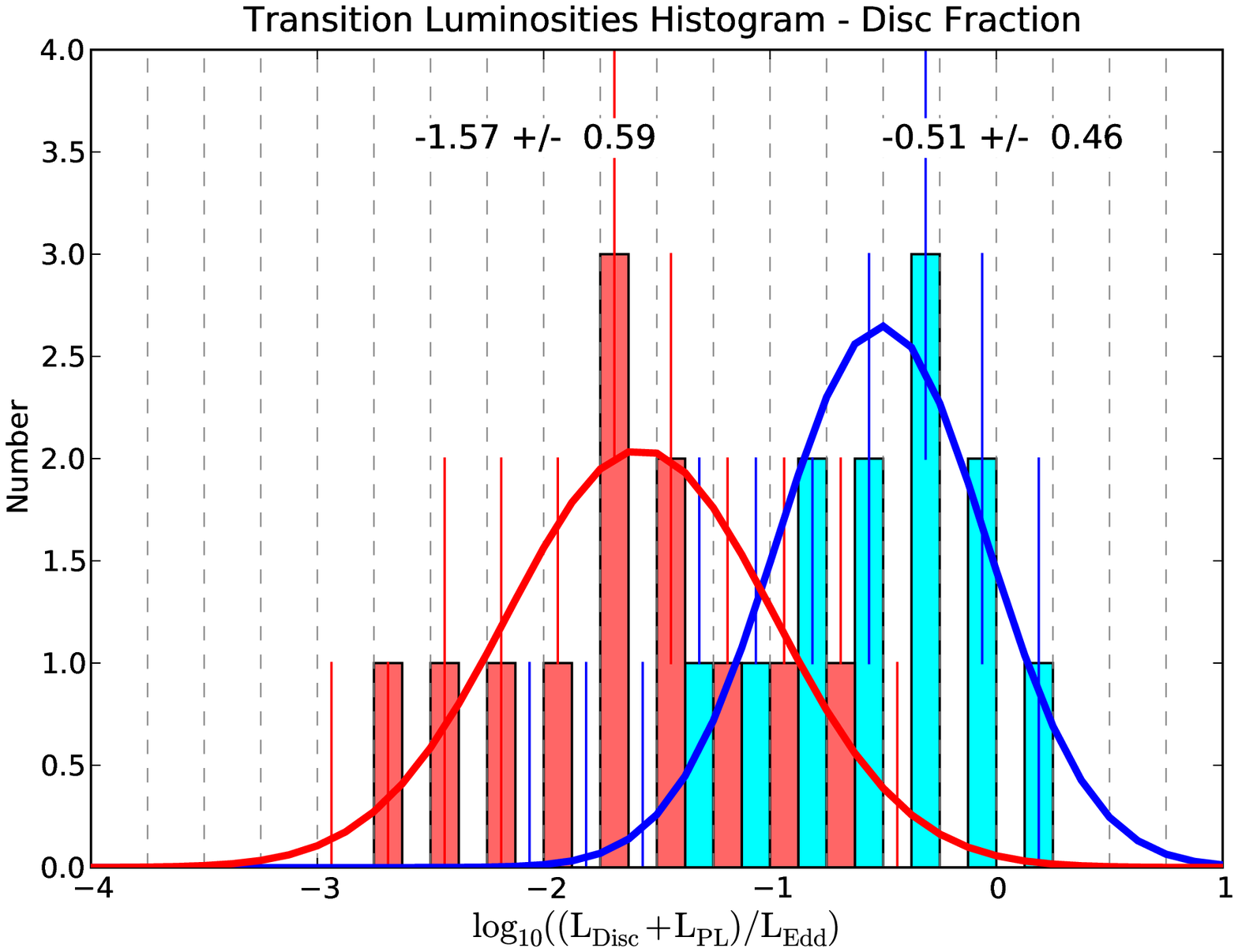}
\caption{\label{fig:transitions} Histograms showing the transition
  luminosities for the outward (blue) and return (red) transitions in
  the HLD and DFLDs for the outburst where this can be clearly done.
  The lines on the bars show the uncertainties in the values of the
  bars as determined by our Monte Carlo routine.  The best fit
  gaussian distributions are shown on the plot.  {\scshape top}:
  transitions from the X-ray colour, {\scshape bottom}: from the disc
  fraction.  Note that the luminosities on the $x$-axis are calculated
in different ways.}
\end{figure}

As not all the binaries have well determined masses and distances we
estimate the effect of these uncertainties on the transition
luminosities in the following way.  The uncertainties in the masses and
distances are propagated through to the $L/L_{\rm Edd}$ values.  We
then do a simple Monte-Carlo simulation on the binning of the
transition values.  Values for $L/L_{\rm Edd}$ are randomly selected
from a gaussian distribution with the $1~\sigma$ values derived from
the error propagation values.  These values are then binned into a
histogram.  We repeat this 1000 times to obtain uncertainties on the
values of the bins, which are shown as the lines on the bars in Fig. \ref{fig:transitions}. 

As a result of the transient nature of the source, many are only
observed by \pca\ after they have already transited into the soft state because
of an extremely rapid rise through the hard state.
Therefore the hard-to-soft transition has been missed.  In order to
fairly compare the distributions of the luminosities where the
transitions occur we show the distributions of only those transitions
where both the egress from- and ingress back into the hard state are determined.

In Fig. \ref{fig:transitions} we the distributions using both the powerlaw fraction
and the X-ray colour as a marker of the transitions.  We also show the
best fit gaussian distributions on top of the histograms.  The X-ray colour
transitions occur at lower values of $L_{\rm Edd}$ than those for the
disc fraction as the former are calculated from the $3-10\kev$
luminosity rather than the total X-ray luminosity.  The two distributions
of the transition luminosities overlap though the peaks are distinct
in both the X-ray colour and disc fraction calculations.

The best fit values to the distributions of the transitions are shown
on Fig. \ref{fig:transitions}.  Using the disc fraction to determine
the transitions shows that most of the transitions from the powerlaw
dominated to the disc dominated state occur between 3 and 100 per cent
$L_{\rm Edd}$.  The luminosities of the transitions from the disc
dominated to the powerlaw dominated states occur between 0.5 and 10
per cent $L_{\rm Edd}$.   

\citet{Kalemci02} compared their powerlaw ratio (PLR, the ratio of the powerlaw to total
flux in $3-25\kev$ band) to changes in the timing behaviour for a
number of black hole binaries.  The changes in short timescale timing
behaviour e.g. the level of the total rms noise, and the frequency of
any quasi-periodical oscillations (QPOs) are frequently used to
determine an accurate transition date between states \citep{vanderKlis06}.  The
value for their PLR is around 0.6 at the time of the transition as
determined from timing analysis.  We construct this PLR for a number
of our data points.  Comparing the two ratios, a value for the PLR of
around 0.6 corresponds to a range in powerlaw fraction of 0.4 to 0.6.

The Powerlaw Fraction presented here has a wider energy band, using values
of $0.8$ to determine the luminosity at the transition is likely to
not be too distant from the true transition point as determined from
changes in the timing properties (see above).  Depending on the
observation rate, there may be a large uncertainty as to whether the
luminosity taken as the transition luminosity is close to the true
value.  When the transition from the hard state into the soft state is
not clearly observed, then it is not clear exactly which observations'
luminosity is closest to the one at transition.  Given the uncertainties in the
masses and distances, the fact that the transitions are not at a
constant luminosity in some sources or the poor sampling, both of
which result in an inaccurate transition luminosity only contribute to
the uncertainties, rather than being the overriding ones.  Further
analysis is required to determine the disc fraction at state
transition, and from these extract the luminosities.

In a study of 10 Neutron and Black Hole X-ray binaries,
\citet{Maccarone03} found that most of the Soft-to-Hard transitions
occurred between 1 and 4 per cent of 
$L_{\rm Edd}$, with a few source's transitions occurring at lower
luminosities.  They also found that the distribution of the
Soft-to-Hard transitions was narrower than that of the Hard-to-Soft.
In Fig. \ref{fig:transitions}, the Soft-to-Hard transition
luminosities calculated using the X-ray colour and also when
calculated using the disc fraction have the same width as the
Hard-to-Soft transition luminosities. In all cases, the 
distributions have been equally affected by broadening caused by the
uncertainties in the distances and masses of the BHXRBs, so there
should still be a difference in the widths of the distributions.  

\begin{figure}
\centering
\includegraphics[width=1.0\columnwidth]{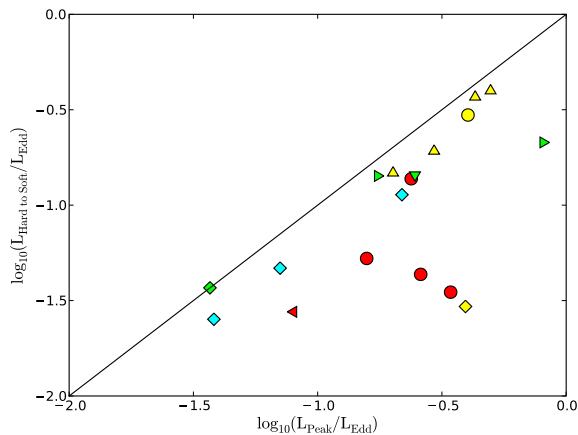}
\caption{\label{fig:peakvstransitions} The luminosity at the Hard to
  Soft transition as a function of the peak luminosity of the
  outburst.  The solid line shows the one-to-one relation.  }
\end{figure}

From Figs. \ref{fig:PeakL} and \ref{fig:transitions} the peak
luminosities and the Hard-to-Soft transition luminosities appear to
occur at the same Eddington fraction.  We show in
Fig. \ref{fig:peakvstransitions} the relation between these two
luminosities.  Only 15 outbursts have a Hard-to-Soft (powerlaw
dominated to disc dominated) transitions determined from the method
outlined above, and most do seem to track (within a factor of three)
the one-to-one relation.  Some from 4U~1630+47 are low, but this BHXRB
has a very complex outburst sequence and some of these may not be true
outbursts; and also one each from
GRO~J1655-40 and XTE~J1550-564, both of which have strong Very High
States.  We note that the luminosities that we extract for the
transition luminosity may not be the same as that determined using the
timing properties of the BHXRB, but they are reasonably close.  There
are only a few BHXRBs (8) and they cover only just over an order 
of magnitude in luminosity but the similarity of the peak and
transition luminosities is intriguing.  When using the X-ray colour to
determine the transitions, the peak luminosity is more likely to occur
during the transition itself.

\section{Radio}\label{sec:radio}

Some of the BHXRBs studied here have been the subject of intensive radio
observations during one or more of their outbursts.  This data comes
from a variety of instruments and at a number of frequencies.  A
detailed and comprehensive study on the radio emission from the BHXRBs is presented in \citet{Fender09}.

We select coincident radio observations by allowing a difference of up
to two days between the X-ray and the radio observation.  We also
allow for upper limits to be linked to X-ray observations where they
are reported.  Both detections and upper limits are listed in Table
\ref{tab:radio}.  Although there are most observations at $15\ghz$
(mainly from XTE~J1859+226 from \citealt{Brocksopp02}), the
observations at $8.4$ and $5\ghz$ are more useful as they span a wider
number of binaries.

\begin{figure*}
\centering
\includegraphics[width=1.0\columnwidth]{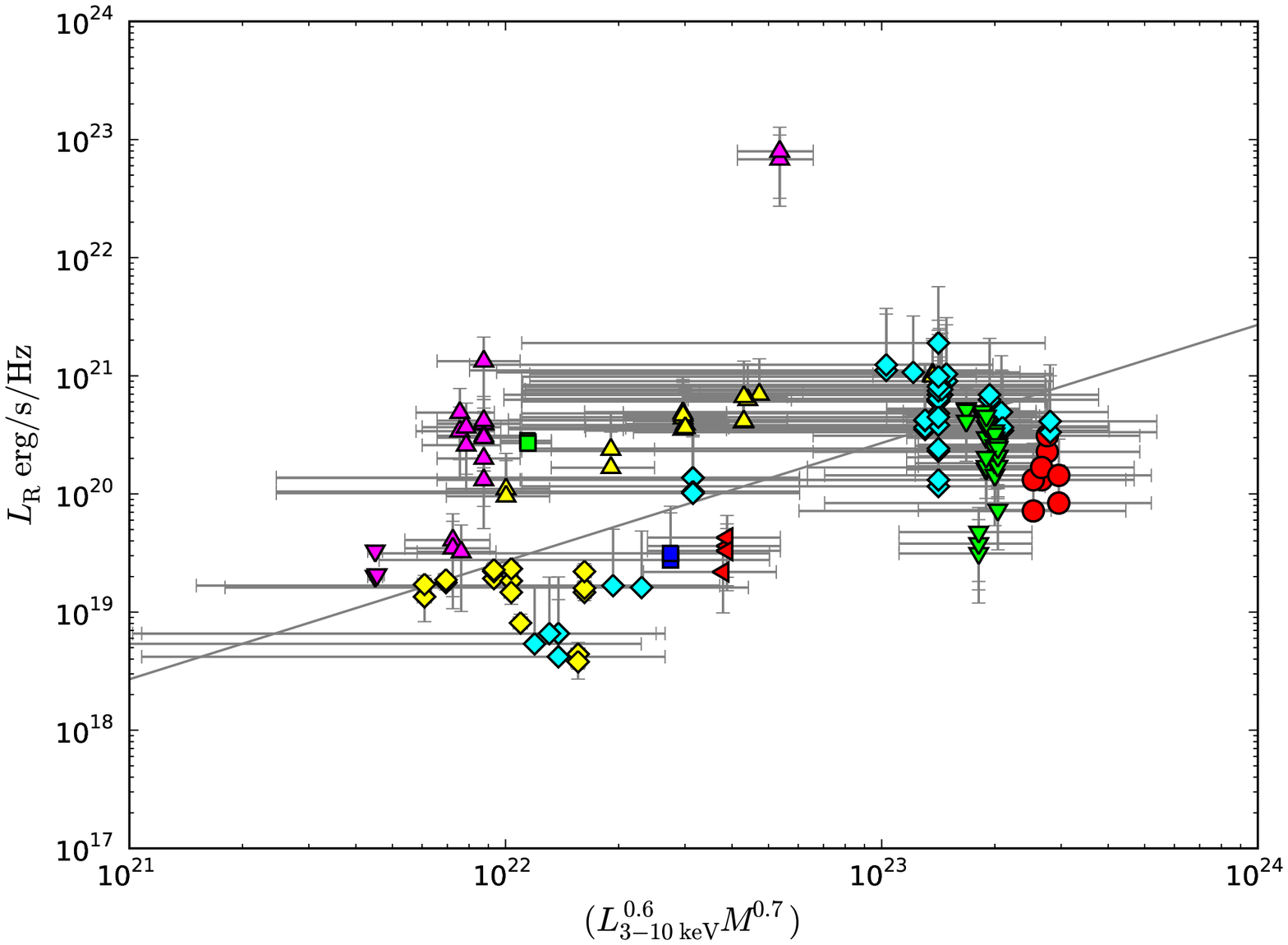}
\includegraphics[width=1.0\columnwidth]{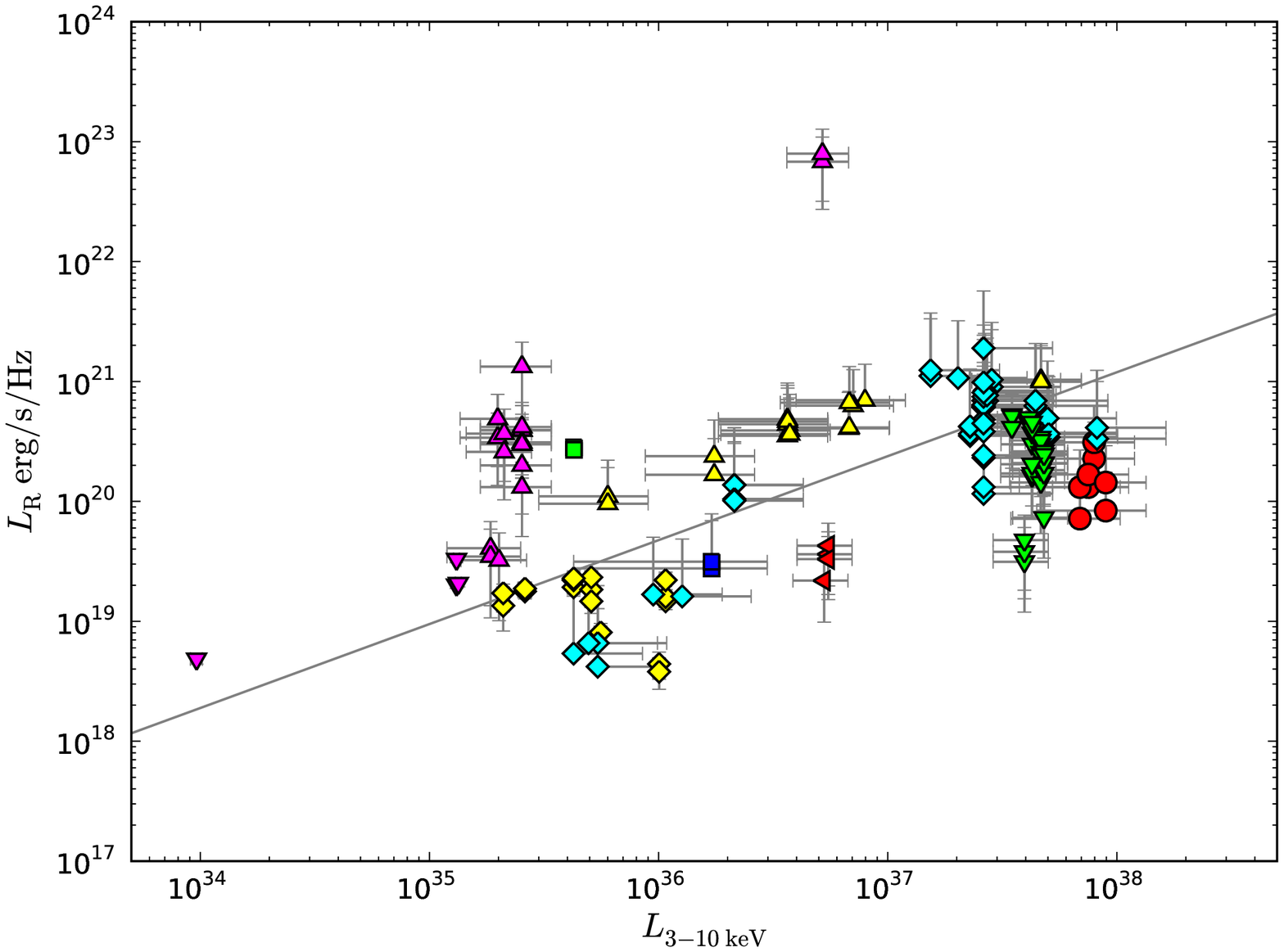}
\includegraphics[width=0.98\textwidth]{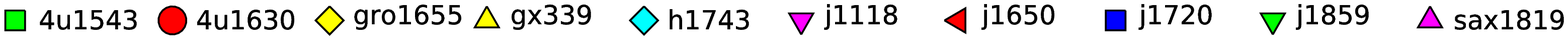}
\caption{\label{fig:LrLx}  The fundamental plane and radio-X-ray correlation for
  the radio observations and objects presented in this work at a radio frequencies
  of $8.4$, $5.0$ and $1.5\ghz$ assuming a spectral index, $\alpha=0.0$.  We restrict the
  observations to those with a Powerlaw Fraction $>0.9$ - the powerlaw
  dominated state.  The
  error bars are derived both from the uncertainties in the fluxes/flux
  densities measured, as well as in the distances and masses of the
  black holes themselves.  {\scshape left}: We show fundamental plane for
  the radio observations, including the masses of the black holes,
  where they are available.  The line is not a fit, but to guide the eye
as to where the relation would lie from \citet{Merloni03} indicating
where $L_{\rm radio} \propto L_{\rm X-ray}^{0.6}M^{0.8}$ would lie.
{\scshape right}: The radio--X-ray luminosity correlation as shown in
e.g. \citet{Corbel03} of $L_{\rm radio} \propto L_{\rm X-ray}^{0.7}$.
The line is also not a fit, but a guide to the eye as to where the
relation would lie.}
\end{figure*}
\begin{figure}
\centering
\includegraphics[width=1.0\columnwidth]{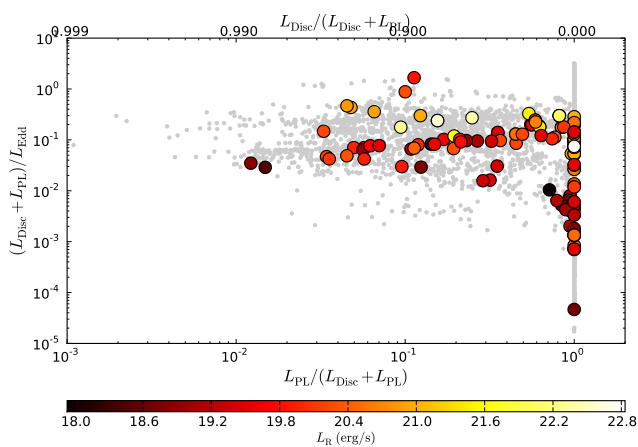}
\caption{\label{fig:DFLDradio} The DLFD with 
the observations which were coincident with $8.4\ghz$ radio observations
highlighted, showing their radio luminosities. }
\end{figure}

Studies of the radio luminosity of GX~339-4 by \citet{Corbel00,
  Corbel03} showed that in the hard state the radio and the X-ray
fluxes are related in a non-linear way.  Using a larger sample of
black hole X-ray binaries \citet{Gallo03} showed this relation to
apply to essentially all binaries.  Shortly thereafter this relation
was extended to AGN by \citet{Merloni03, Falcke04} independently.
This fundamental plane of black hole activity includes the mass of the
black hole as
\[
L_{\rm radio} \propto L_{\rm X-ray}^{0.6}M^{0.7}.
\]
\noindent This relation is applicable to BHXRBs in their hard state.  Using the
disc fraction to restrict which points to plot is in this case
problematic.  Choosing radio observations taken in the powerlaw
dominated state would include hard state observations, but also
hard-intermediate states.  However, \citet{Fender04} show that the
radio flares tend to occur on the transition between the soft and hard
states.  We therefore do select the radio observations from the
powerlaw dominated state only.

As our sample covers a relatively small range in $L_{\rm radio}$,
$L_{\rm X-ray}$ and $M$ rather than try and obtain the best fit for
the fundamental plane, we plot a projection of the best fit plane
parameters as obtained from larger samples across a wide range of
black hole masses in Fig. \ref{fig:LrLx}.  We have plotted points from
a number of radio frequencies on the same diagram in the assumption
that the spectral index\footnote{We define the radio spectral index as
  $S_{\nu}\propto \nu^{\alpha}$.}, $\alpha=0.0$.  This was done to increase the number of points on
the diagram, and the assumption of $\alpha=0.0$ is reasonable as the
jet should be self-absorbed in the powerlaw state.  There is a large
scatter around the expected relation, however with the uncertainties
in the masses and distances of the black holes this is expected.  The high radio
luminosity points from SAX J1819.3-2525 is that from the exceptional 
radio flare reported in \citet{ATel172}, the like of which has not
been observed since from any other system.  The radio observations of
GX~339-4 clearly follow the fundamental plane relation.  For most of
the other objects the variation in X-ray and/or radio luminosity is
too limited to determine any result.  The uncertainties in the masses
and distances effect the clarity of any underlying relation across
this very small sample.

The development of the fundamental plane built on earlier
investigations into a correlation between the radio and X-ray
luminosities of the BHXRBs in the hard state.  
In Fig. \ref{fig:LrLx} we also show the correlation between $L_{\rm
  radio}$ and $L_{\rm X-ray}$ as shown for GX~339-4 by
e.g. \citet{Corbel00,Corbel03}
\[
L_{\rm radio} \propto L_{\rm X-ray}^{0.7}.
\]
As the masses for many objects are
not well determined, and are very similar, this correlation is as
clear as the fundamental plane, but the error bars are much reduced.
In this case the considerable scatter still remains even though there
are few parameters which are uncertain.

On first sight, the radio/X-ray correlation of our sources is not very
tight. However, this is mainly do to our statistical approach. In
Fig. \ref{fig:LrLx} we show all observations, which do have a radio detection
and have a powerlaw fraction above 0.9. While this selects mainly
hard state observations, it cannot take peculiarities into
account. For example, SAX~J1819.3-2525 has nearly always a powerlaw
fraction of
around 1 (see Appendix, Fig. \ref{fig:remainder}), but the observed radio emission originates in
an optically thin flare \cite{ATel175}. As the radio emission
therefore does not originate in an optically thick compact jet it
should not follow the radio/X-ray correlation found for hard state
XRBs. Also GRO~J1655-40 shows some points far away from the
correlation. These points correspond to a time shortly after the state
transition, where the compact jet is just restarting. For restarting
jets see e.g., \citet{Russell07,Fender09}. This explains
the 'inverted dependency' (lower X-ray luminosities show higher radio
fluxes) of those points. In fact, detailed inspection of most outlying
points indicate that the radio emission does not originate in a
compact jet. In addition the correlation shown here combines several
radio frequencies to obtain enough data-points. The necessary
extrapolation further increases the scatter around the intrinsic
correlation. As the radio/X-ray correlation is not a central theme
here and one would need careful inspection of each observation, we do
not proceed further in this analysis. 

We also show the $8.4\ghz$ radio luminosity as a function of position on the
DFLD (Fig. \ref{fig:DFLDradio}).  The radio luminosity from an BHXRB
peaks (sometimes with an 
extremely bright radio flare) during the transition from the powerlaw to
the disc dominated state, as seen on an HID.  Showing the data on a DFLD
compresses the hard-intermediate state onto the powerlaw dominated
branch and so the radio flare is expected to be seen at the top of the
``stalk'' of the DFLD.  However, as not every object is observed in
the radio, not every radio flare is observed, and, from Section
\ref{sec:Trans_L}, the transitions from the powerlaw dominated to the
disc dominated state do not
occur at very similar total luminosities, the distribution is not very clear
in Fig. \ref{fig:LrLx}.  

\begin{table*}
\centering
\caption{\label{tab:radio} {\sc Radio Data}}
\begin{tabular}{lllllll}
\hline
\hline
Object &\multicolumn{5}{c}{Number of observations}& Reference\\
&$15\ghz$&$8.4\ghz$&$5\ghz$&$2.3\ghz$&$1.5\ghz$\\
\hline
4U 1543-47    		&&2&2&&2&(1,2)\\
4U 1630-47		&&20&22&&&(3)\\
4U 1957+115		&&&&&&-\\
GRO J1655-40		&&10&15&&3&(4,5,6,7,8,9,10,11,12,13)\\
GRS 1737-31		&&&&&&-\\
GRS 1739-278		&&&&&&-\\
GRS 1758-258		&&&&&&(14)\\
GS 1354-644		&&&&&&-\\
GS 2023+338		&&&&&&-\\
GX 339-4		&&17&15&1&3&(15,16)\\
H 1743-322		&7&59&44&&23&(17,18,19,20,21,22,23,24,25)\\
XTE J1118+480		&30&2&2&&&(26,27,28,29,30,31,32$^1$)\\
XTE J1550-564		&&&&&&-\\
XTE J1650-500		&&5&5&2&2&(33)\\
XTE J1720-318		&&8&9&&&(34)\\
XTE J1748-288		&&8&12&5&6&(35,36,37)\\
XTE J1755-324		&&&&&&(38)\\
XTE J1817-330		&&1&3&&1&(39,40)\\
XTE J1859+226		&225&27&21&5&90&(41)\\
XTE J2012+381		&2&1&2&&2&(42,43,44)\\
LMC X1			&&&&&&-\\
LMC X3			&&1&&&&(45)\\
SAX 1711.6-3808		&&&&&&-\\
SAX 1819.3-2525		&&11&7&&&(46,47,48,49,50,51,52)\\
SLX 1746-331		&&&&&&-\\
\hline
Totals&264&171&160&13&132\\
\hline
\end{tabular}
\begin{quote}
The number of radio observations which were taken within 2 days of an
X-ray observation.  In some cases radio observations have been found,
but none were taken within 2 days of an X-ray observation.\\
(1) \citet{Park04}
, (2) \citet{Kalemci05}
, (3) \citet{Hjellming99}
, (4) \citet{IAUC6410}
, (5) \citet{ATel419}
, (6) \citet{ATel425}
, (7) \citet{ATel434}
, (8) \citet{ATel437}
, (9) \citet{ATel441}
, (10) \citet{ATel443}
, (11) \citet{ATel489}
, (12) \citet{ATel609}
, (13) \citet{ATel612}
, (14) \citet{Lin00}
, (15) \citet{Corbel00}
, (16) \citet{Gallo04}
, (17) \citet{McClintock07}
, (18) \citet{ATel142}
, (19) \citet{ATel304}
, (20) \citet{ATel314}
, (21) \citet{ATel575}
, (22) \citet{ATel1349}
, (23) \citet{ATel1352}
, (24) \citet{ATel1378}
, (25) \citet{ATel1384}
, (26) \citet{ATel385}
, (27) \citet{ATel387}
, (28) \citet{ATel400}
, (29) \citet{ATel404}
, (30) \citet{ATel420}
, (31) \citet{McClintock01}
, (32) \citet{Pooley01}
, (33) \citet{Corbel04}
, (34) \citet{Brocksopp03}
, (35) \citet{Brocksopp07}
, (36) \citet{IAUC6937}
, (37) \citet{IAUC6934}
, (38) \citet{IAUC6726}
, (39) \citet{ATel717}
, (40) \citet{ATel721}
, (41) \citet{Brocksopp02}
, (42) \citet{IAUC6926}
, (43) \citet{IAUC6924}
, (44) \citet{IAUC6932}
, (45) \citet{ATel1138}
, (46) \citet{ATel61}
, (47) \citet{ATel105}
, (48) \citet{ATel172}
, (49) \citet{ATel175}
, (50) \citet{ATel296}
, (51) \citet{ATel303}
, (52) \citet{ATel315}.  
$^1$ We thank Guy Pooley for providing the radio data on XTE~J1118+480.
\end{quote}
\end{table*}

\section{Discussion}

We have completed a complete and comprehensive analysis of a large
fraction of the publicly available data on low mass black hole X-ray
binaries present in the \rxte\ archive.  Using the best fitting of one
of three simple models for the continuum emission we have been able to
investigate the global properties of the BHXRB outbursts.

\subsection{Diagnostic Diagrams}

In the study of the global properties of the outbursts of BHXRBs we
have presented two diagnostic diagrams, the Hardness-Luminosity Diagram
and the Disc Fraction Luminosity Diagram.  These two diagrams are both
useful when studying BHXRBs but are best suited to different aspects
of the investigations.

The HLD is a simple diagram to construct, requiring in the simplest
case only the ratio of
the detected counts in two bands.  This means that it can be quickly
constructed to follow an in-progress outburst, but will also work for
short observations or faint sources or states, when the number of
counts is low and reliable spectral studies are difficult.  However,
by only using the detected counts, the X-ray colours determined are
affected by any absorption along the line of sight.  This makes
comparing the behaviour between sources non-trivial, as each BHXRB has a
different $N_{\rm H}$ column. 

The DFLD returns more physical information on the parameters of the
XRB system, within the limitations of the models fitted.  With the
simple set of models fitted in our analysis presented in this paper,
the behaviour and variation of the powerlaw and the disc throughout
the outburst can easily be traced.  The result of the analysis, if not
the diagram, returns more information than the HLD.  However, the
requirement to fit spectra and determine model parameters limits the
type of observations which can be investigated, ruling out low counts
(short or faint) observations.  

The limitations of the \rxte\ satellite also restrict the usefulness
of the DFLD in the study of the 
powerlaw and intermediate states (as identified from the HLD).  The
lack of low energy sensitivity means that the detection of
non-dominant discs is not always successful.  This compresses a large
fraction of the intermediate states on top of the hard states (without
any disc).  However, as the disc parameters are extracted from the
model fits to the spectra, the investigation into the the disc
parameters and their behaviour throughout the BHXRB outbursts is more
accurate (to be presented in a forthcoming paper).

As spectral models are fit to the observations, the effects of the
absorption column can be removed, allowing for an easier comparison
between sources.  This allows the variation of the BHXRB population to
be easily studied.  However the compression of the intermediate states
onto the powerlaw dominated ``stalk'' means that the HLD is more
useful for studying an individual source or outburst, whereas the DFLD
is useful for studying the outburst evolution for a population of BHXRBs.

The \rxte\ satellite is currently the only instrument which allows the
work done in this study to be carried out.  The large archive contains
a large number of observations of a wide variety of sources.  The
coverage at high energies is vital for constraining the slope of the
powerlaw emission in the soft state, where the disc emission dominates
the signal from the \pca.  However, the limited low energy coverage is
not ideal as it does not allow accurate fitting of the disc properties
when the disc does not completely dominate the low-energy spectrum.  Also the $N_{\rm
  H}$ cannot be determined from the observations themselves and we are
reliant on values from more detailed studies.

Until an observatory with a greater sensitivity and wider (at least at
the low energy end) spectral coverage has created an archive of
observations to match those from \rxte\ presented here, the
combination of both the Disc Fraction information and the X-ray colour
may be necessary to fully determine the state of the BHXRB.
ASTROSAT\footnote{\url{http://meghnad.iucaa.ernet.in/~astrosat/}} and MAXI\footnote{\url{http://kibo.jaxa.jp/en/experiment/ef/maxi/}} fit
these specifications and hopefully over time will build up the archive
required.  This increase in the low energy sensitivity should allow
the disc to be fit in cases where it is not dominant, therefore
allowing the intermediate states to be investigated by the DFLD.

The individual DFLDs for the BHXRBs show a great deal of
similarity between each other - which is then reflected in the
combined DFLD in Section \ref{sec:pop:dfld}.  All BHXRBs where the
outbursts are well sampled and not ``messy'' show the
simultaneous change in disc fraction and luminosity on the
return to the hard powerlaw dominated state at the end of an
outburst, moving diagonally in the DFLD.  Their equivalent motion in
the HLD tends to be horizontally during the transition to the hard state.

A number of BHXRBs also show this diagonal motion in the DFLD at the
start of their outbursts - GX~339-4 is a notable exception.  These two
trends indicate that, whereas in most HLDs the transitions to the soft
state are at a constant luminosity and very quick, the DFLD compresses
these intermediate states into the powerlaw dominated state.  Only the
softest states exhibit a powerlaw fraction $<1$, and so the decay
through the disc dominated state is clearly seen in the DFLD.  The
most disc dominated states appear mid-way through the outburst, rather
than at the highest luminosity disc dominated states.  This may result
from the limitations of the {\scshape diskbb} model outlined in
Section \ref{sec:model}.

\subsection{Outbursts}

We used an initial version of the results to identify periods when the
XRBs were in outburst.  Given the range in outburst forms and shapes in
the HLD and DLFD this allowed the restriction of the ranges in which
to search for transitions and peak luminosities.  There are only a few
objects which have periods when it is not exceedingly clear where to
split up the light curve.

In Fig. \ref{fig:curves}, many of the outbursts show a hardening
after the first excursion into the disc dominated state, which is sometimes associated with an increase in
luminosity (see also \citealp{Casella04,Kubota04, Belloni05, Belloni06,Dunn08}).  If
the increase in luminosity occurs, then in some cases 
the track in the DFLD is the same - there is no hysteresis.  In
others the luminosity does not rise, or even continues to fall during
the excursion towards the hard state.  The cause of this
excursion is not clear.  

\begin{figure*}
\centering
\includegraphics[width=0.43\textwidth]{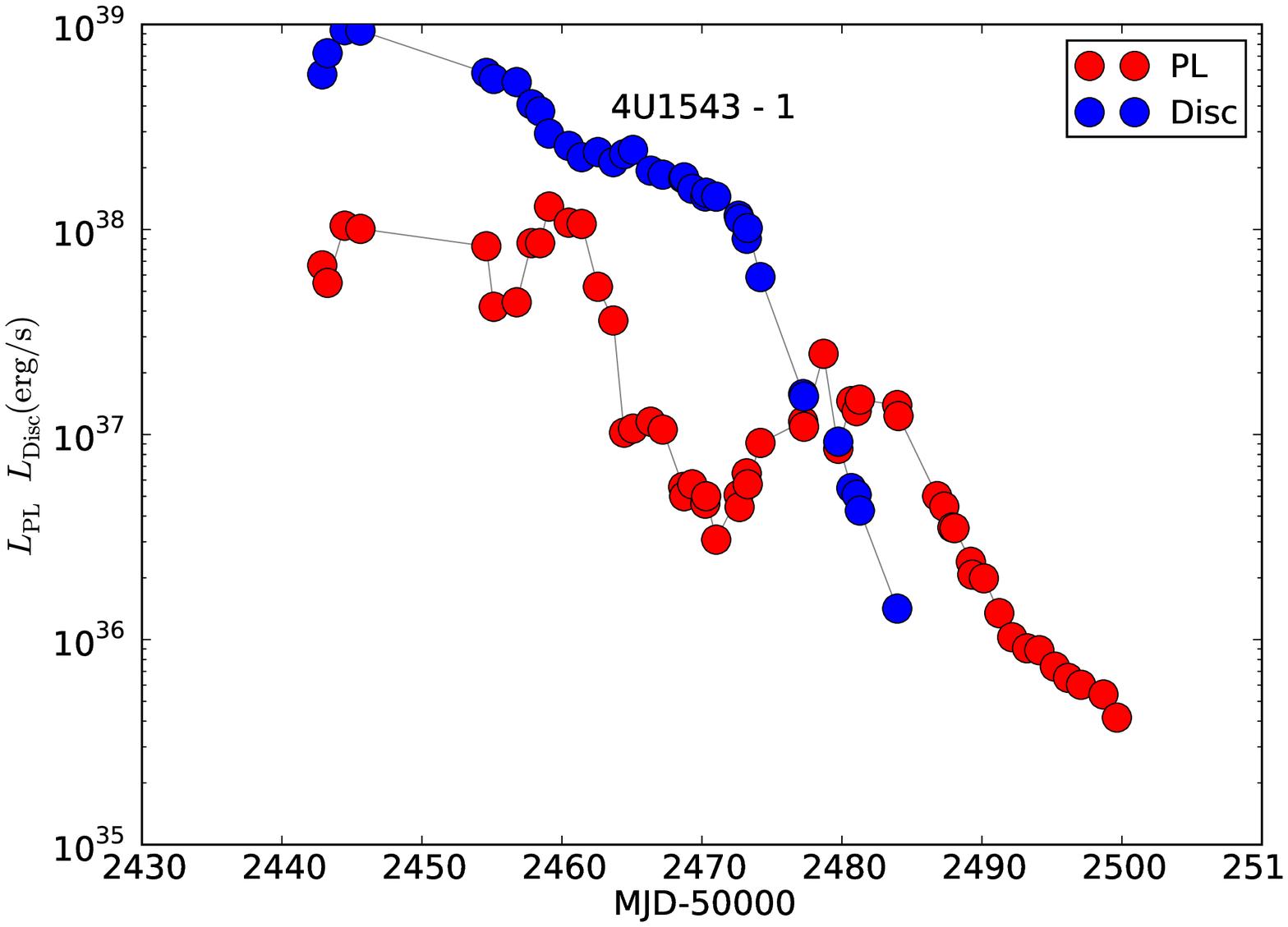}
\includegraphics[width=0.43\textwidth]{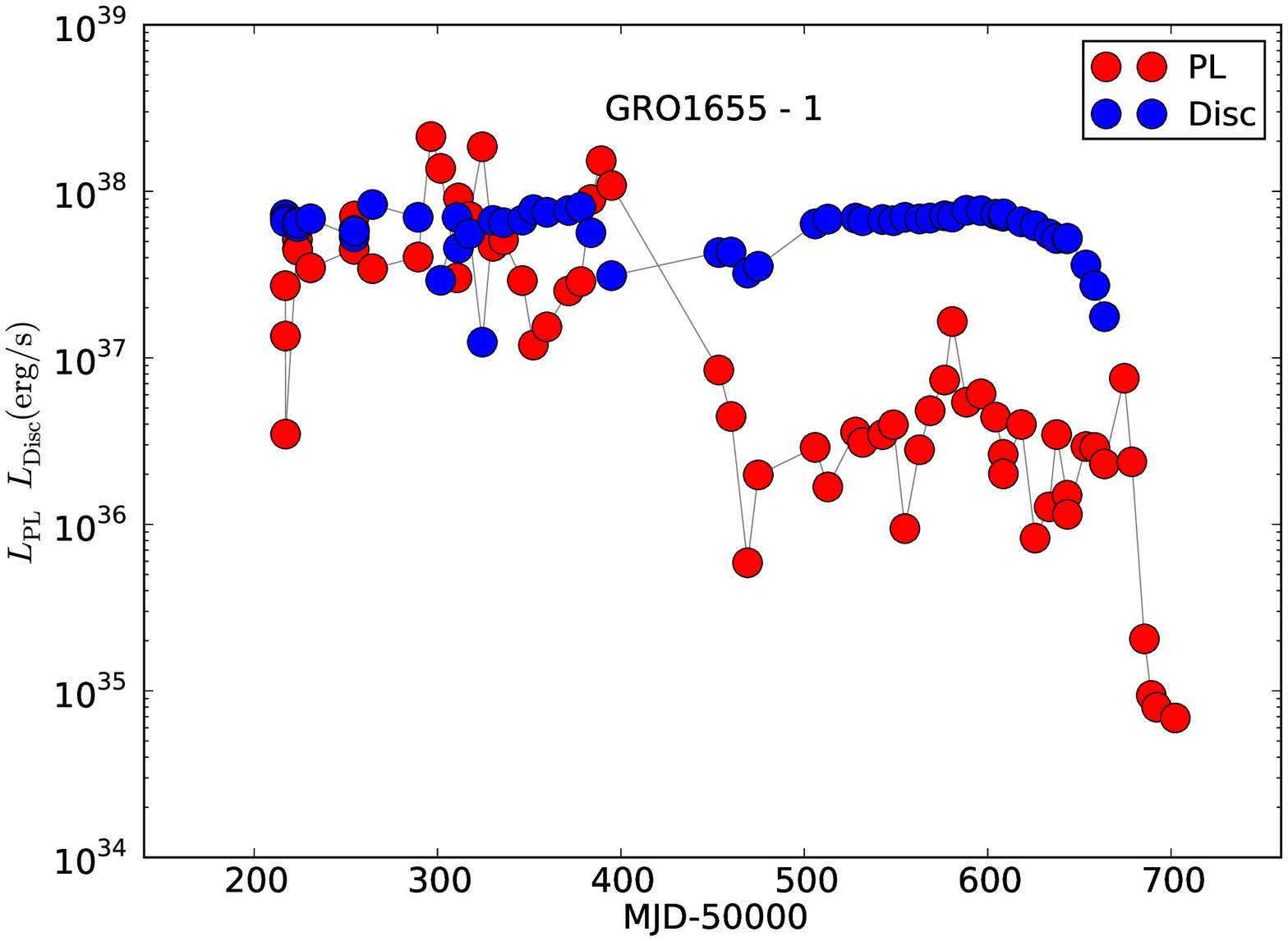}
\includegraphics[width=0.43\textwidth]{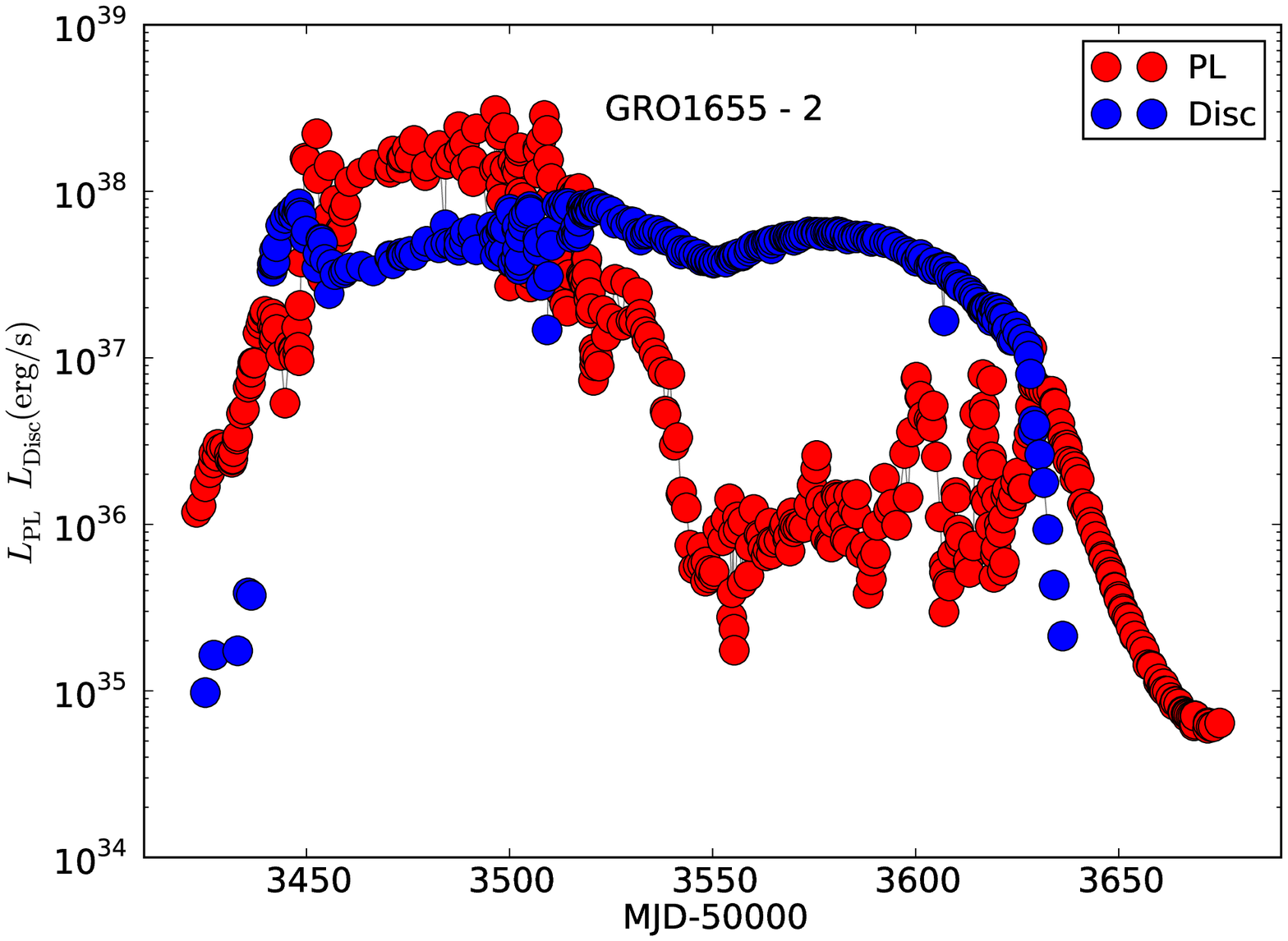}
\includegraphics[width=0.43\textwidth]{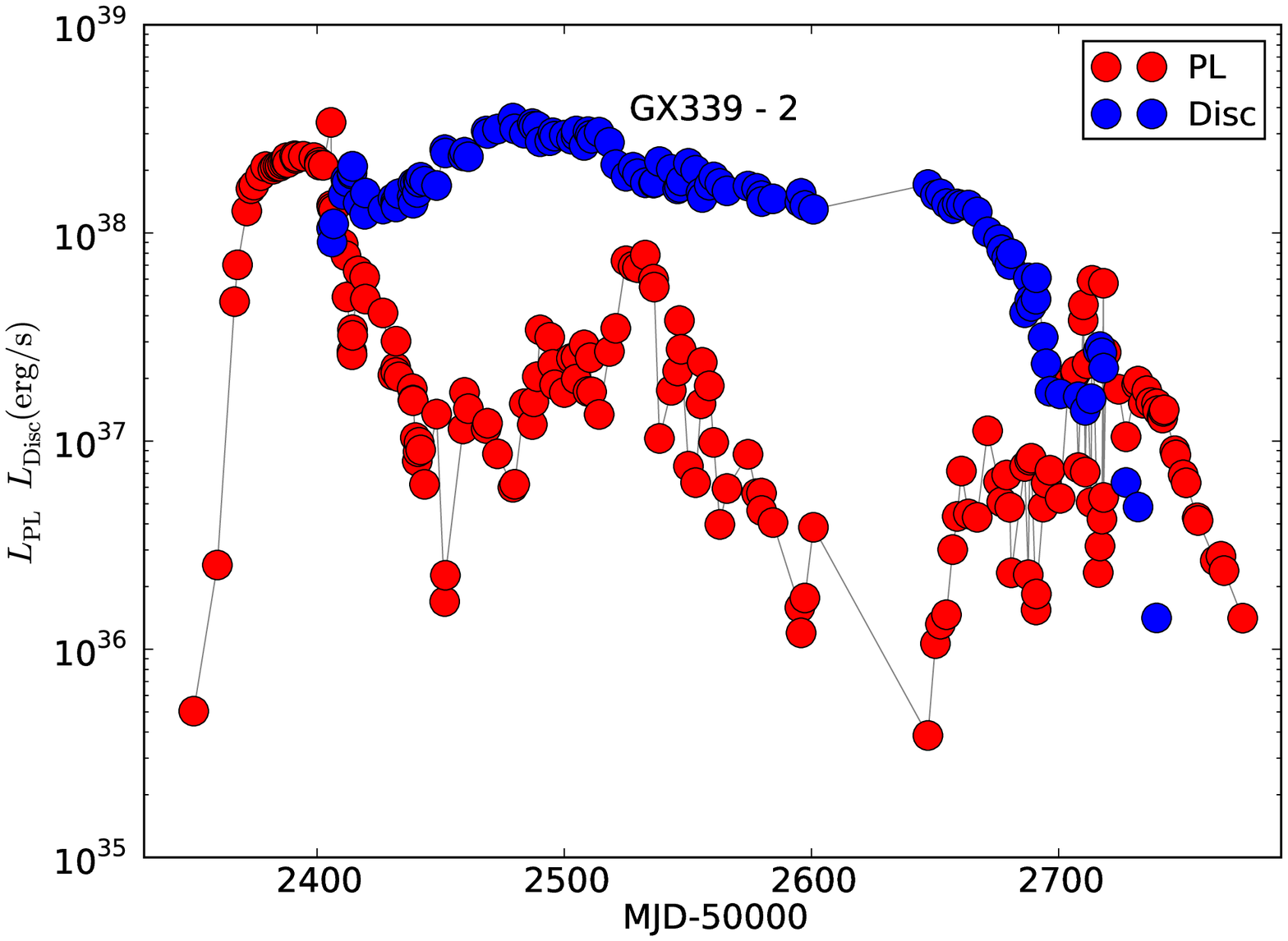}
\includegraphics[width=0.43\textwidth]{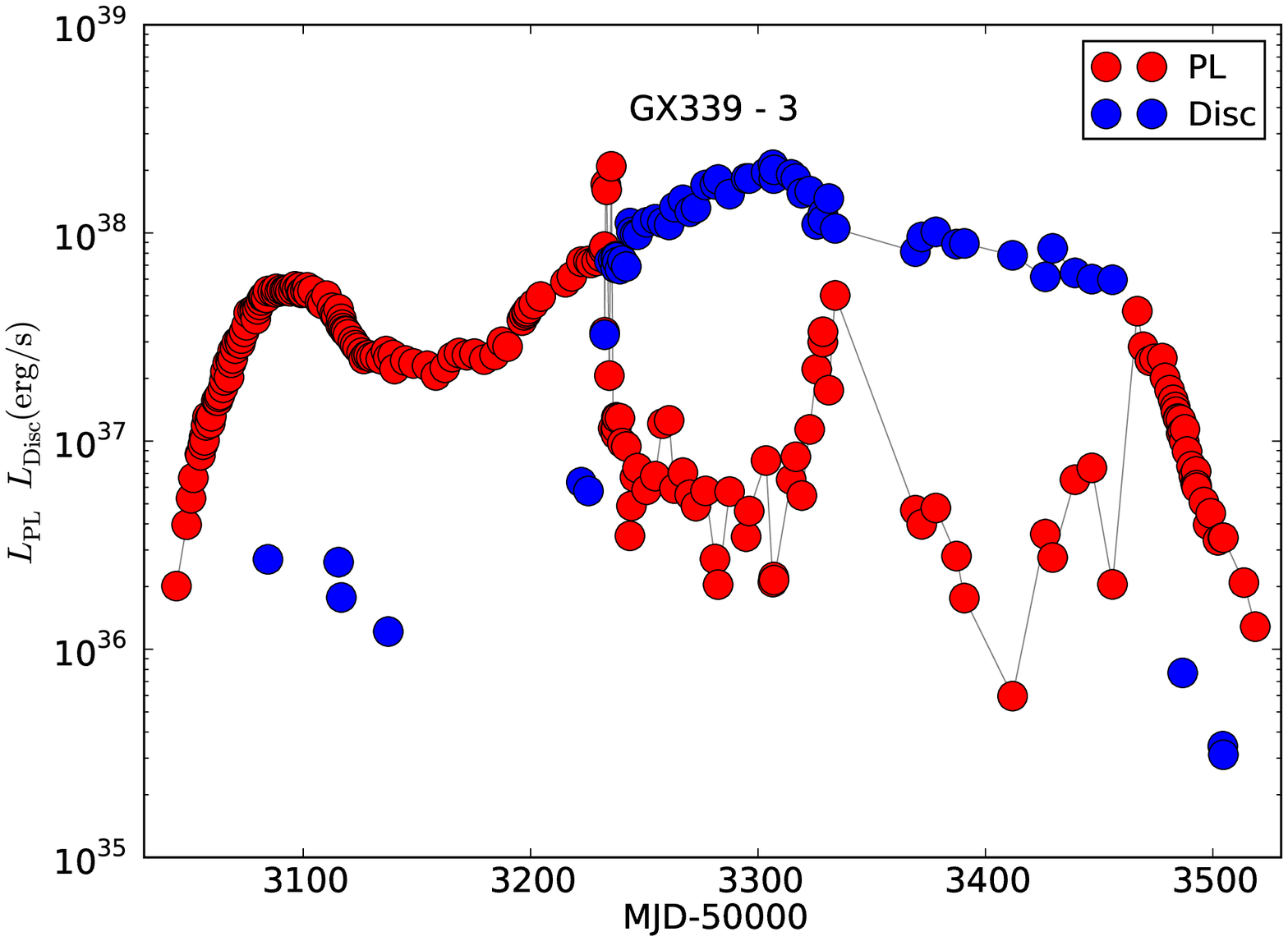}
\includegraphics[width=0.43\textwidth]{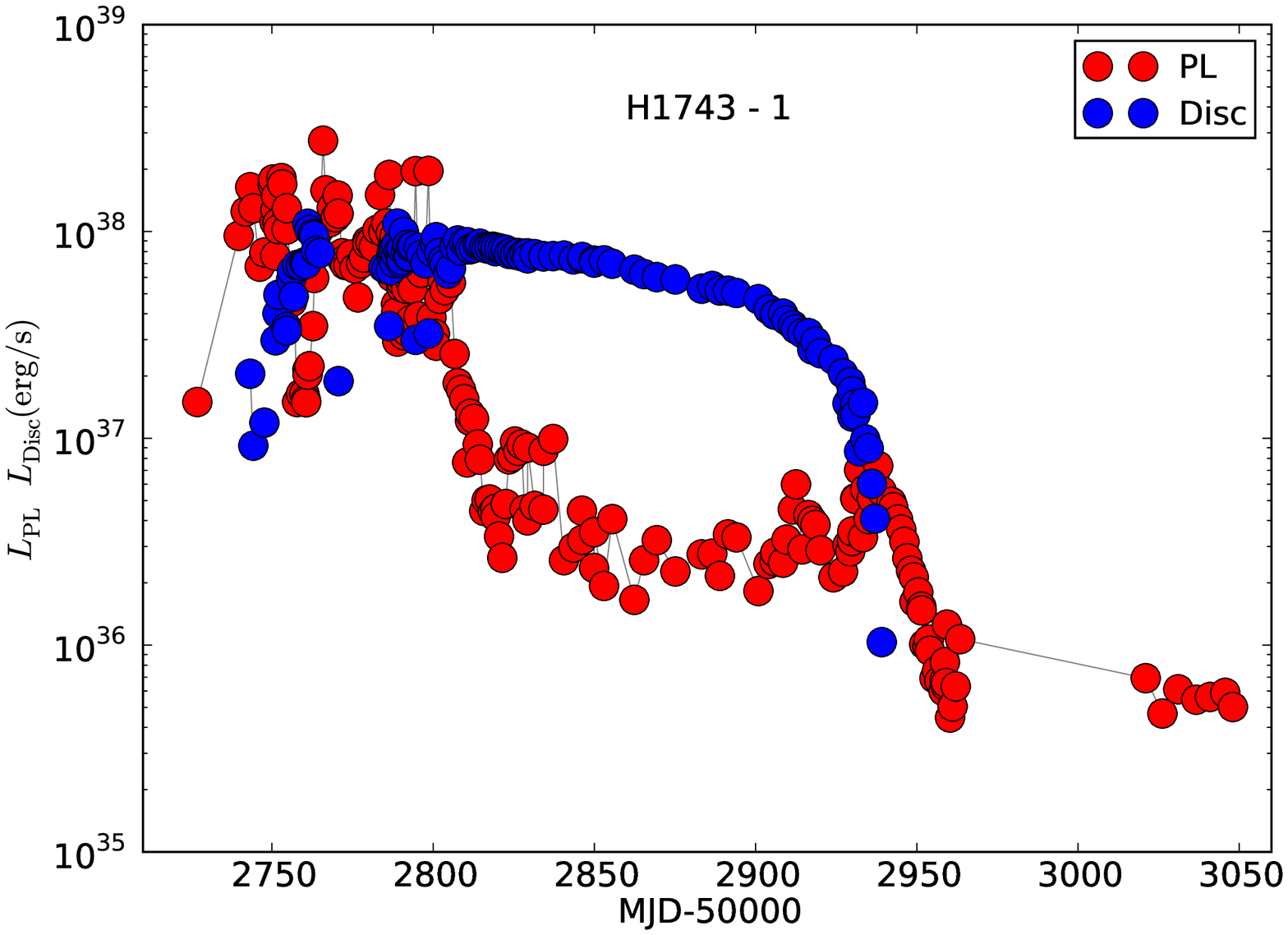}
\includegraphics[width=0.43\textwidth]{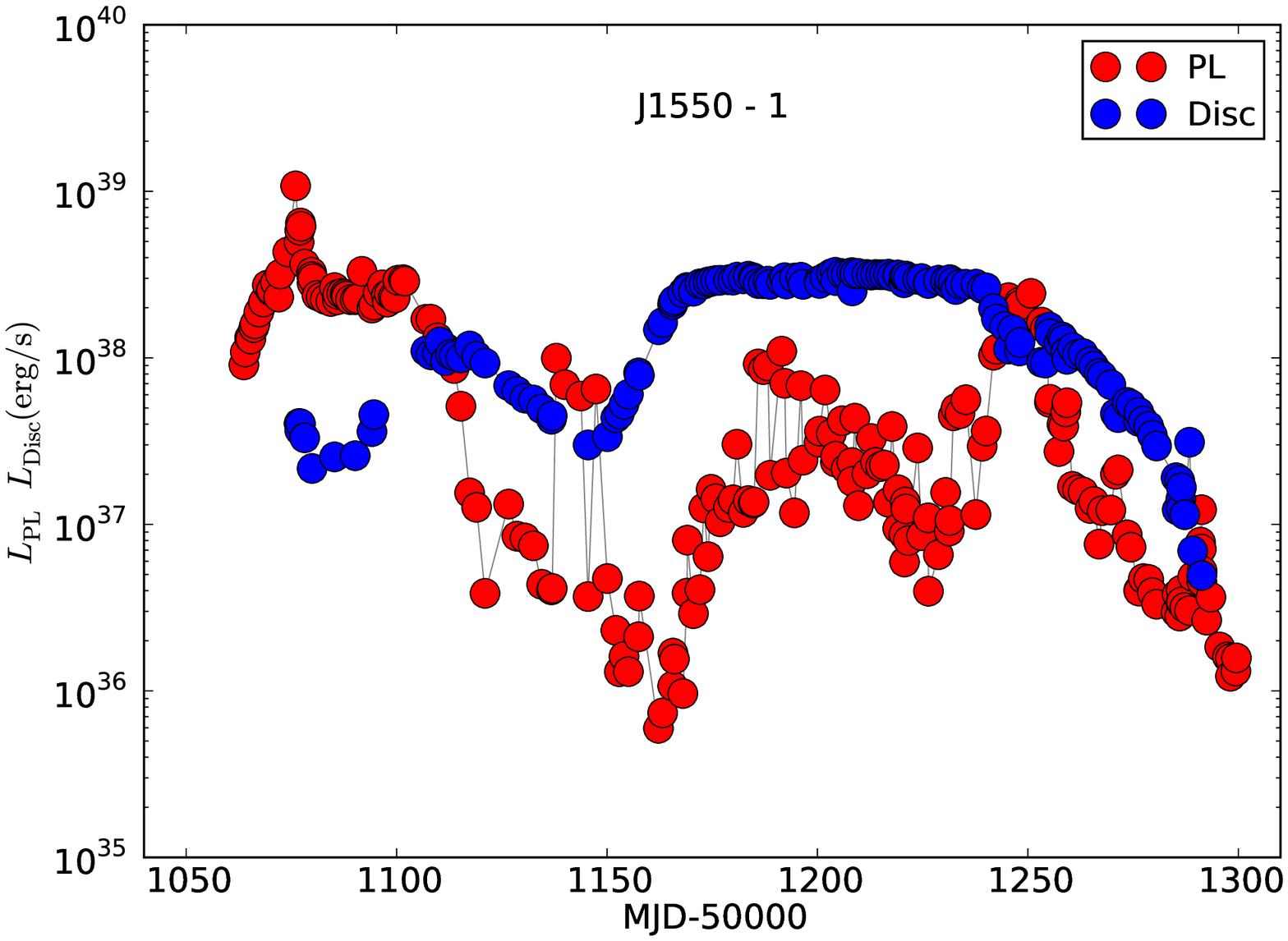}
\includegraphics[width=0.43\textwidth]{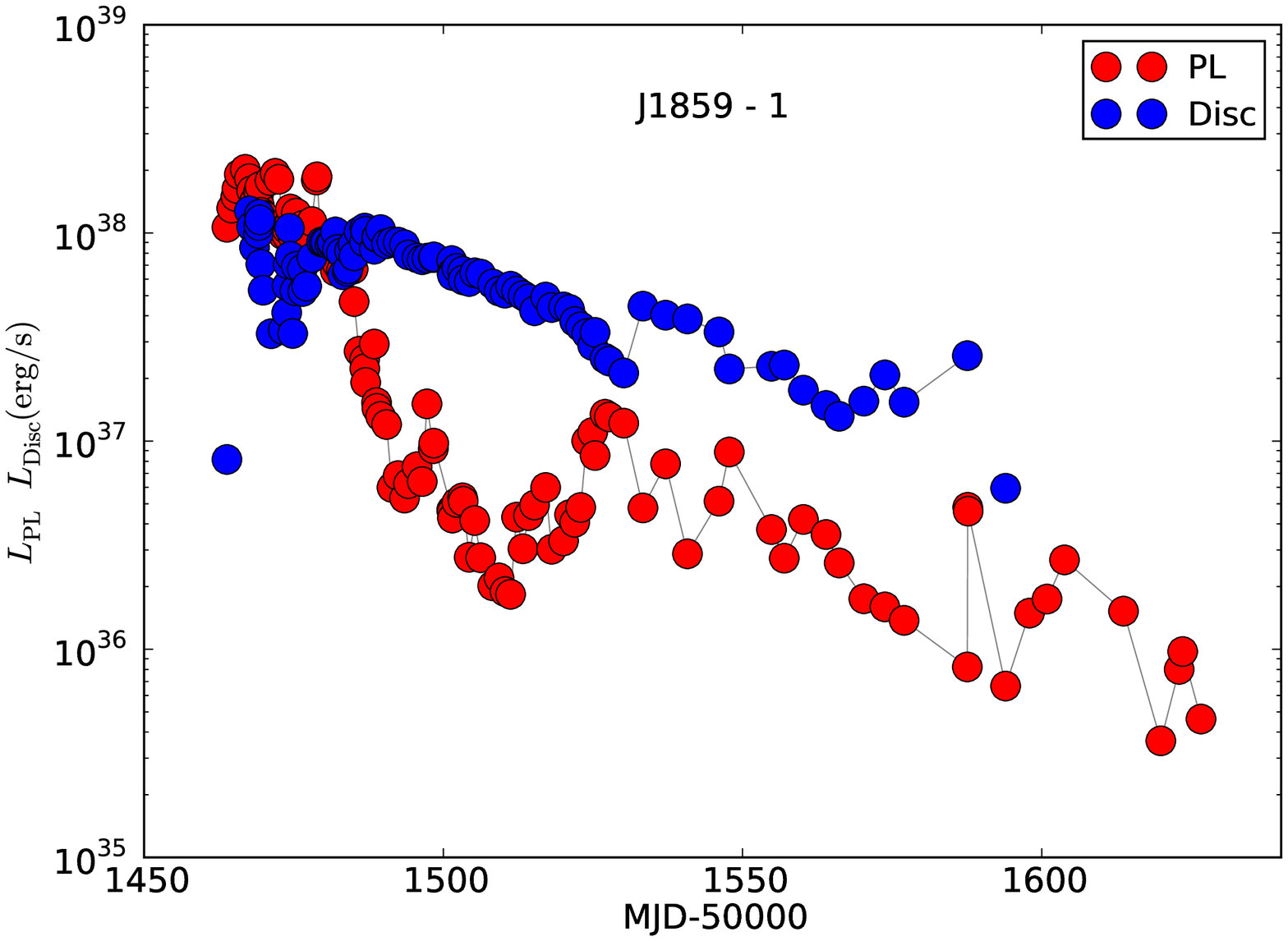}
\caption{The powerlaw and disc luminosity curves through eight of the
  outbursts covered in this work. \label{fig:PlDiscCurve}}
\end{figure*}

The relative variation of the luminosities of the powerlaw and disc
components determine whether the excursion runs along the same track.
In Fig. \ref{fig:PlDiscCurve} we show the relative variation of these
luminosities.  Where the powerlaw and disc luminosities are
approximately equal at the beginning of the outburst, the result will
be overlapping tracks in the DFLD (GRO~J1655-40, H~1743-322 and XTE~J1859+226).
These all occur before the powerlaw luminosity falls drastically away
from the disc luminosity once the BHXRB is fully in the disc
dominated state, and therefore are in the transition stage across to
the disc dominated state, and possibly indicative of a very high state.

If the powerlaw luminosity increases
mid-outburst, with little change in the disc luminosity (4U~1543-47,
GX~339-4 and XTE~J1859+226) then the behaviour is different.
Depending on the rate of 
decay of the disc, the total luminosity decline slows or even
temporarily reverses.  The powerlaw fraction increases, but the track
of the BHXRB through the DFLD does not lie on the track of the
entry into the disc dominated state.

Although distributions of the transition luminosities, as shown in
Fig. \ref{fig:transitions}, are easily determined, it is difficult to
conclude whether the distribution for one transition is broader than
the other or not.  When using the X-ray colour to determine the
transitions, the Hard-to-Soft transition luminosity distribution is
broader than the Soft-to-Hard transition luminosity distribution.
However, when using the disc fraction, the two distributions are
almost equally broad.  There are 8 outbursts in the X-ray colour and
11 outbursts in the disc fraction plots\footnote{These have been
  blurred by the Monte-Carlo process used to determine the
  uncertainties on the distributions.}.  Given these relatively small
numbers of transitions and the fact that differences in the
distribution widths are not present in both methods of calculation, we
cannot say whether our calculations support the work of
\citet{Maccarone03} or not.  Future determinations of the transitions
(from e.g. timing analysis) and improved measurements of the distances
and masses will allow a more detailed analysis.

\section{Conclusions}

We have performed a consistent spectral analysis on a large sample of
well known black hole X-ray binaries.  Using a set of simple models we
fit the spectra to study the evolution of the outbursts.  Two standard
diagnostic diagrams show that the level of similarity between the
outbursts is very high.  Further improvements in the overlap and the comparison of the sources
will result with more accurately determined distances and masses.  A
critical analysis of the two diagrams shows that for individual
sources or outbursts, the HLD is best suited,
being able to separate the different states (even if not being able to
identify them).  The DFLD, being independent from the effect of
$N_{\rm H}$ and a more physical description of the characteristics of
the system, is more useful when studying the population as a whole.
However, currently the spectral limitations of the \rxte\ satellite
restrict the diagnostic ability of the DFLD to the soft state.  The
overlap in the DFLD is very striking, indicating that in at least the
states where the disc is strongest, the BHXRBs behave in a very
similar way.

The rates of motion through the DFLD show
surprising uniformity, however currently this only reflects the motion
through the soft state, rather than at the true transition periods.
There is a region of increased rate of motion on the transition
between the disc dominated and the powerlaw dominated states, and
another on the lower luminosity part of the disc dominated states. 
Using the change in the disc fraction, as well as the X-ray colour, we
determine the dates and luminosities of the transitions to- and from
the soft or disc dominated states.  The distributions of these
luminosities are both broad and overlap.  However, the uncertainty in
the masses and especially distances comes through to this analysis and
we are not able to clearly say whether one distribution is
significantly broader than the other, especially as there are only a
comparatively small number of outbursts which show both transitions.

We investigate the location of the radio emission on top of the global
DFLD.  However, the very different observation rate make it difficult
to determine if there is any pattern from this analysis alone.  The
small range in black hole masses also limits our ability to study the
$L_{X}-L_{R}$ correlation.

\section*{Acknowledgements}
RJHD acknowledges support from the Alexander von Humboldt Foundation.
EGK acknowledges funding form a Marie Curie Intra-European fellowship 
under contract Nr. MEIF-CT-2006-024668.  TMB acknowledges support ASI via
contract I/088/06/0  and thanks the International Space Science
Institute (ISSI).  This research was supported
by the DFG cluster of excellence ‘Origin and Structure of the
Universe’ (www.universe-cluster.de).  We thank the anonymous referee
and Andrea Merloni for  
suggestions which improved this manuscript and Guy Pooley for the
radio data for XTE~J1118+480.  This research has made use of data
obtained through the High Energy Astrophysics Science Archive Research
Center Online Service, provided by the NASA/Goddard Space Flight
Center. 

\bibliographystyle{mn2e} 
\bibliography{mn-jour,./dunn}

\section*{Appendix}


In Fig. \ref{fig:curves} we plot all the HIDs, DFLDs, the X-ray colour and Disc Fraction
curves for the outbursts where a transition as determined from the
automated routine is detected in either the X-ray colour or Disc
Fraction curves.  The source 4U~1630-47 shows a complicated outburst
structure between MJD 52500 and 53500.  We have split this outburst
structure into two separate components as the final rise in flux is
very similar to stand-alone outbursts from other well known BHXRBs.

The colour schemes are as follows; for the HIDs and DFLDs it shows the
time since the beginning of the current outburst, starting at black
and ending at white.  For the X-ray colour
and Disc Fraction curves it shows the
flux ($3-10\kev$)/luminosity ($L_{\rm PL} + L_{\rm Disc}$) of the
source, white being bright, and black being faint.  The normalisation
is such to use the full range for each diagram.  

In Fig. \ref{fig:remainder} we show the HIDs and DFLDs of the remaining BHXRBs
which are not shown in Fig. \ref{fig:curves}

 \renewcommand{\thefigure}{A.\arabic{figure}}
 \setcounter{figure}{0}

\begin{figure*}
\centering
\includegraphics[width=0.41\textwidth]{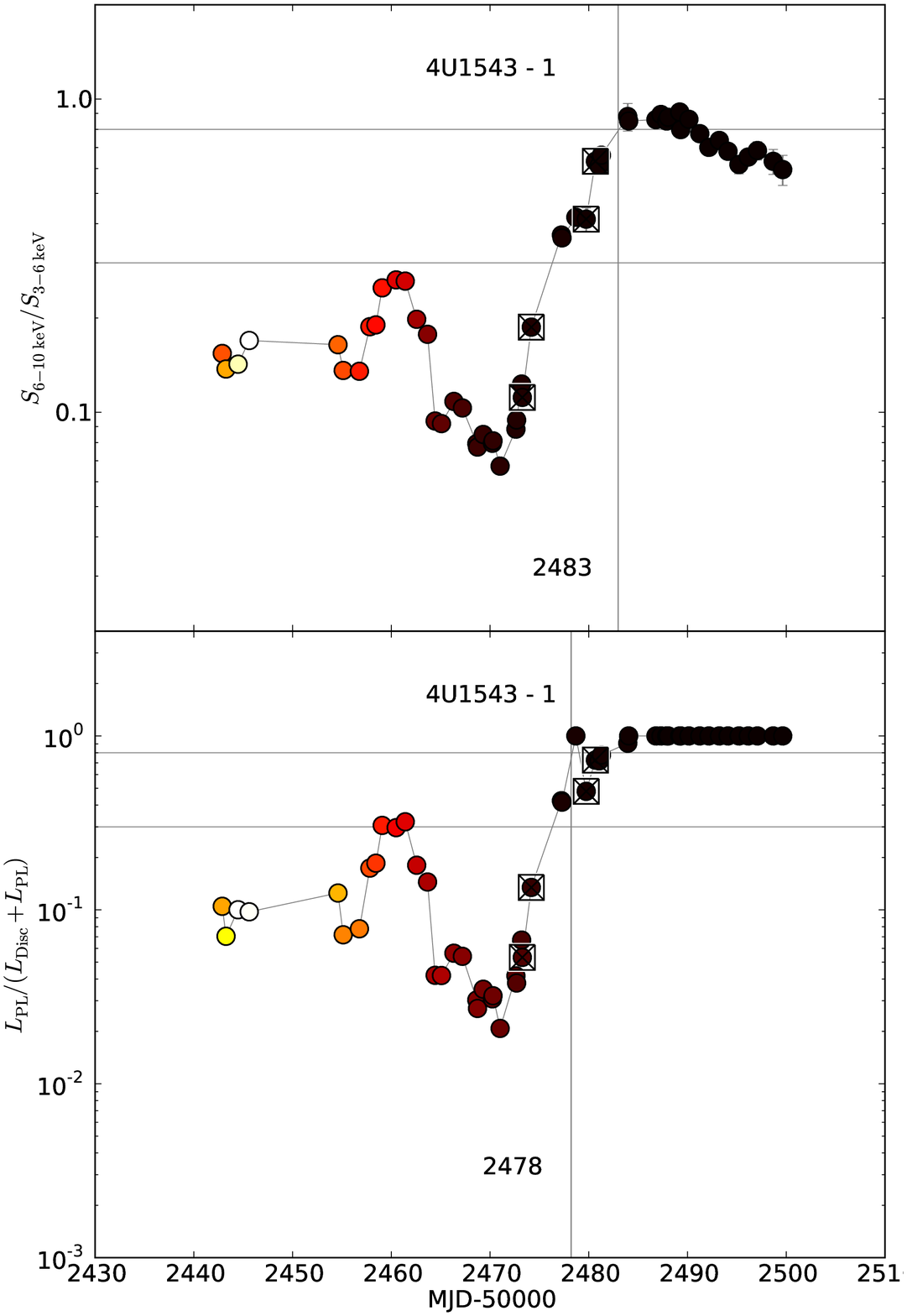}
\includegraphics[width=0.41\textwidth]{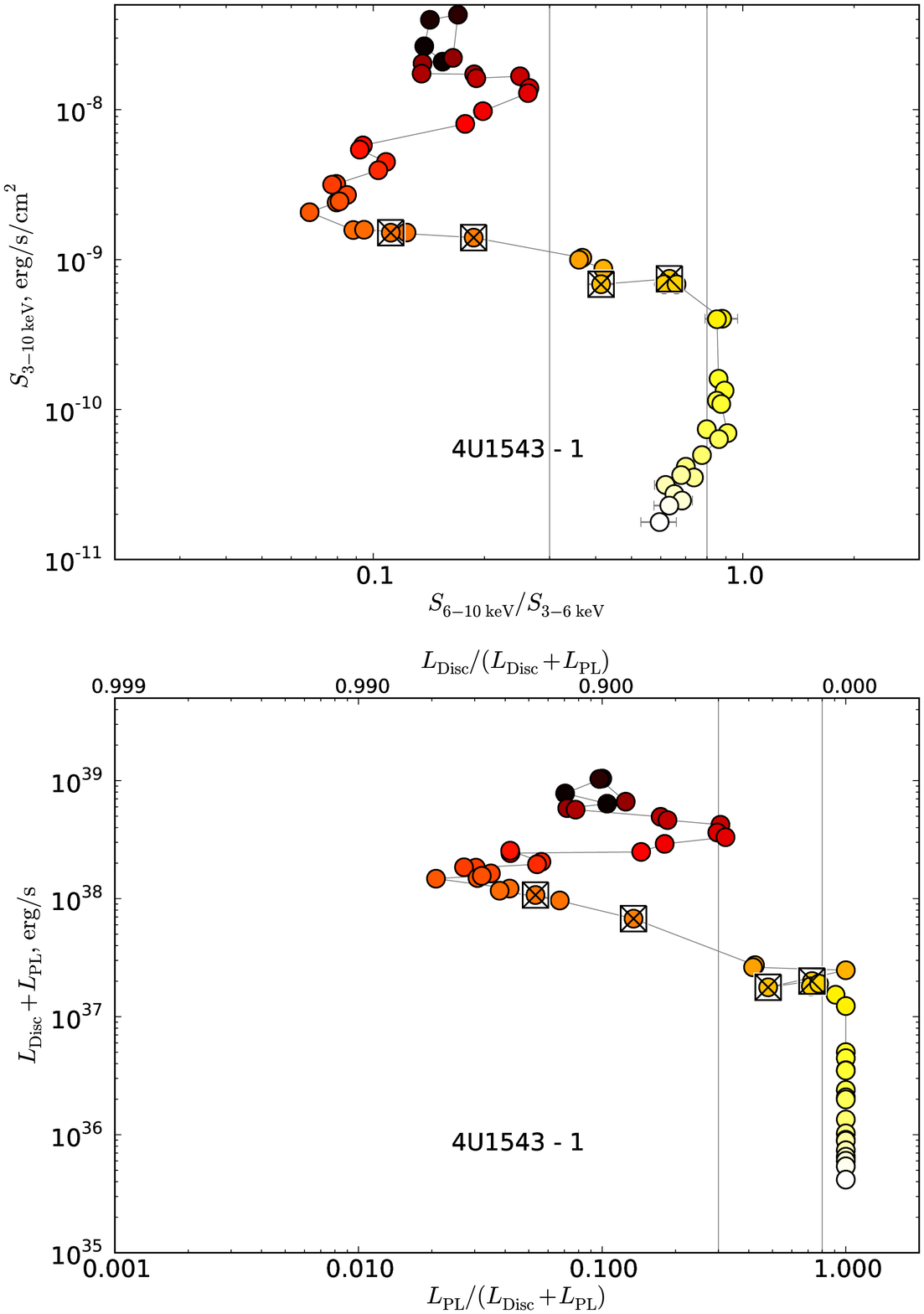}
\caption{\label{fig:curves} The curves of X-ray colour ({\scshape top left}) and
  powerlaw fraction ({\scshape bottom left}) against time, and the HID ({\scshape top
    right}) and DFLD ({\scshape bottom right}) for 4U 1543-47 Outburst 1.
  The observations from which the transition
  dates have been determined from timing properties are shown with the
crossed box (see Table \ref{tab:transitions} for the dates and
references).  In the HID and DFLD, the two
vertical lines are the Powerlaw Fractions and X-ray colours used when
calculating the transitions.  For the X-ray colour and Powerlaw
fraction curves, the two horizontal lines show the values used for
calculating the transition.  The vertical lines show the dates of the
transitions as calculated in Section \ref{sec:Trans_L}..}
\end{figure*}
\addtocounter{figure}{-1}
\begin{figure*}
\centering
\includegraphics[width=0.41\textwidth]{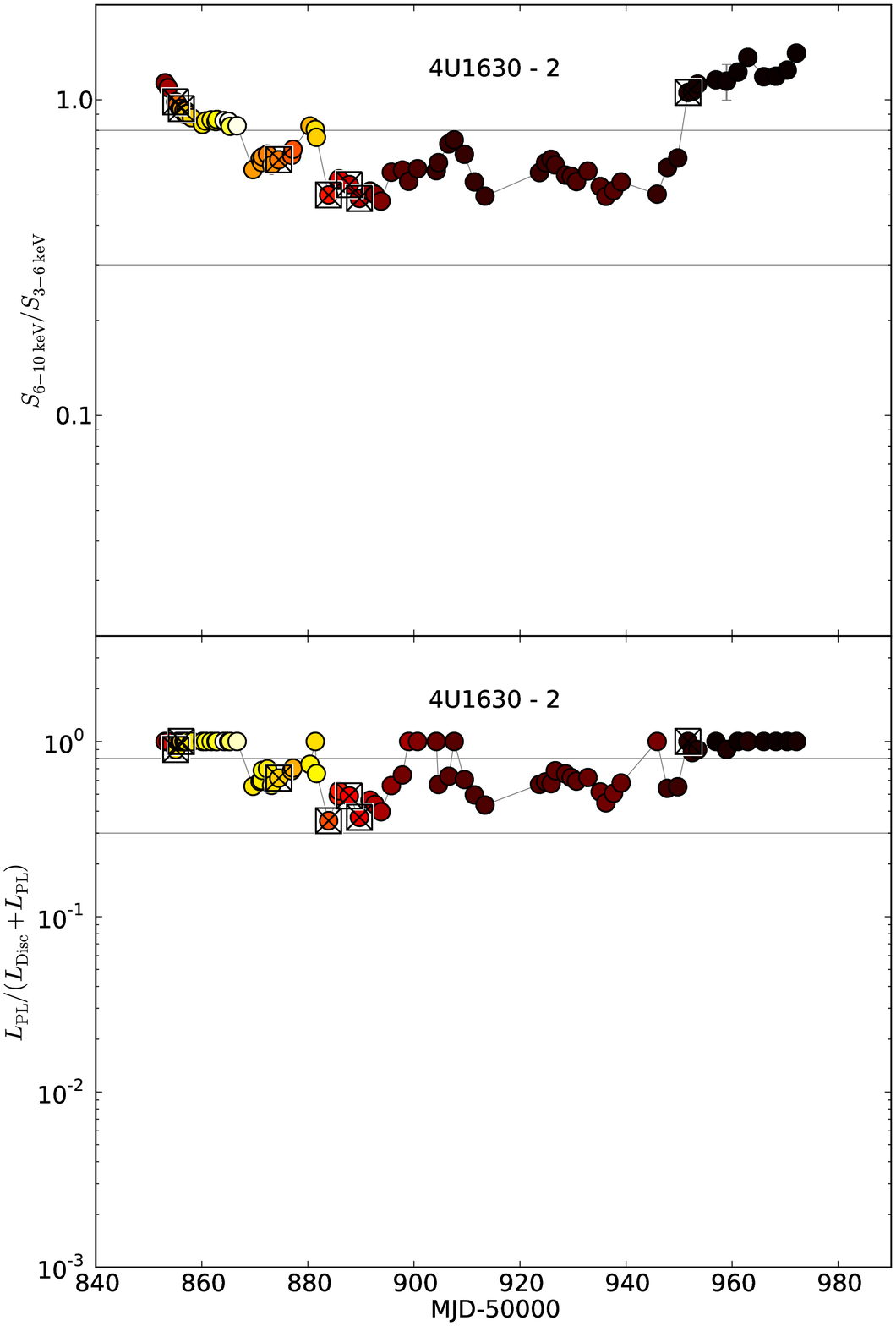}
\includegraphics[width=0.41\textwidth]{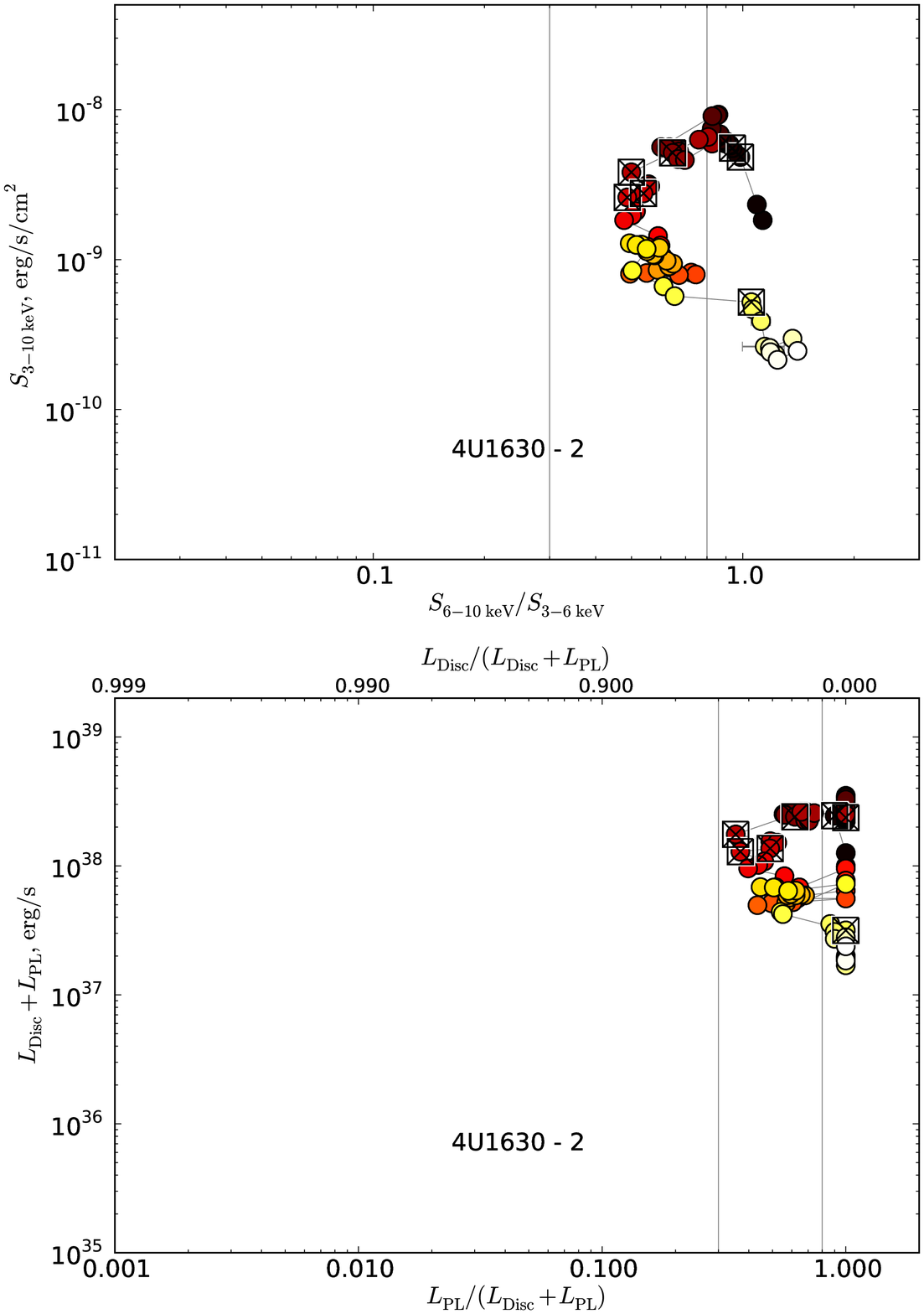}
\caption{(cont) X-ray colour curves and diagnostic diagrams for the
  outbursts observed - 4U 1630-47 Outburst 2.}
\end{figure*}
\addtocounter{figure}{-1}
\begin{figure*}
\centering
\includegraphics[width=0.41\textwidth]{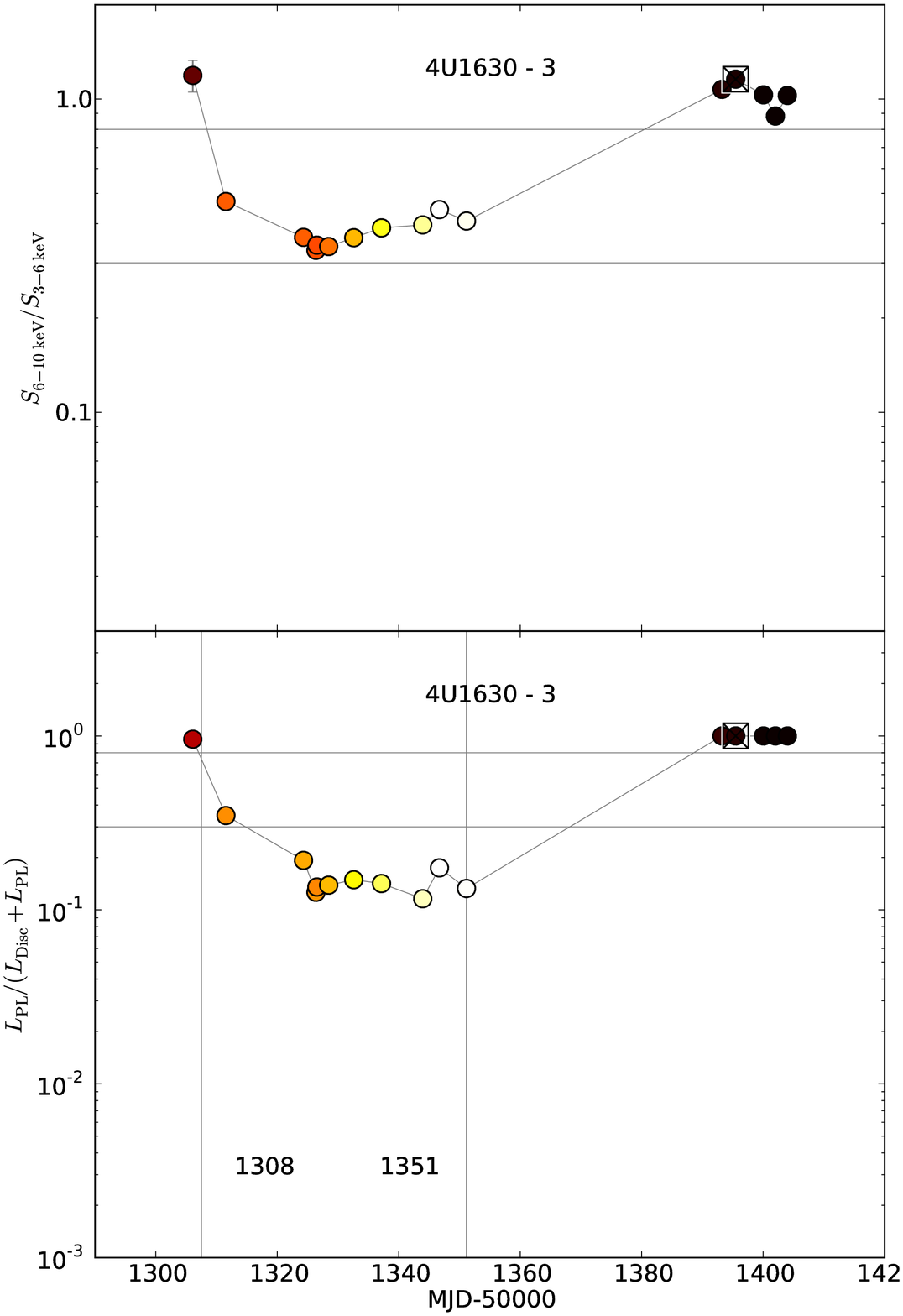}
\includegraphics[width=0.41\textwidth]{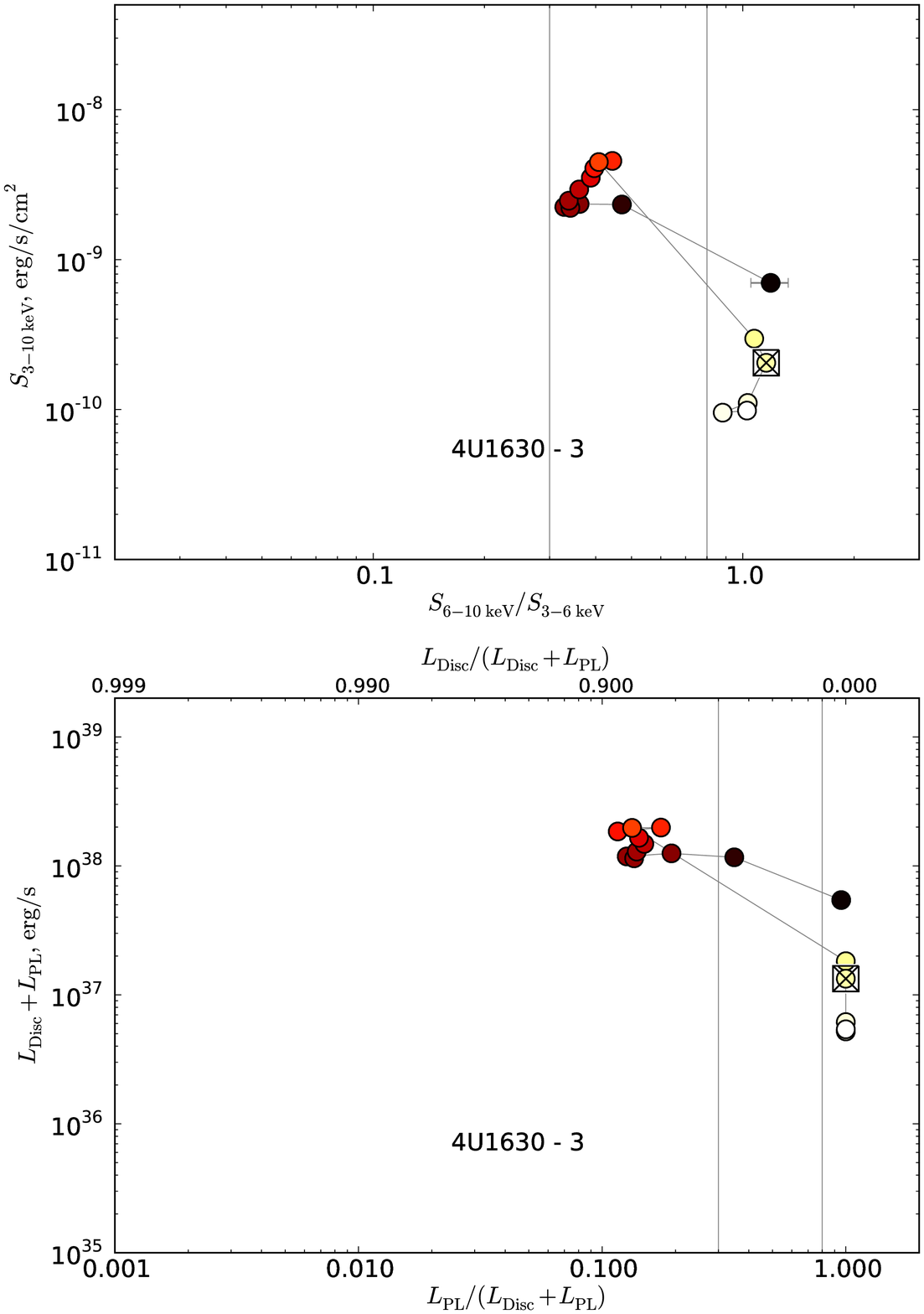}
\caption{(cont) X-ray colour curves and diagnostic diagrams for the
  outbursts observed - 4U 1630-47 Outburst 3.}
\end{figure*}
\addtocounter{figure}{-1}
\begin{figure*}
\centering
\includegraphics[width=0.41\textwidth]{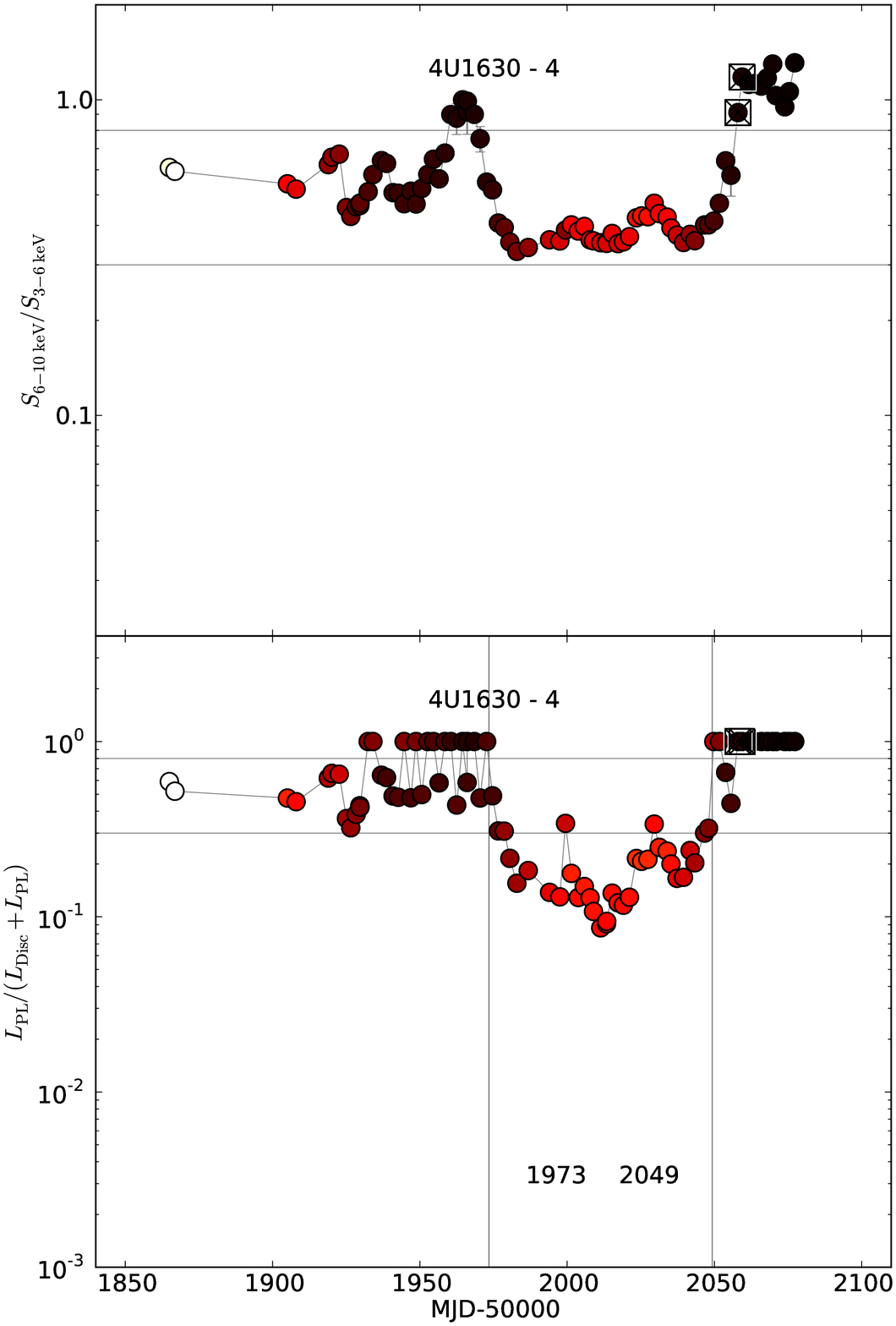}
\includegraphics[width=0.41\textwidth]{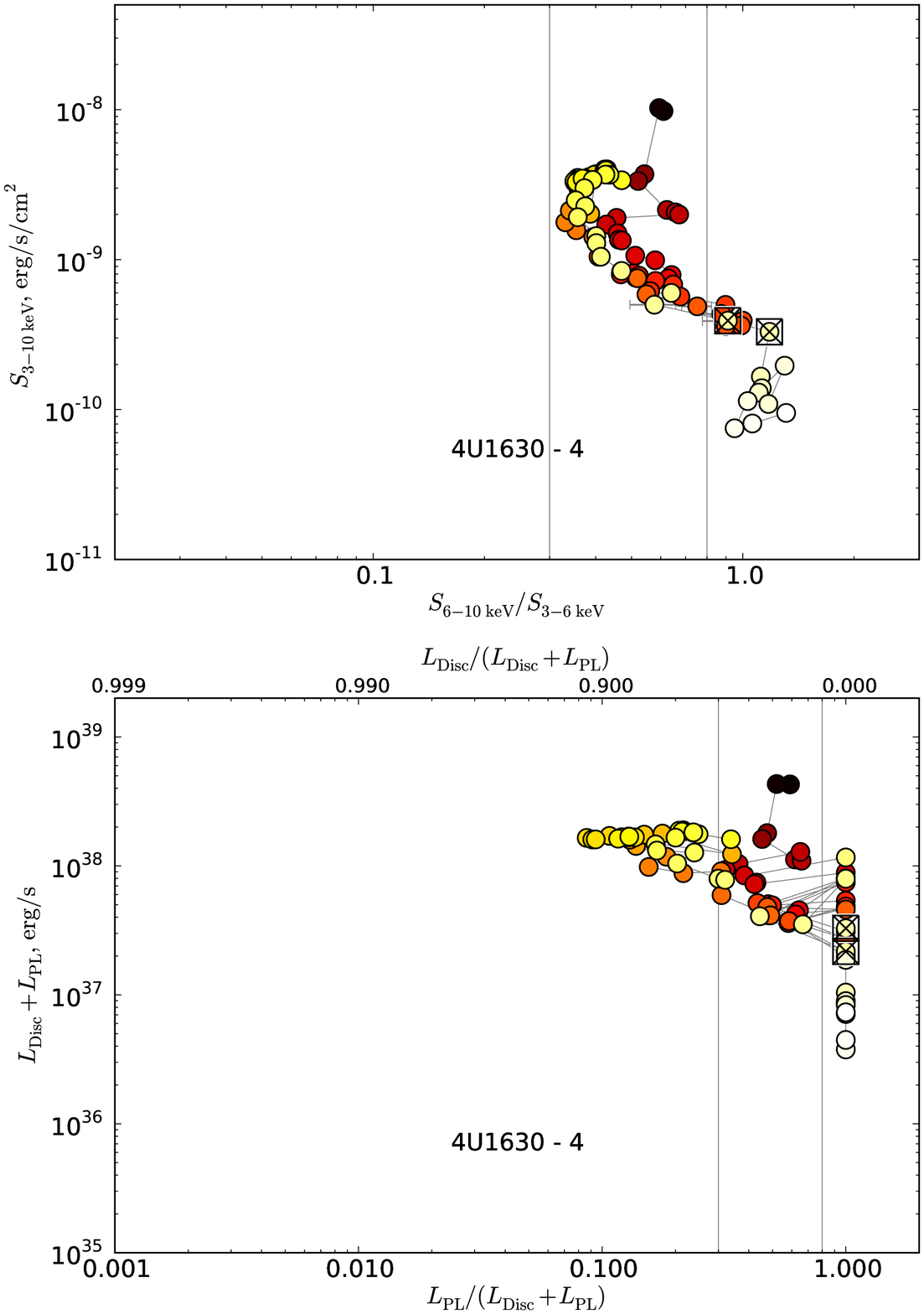}
\caption{(cont) X-ray colour curves and diagnostic diagrams for the
  outbursts observed - 4U 1630-47 Outburst 4.}
\end{figure*}
\addtocounter{figure}{-1}
\begin{figure*}
\centering
\includegraphics[width=0.41\textwidth]{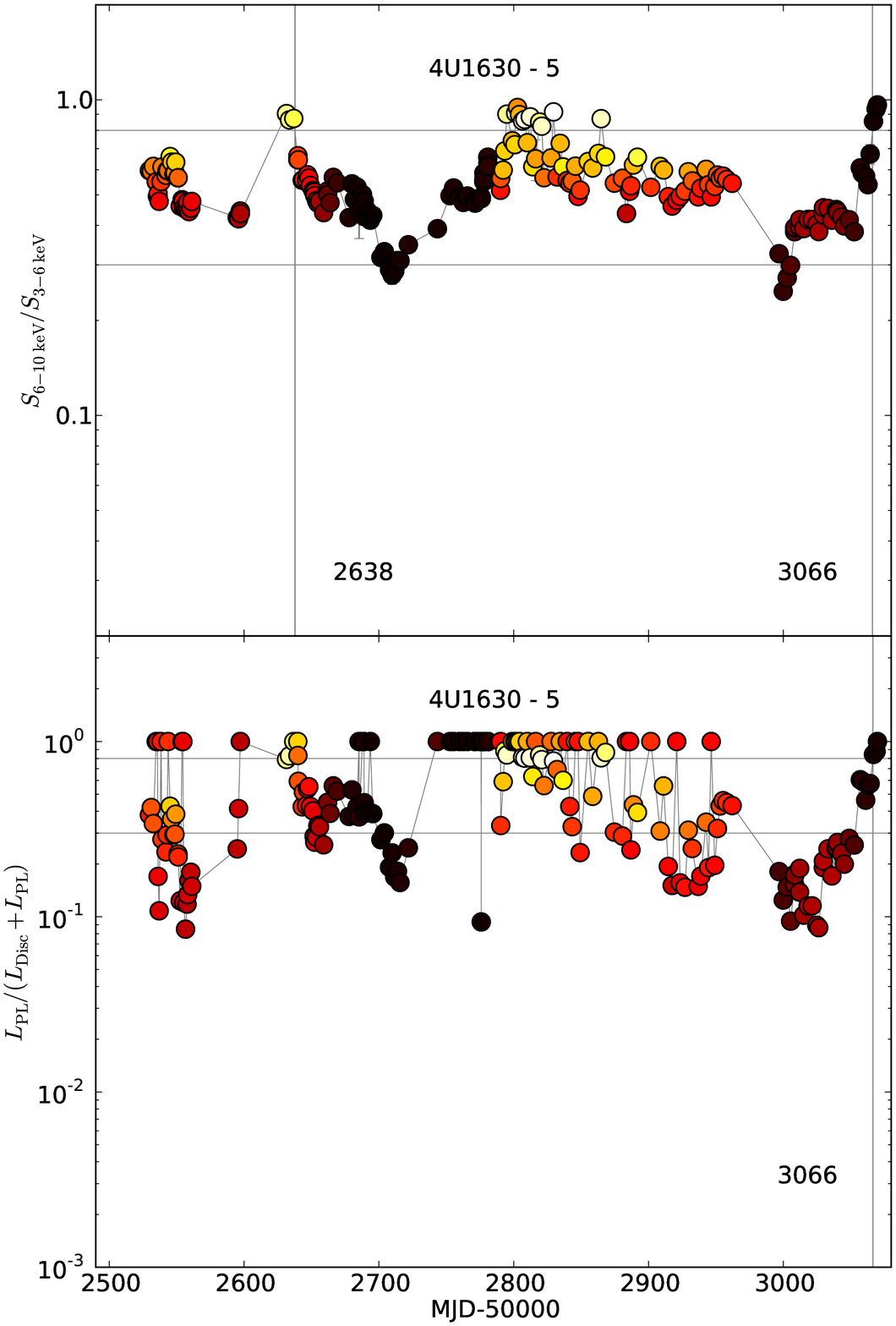}
\includegraphics[width=0.41\textwidth]{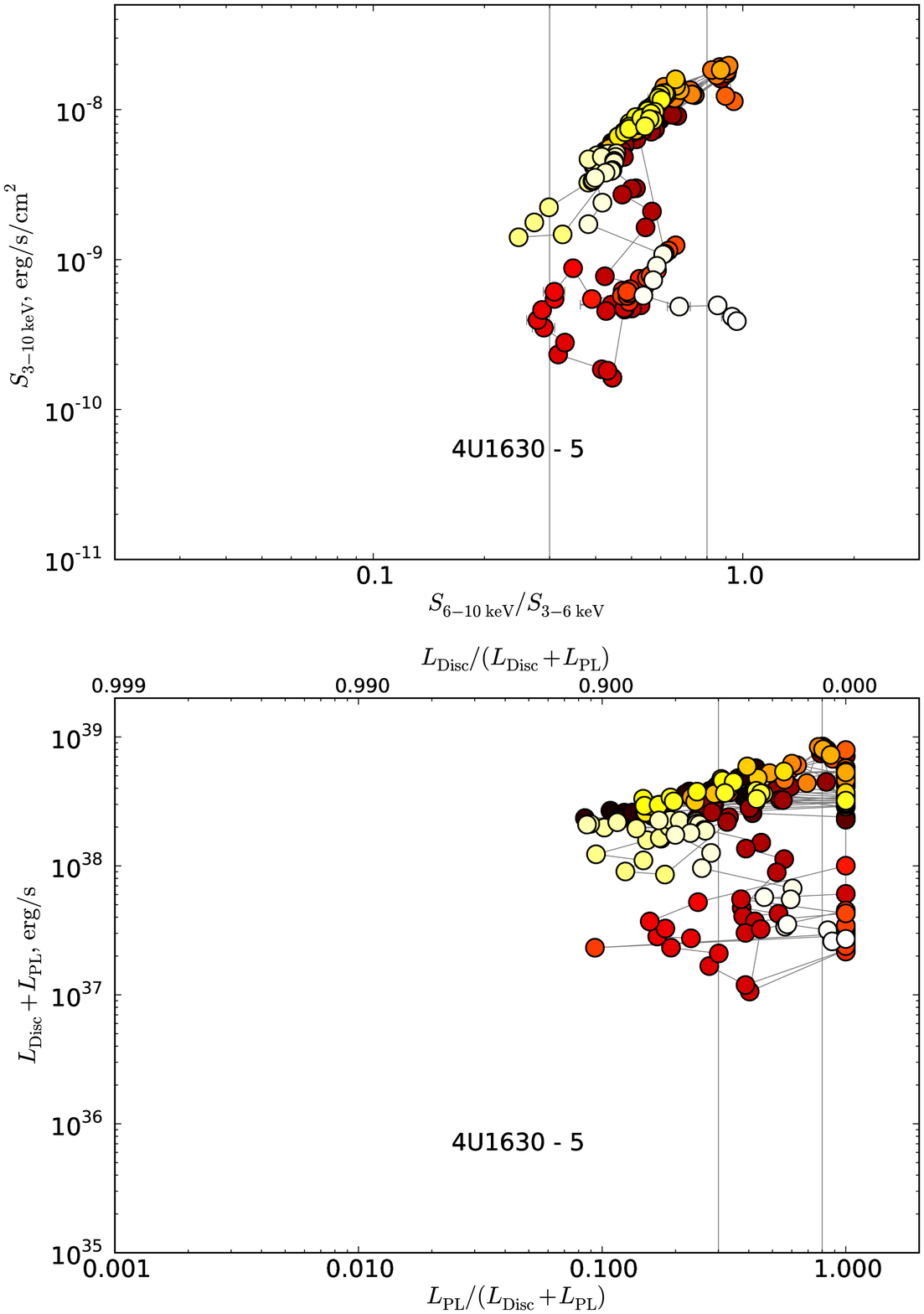}
\caption{(cont) X-ray colour curves and diagnostic diagrams for the
  outbursts observed - 4U 1630-47 Outburst 5.}
\end{figure*}
\addtocounter{figure}{-1}
\begin{figure*}
\centering
\includegraphics[width=0.41\textwidth]{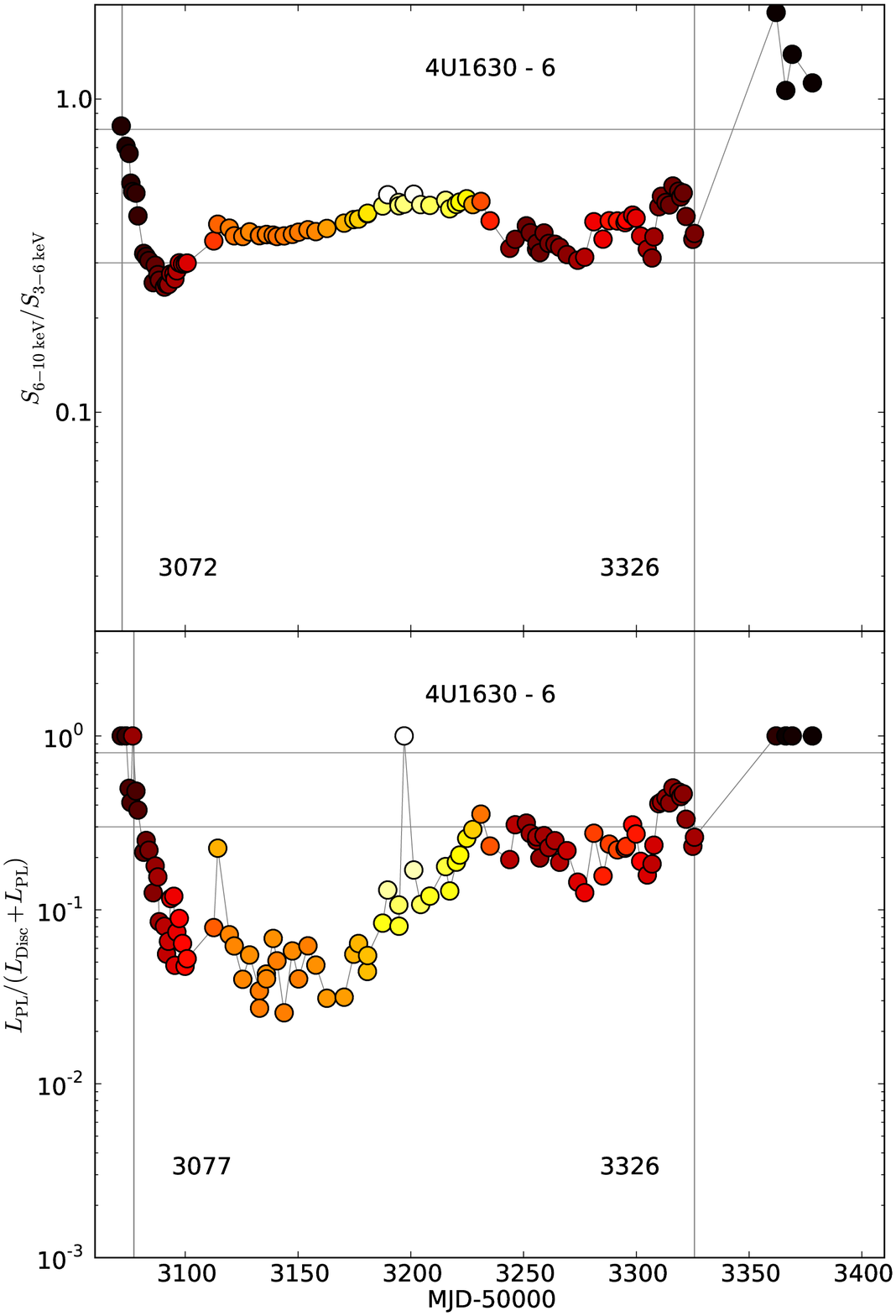}
\includegraphics[width=0.41\textwidth]{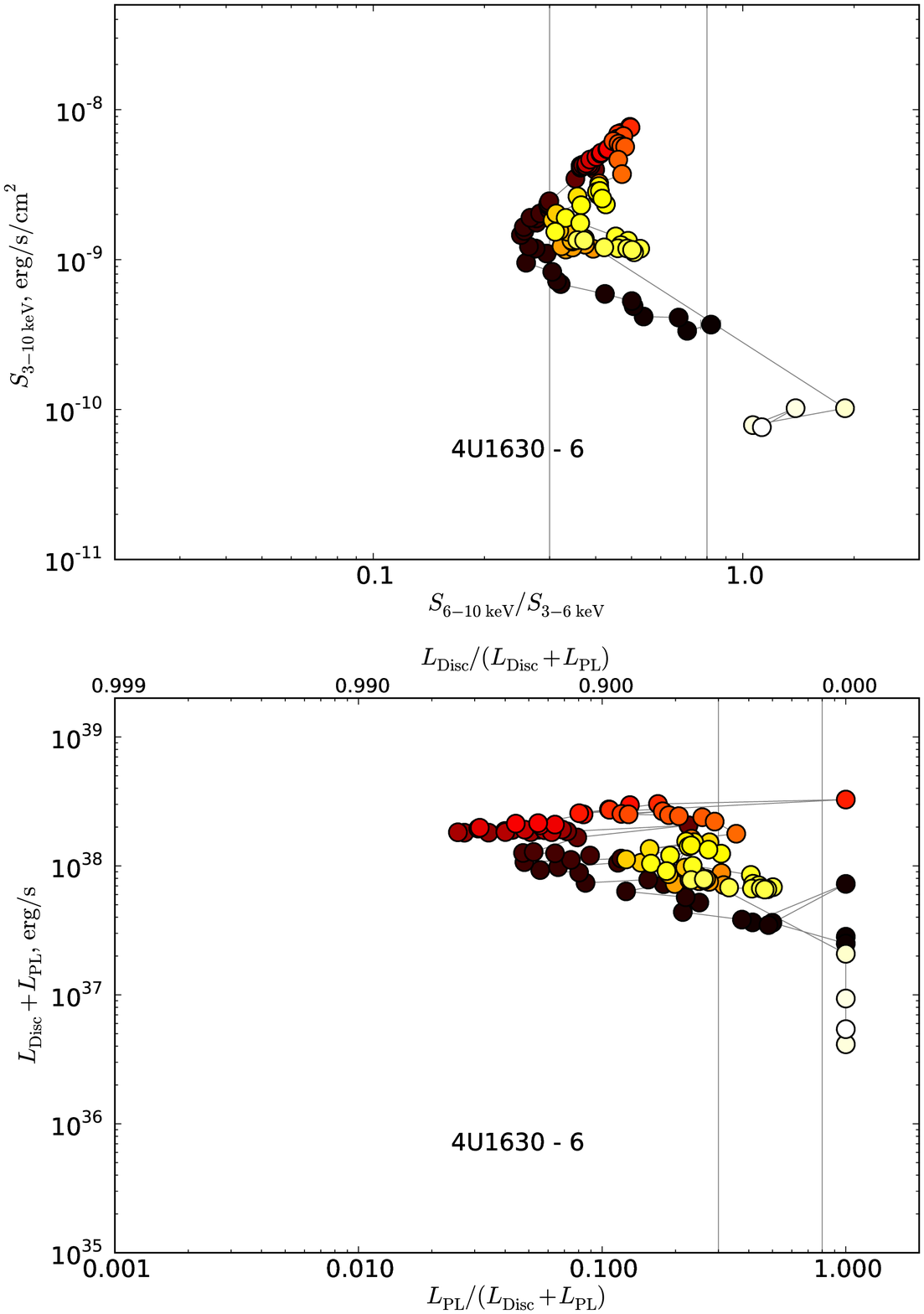}
\caption{(cont) X-ray colour curves and diagnostic diagrams for the
  outbursts observed - 4U 1630-47 Outburst 6.}
\end{figure*}
\addtocounter{figure}{-1}
\begin{figure*}
\centering
\includegraphics[width=0.41\textwidth]{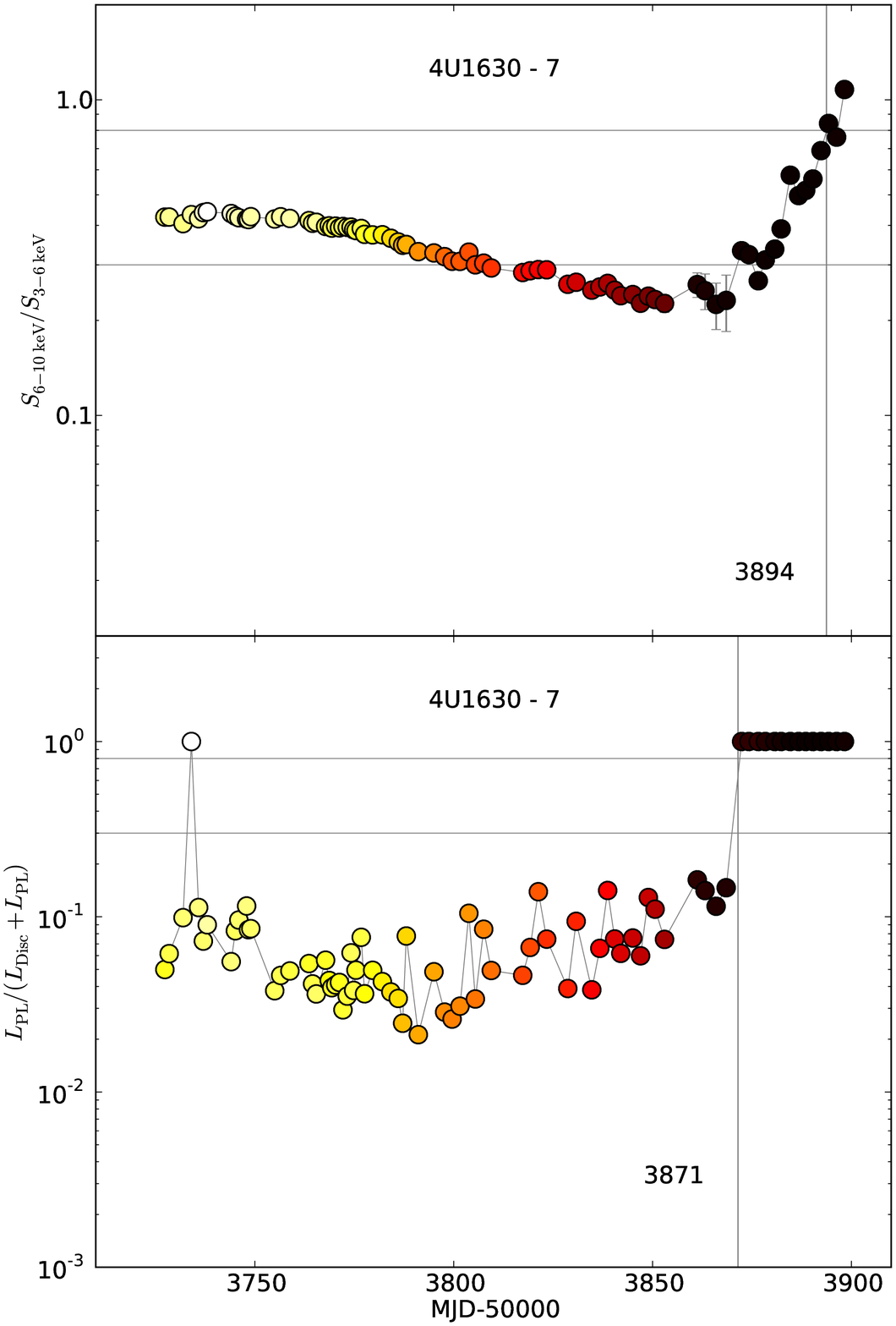}
\includegraphics[width=0.41\textwidth]{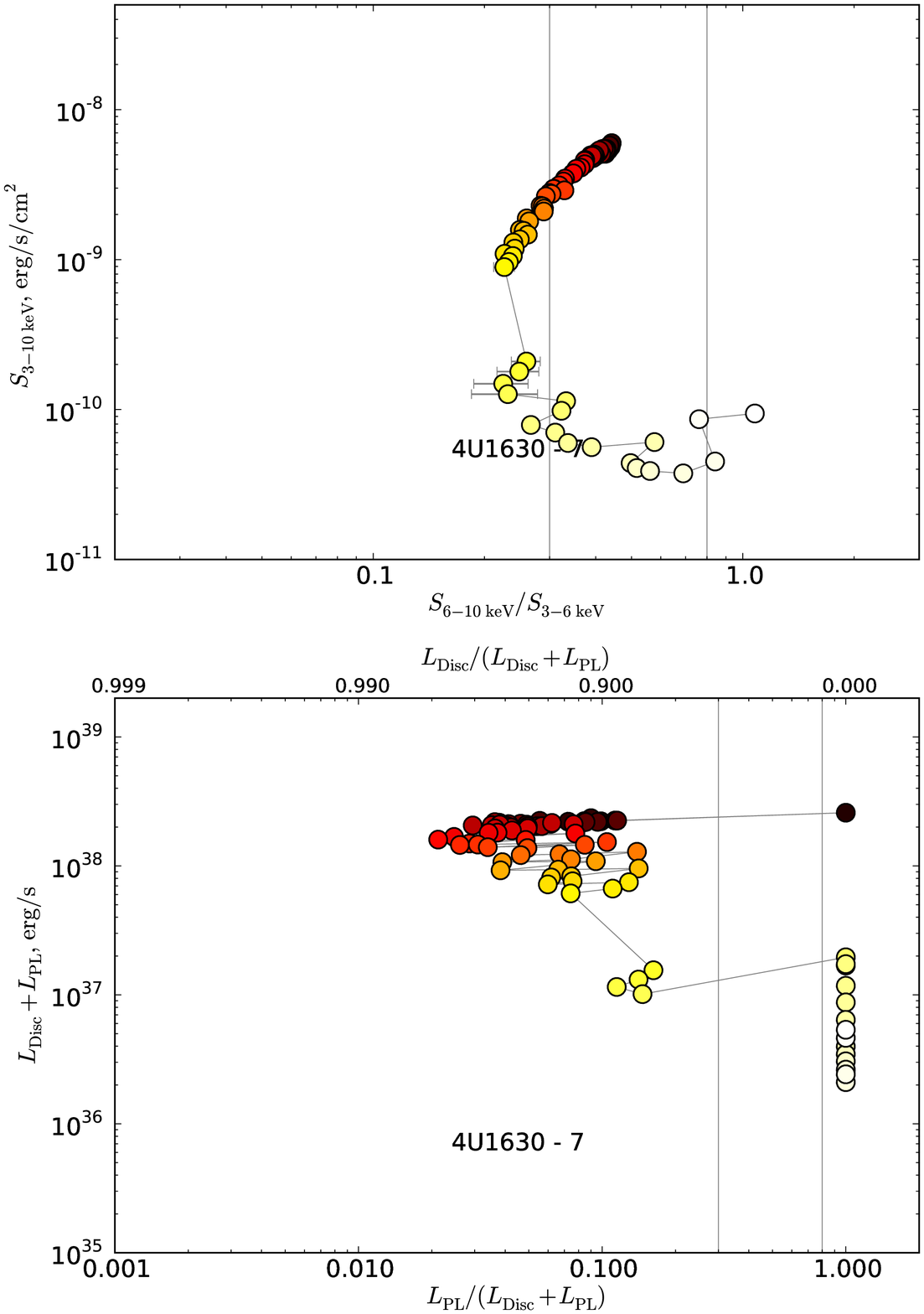}
\caption{(cont) X-ray colour curves and diagnostic diagrams for the
  outbursts observed - 4U 1630-47 Outburst 7.}
\end{figure*}
\addtocounter{figure}{-1}
\begin{figure*}
\centering
\includegraphics[width=0.41\textwidth]{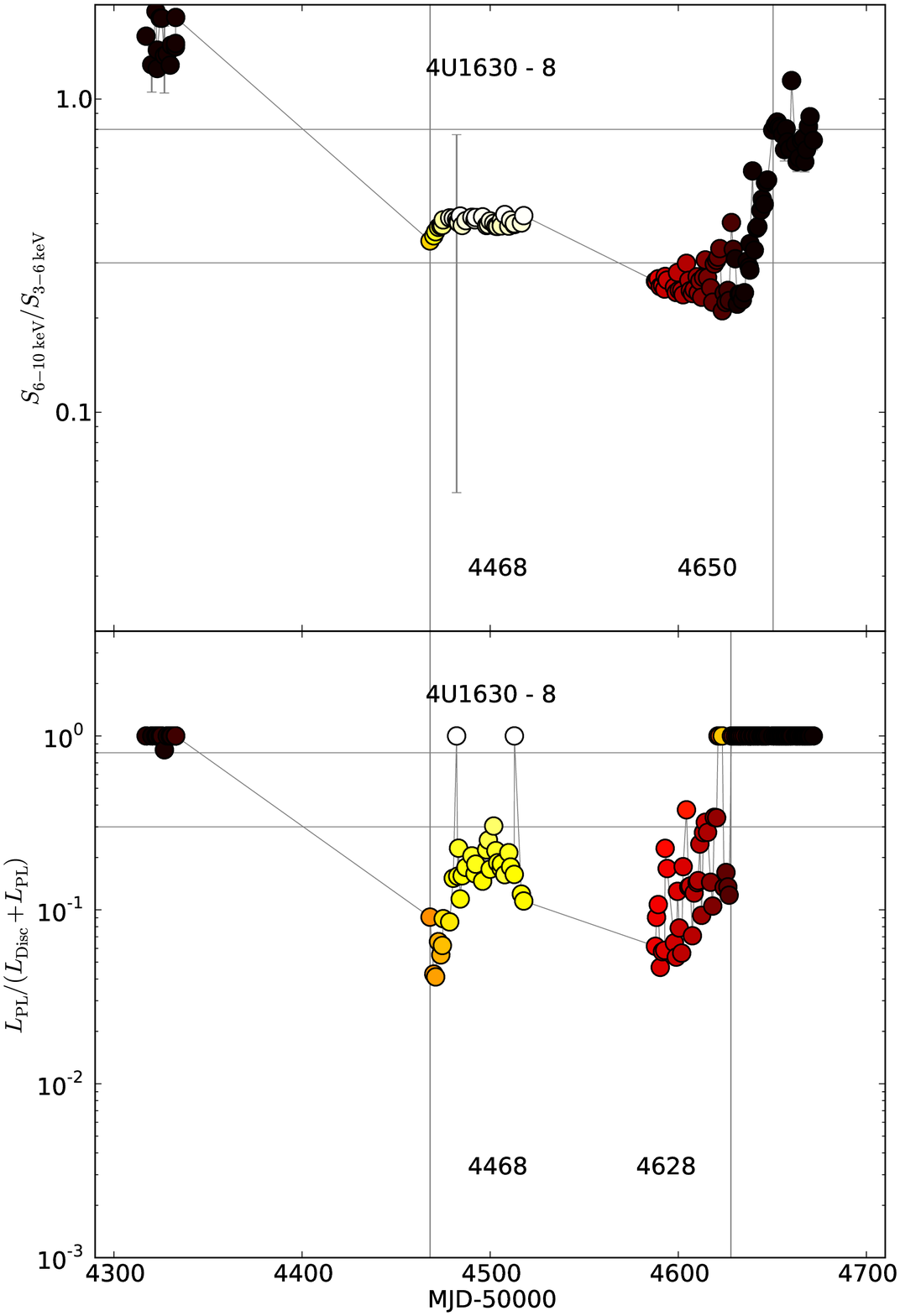}
\includegraphics[width=0.41\textwidth]{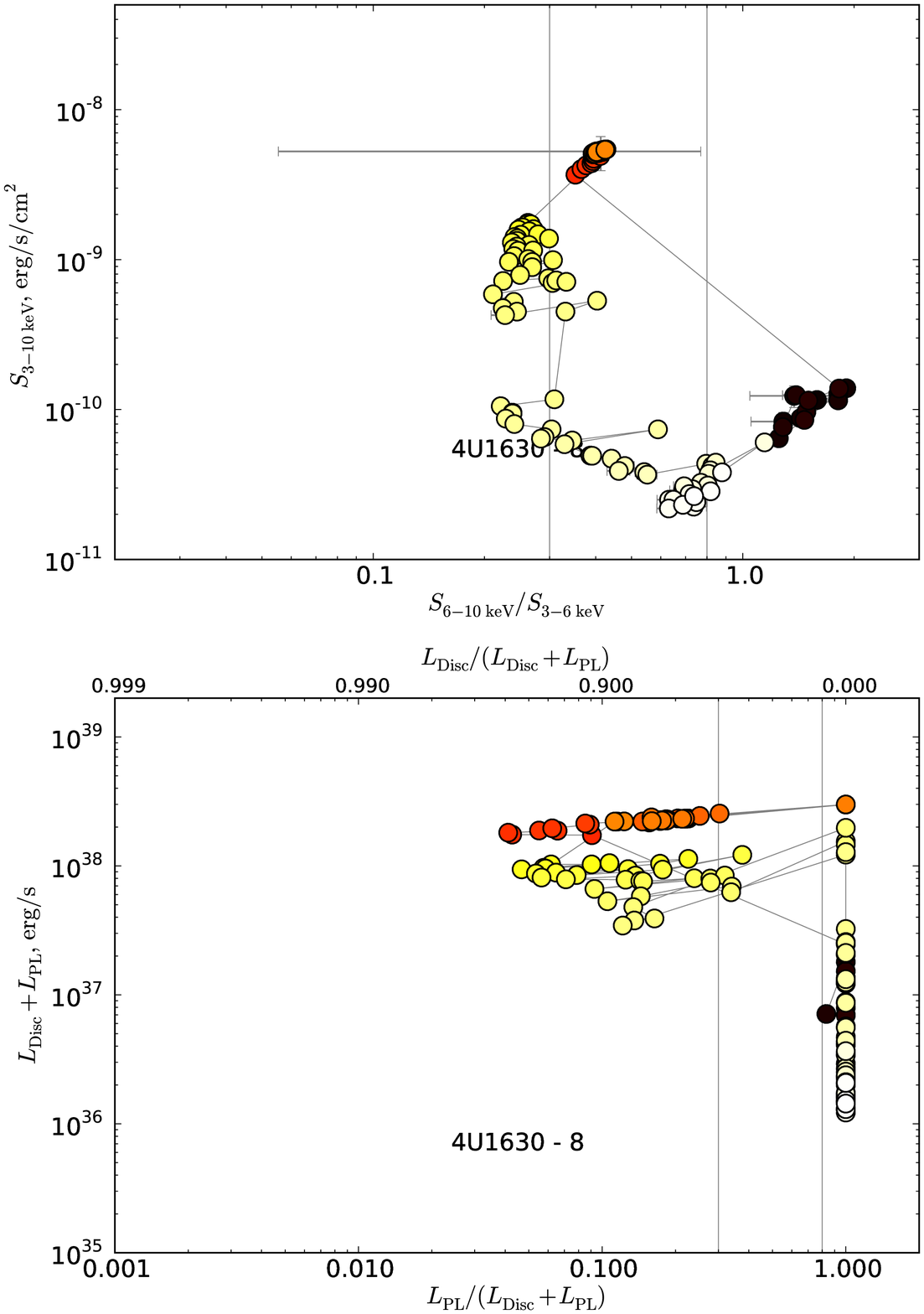}
\caption{(cont) X-ray colour curves and diagnostic diagrams for the
  outbursts observed - 4U 1630-47 Outburst 8.}
\end{figure*}
\addtocounter{figure}{-1}
\begin{figure*}
\centering
\includegraphics[width=0.41\textwidth]{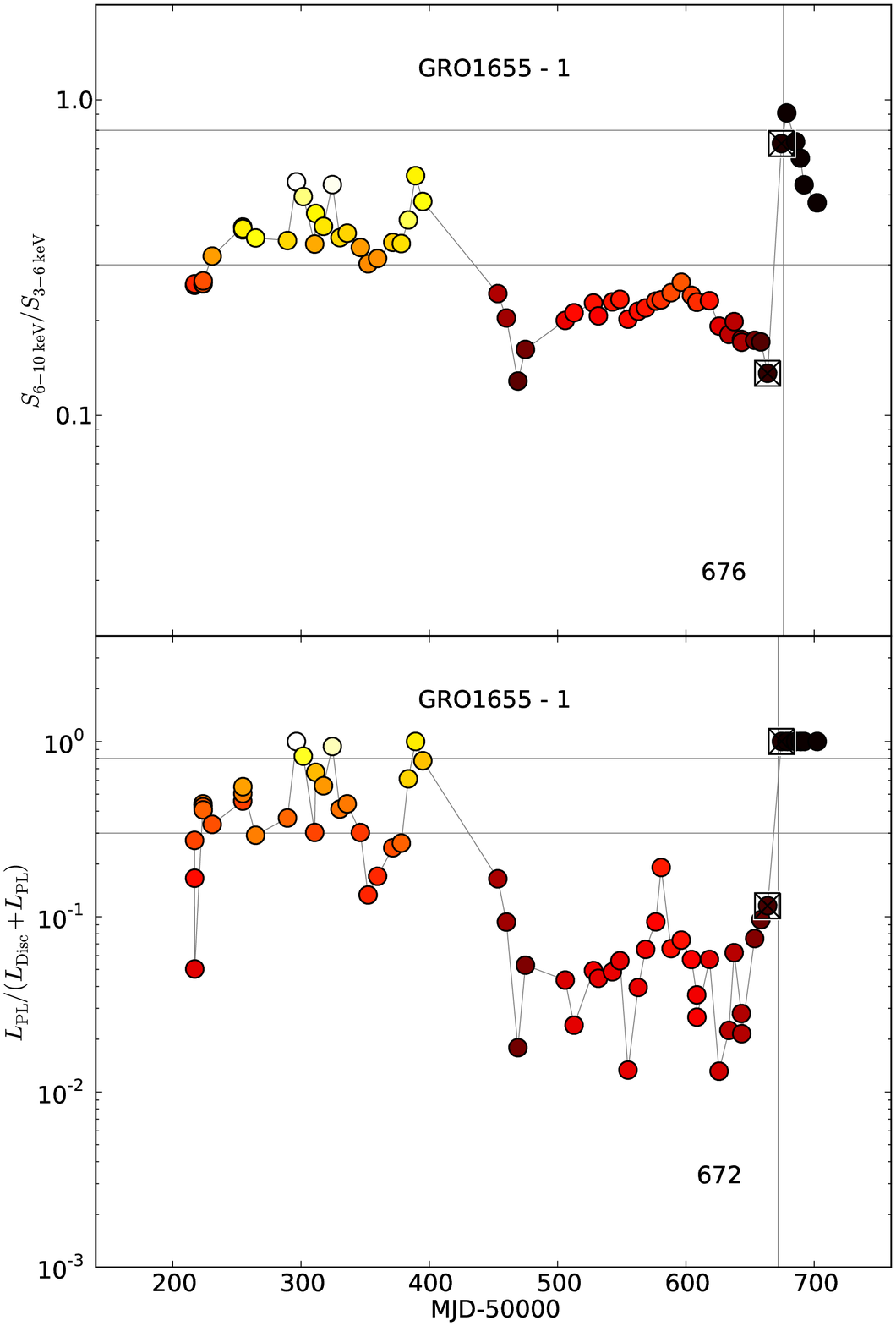}
\includegraphics[width=0.41\textwidth]{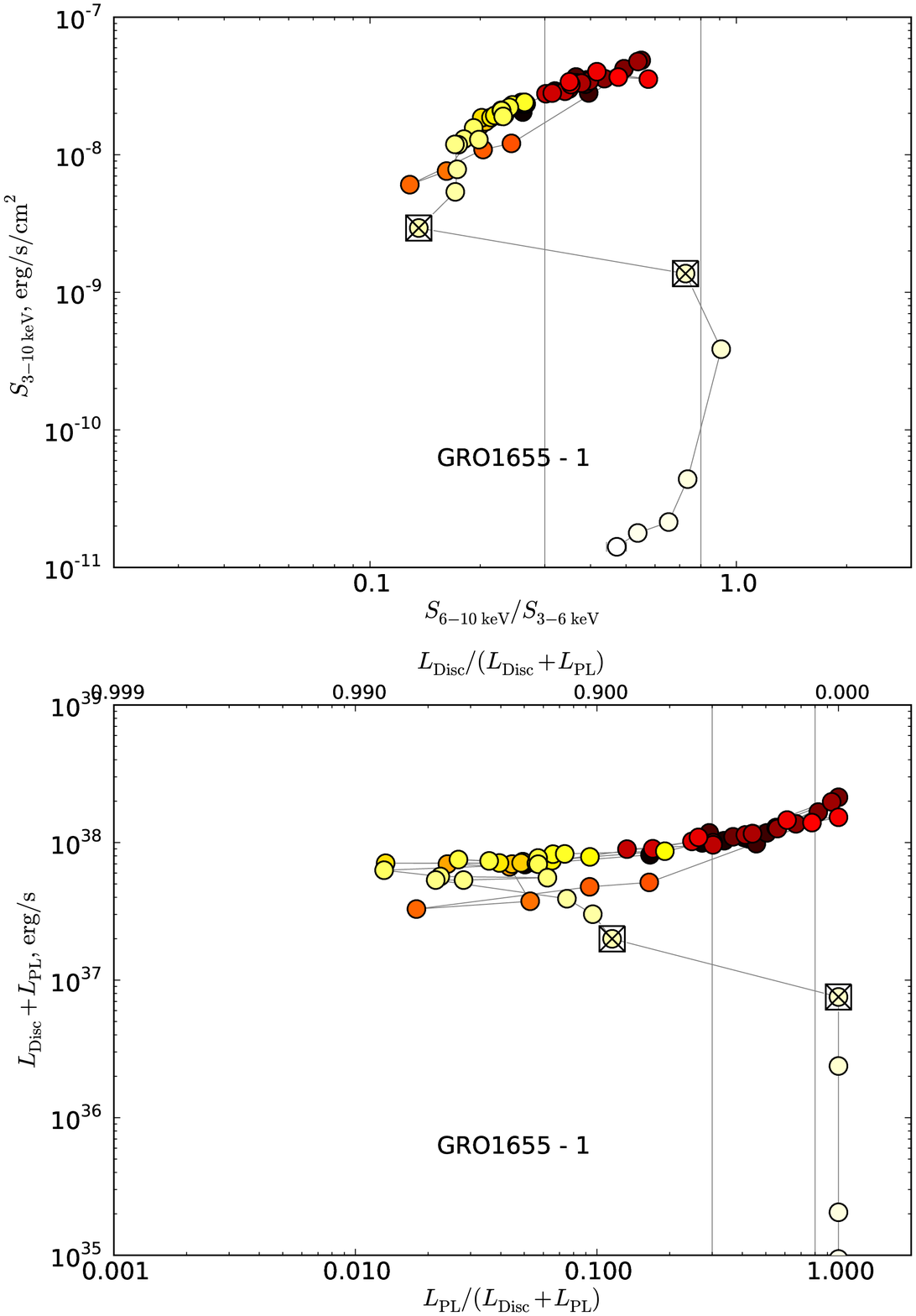}
\caption{(cont) X-ray colour curves and diagnostic diagrams for the
  outbursts observed.}
\end{figure*}
\addtocounter{figure}{-1}
\begin{figure*}
\centering
\includegraphics[width=0.41\textwidth]{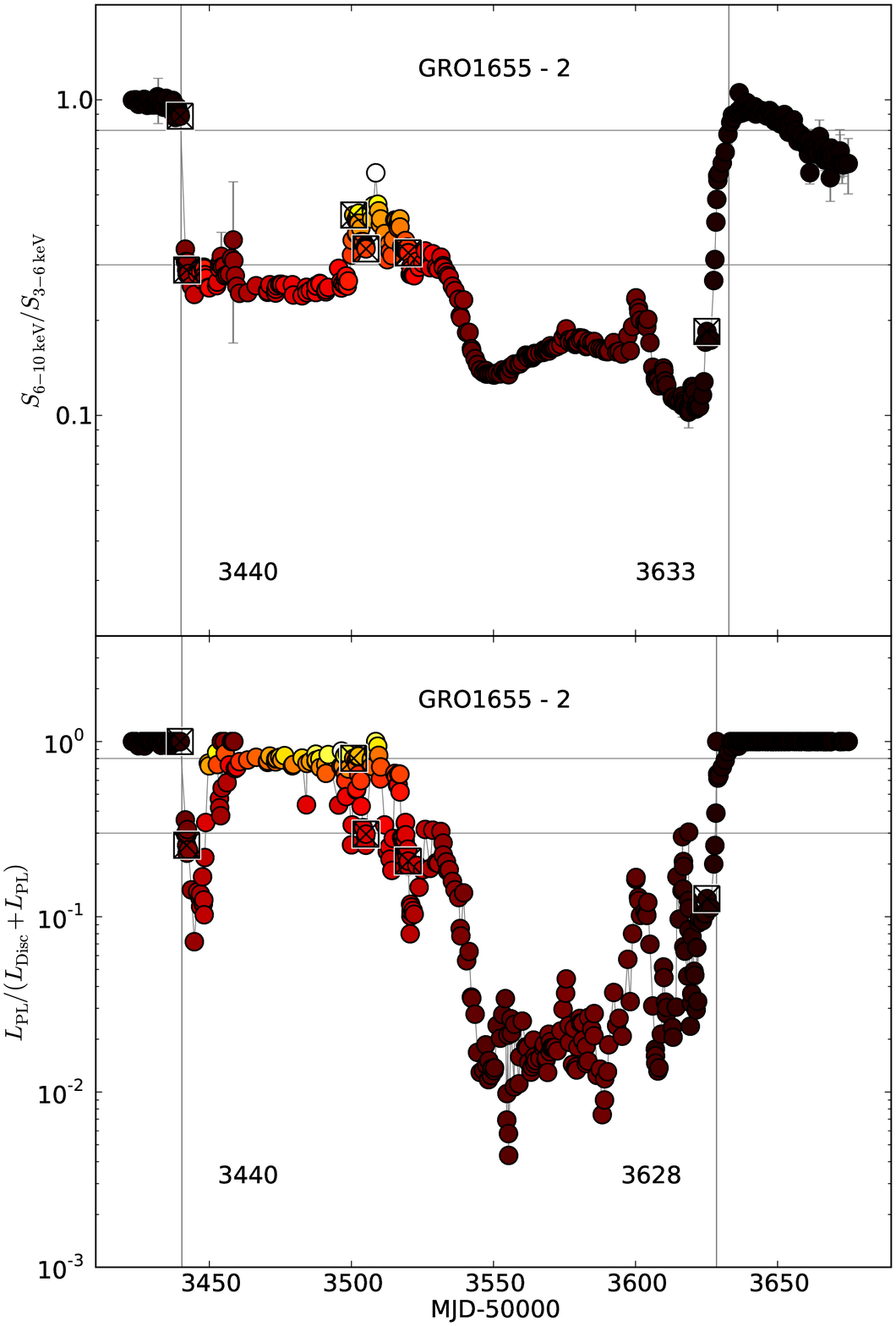}
\includegraphics[width=0.41\textwidth]{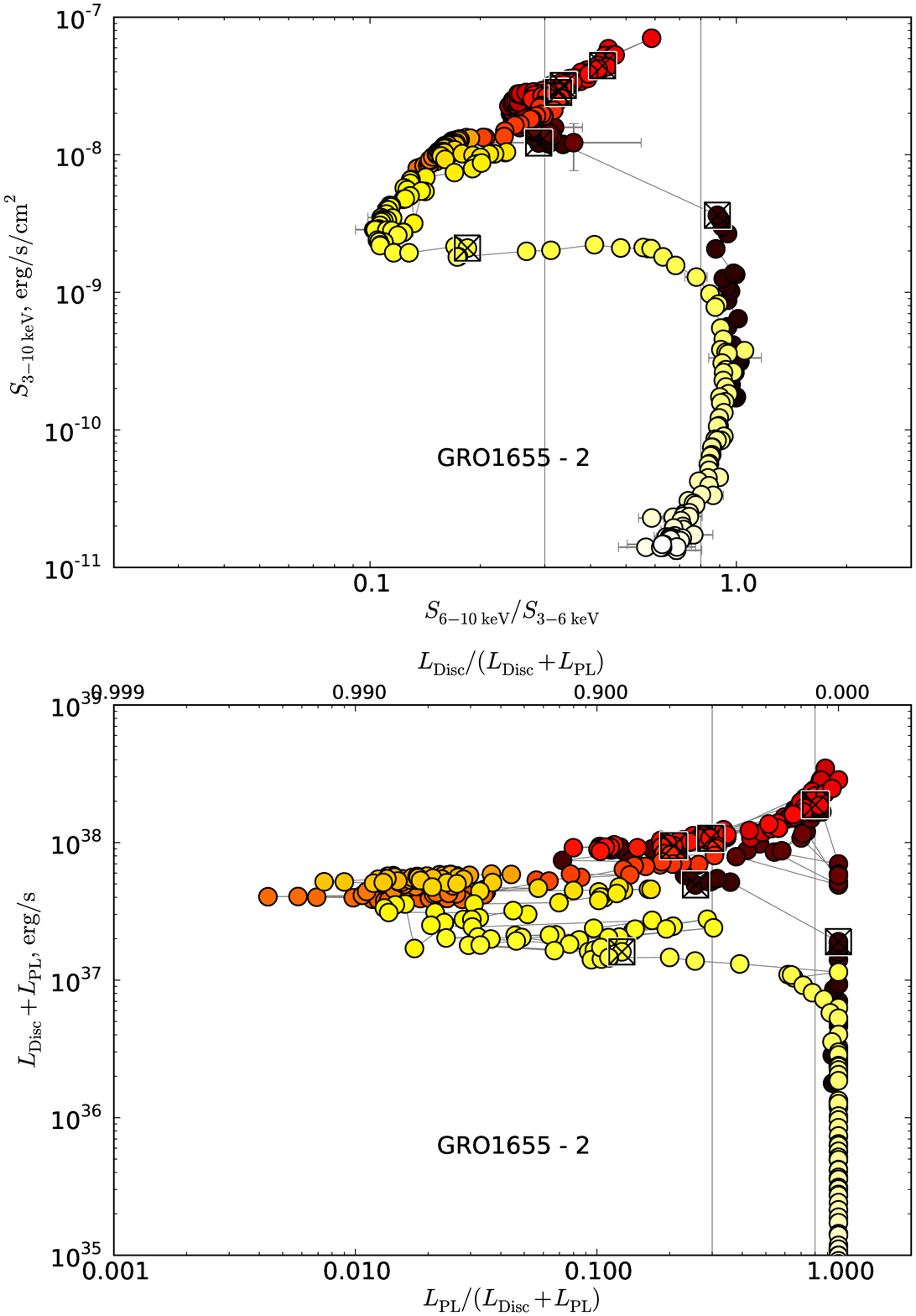}
\caption{(cont) X-ray colour curves and diagnostic diagrams for the
  outbursts observed.}
\end{figure*}
\addtocounter{figure}{-1}
\begin{figure*}
\centering
\includegraphics[width=0.41\textwidth]{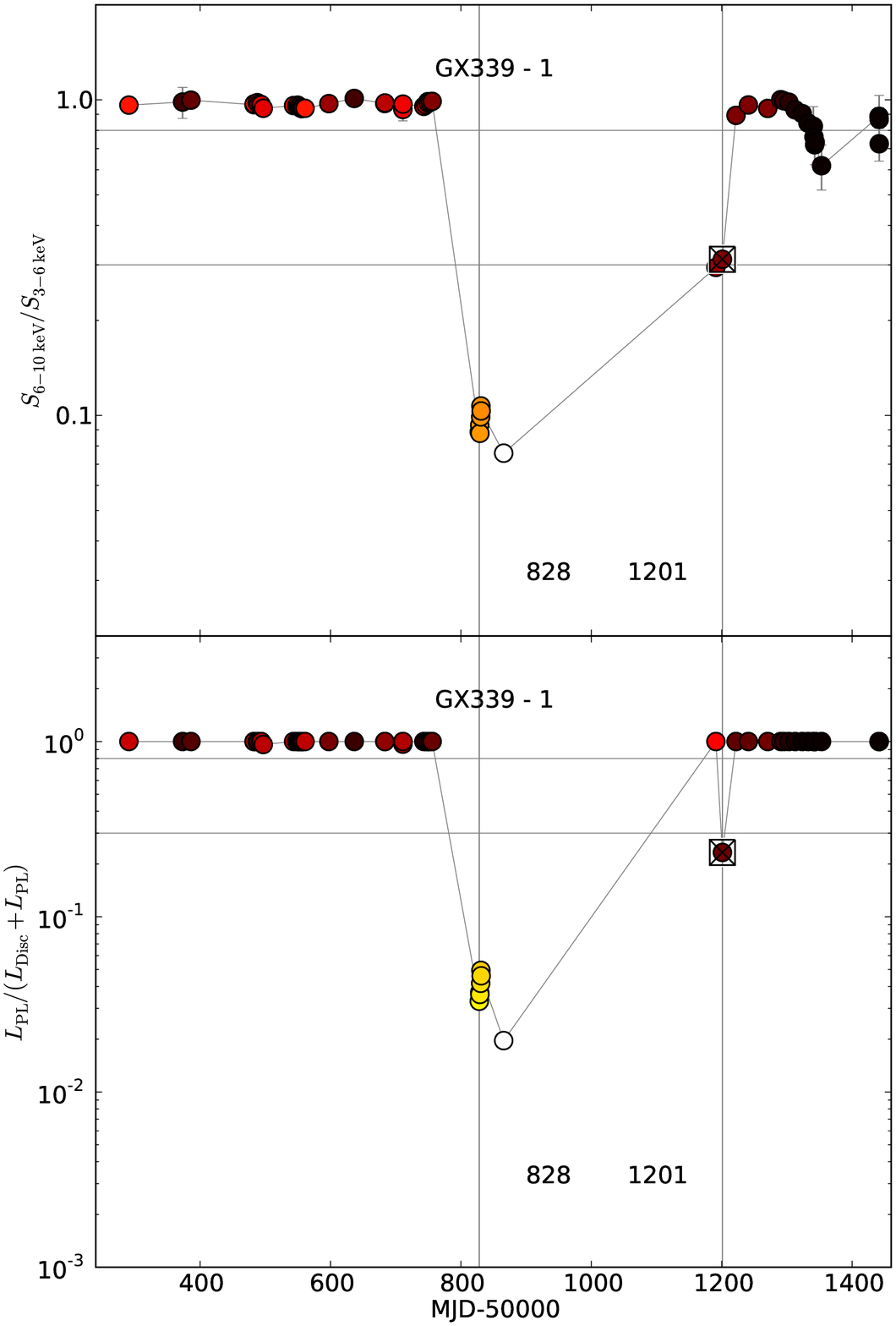}
\includegraphics[width=0.41\textwidth]{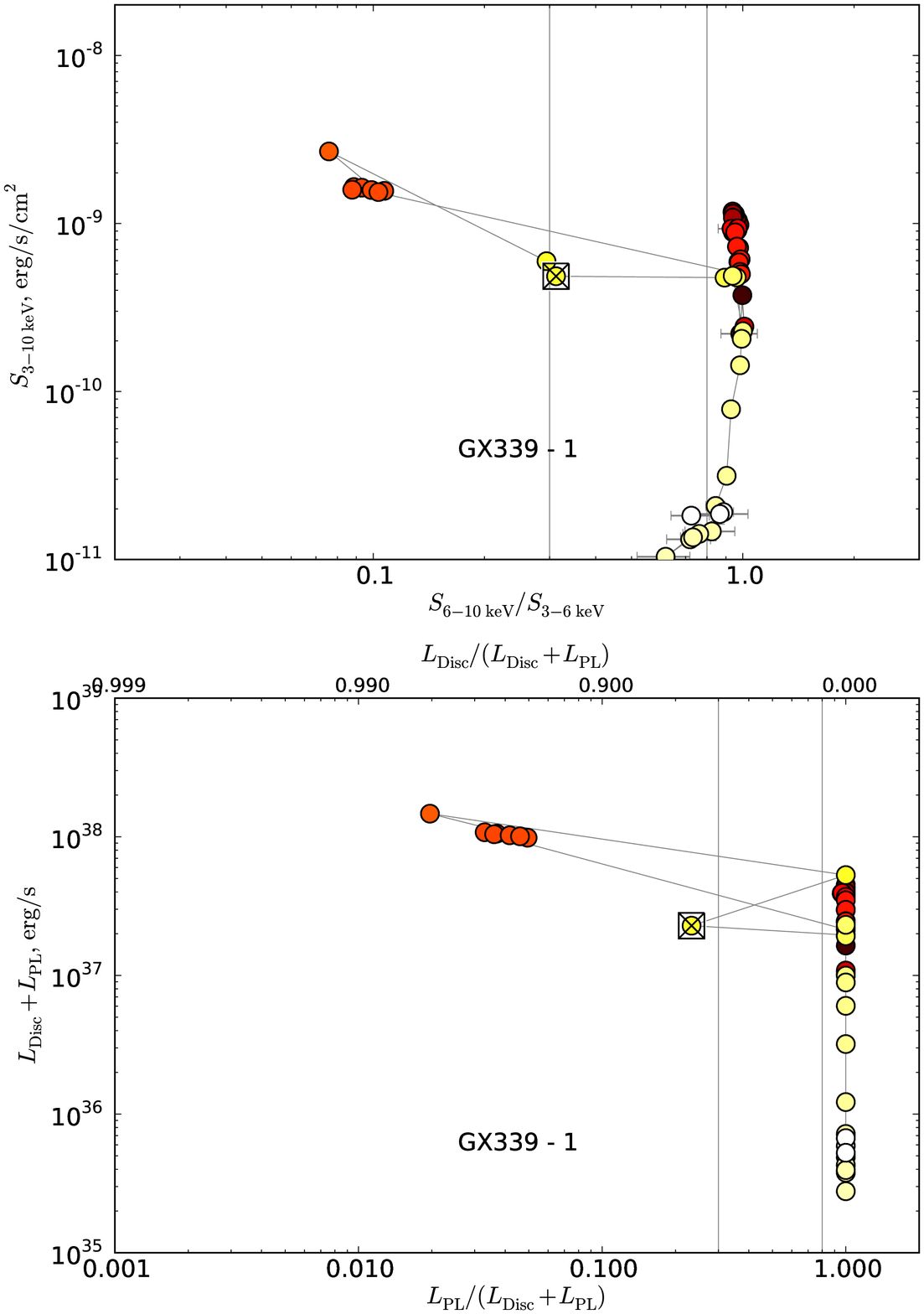}
\caption{(cont) X-ray colour curves and diagnostic diagrams for the
  outbursts observed - GX 339-4 Outburst 1.}
\end{figure*}
\addtocounter{figure}{-1}
\begin{figure*}
\centering
\includegraphics[width=0.41\textwidth]{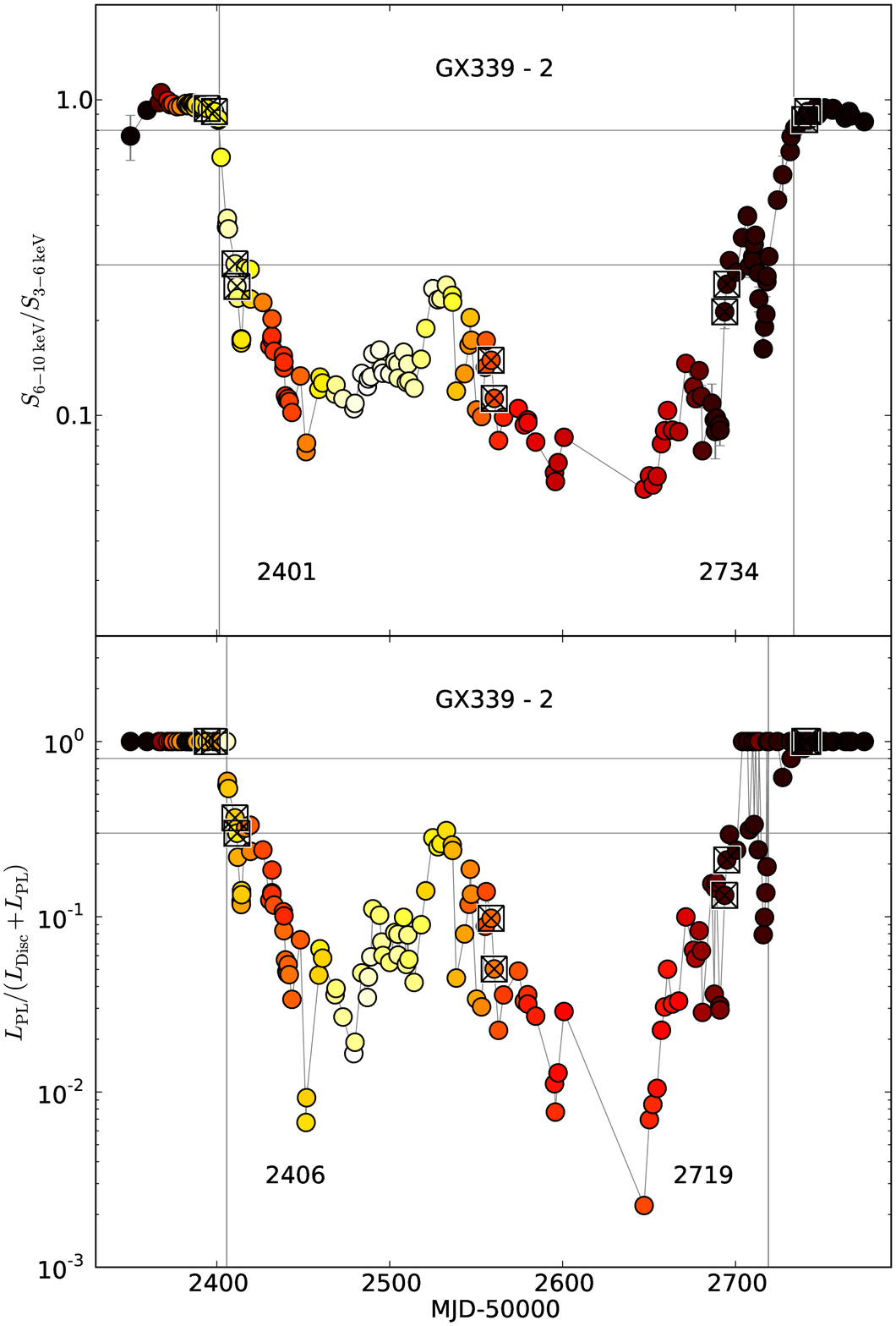}
\includegraphics[width=0.41\textwidth]{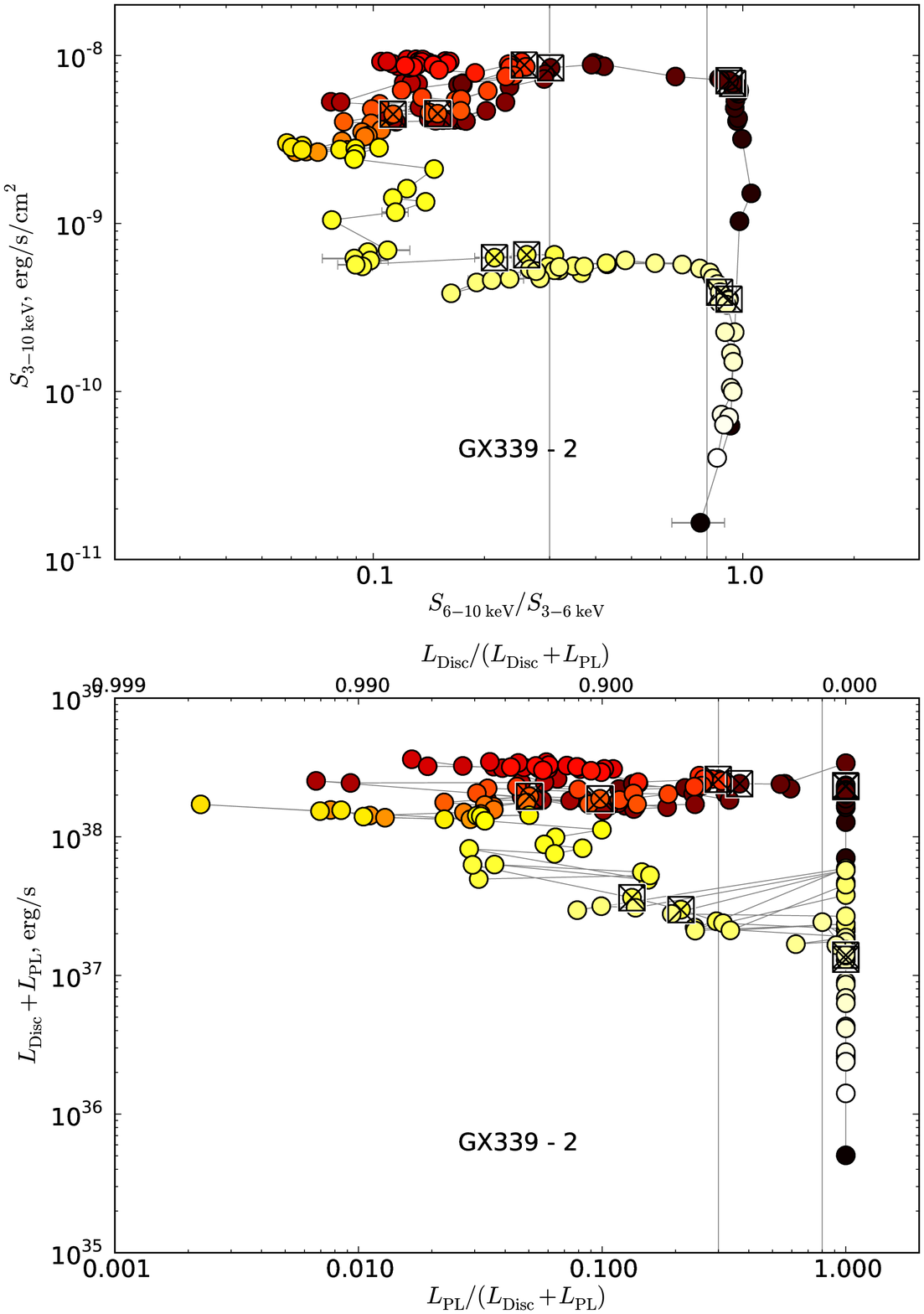}
\caption{(cont) X-ray colour curves and diagnostic diagrams for the
  outbursts observed - GX 339-4 Outburst 2.}
\end{figure*}
\addtocounter{figure}{-1}
\begin{figure*}
\centering
\includegraphics[width=0.41\textwidth]{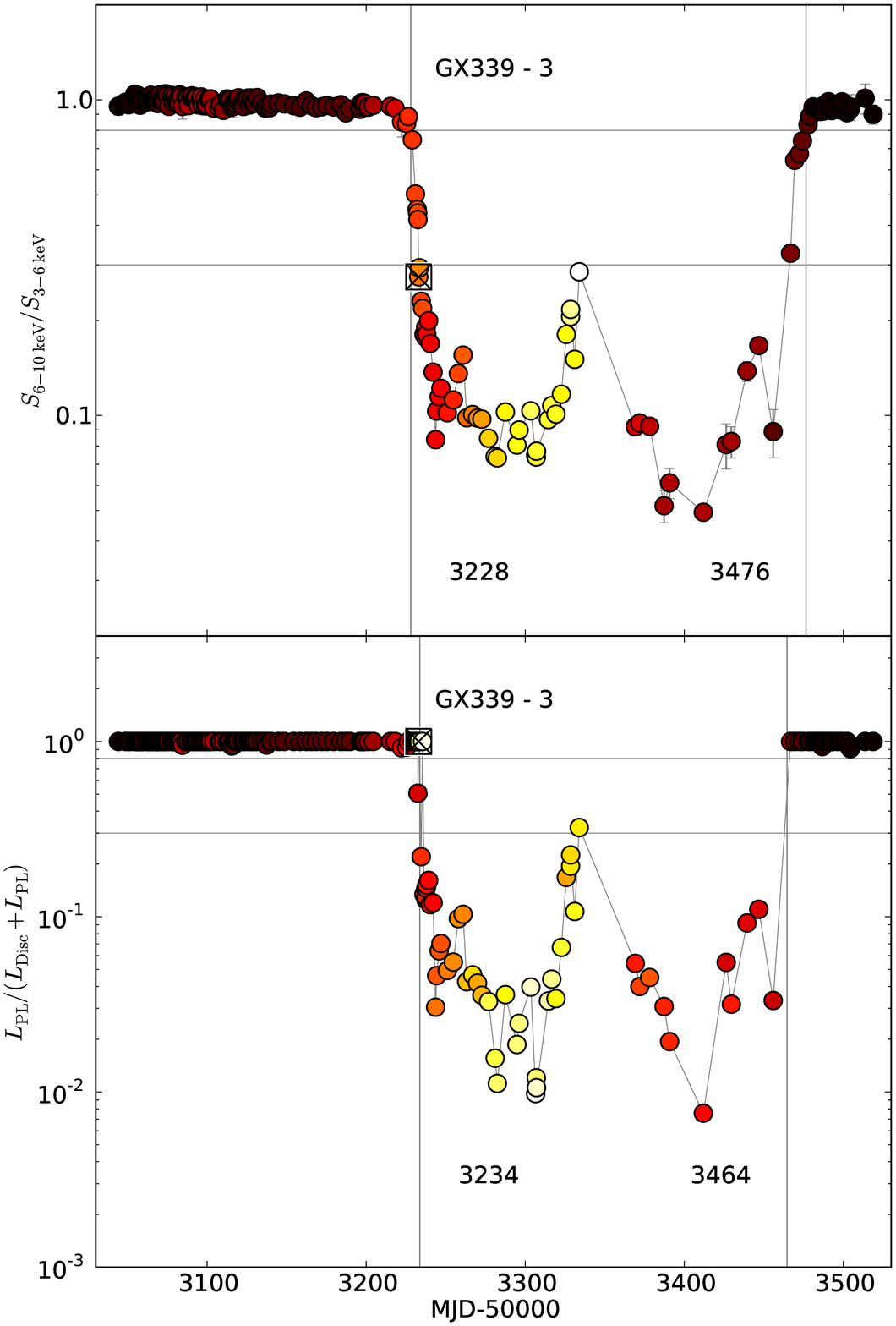}
\includegraphics[width=0.41\textwidth]{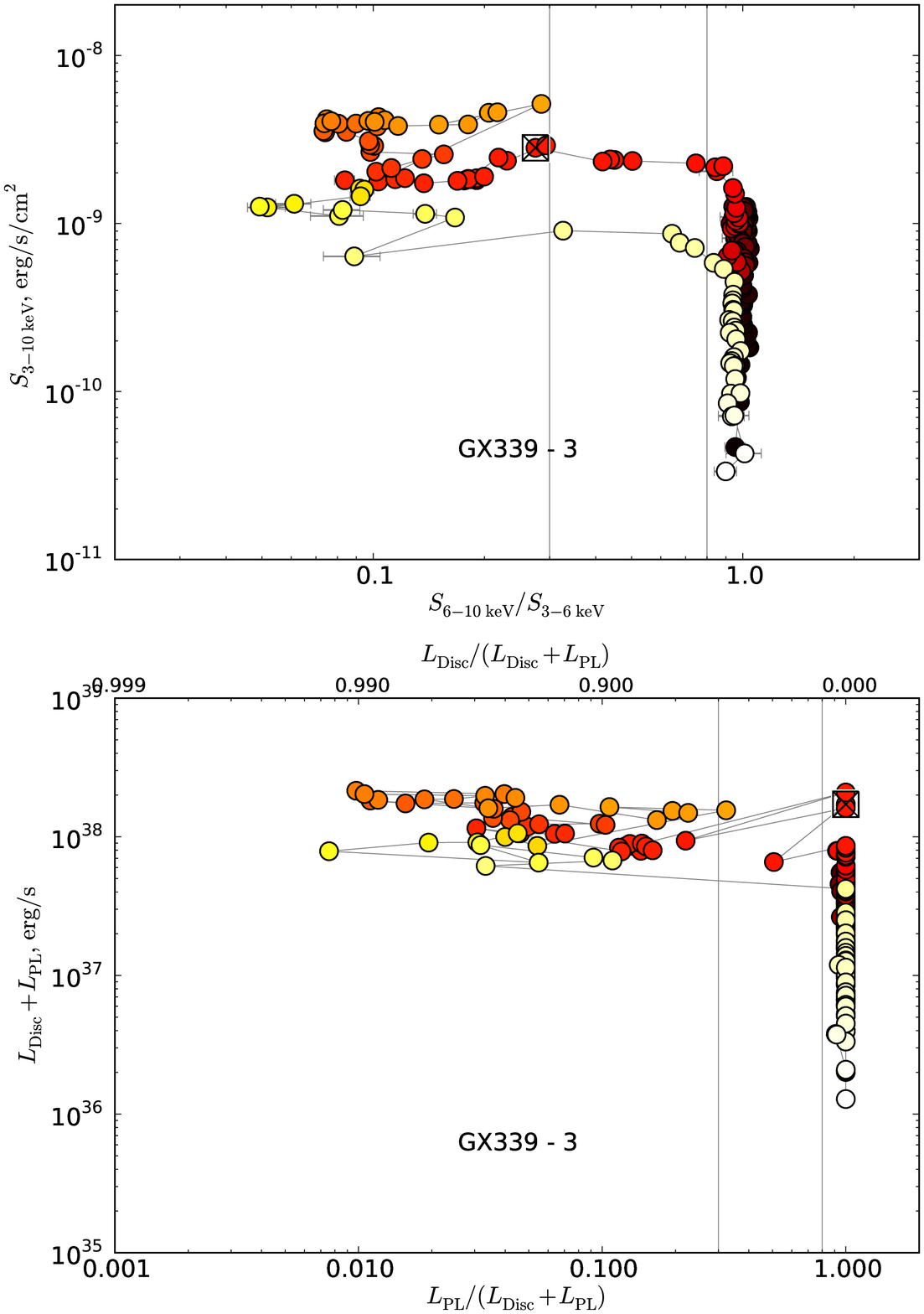}
\caption{(cont) X-ray colour curves and diagnostic diagrams for the
  outbursts observed - GX 339-4 Outburst 3.}
\end{figure*}
\addtocounter{figure}{-1}
\begin{figure*}
\centering
\includegraphics[width=0.41\textwidth]{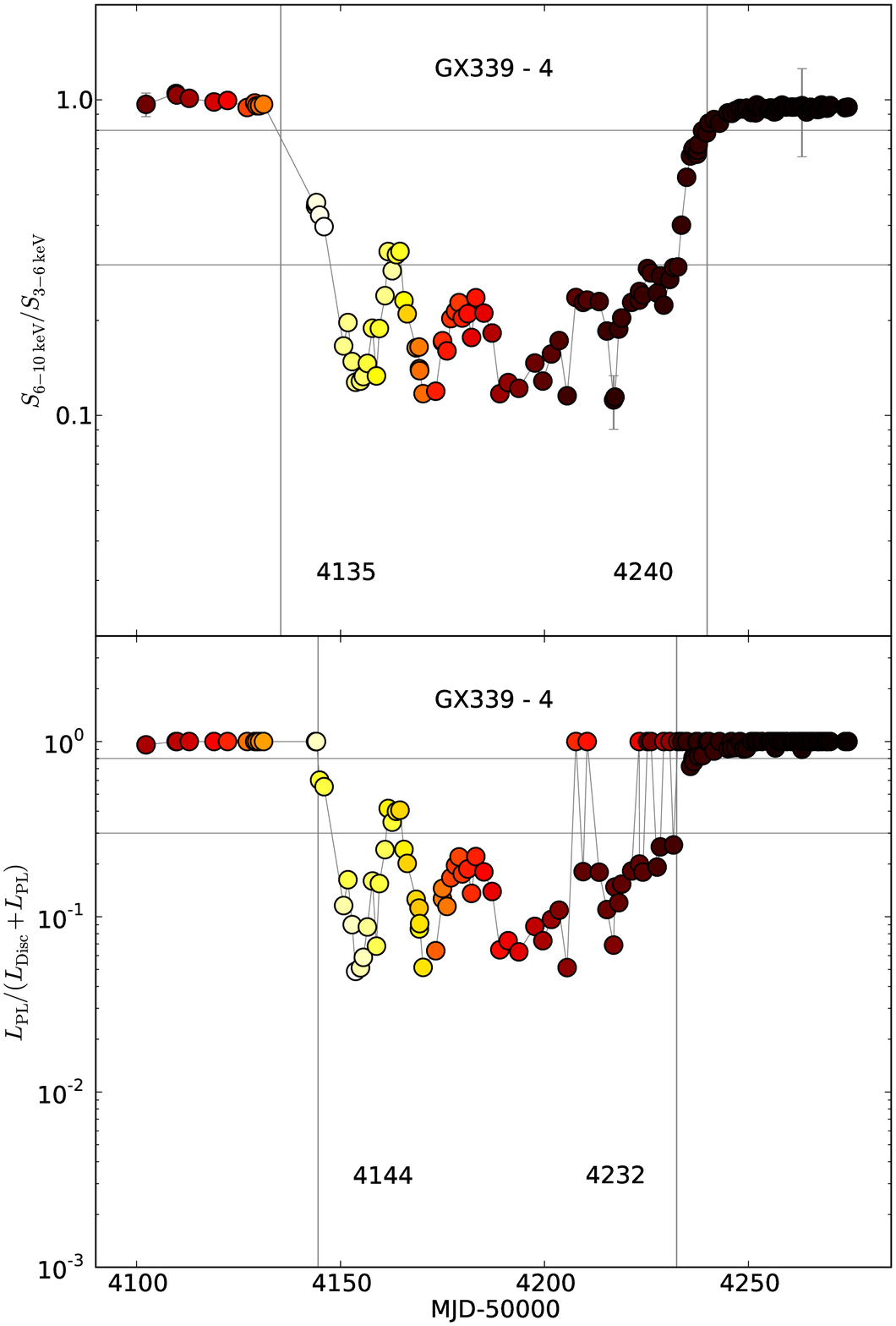}
\includegraphics[width=0.41\textwidth]{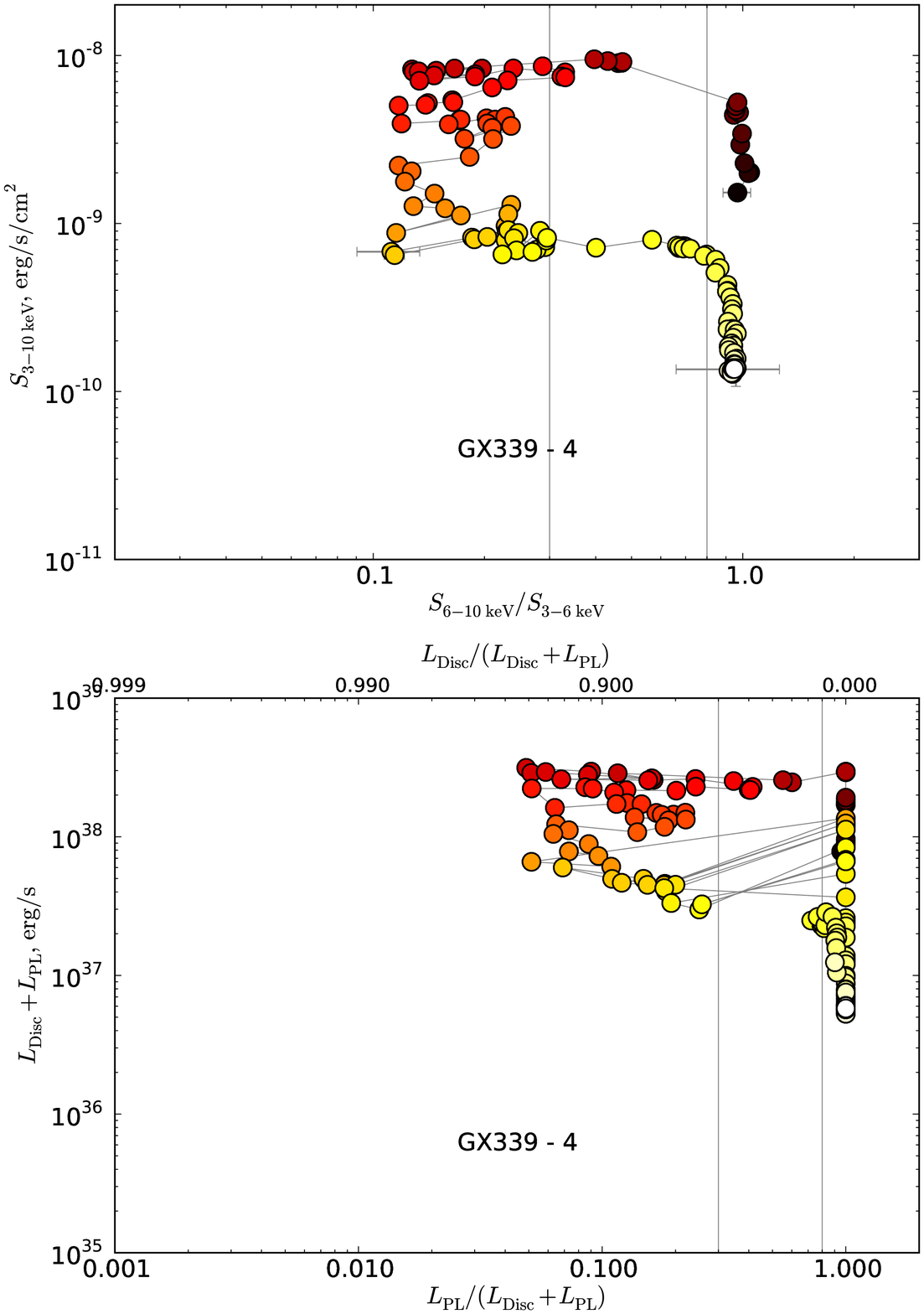}
\caption{(cont) X-ray colour curves and diagnostic diagrams for the
  outbursts observed - GX 339-4 Outburst 4.}
\end{figure*}
\addtocounter{figure}{-1}
\begin{figure*}
\centering
\includegraphics[width=0.41\textwidth]{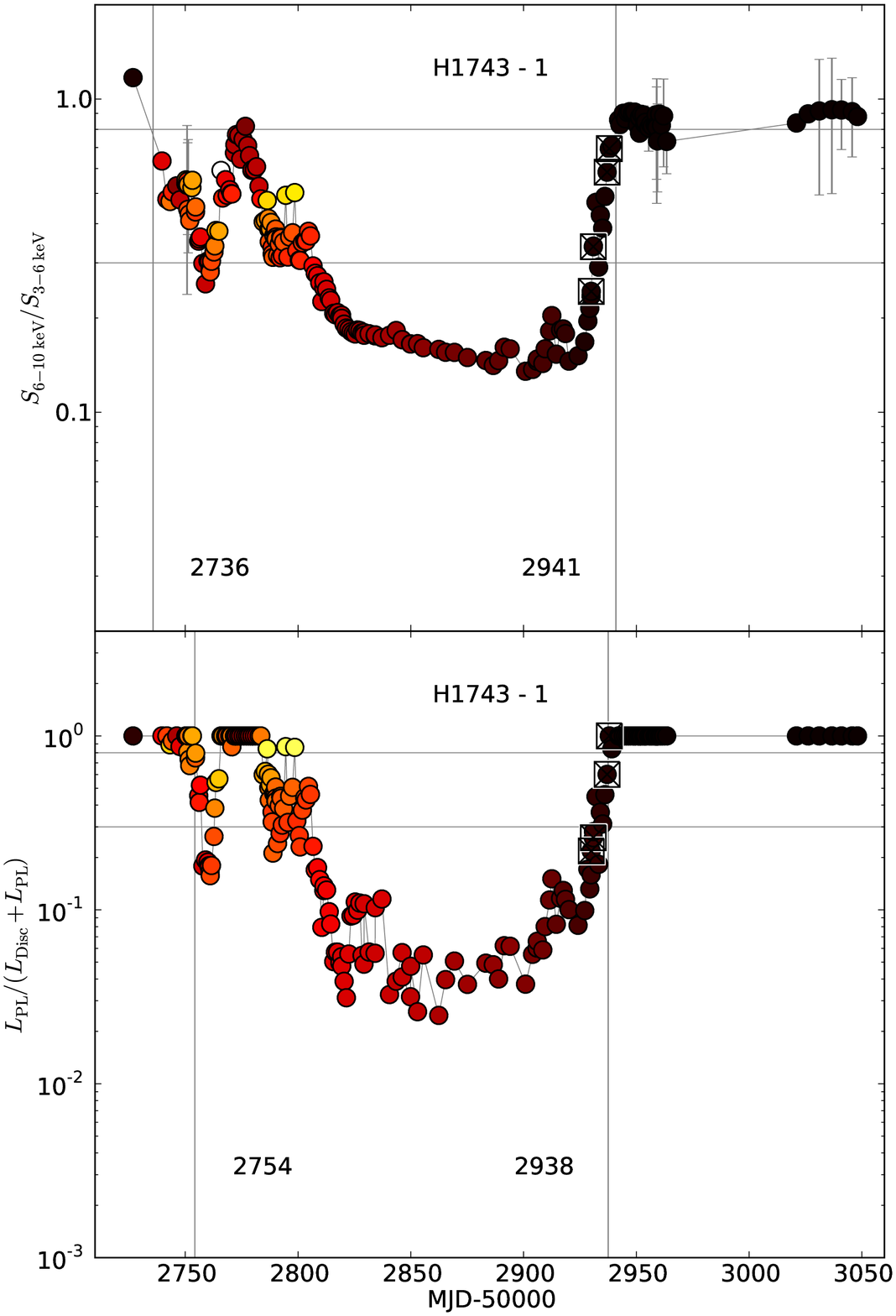}
\includegraphics[width=0.41\textwidth]{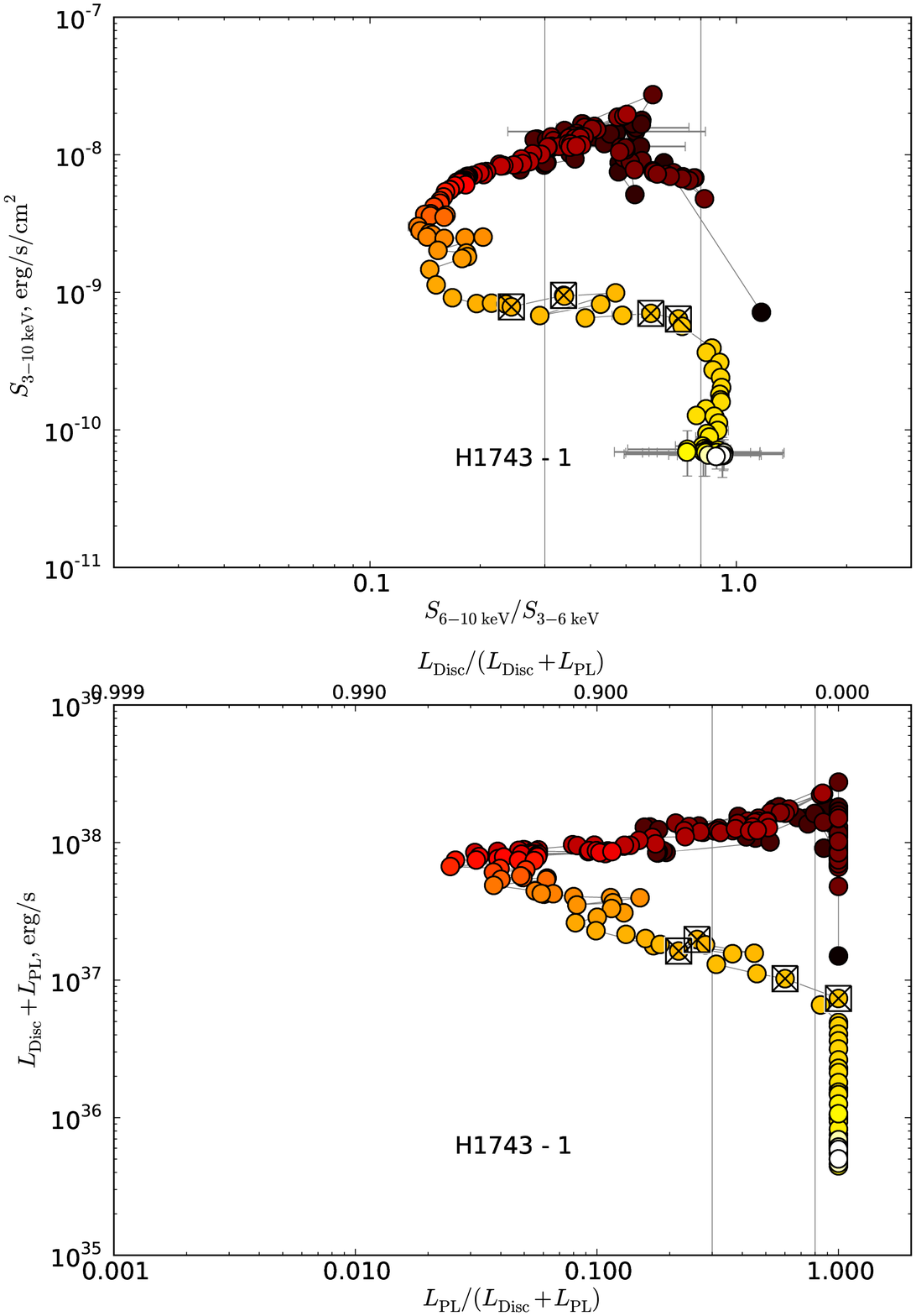}
\caption{(cont) X-ray colour curves and diagnostic diagrams for the
  outbursts observed - H1743-332 Outburst 1.}
\end{figure*}
\clearpage
\addtocounter{figure}{-1}
\begin{figure*}
\centering
\includegraphics[width=0.41\textwidth]{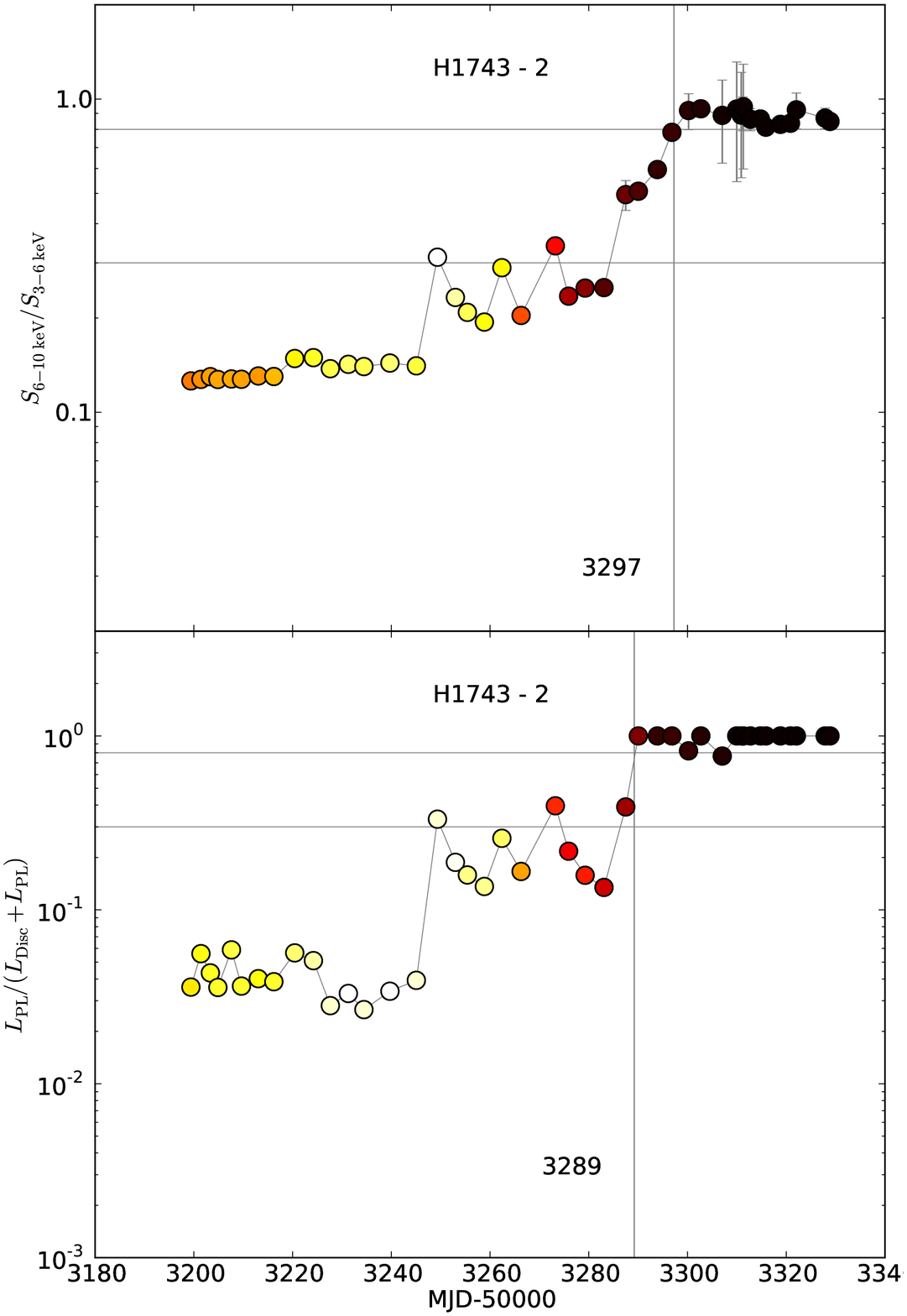}
\includegraphics[width=0.41\textwidth]{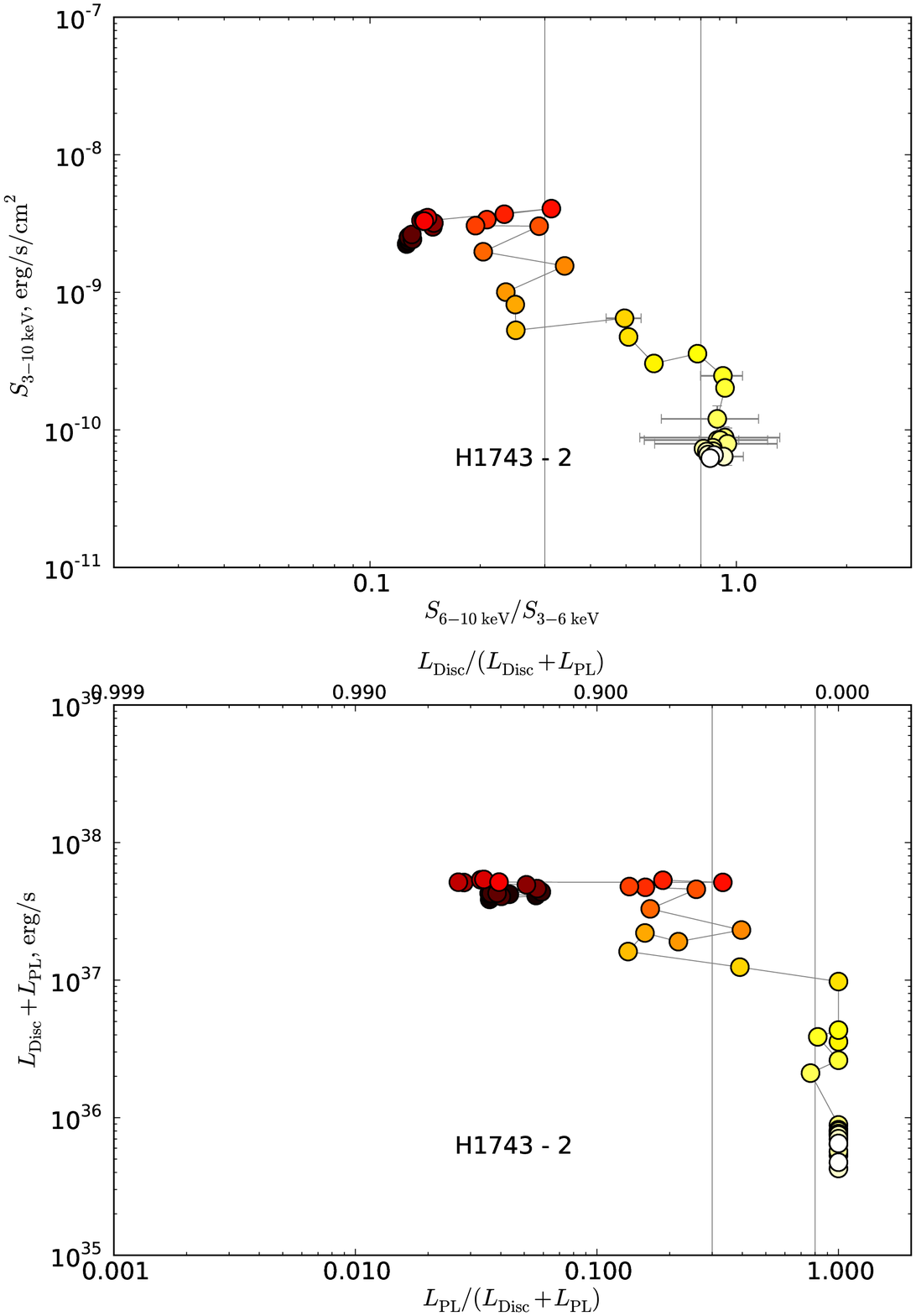}
\caption{(cont) X-ray colour curves and diagnostic diagrams for the
  outbursts observed - H1743-332 Outburst 2.}
\end{figure*}
\addtocounter{figure}{-1}
\begin{figure*}
\centering
\includegraphics[width=0.41\textwidth]{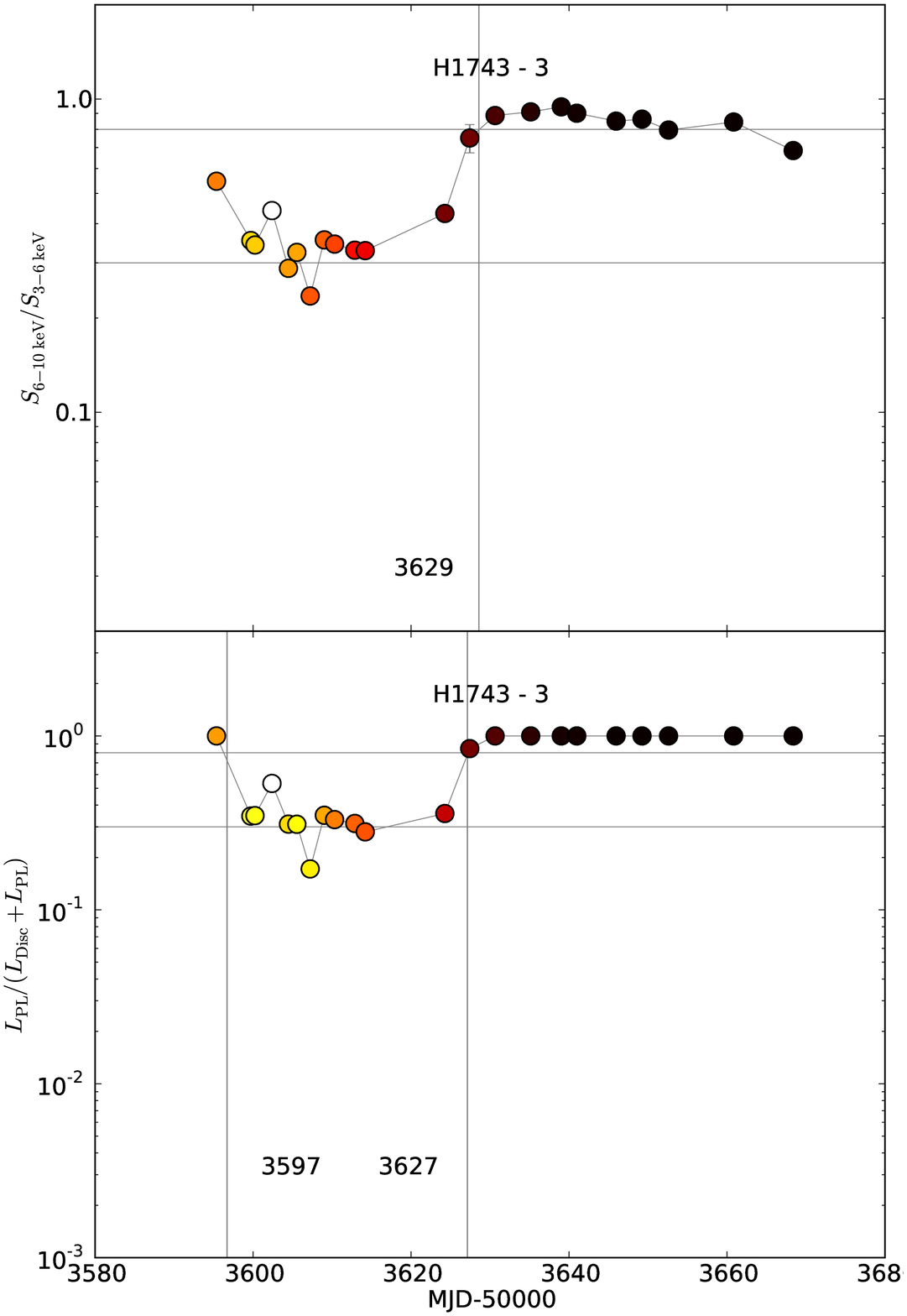}
\includegraphics[width=0.41\textwidth]{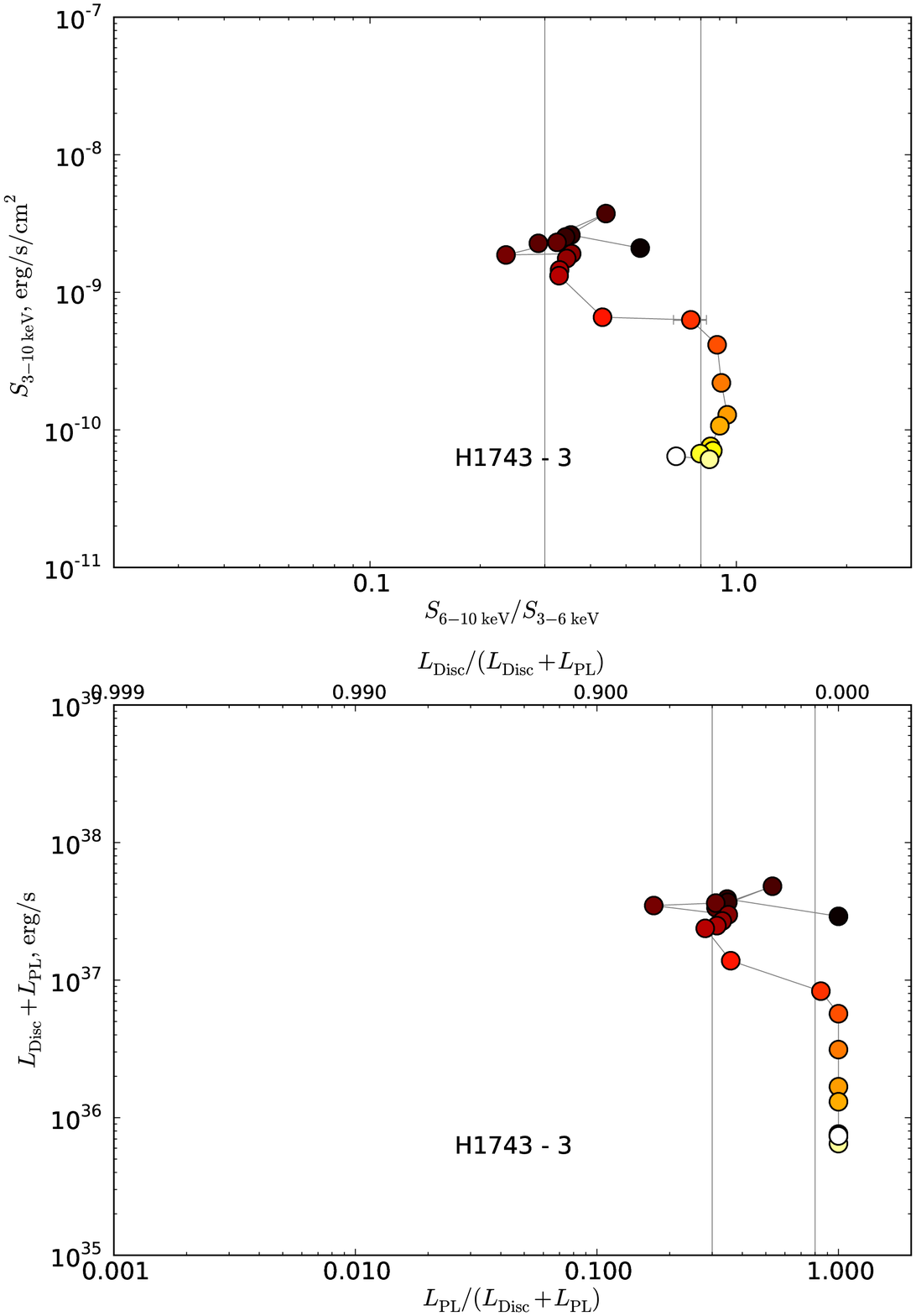}
\caption{(cont) X-ray colour curves and diagnostic diagrams for the
  outbursts observed - H1743-332 Outburst 3.}
\end{figure*}
\addtocounter{figure}{-1}
\begin{figure*}
\centering
\includegraphics[width=0.41\textwidth]{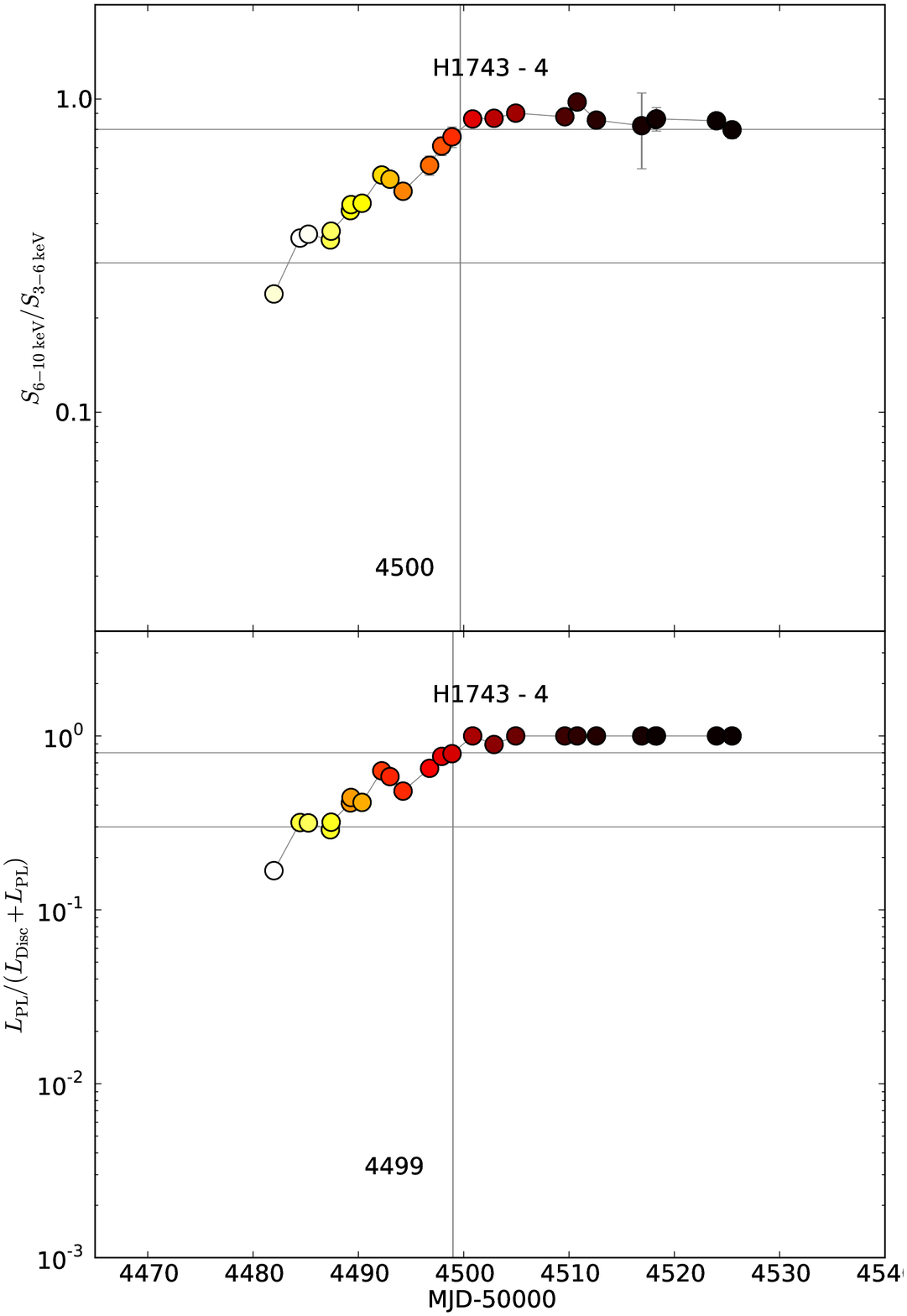}
\includegraphics[width=0.41\textwidth]{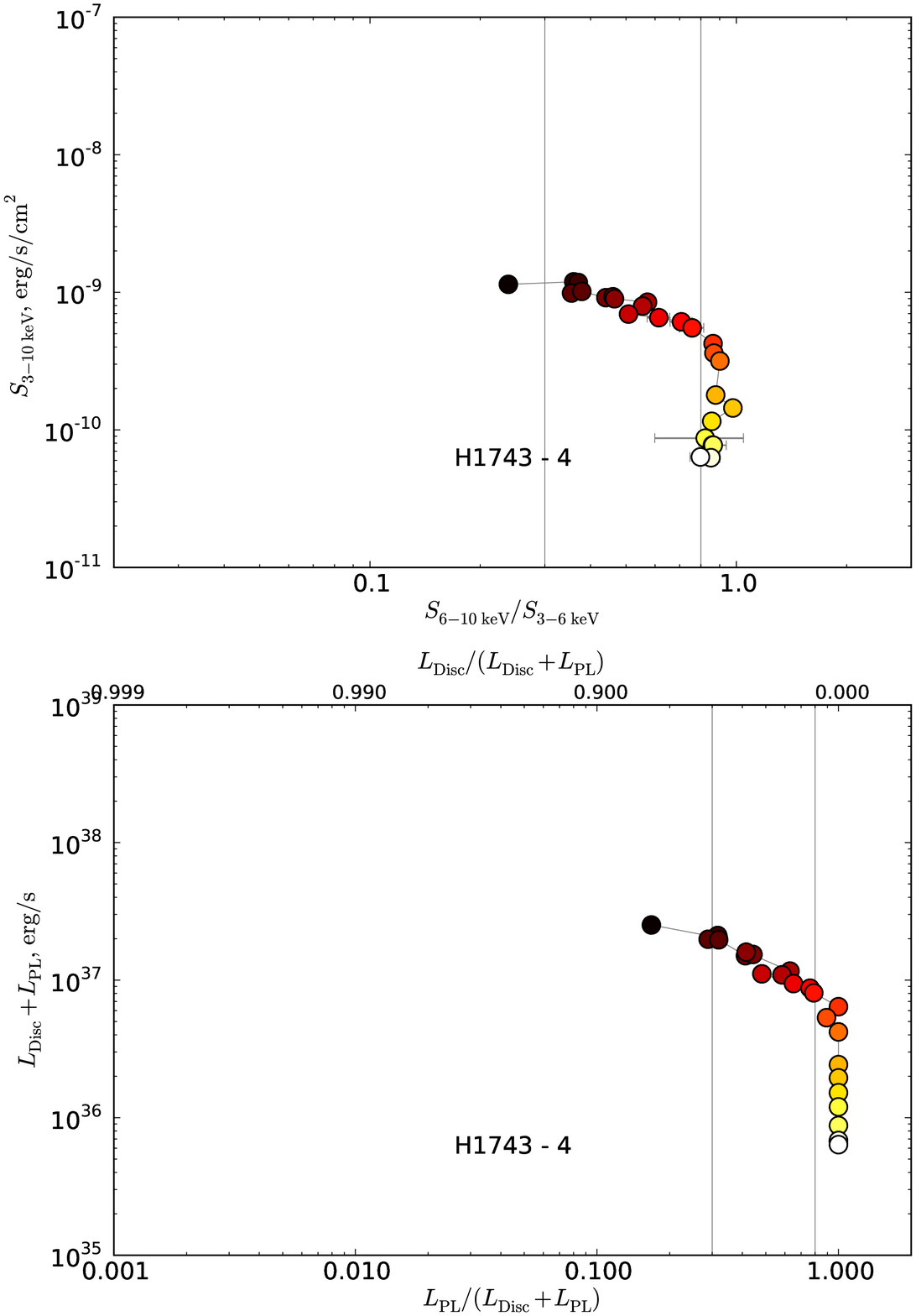}
\caption{(cont) X-ray colour curves and diagnostic diagrams for the
  outbursts observed - H1743-332 Outburst 4.}
\end{figure*}
\addtocounter{figure}{-1}
\begin{figure*}
\centering
\includegraphics[width=0.41\textwidth]{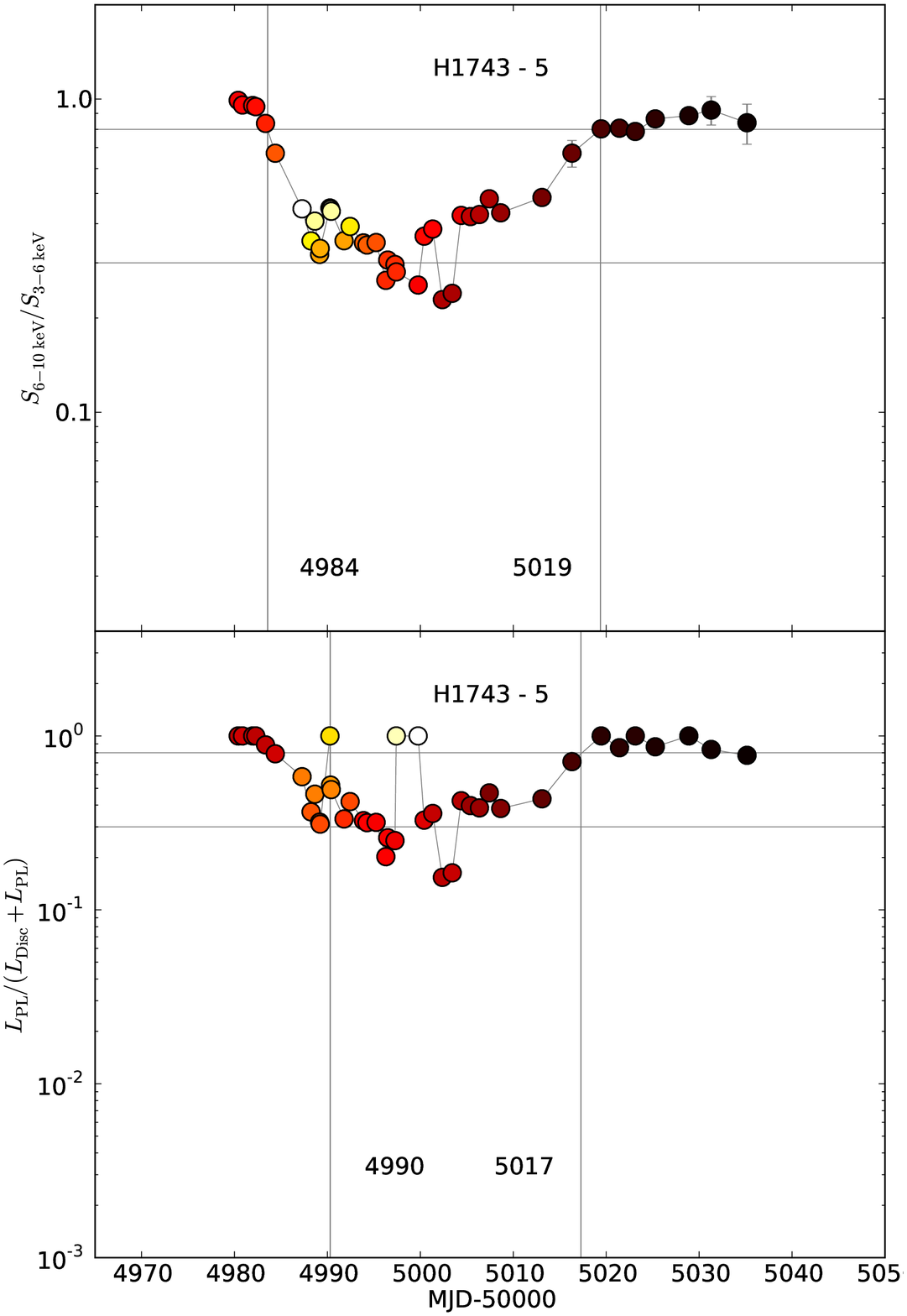}
\includegraphics[width=0.41\textwidth]{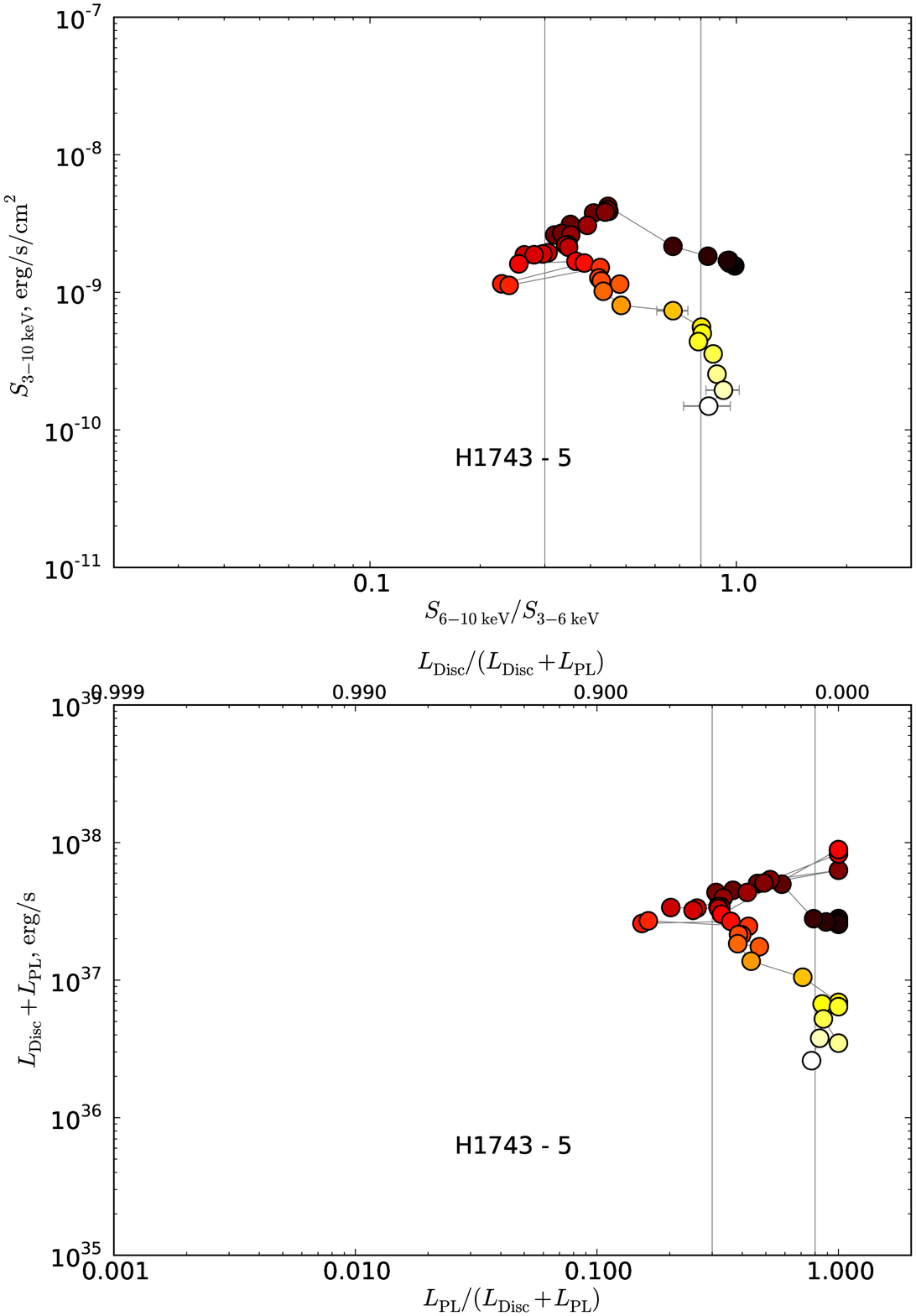}
\caption{(cont) X-ray colour curves and diagnostic diagrams for the
  outbursts observed - H1743-332 Outburst 5.}
\end{figure*}
\addtocounter{figure}{-1}
\begin{figure*}
\centering
\includegraphics[width=0.41\textwidth]{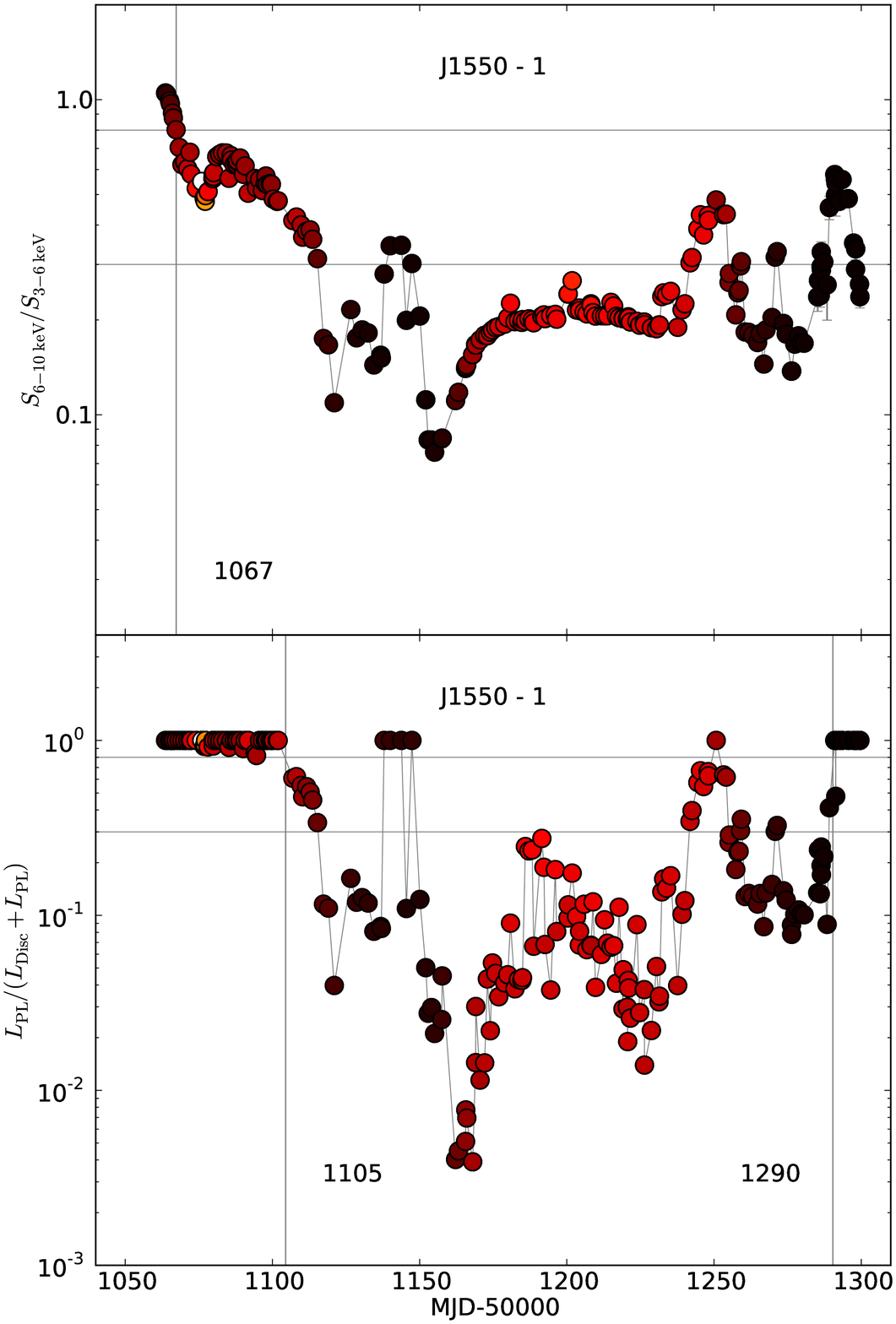}
\includegraphics[width=0.41\textwidth]{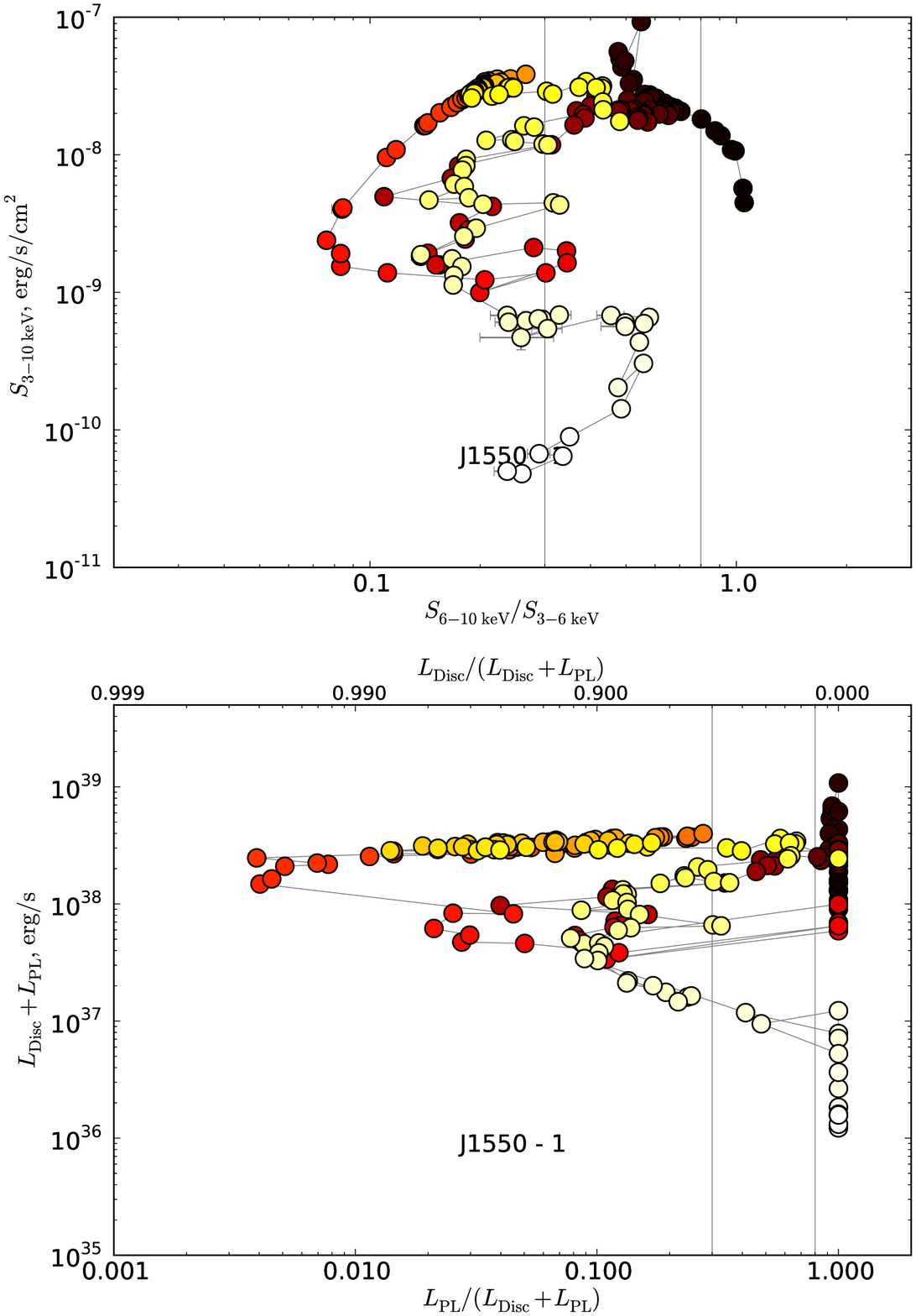}
\caption{(cont) X-ray colour curves and diagnostic diagrams for the
  outbursts observed - XTE J1550-564 Outburst 1.}
\end{figure*}
\addtocounter{figure}{-1}
\begin{figure*}
\centering
\includegraphics[width=0.41\textwidth]{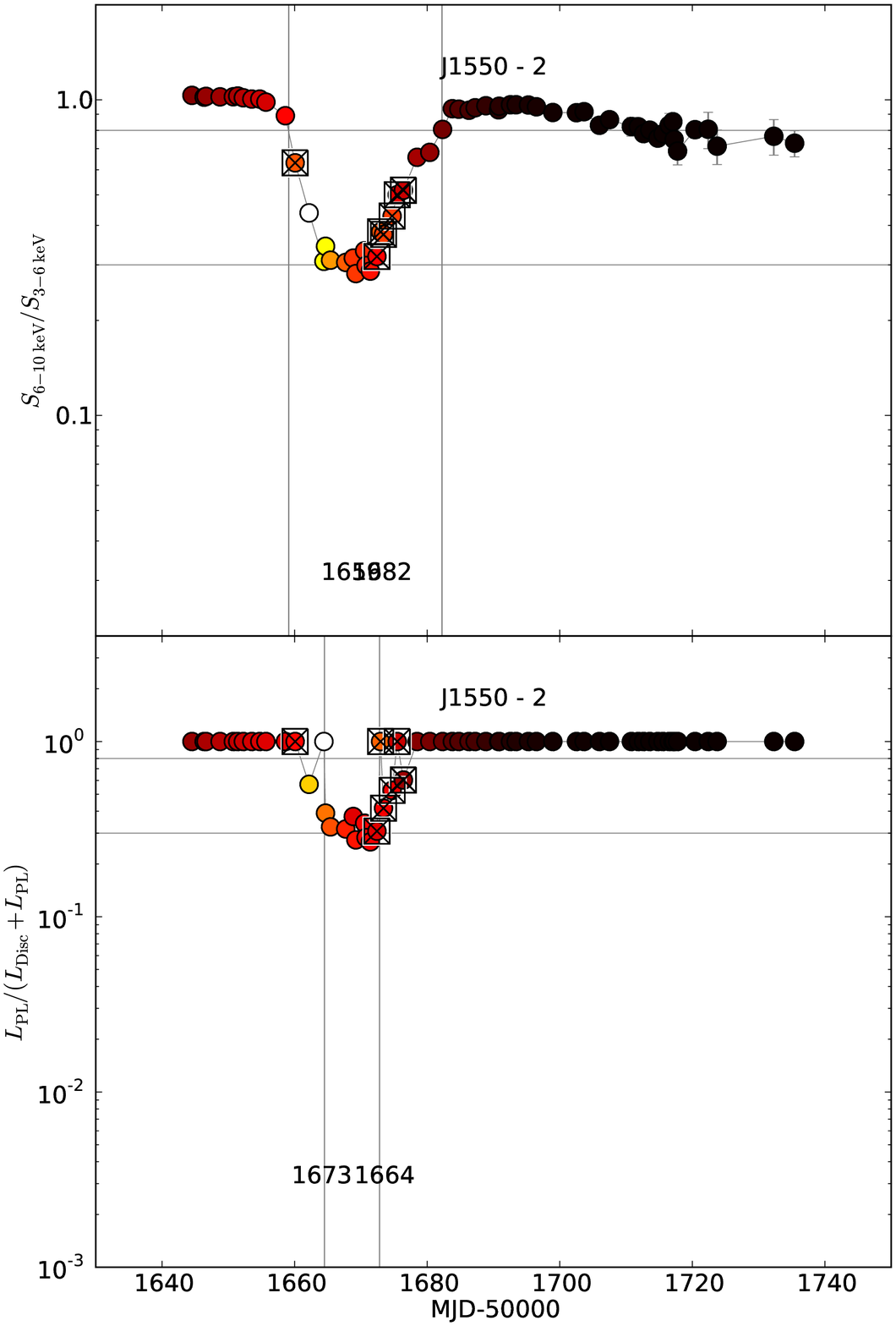}
\includegraphics[width=0.41\textwidth]{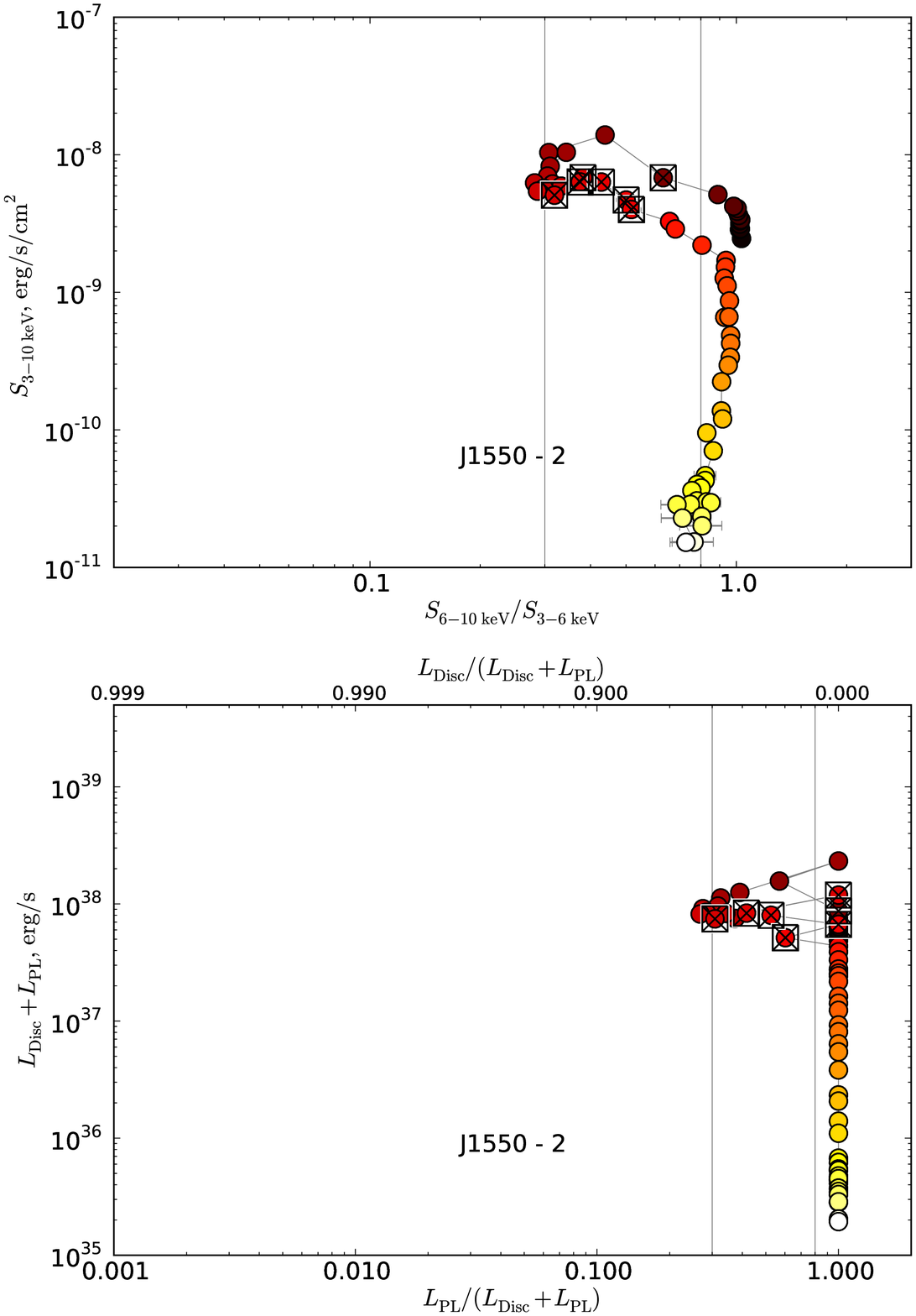}
\caption{(cont) X-ray colour curves and diagnostic diagrams for the
  outbursts observed - XTE J1550-564 Outburst 2.}
\end{figure*}
\addtocounter{figure}{-1}
\begin{figure*}
\centering
\includegraphics[width=0.41\textwidth]{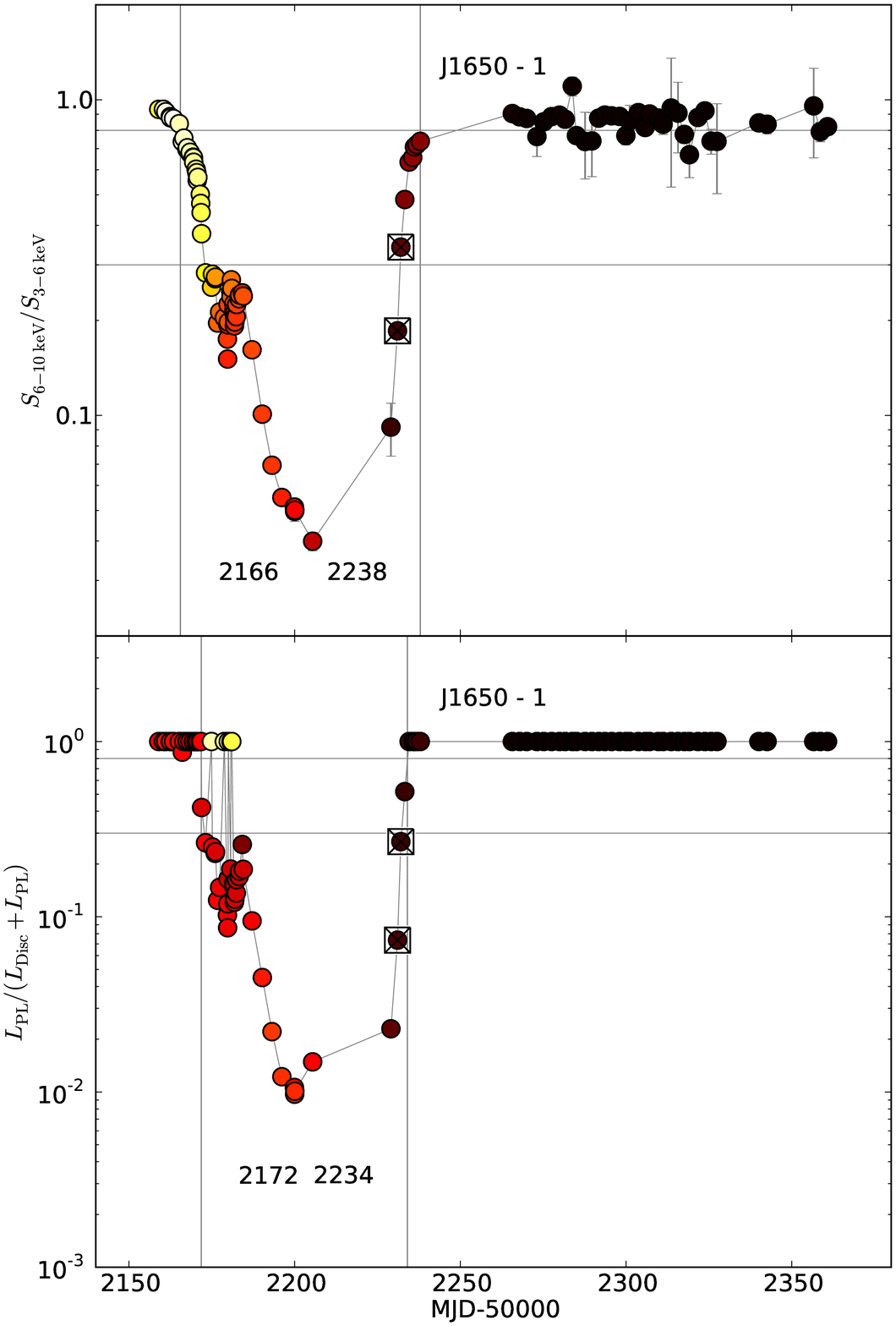}
\includegraphics[width=0.41\textwidth]{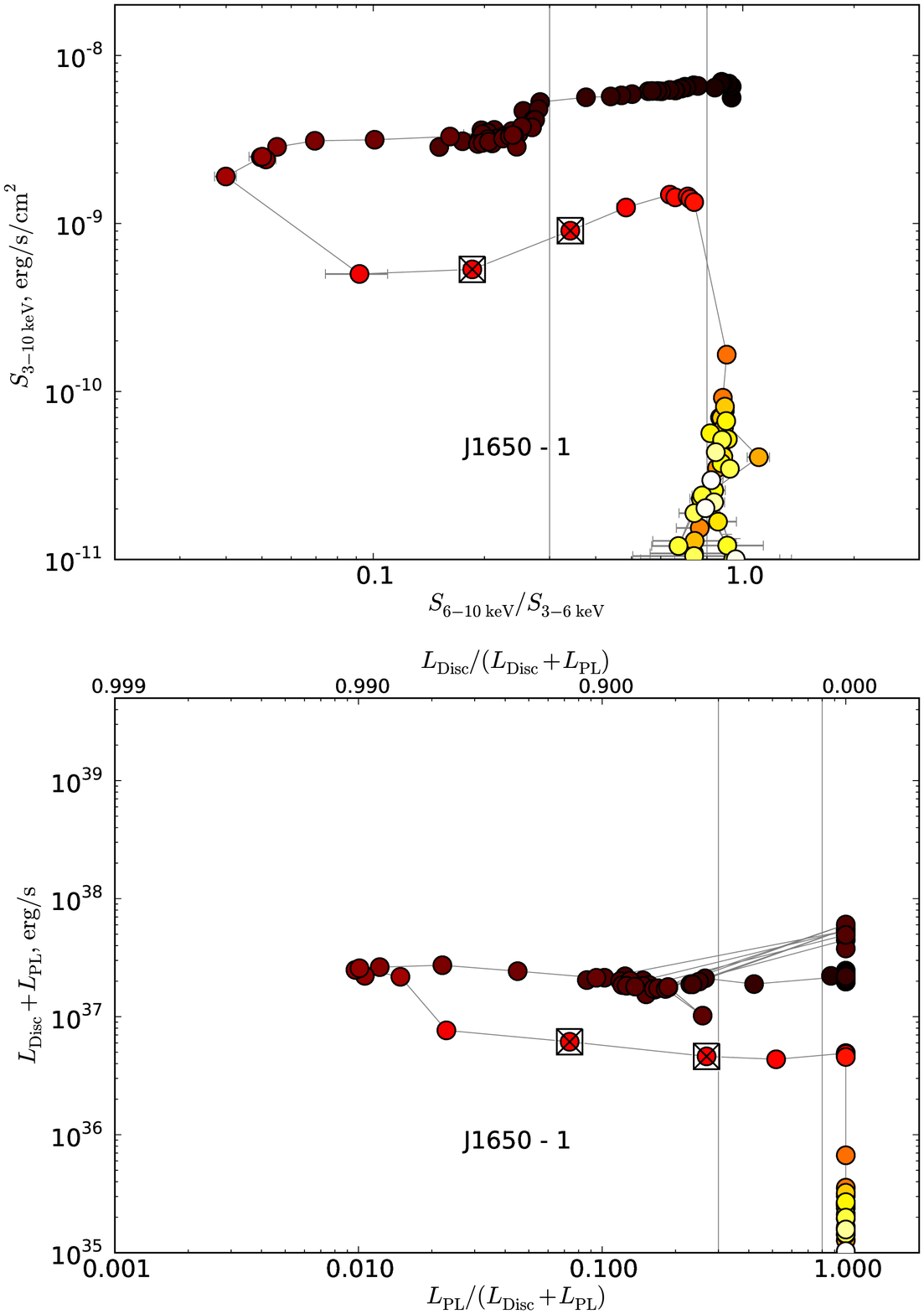}
\caption{(cont) X-ray colour curves and diagnostic diagrams for the
  outbursts observed - XTE J1650-500 Outburst 1.}
\end{figure*}
\addtocounter{figure}{-1}
\begin{figure*}
\centering
\includegraphics[width=0.41\textwidth]{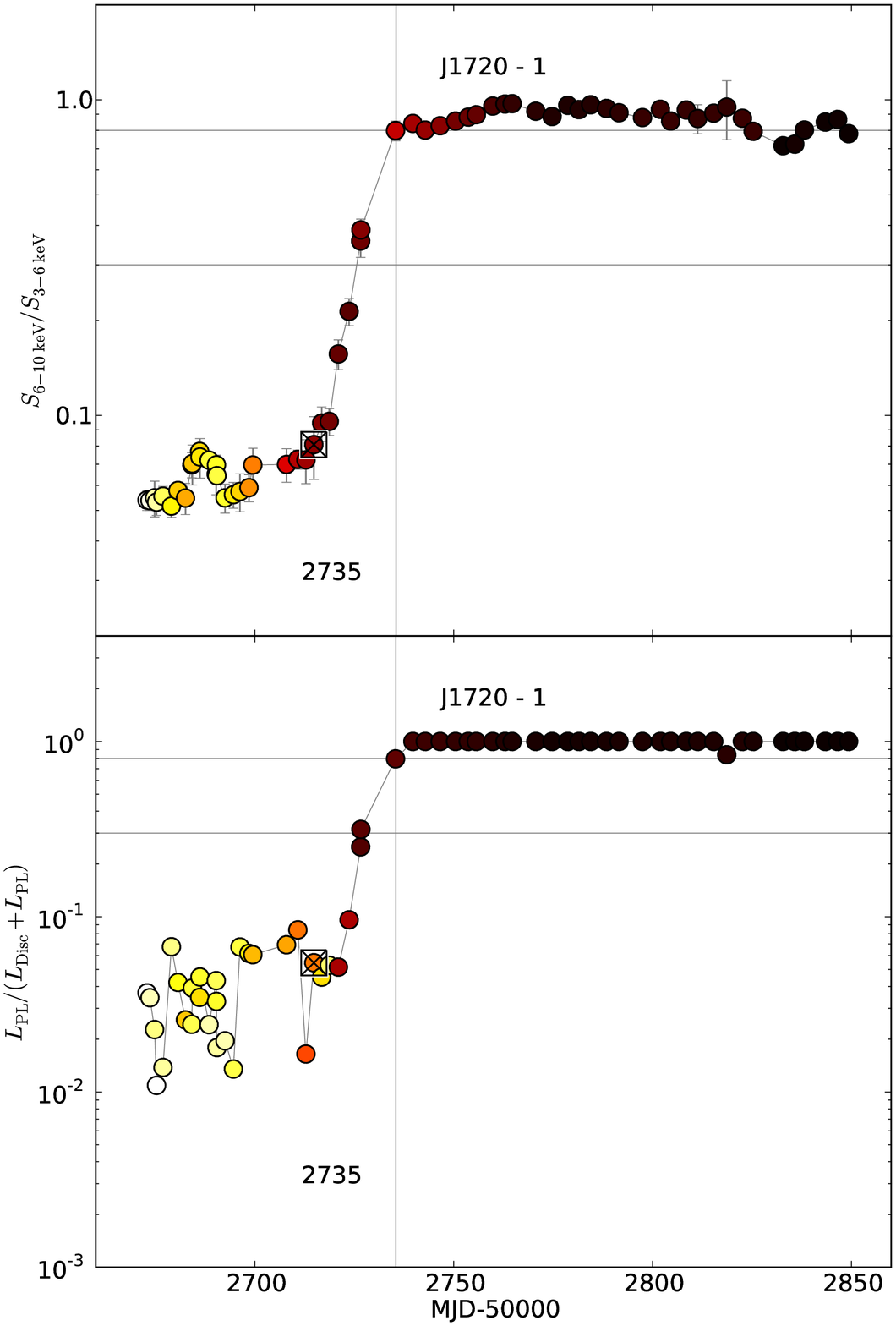}
\includegraphics[width=0.41\textwidth]{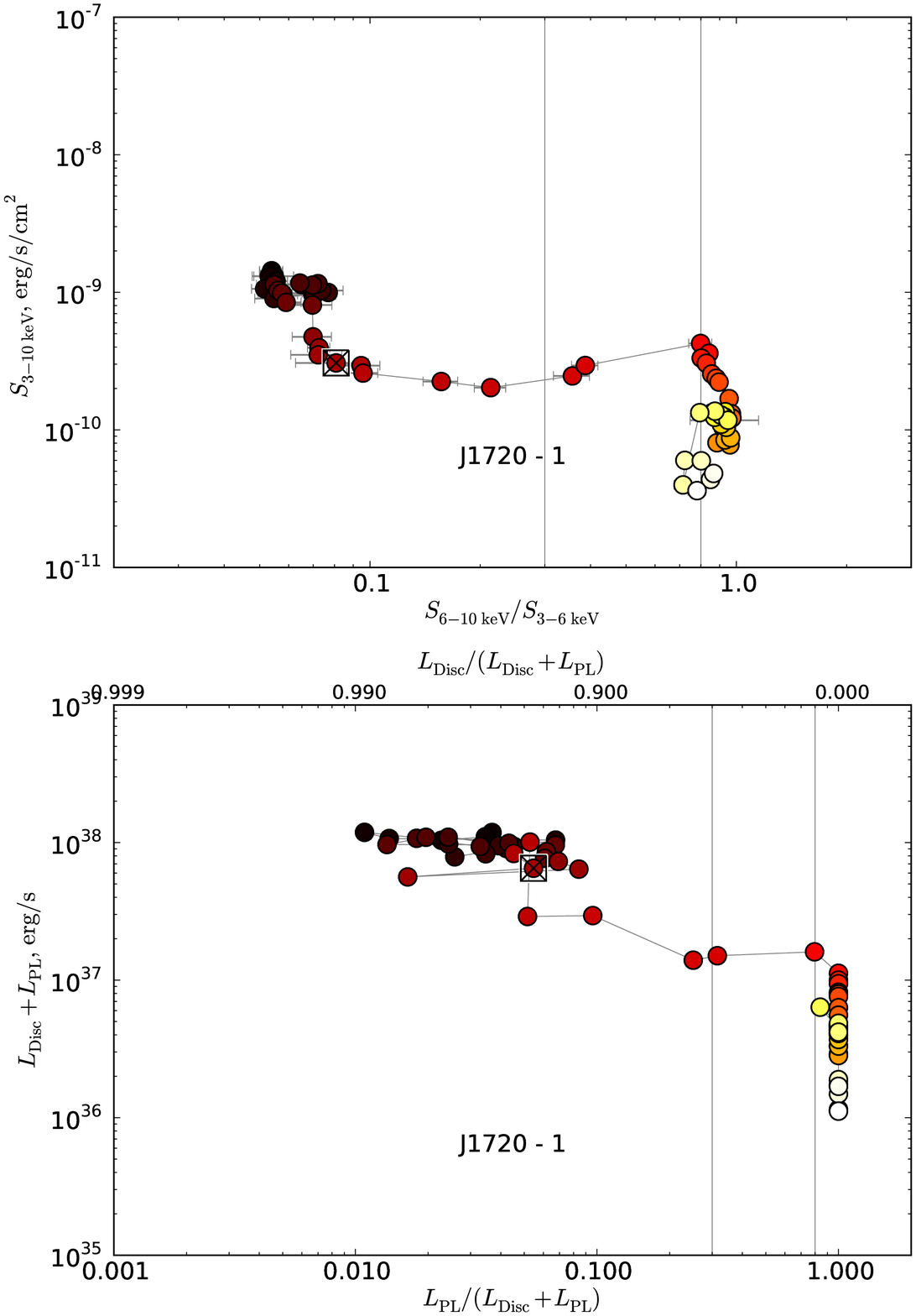}
\caption{(cont) X-ray colour curves and diagnostic diagrams for the
  outbursts observed - XTE J1720-318 Outburst 1.}
\end{figure*}
\addtocounter{figure}{-1}
\begin{figure*}
\centering
\includegraphics[width=0.41\textwidth]{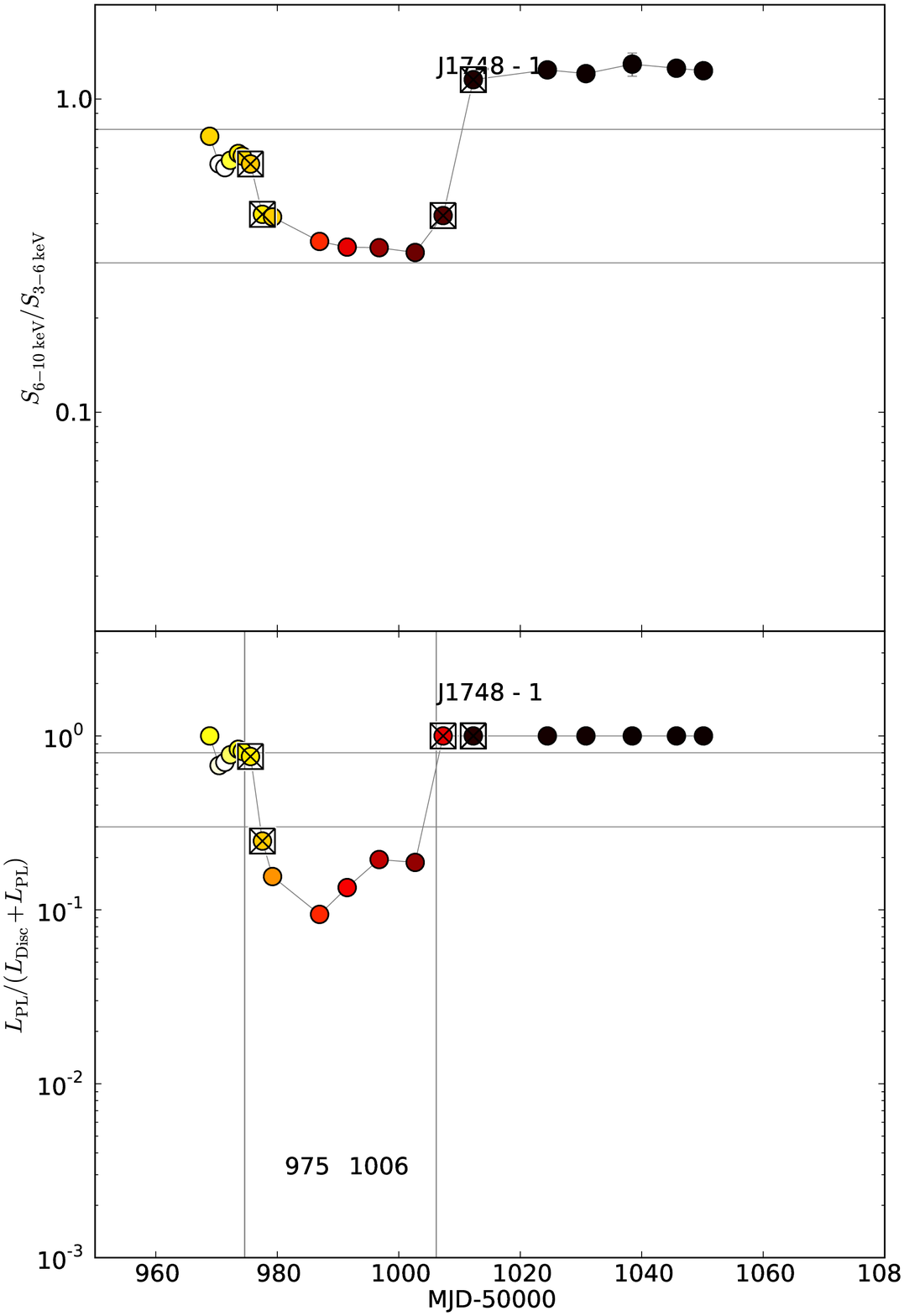}
\includegraphics[width=0.41\textwidth]{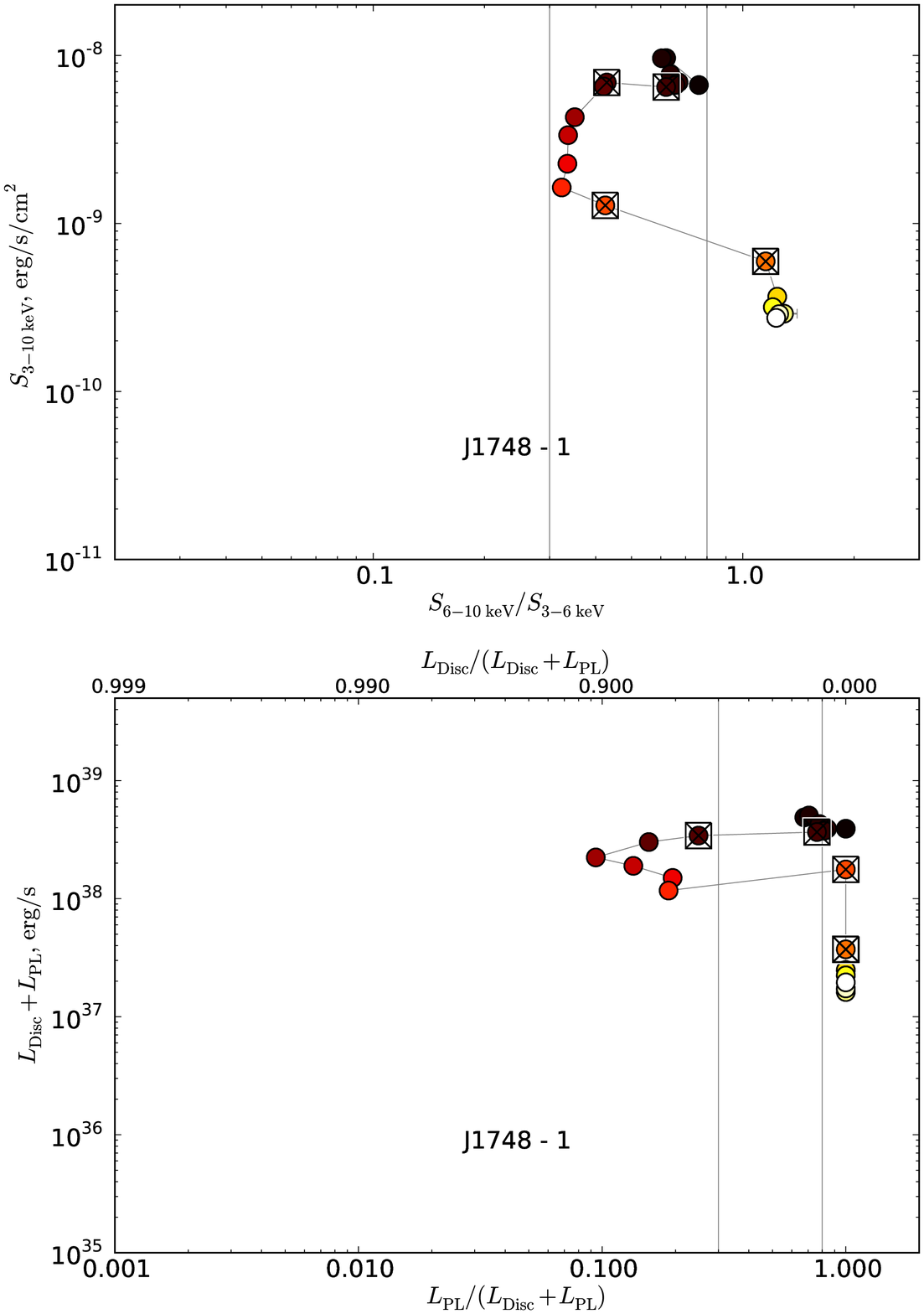}
\caption{(cont) X-ray colour curves and diagnostic diagrams for the
  outbursts observed - XTE J1748-2888 Outburst 1.}
\end{figure*}
\addtocounter{figure}{-1}
\begin{figure*}
\centering
\includegraphics[width=0.41\textwidth]{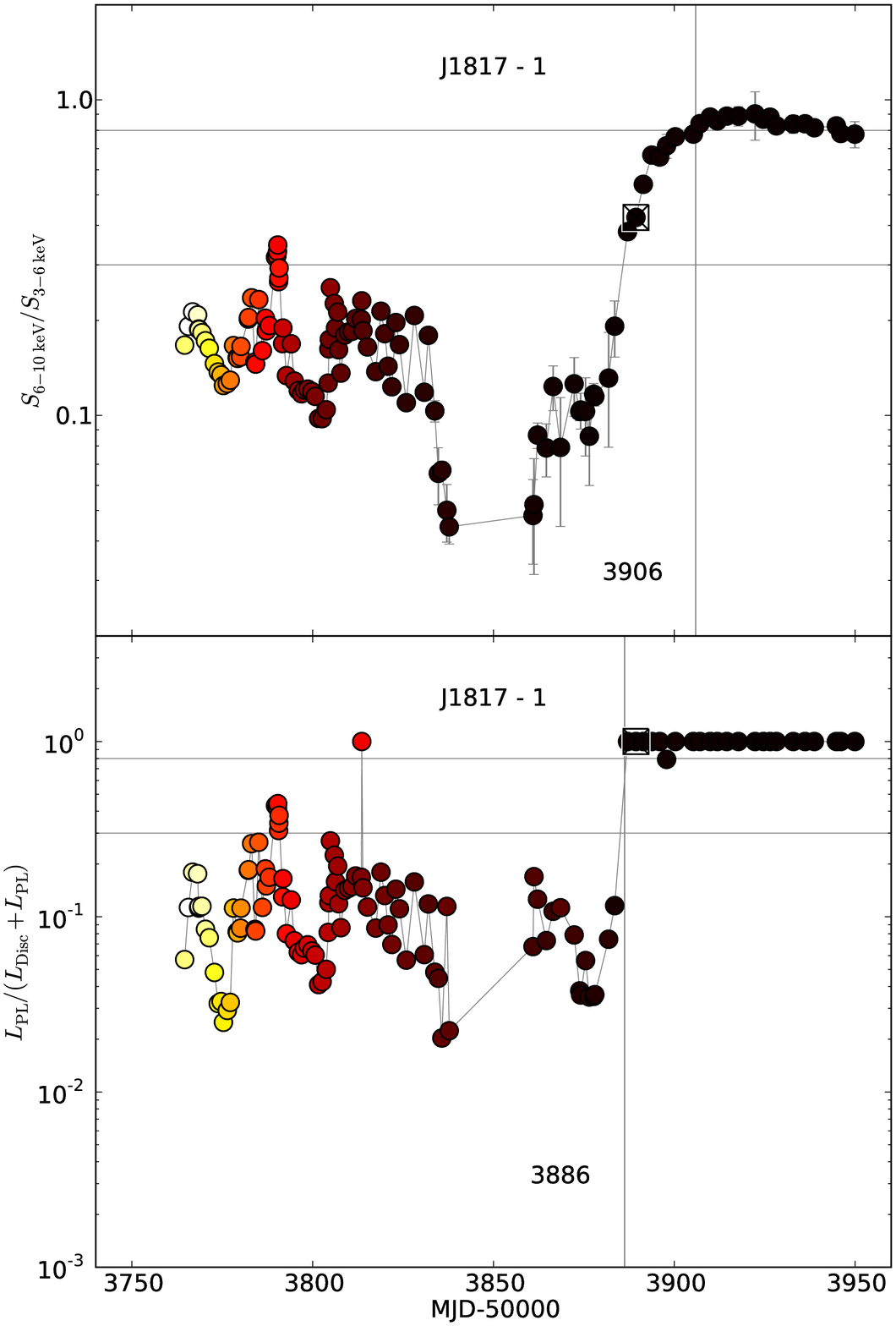}
\includegraphics[width=0.41\textwidth]{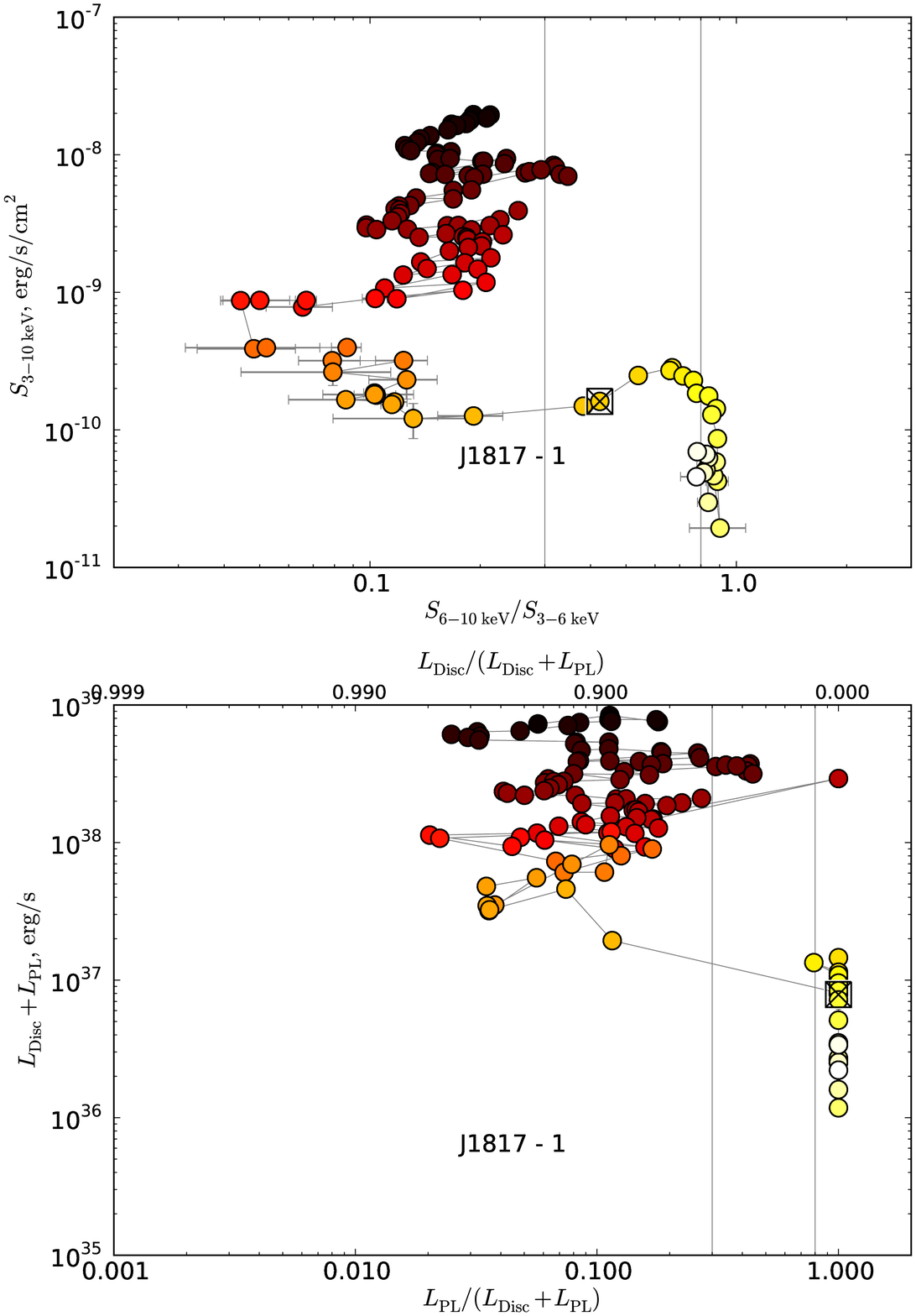}
\caption{(cont) X-ray colour curves and diagnostic diagrams for the
  outbursts observed - XTE J1817-330 Outburst 1.}
\end{figure*}
\addtocounter{figure}{-1}
\begin{figure*}
\centering
\includegraphics[width=0.41\textwidth]{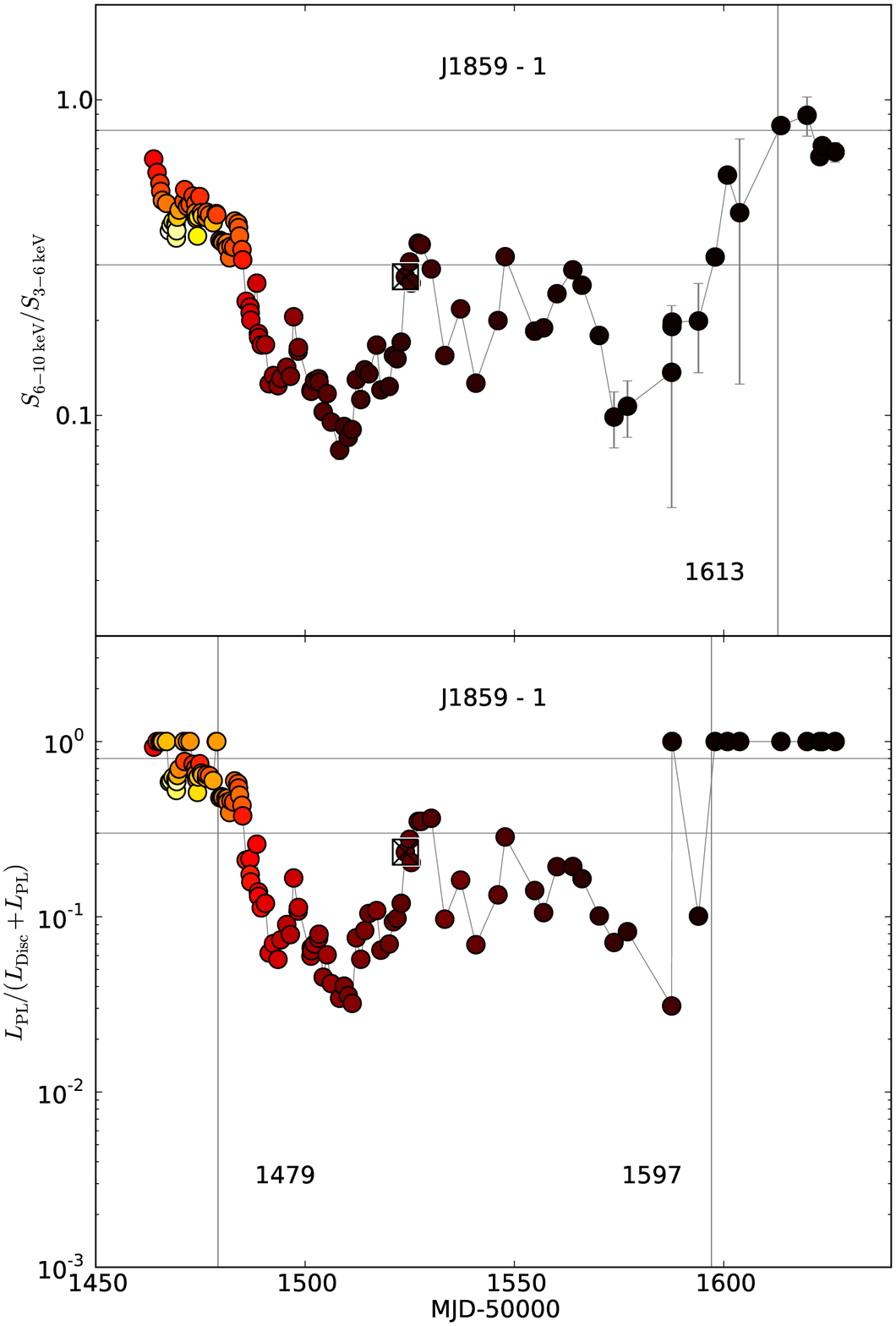}
\includegraphics[width=0.41\textwidth]{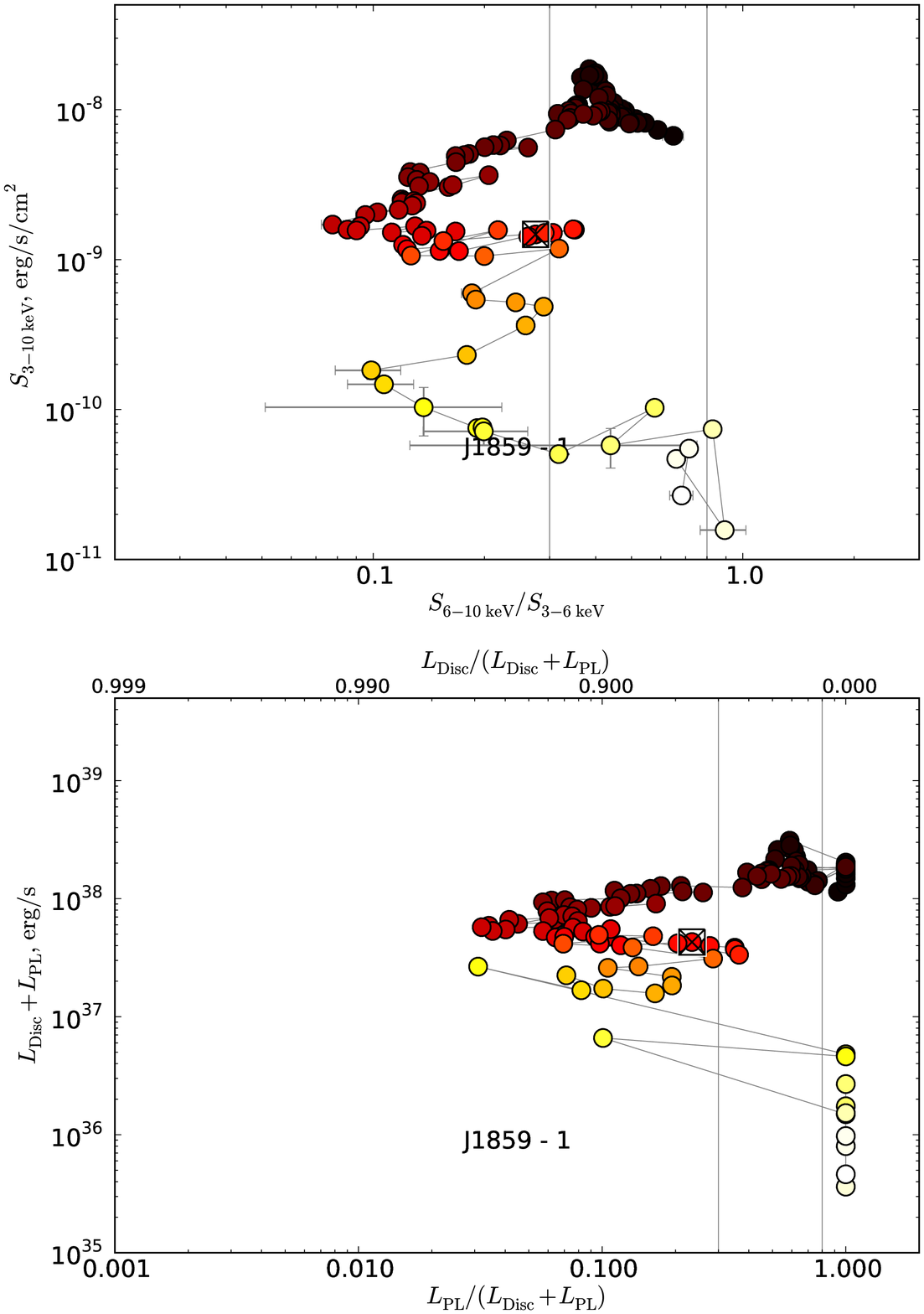}
\caption{(cont) X-ray colour curves and diagnostic diagrams for the
  outbursts observed - XTE J1859-226 Outburst 1.}
\end{figure*}
\addtocounter{figure}{-1}
\begin{figure*}
\centering
\includegraphics[width=0.41\textwidth]{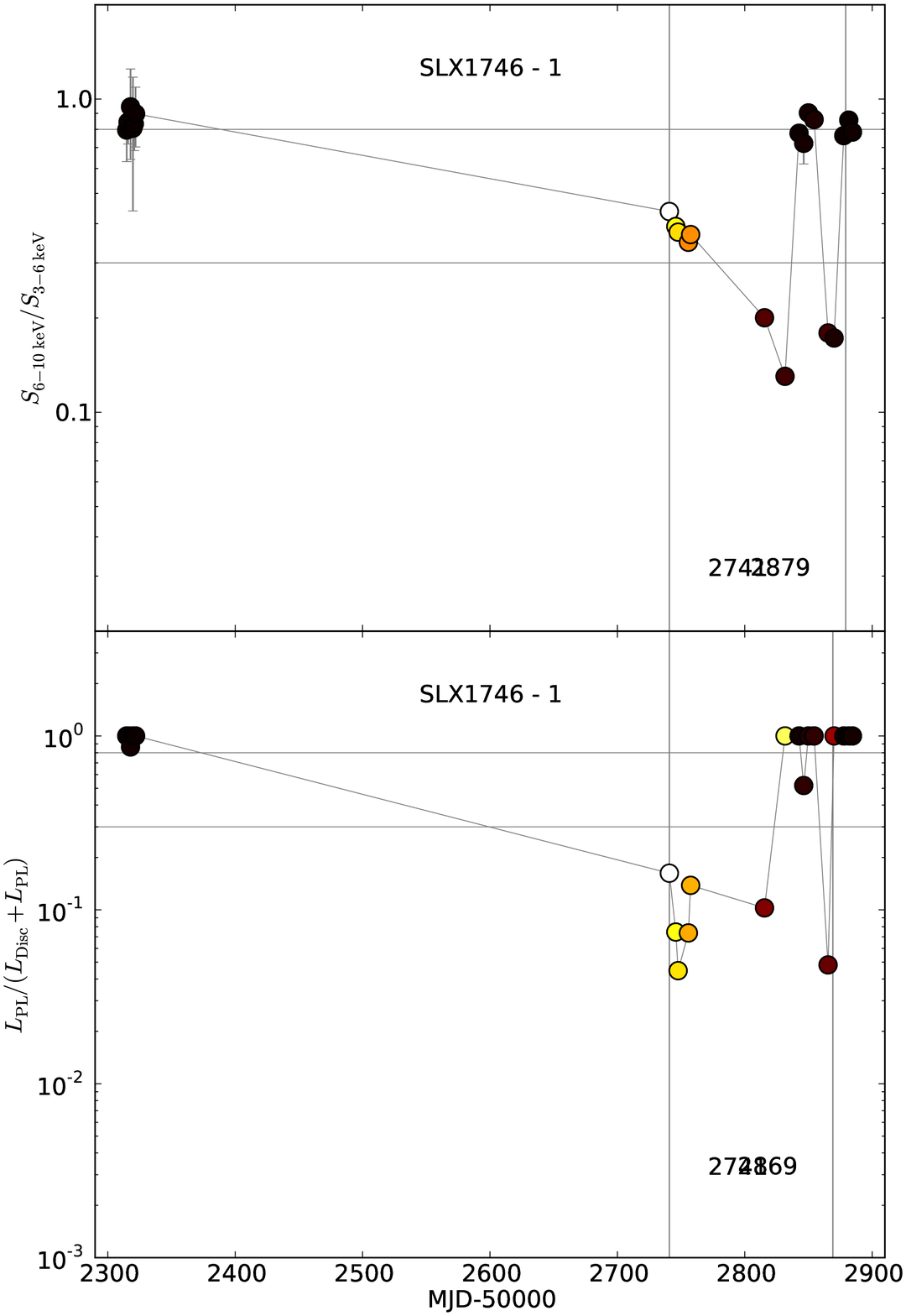}
\includegraphics[width=0.41\textwidth]{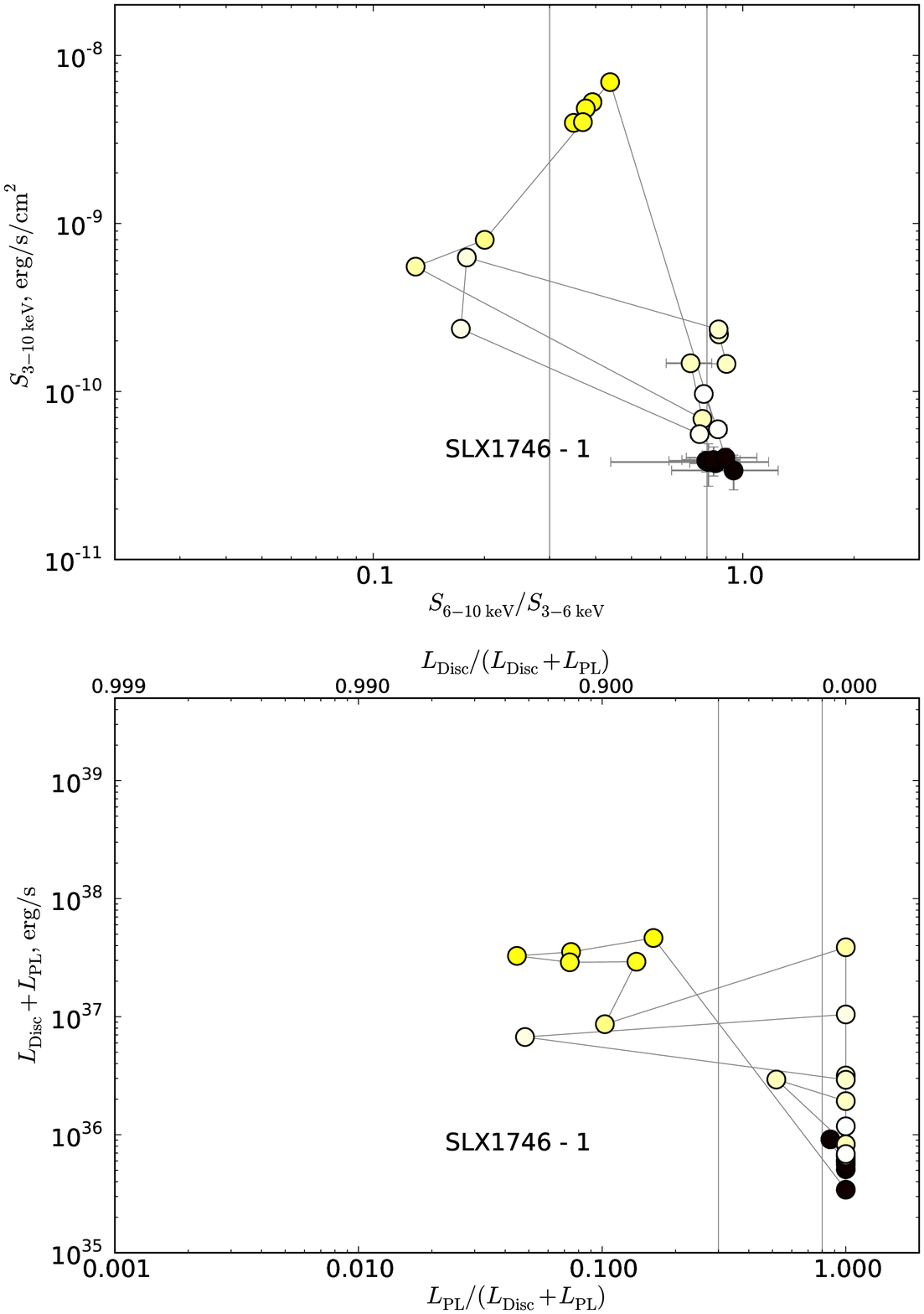}
\caption{(cont) X-ray colour curves and diagnostic diagrams for the
  outbursts observed - SLX 1746-331 Outburst 1.}
\end{figure*}

\begin{figure*}
\centering
\includegraphics[width=0.45\textwidth]{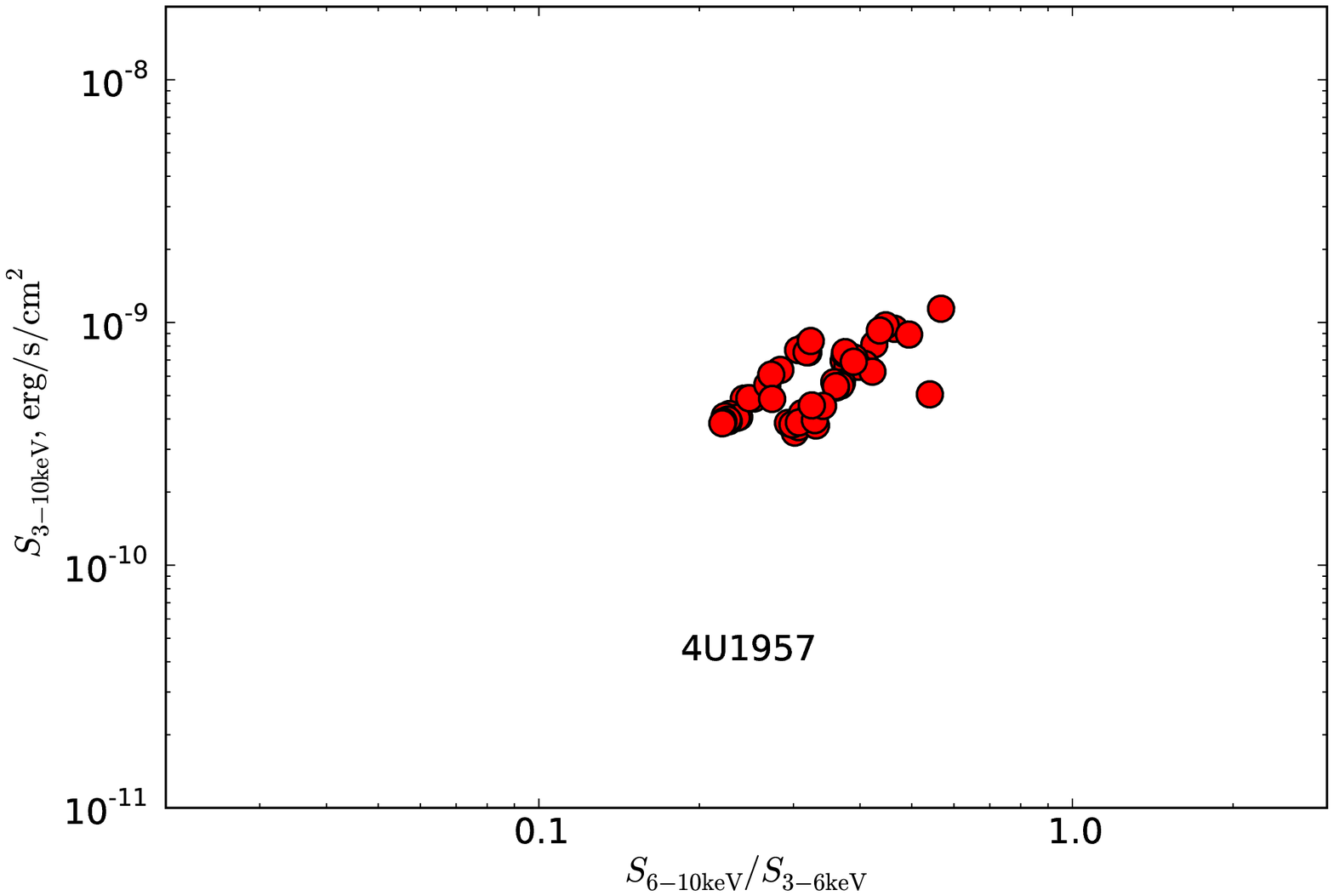}
\includegraphics[width=0.45\textwidth]{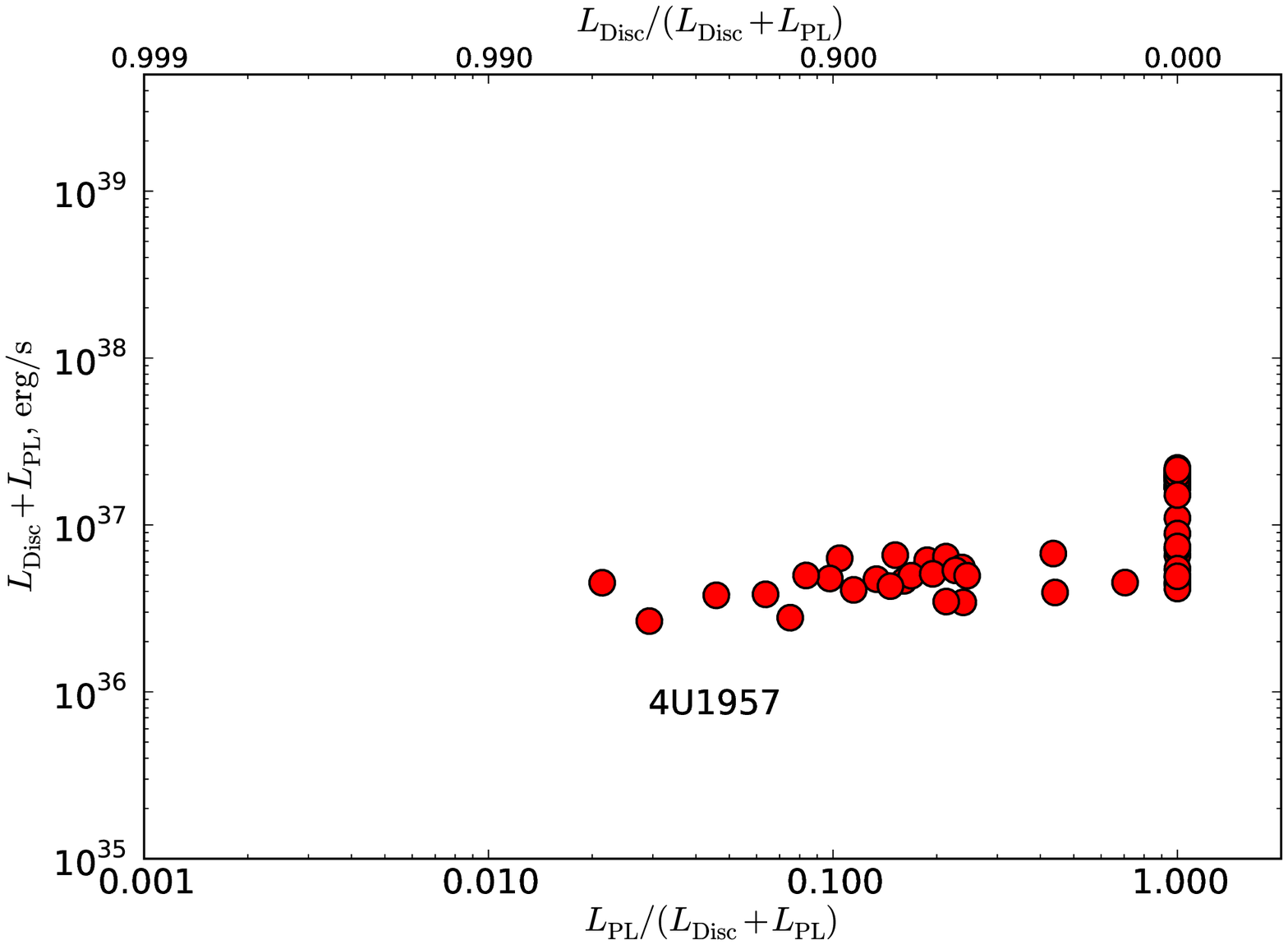}

\includegraphics[width=0.45\textwidth]{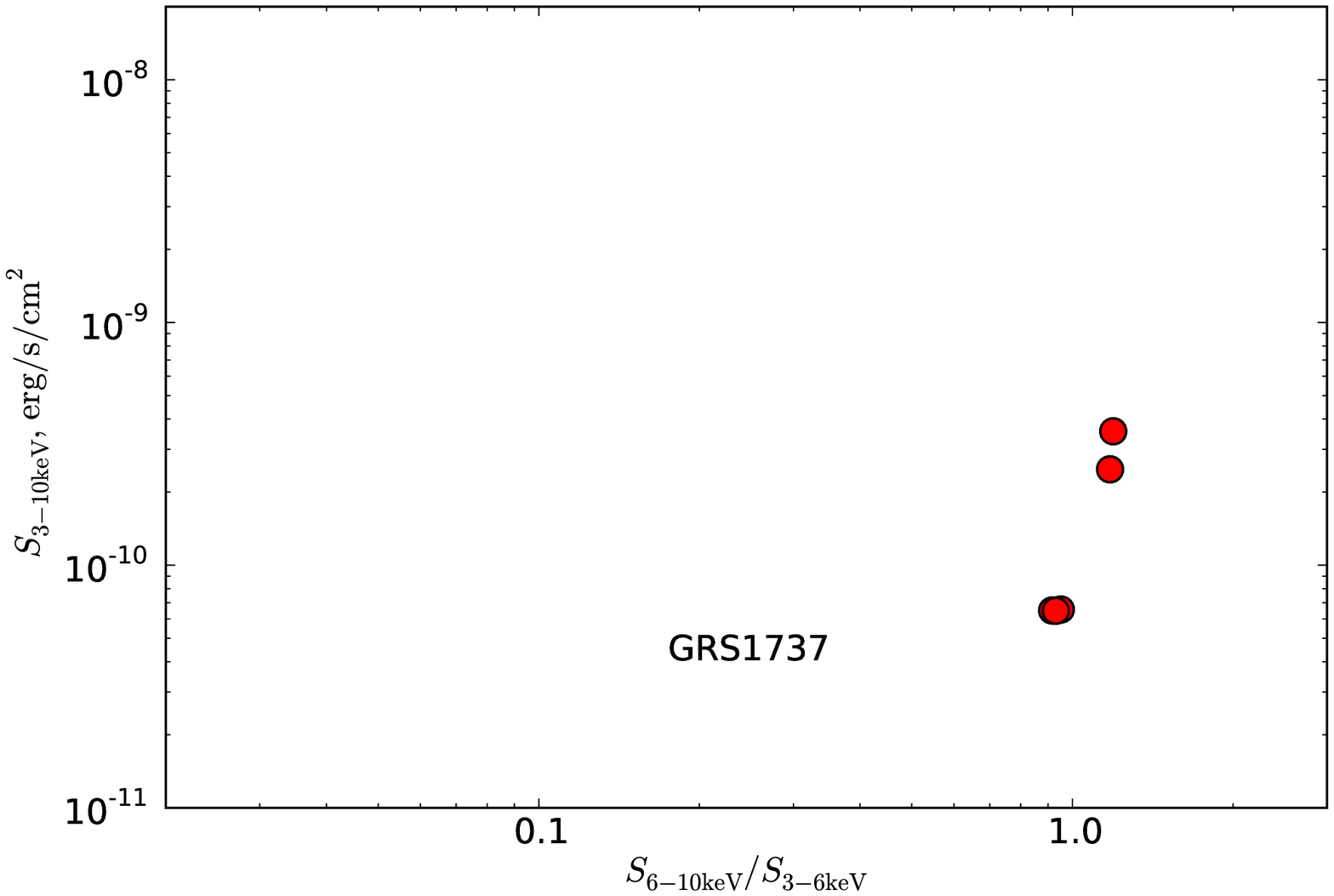}
\includegraphics[width=0.45\textwidth]{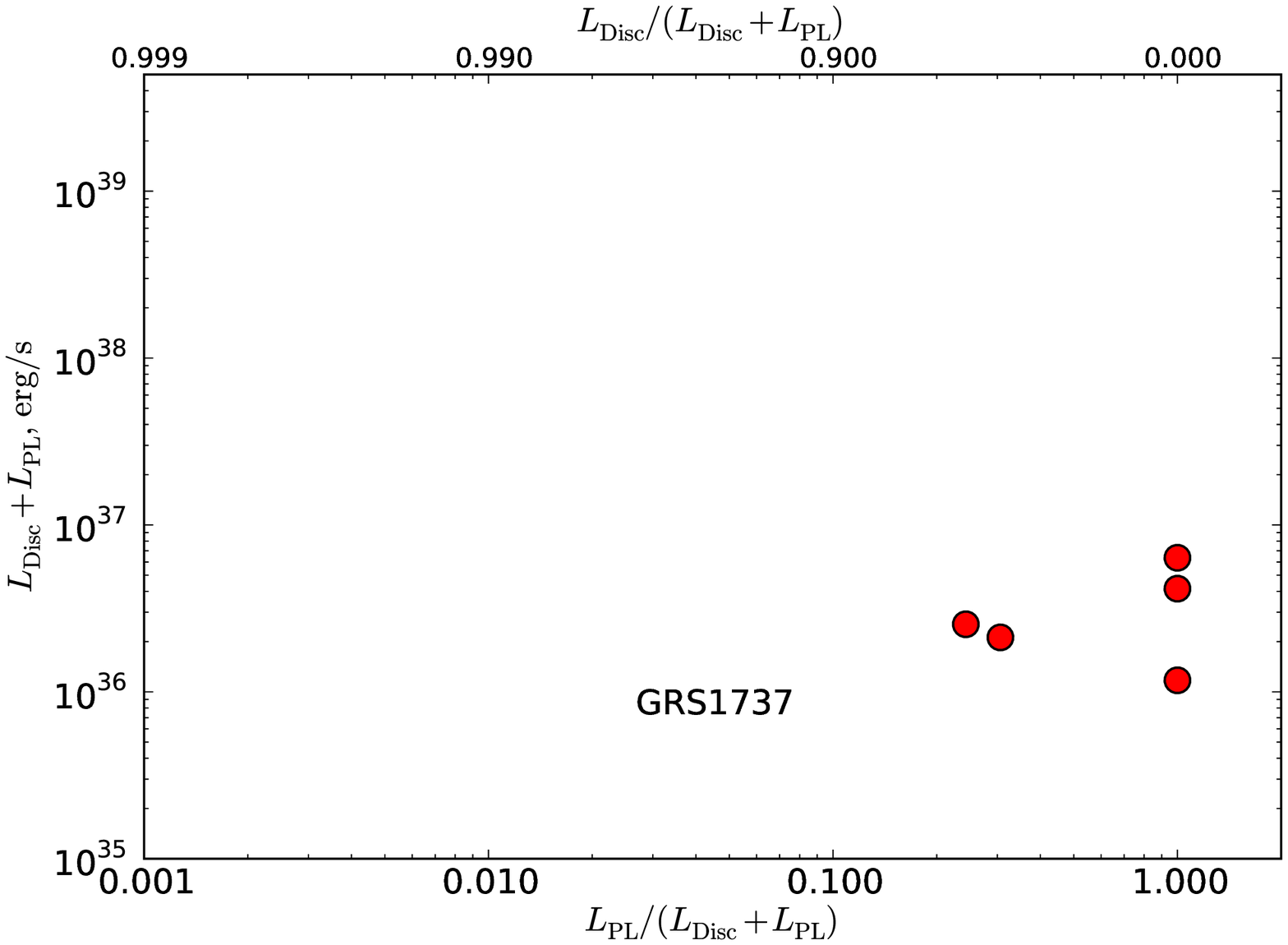}

\includegraphics[width=0.45\textwidth]{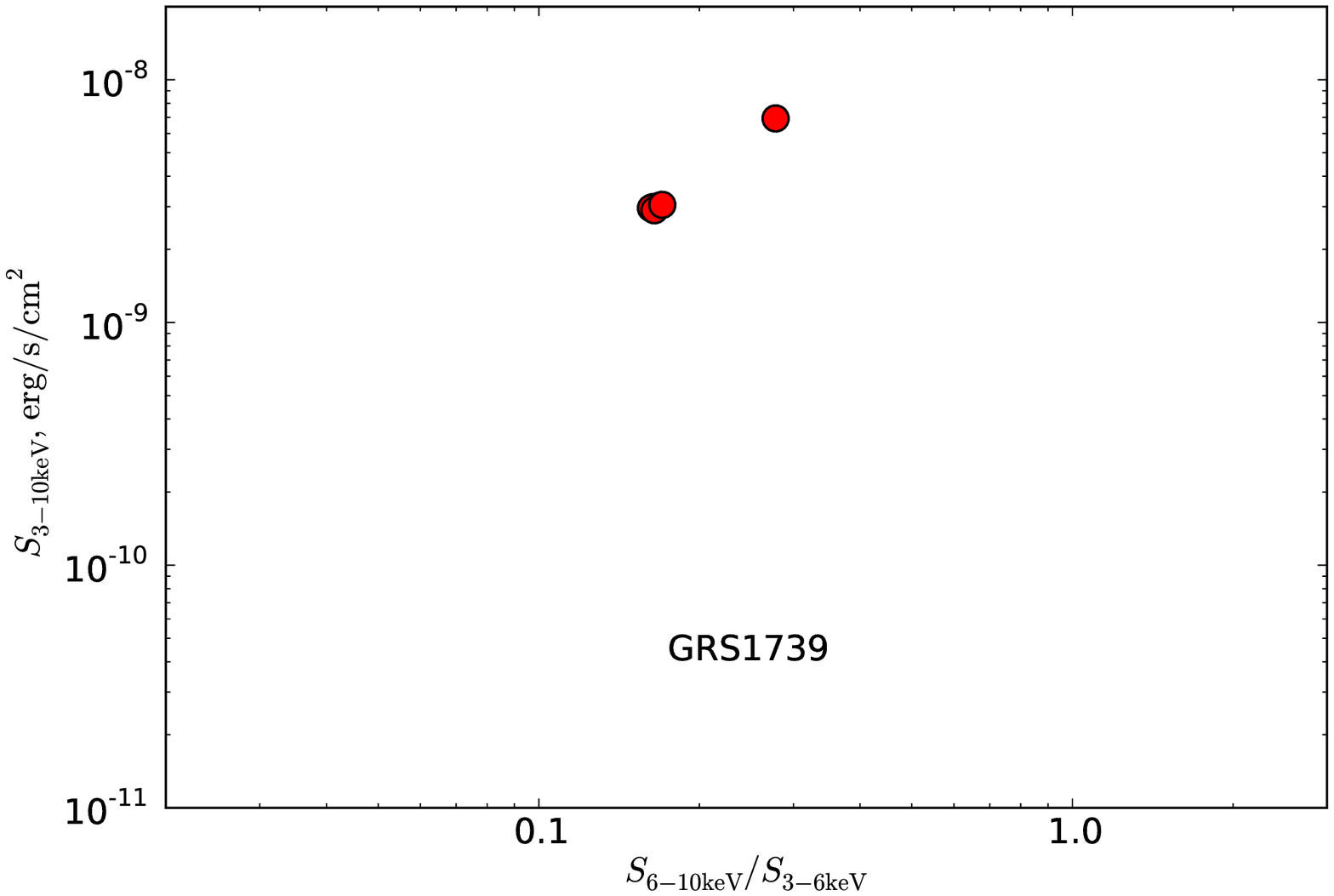}
\includegraphics[width=0.45\textwidth]{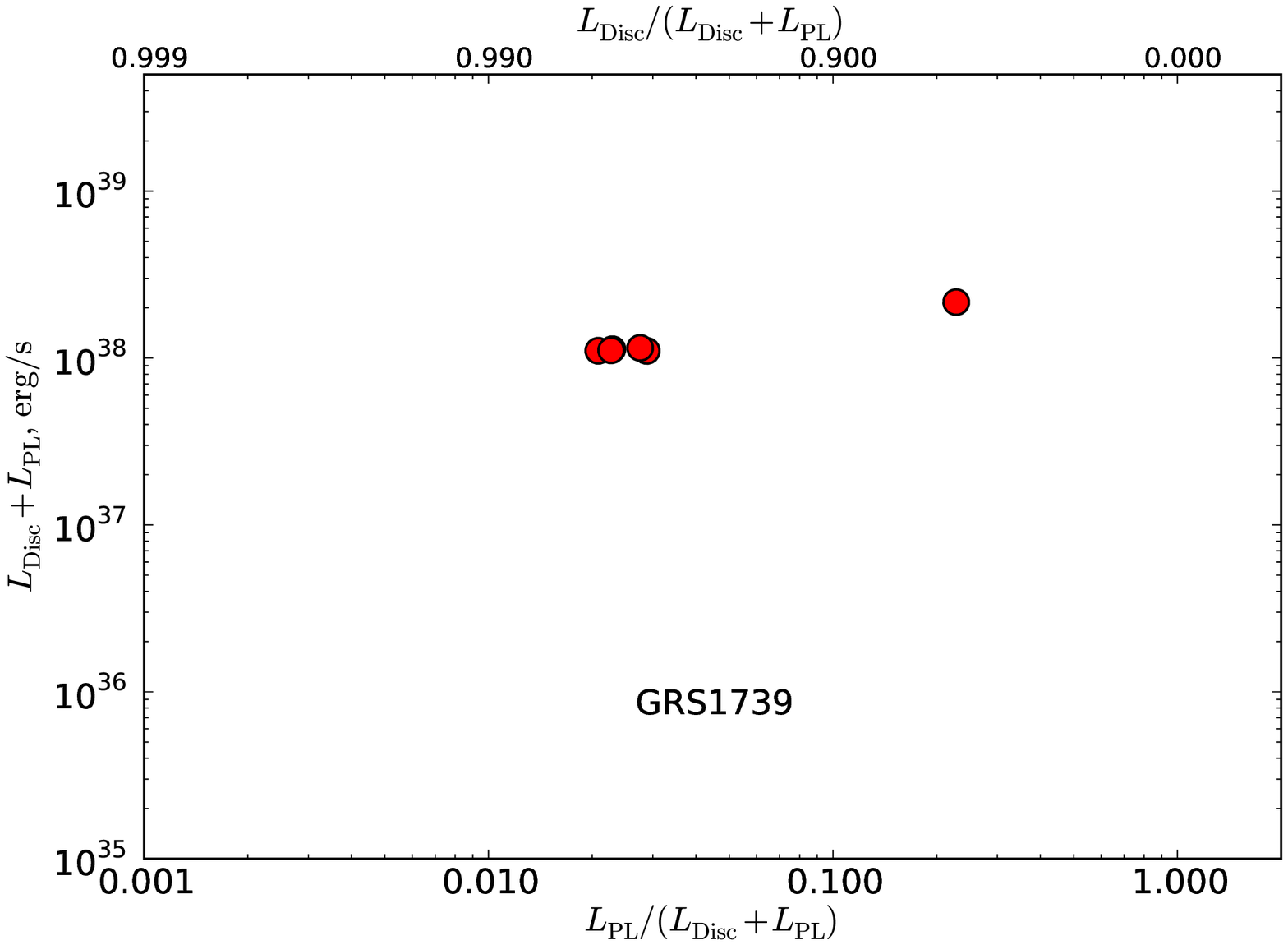}

\includegraphics[width=0.45\textwidth]{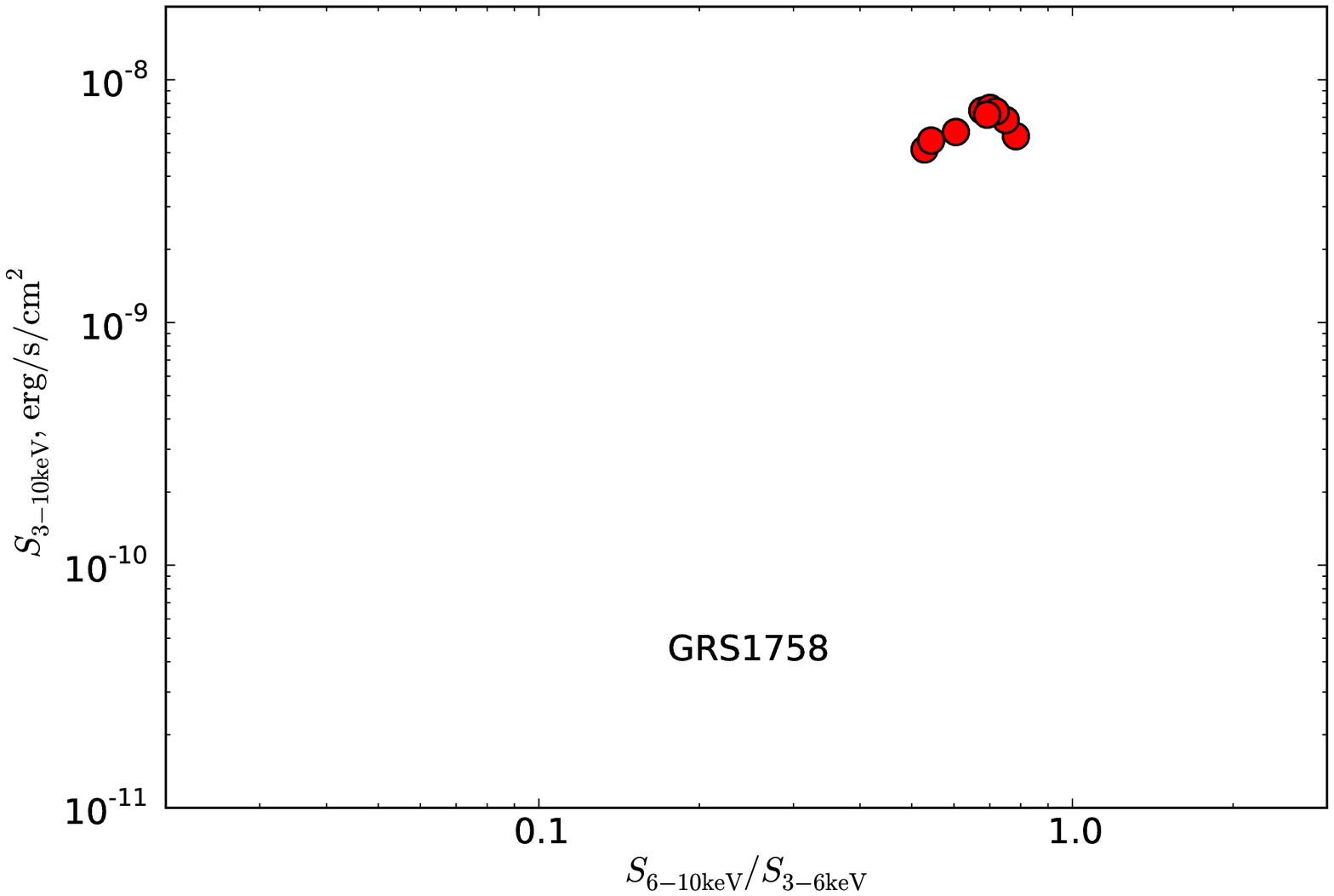}
\includegraphics[width=0.45\textwidth]{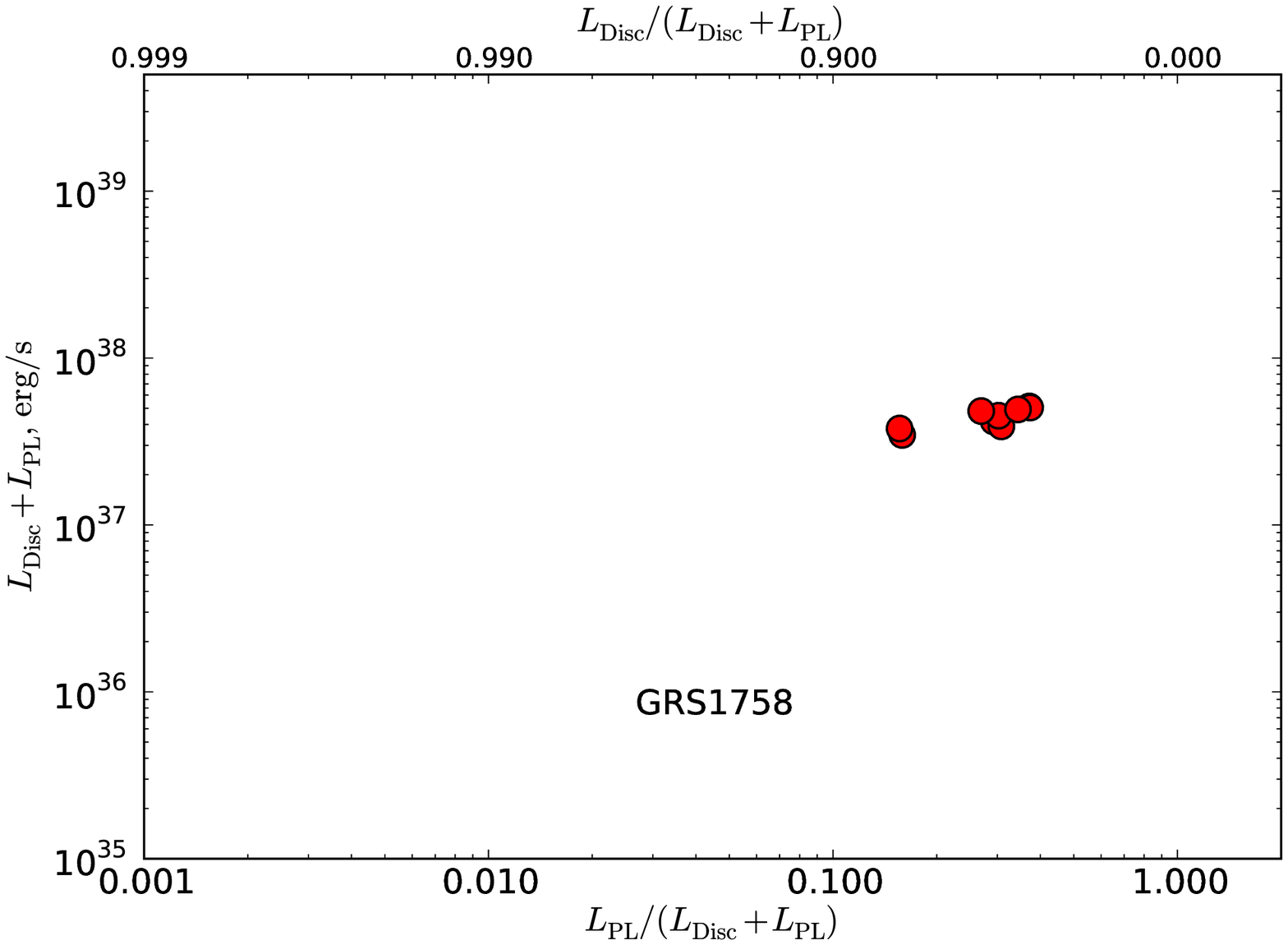}
\caption{\label{fig:remainder} We show the HID and the DFLD for the
  remainder of the BHXRBs not shown in Fig. \ref{fig:curves}.}
\end{figure*}
\addtocounter{figure}{-1}

\begin{figure*}
\centering
\includegraphics[width=0.45\textwidth]{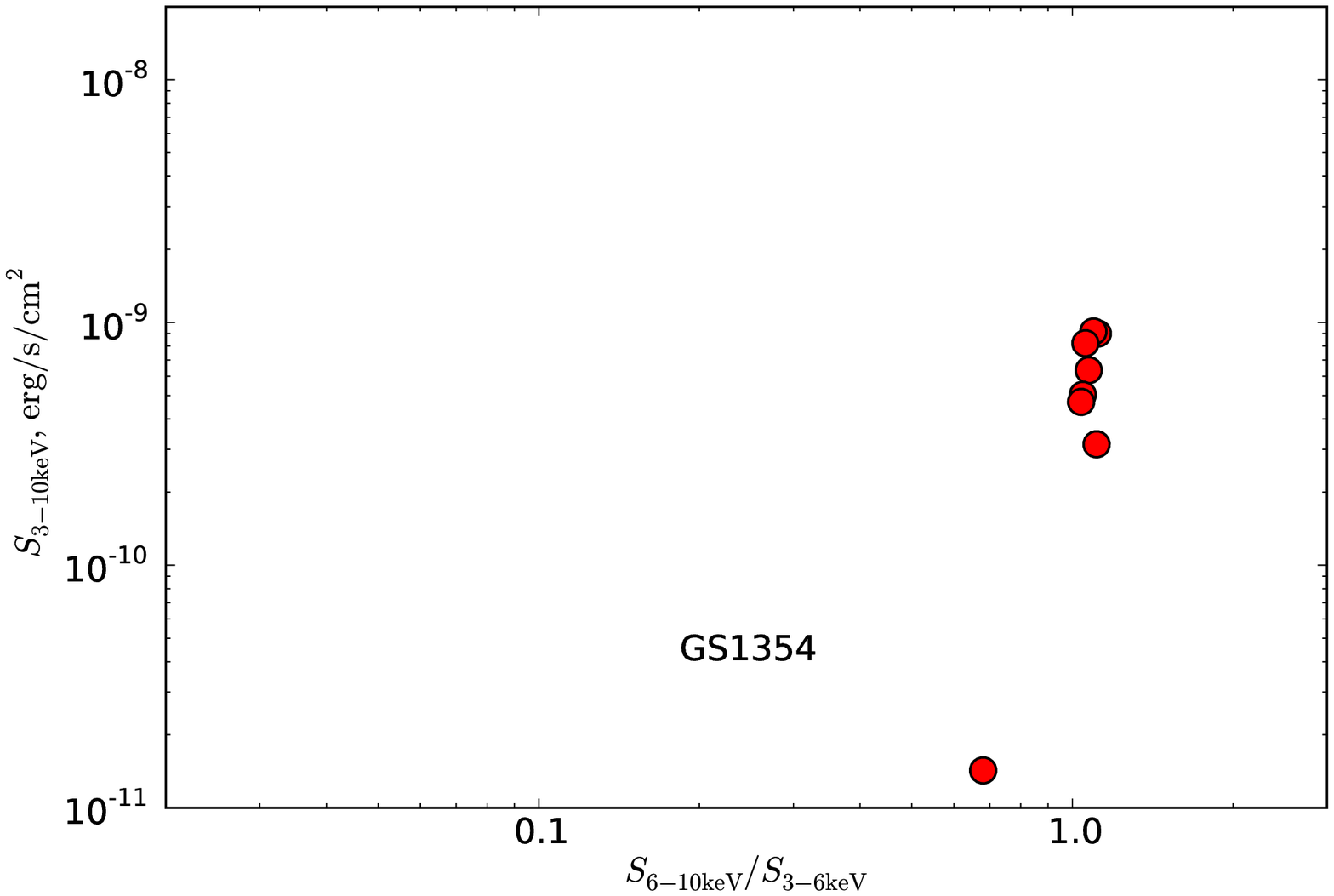}
\includegraphics[width=0.45\textwidth]{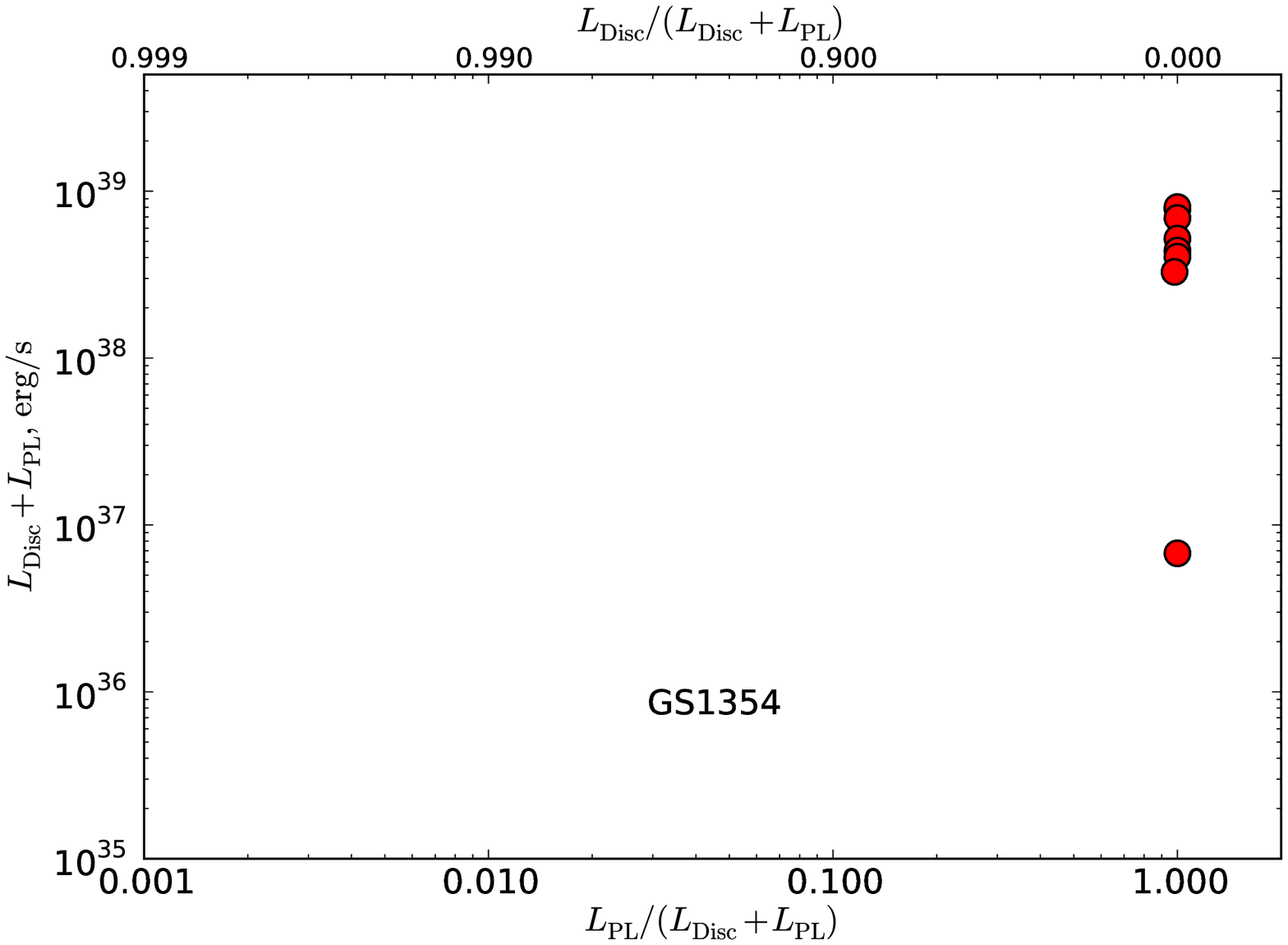}

\includegraphics[width=0.45\textwidth]{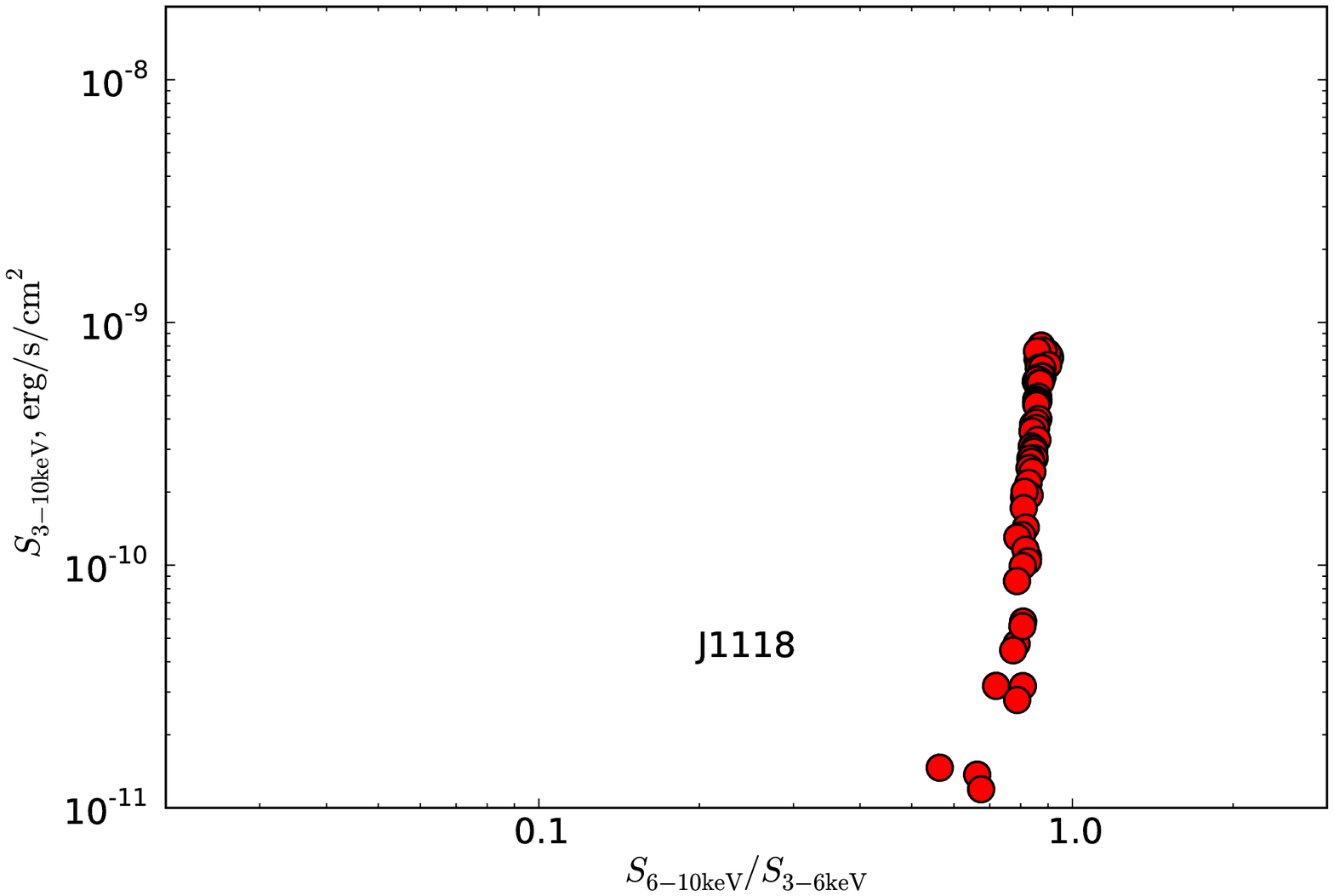}
\includegraphics[width=0.45\textwidth]{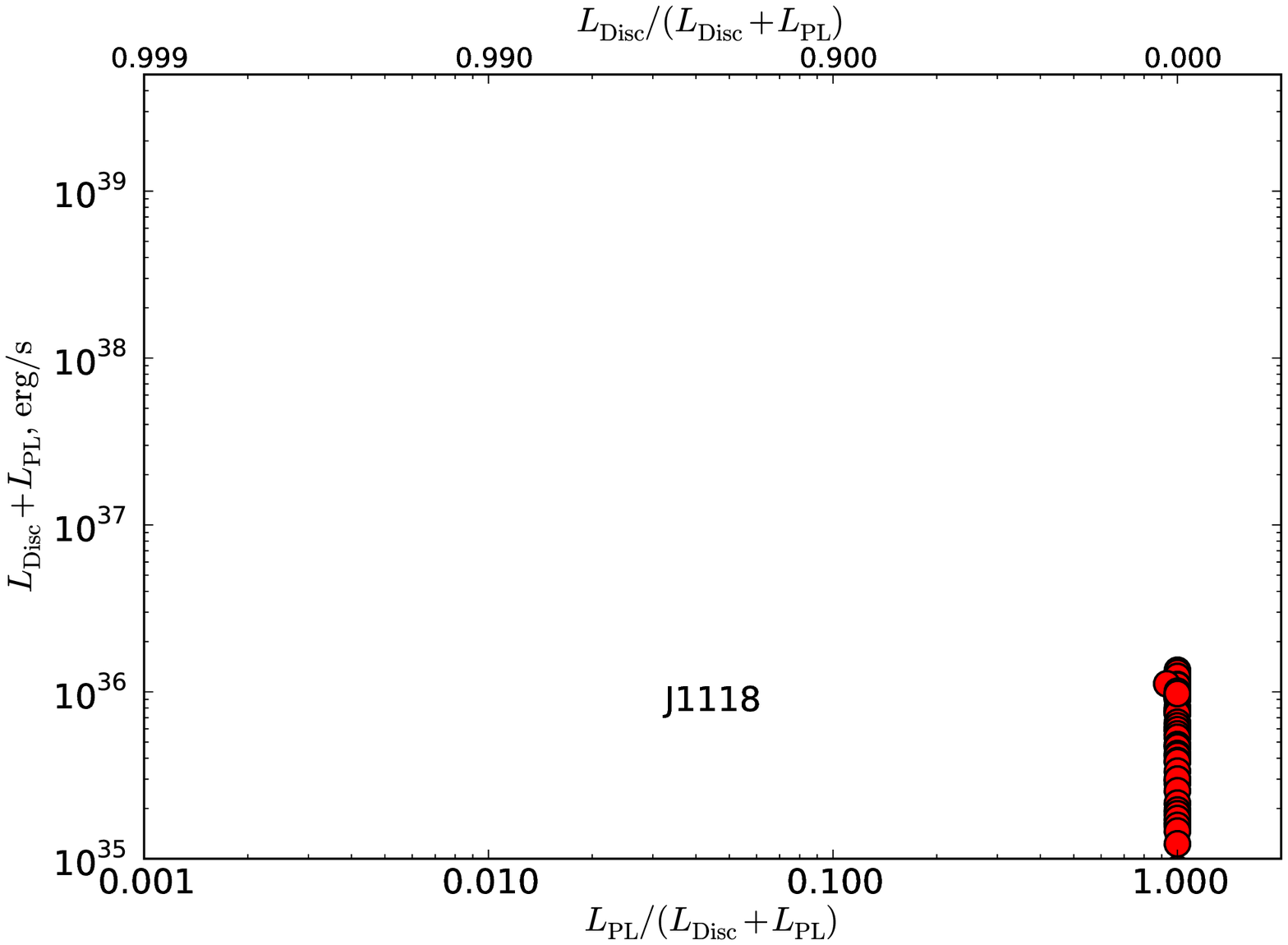}

\includegraphics[width=0.45\textwidth]{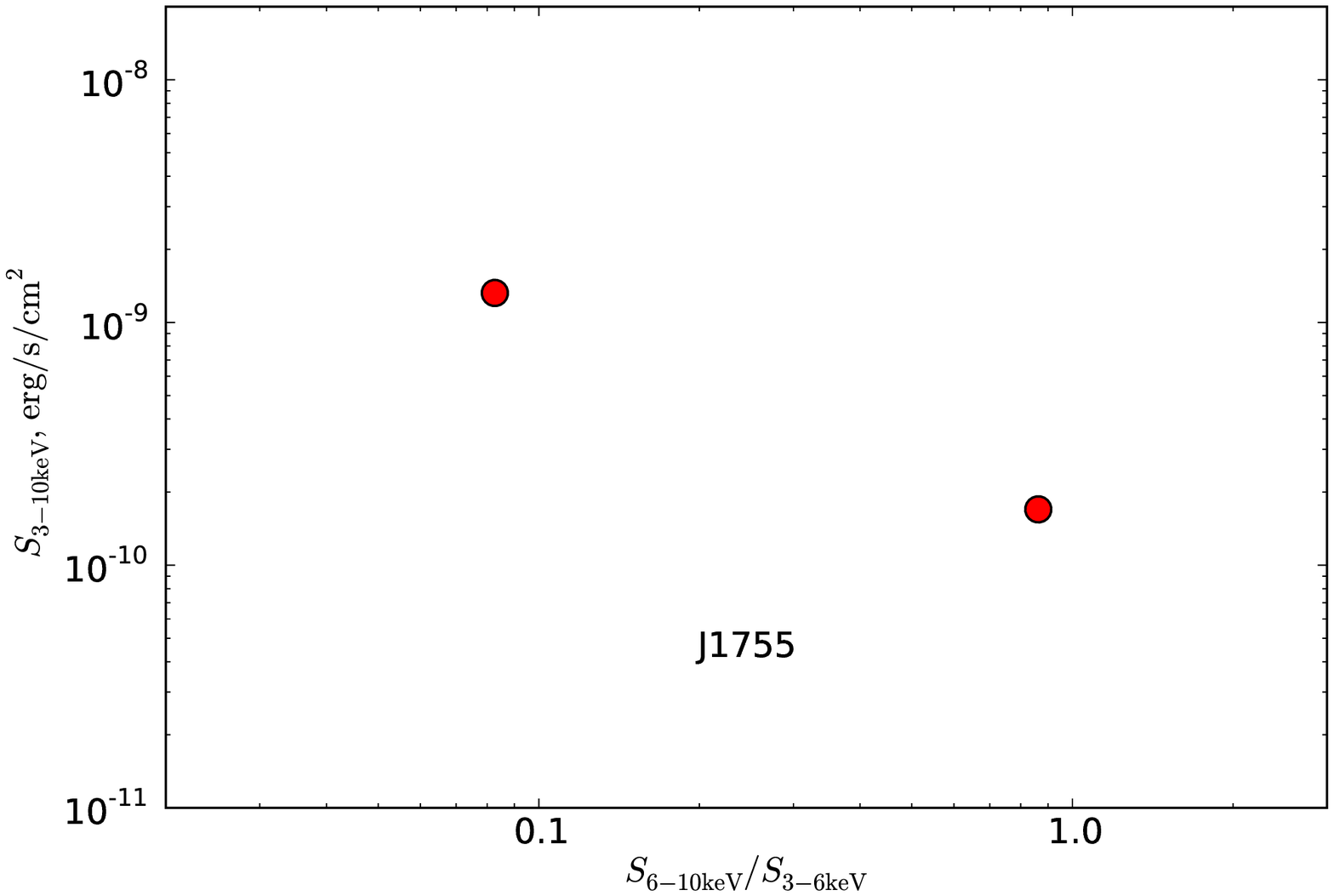}
\includegraphics[width=0.45\textwidth]{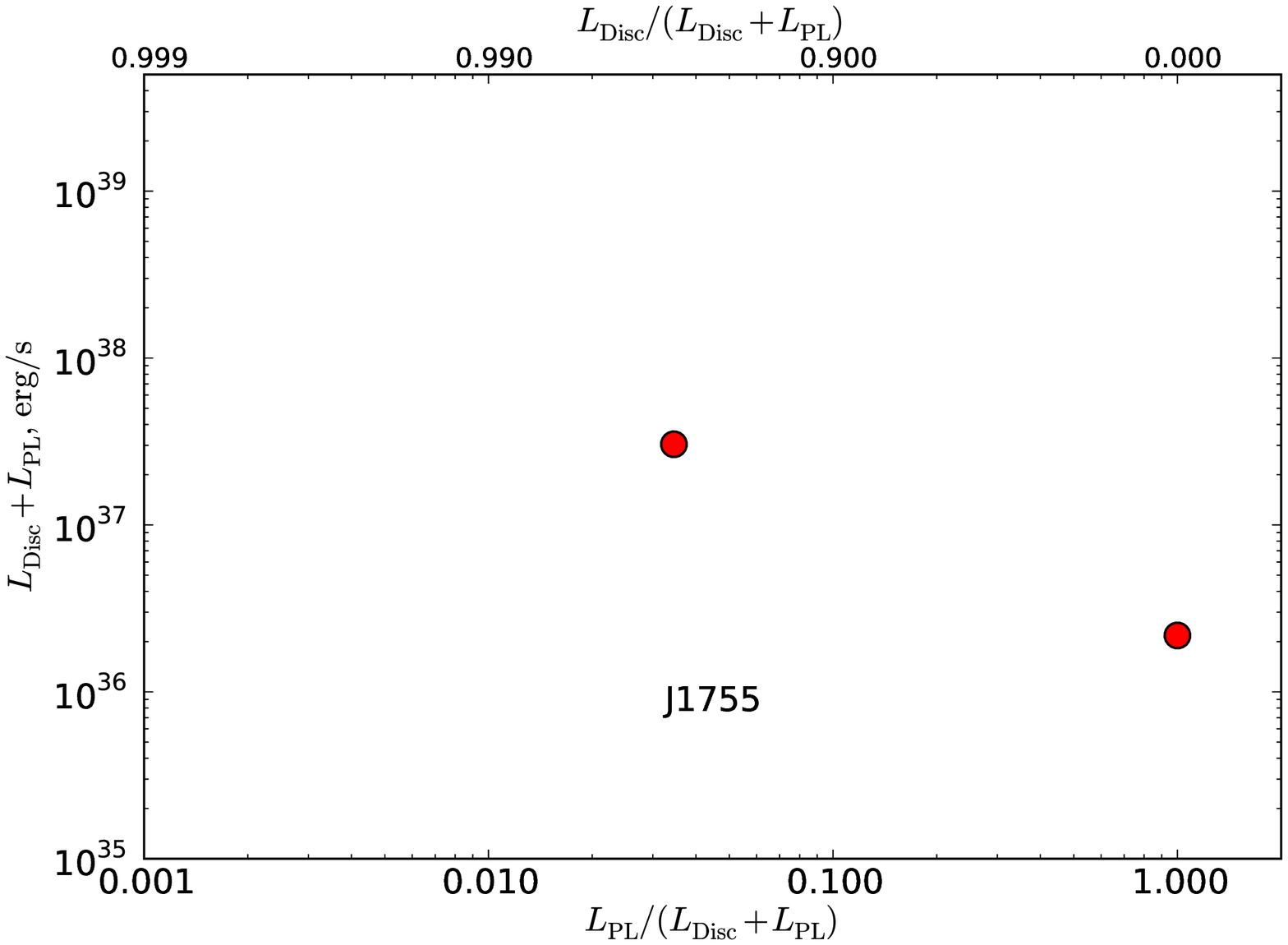}

\includegraphics[width=0.45\textwidth]{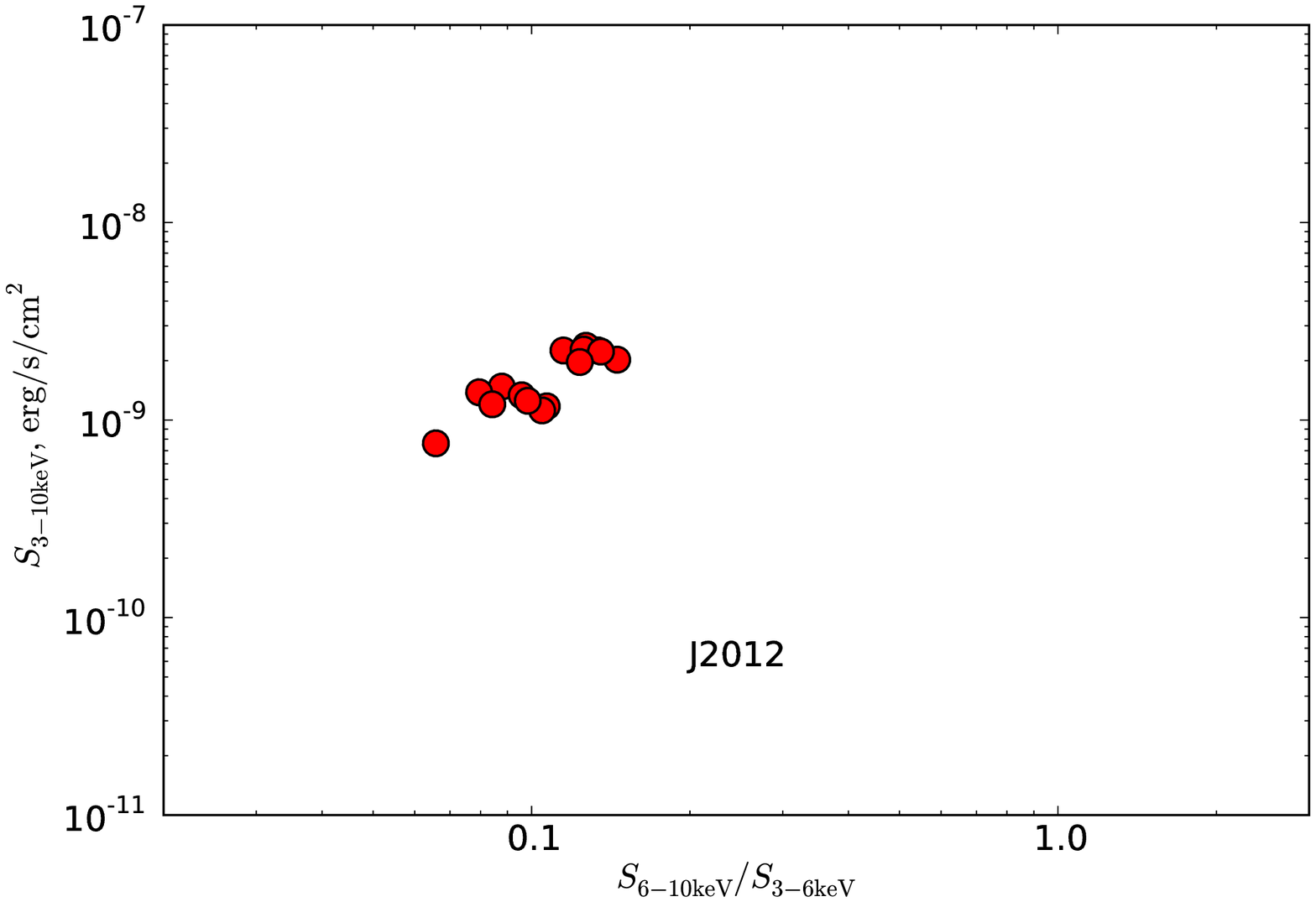}
\includegraphics[width=0.45\textwidth]{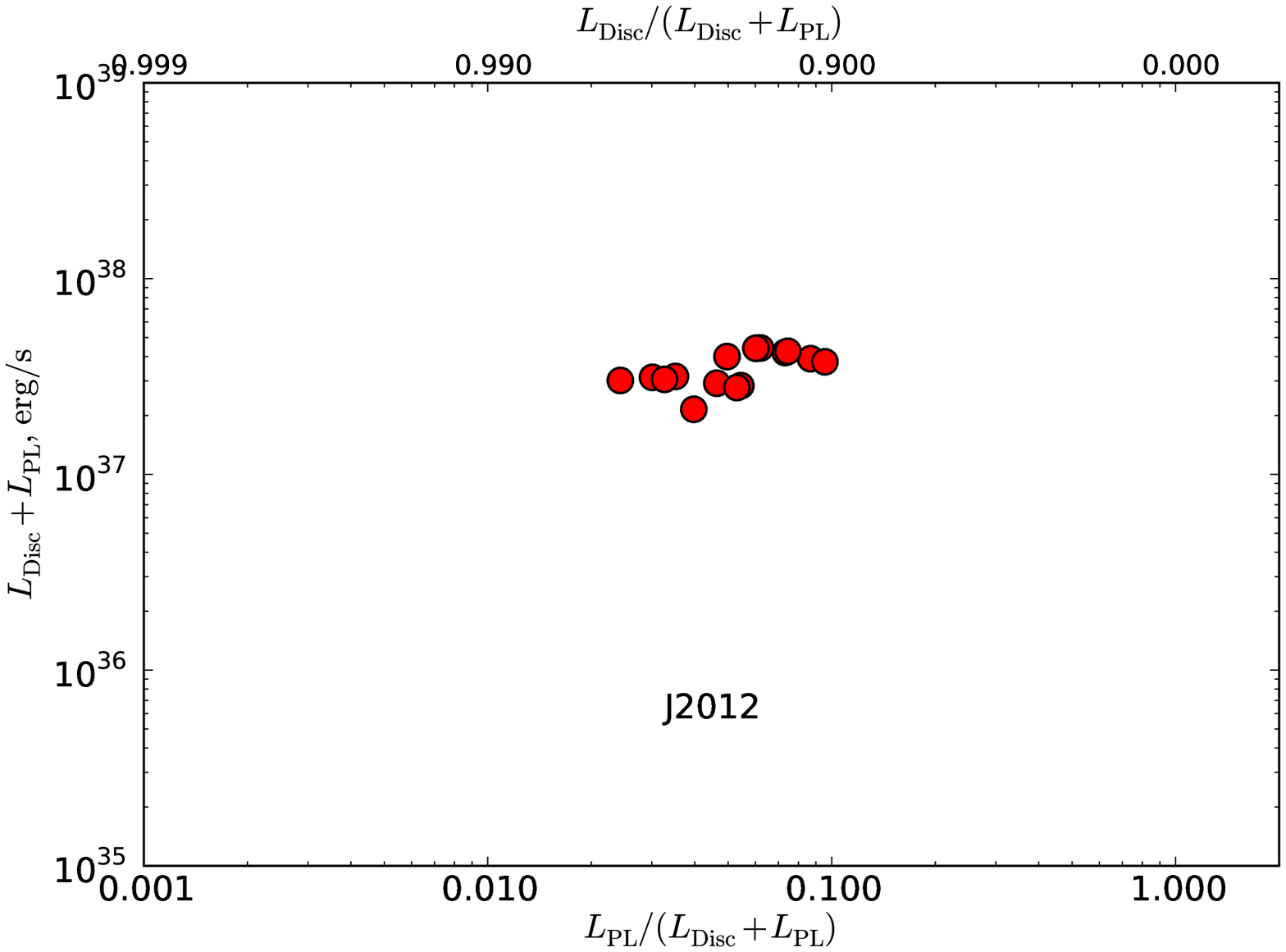}
\caption{(cont) HID and DFLDs for the remainder of the BHXRBs.}
\end{figure*}
\addtocounter{figure}{-1}

\begin{figure*}
\centering
\includegraphics[width=0.45\textwidth]{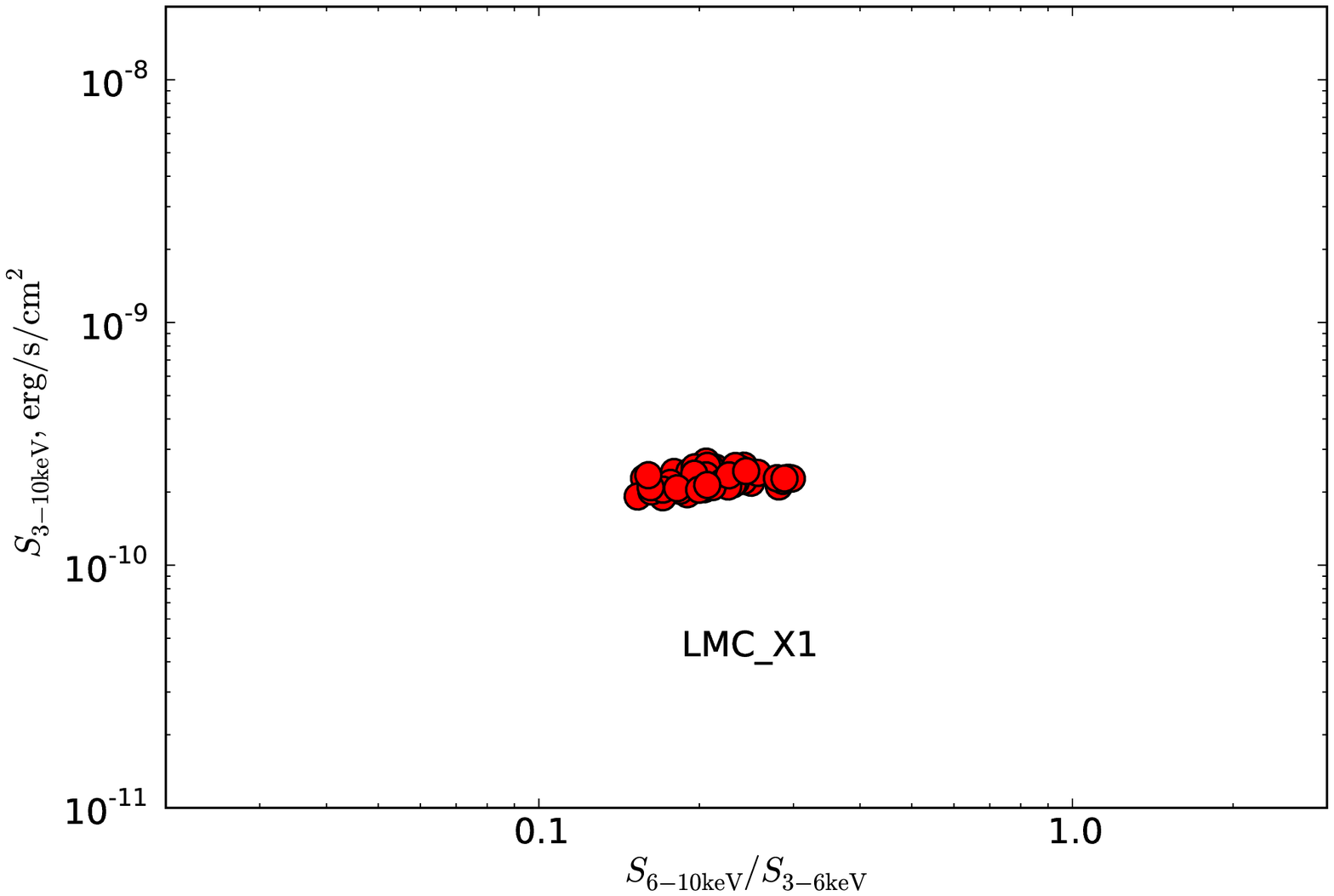}
\includegraphics[width=0.45\textwidth]{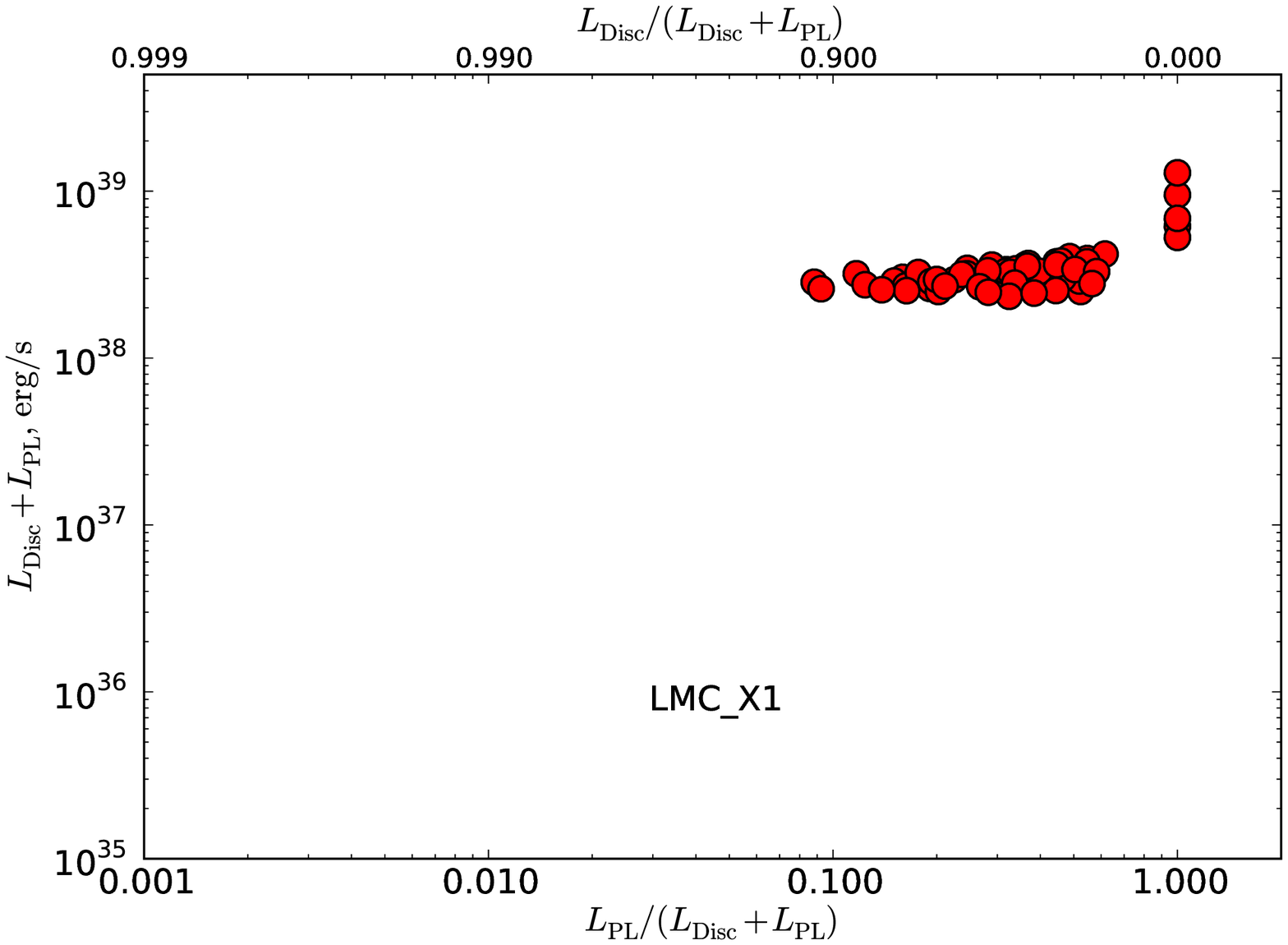}

\includegraphics[width=0.45\textwidth]{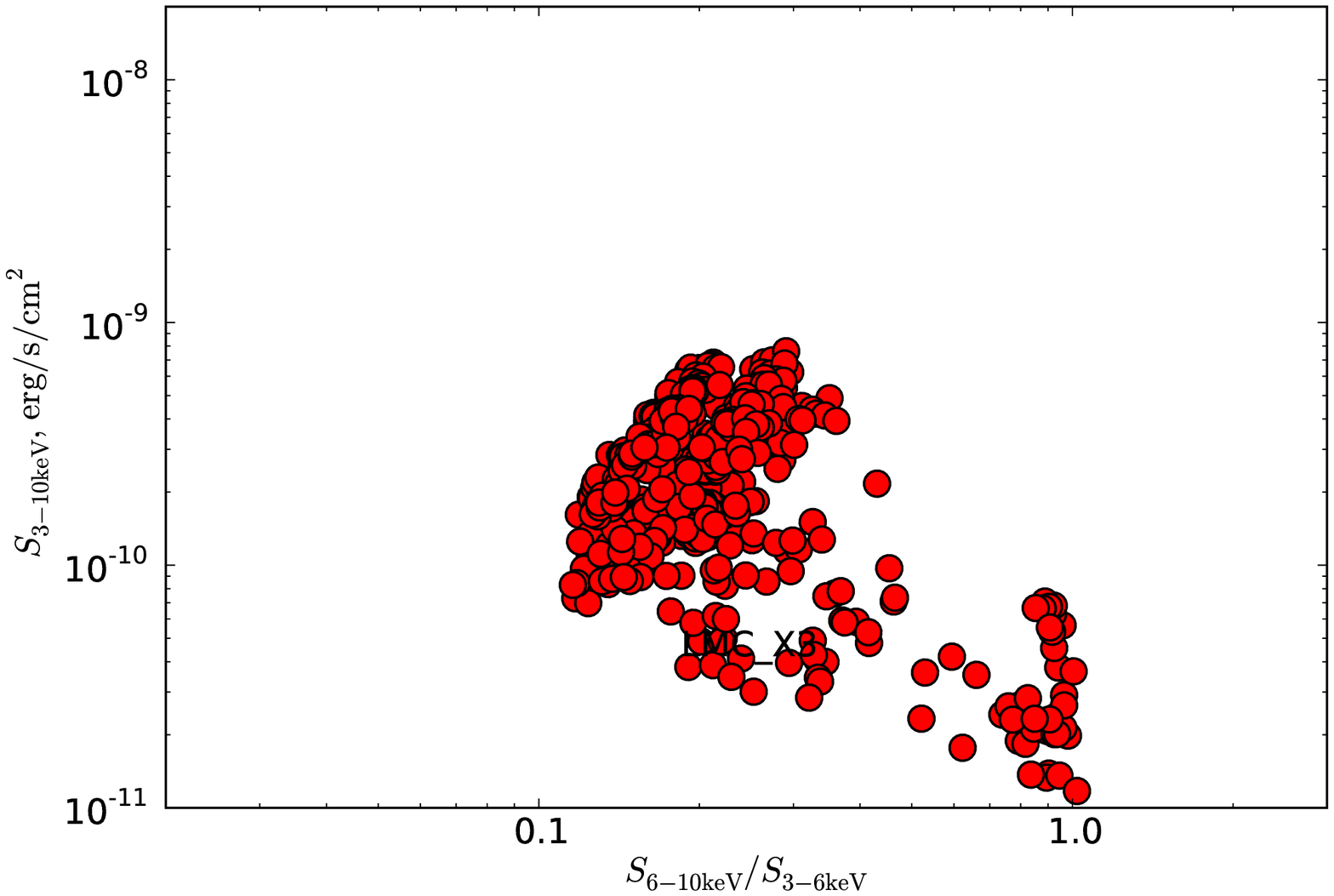}
\includegraphics[width=0.45\textwidth]{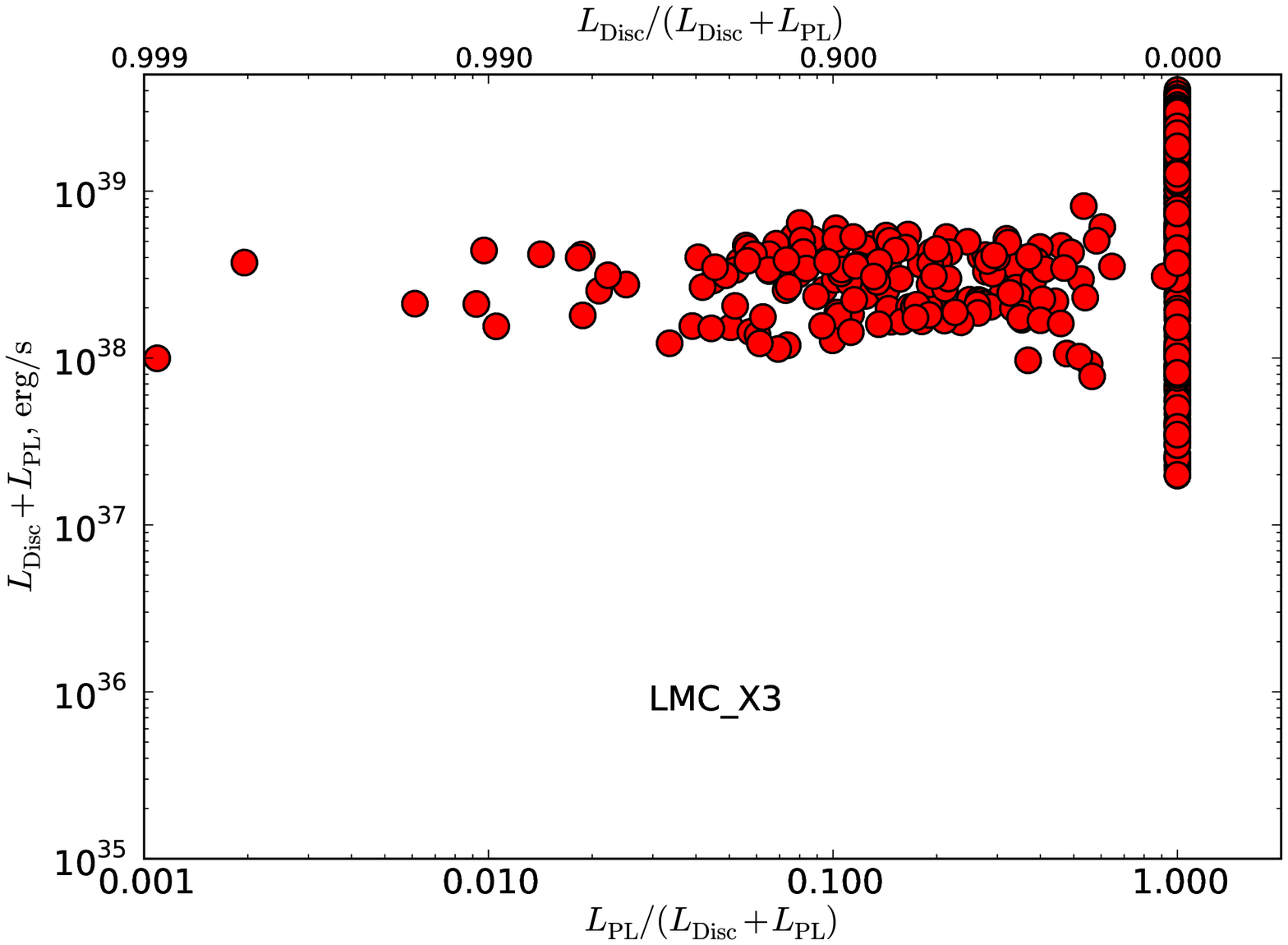}

\includegraphics[width=0.45\textwidth]{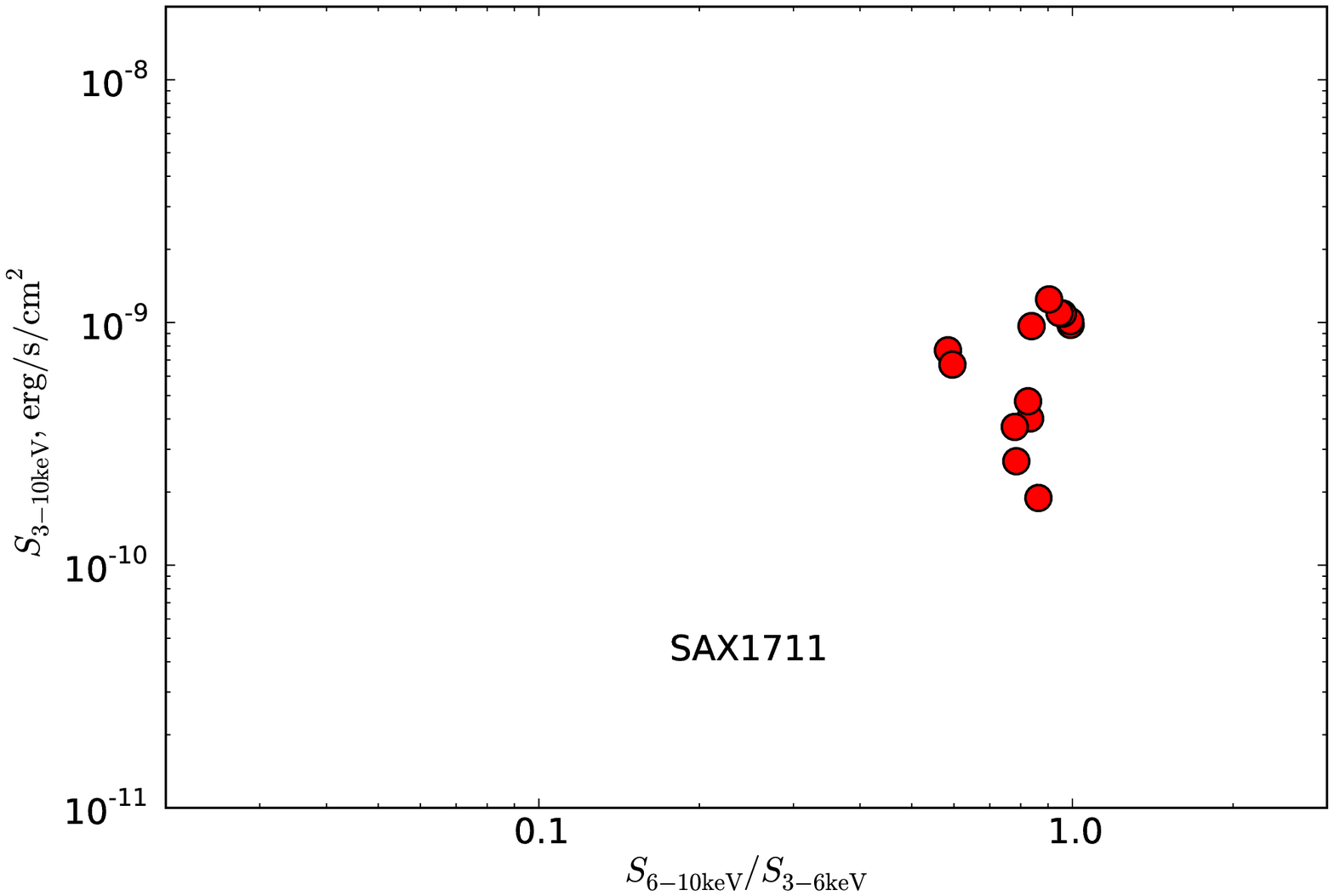}
\includegraphics[width=0.45\textwidth]{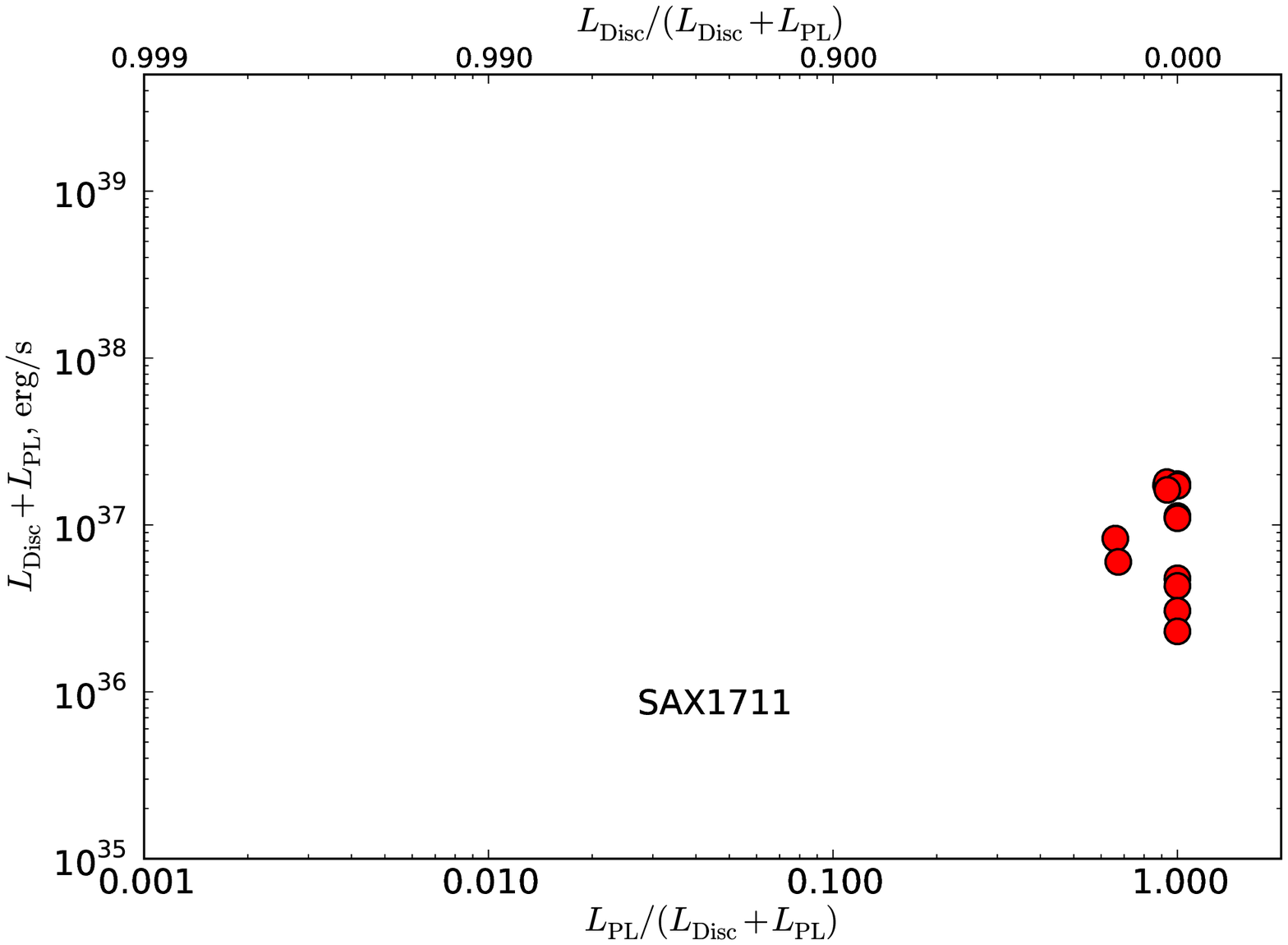}

\includegraphics[width=0.45\textwidth]{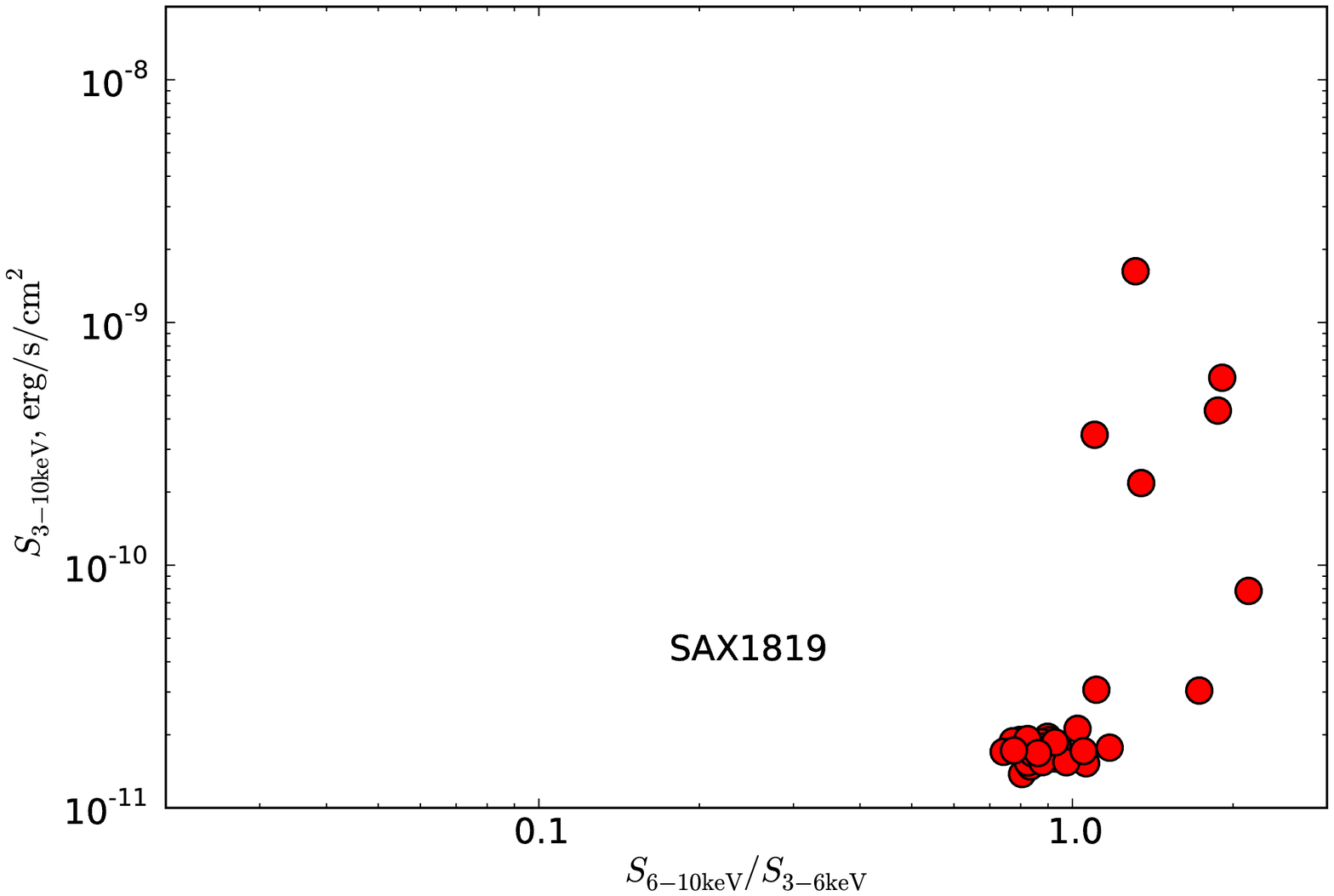}
\includegraphics[width=0.45\textwidth]{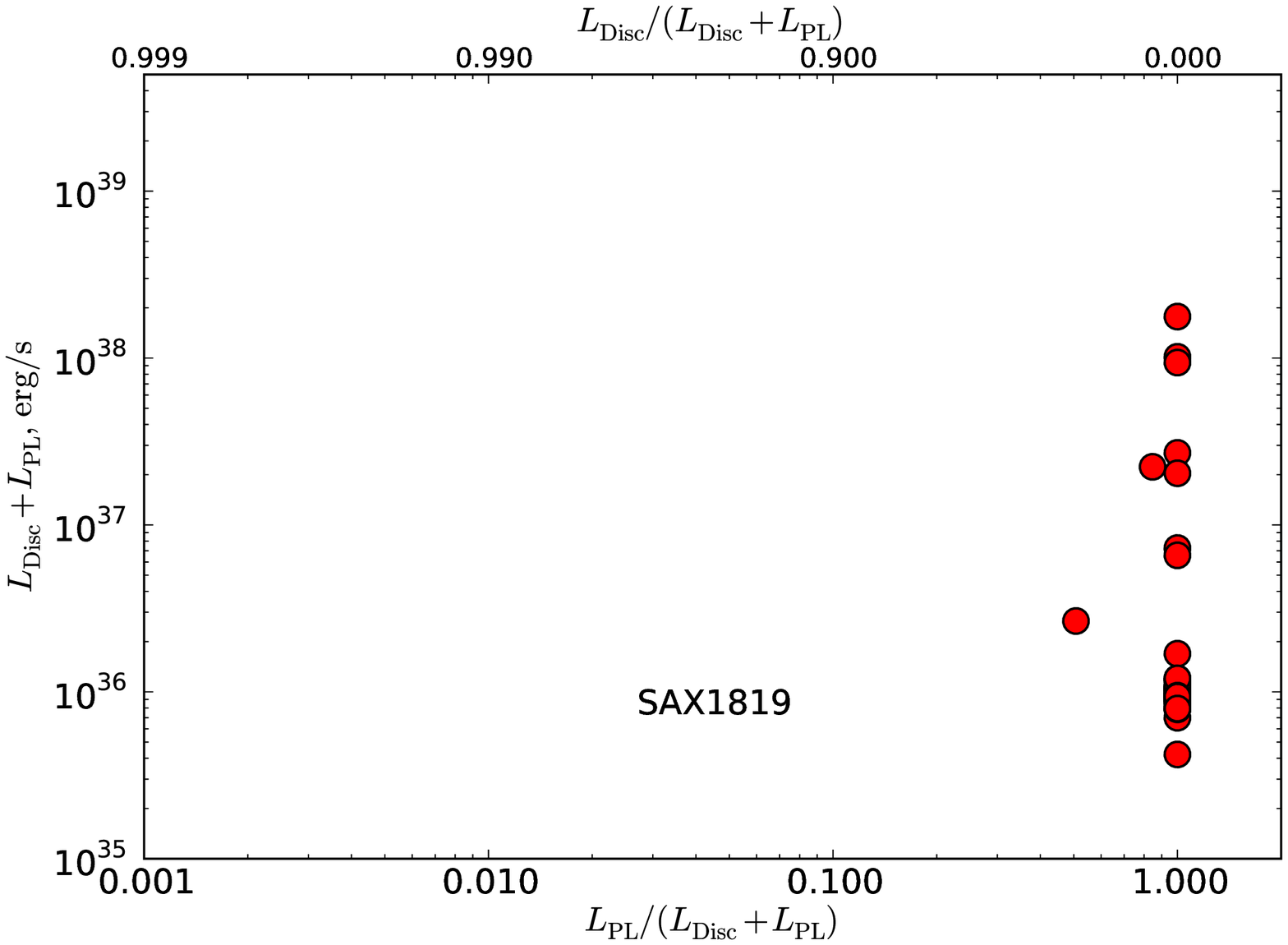}
\caption{(cont) HID and DFLDs for the remainder of the BHXRBs.}
\end{figure*}

\end{document}